\documentclass[11pt, phd]{osudissert96}

\usepackage{amsmath, amssymb, amsthm,mathtools}
\usepackage{graphicx}
\usepackage{hyperref}
\usepackage{array}
\usepackage{xspace}
\usepackage{nicefrac}
\usepackage{overpic}
\usepackage{float}

\usepackage{tikz}
\usepackage{rotating} 
\usepackage{multirow} 
\usepackage{caption}
\usepackage{subcaption}

\usepackage{bm,bbm} 
\usepackage{booktabs} 
\usepackage{slashed} 
\usepackage{dsfont}
\usepackage{bbm}
\usepackage{environ}

\usepackage[silent,margin]{fixme}

\usepackage{bookmark} 

\hypersetup{colorlinks=true,linkcolor=blue}

\usepackage{cite} 

\usepackage[xindy,acronym, section=chapter]{glossaries}
\glsdisablehyper
\makeglossaries

\graphicspath{{figures/}}

\newcommand{\fm}{\,\text{fm}}

\newcommand{\fmi}{\,\text{fm}^{-1}}
\newcommand{\MeV}{\,\text{MeV}}

\newcommand{\LamUV}{\Lambda_{\text{UV}}}

\newcommand{\bea}{\begin{eqnarray}}
\newcommand{\eea}{\end{eqnarray}}
\newcommand{\beq}{\begin{equation}}
\newcommand{\eeq}{\end{equation}}
\newcommand{\ben}{\begin{equation*}}
\newcommand{\een}{\end{equation*}}

\newcommand{\lp}{{l^{{\prime}}}}
\newcommand{\lpp}{{l^{{\prime \prime}}}}

\newcommand{\bra}[1]{\ensuremath{{\langle#1|}}}
\newcommand{\ket}[1]{\ensuremath{{|#1\rangle}}}

\newcommand{\braket}[2]{\ensuremath{{\langle#1|#2\rangle}}}

\newcommand{\hw}{\ensuremath{\hbar\Omega}}

\renewcommand{\imath}{\ensuremath{i}}

\newcommand{\Einf}{\ensuremath{E_{\infty}}}
\newcommand{\Ainf}{\ensuremath{A_{\infty}}}
\newcommand{\kinf}{\ensuremath{k_{\infty}}}
\newcommand{\kappainf}{\ensuremath{\kappa_{\infty}}}
\newcommand{\Nmax}{{N_{\mathrm{max}}}}
\newcommand{\LA}{L'_0}  
\newcommand{\ANC}{\gamma_\infty}
\newcommand{\Jost}{\mathfrak{f}}
\newcommand{\LNLO}{\text{L-NLO}}
\newcommand{\rsqav}{\langle r^2 \rangle}
\newcommand{\rsqinf}{\ensuremath{\rsqav_{\infty}}}
\newcommand{\Lameff}{\Lambda_{\rm eff}}
\newcommand{\Leff}{L_{\rm eff}}

\newcommand{\ee}{\mathrm{e}}
\newcommand{\ii}{\mathrm{i}}

\newcommand{\wt}{\widetilde}
\newcommand{\wh}{\widehat}
\newcommand{\dd}{\mathrm{d}}

\newcommand{\PT}{\text{PT}}
\newcommand{\VPT}{V_\PT}

\newcommand{\ie}{\textit{i.e.}}
\newcommand{\eg}{\textit{e.g.}}

\newcommand{\mbraket}[3]{\langle #1|#2|#3\rangle}

\newcommand{\etal}{\textit{et al.}\xspace}

\newcommand{\mbf}{\mathbf}
\newcommand{\pp}{p^\prime}
\newcommand{\ppp}{p^{\prime \prime}}
\newcommand{\Ep}{E^\prime}

\newcommand{\GreensFn}{G_0}
\newcommand{\ampT}{\mathcal{T}}
\newcommand{\fL}{f_L}
\newcommand{\thetacm}{\theta^{\prime}}
\newcommand{\phicm}{\varphi^{\prime}}
\newcommand{\thetacprime}{\alpha^\prime}
\newcommand{\thetacdoubleprime}{\alpha^{\prime \prime}}
\newcommand{\msf}{m_{s_f}}
\newcommand{\mJd}{m_{J_d}}
\newcommand{\mst}{\wt{m}_s}
\newcommand{\dcostheta}{\dd \! \cos \theta}
\newcommand{\mbfq}{\mbf{q}}
\newcommand{\la}{\langle}
\newcommand{\ra}{\rangle}
\newcommand{\bi}{\begin{itemize}}
\newcommand{\ei}{\end{itemize}}
\newcommand{\li}{\item}
\newcommand{\CG}[6]{\ensuremath{\braket{#1\,#2\,#3\,#4}{#5\,#6}}}

\NewEnviron{subalign}[1][]{%
\begin{subequations}\begin{align}
  \BODY
\end{align}\label{#1}\end{subequations}
}

\NewEnviron{spliteq}{%
\begin{equation}\begin{split}
  \BODY
\end{split}\end{equation}
}

\newlength{\movieplotwidth}
\newlength{\phaseplotwidth}
\newlength{\universalityplotwidth}
\newlength{\wideplotwidth}

\newacronym{ANC}{ANC}{Asymptotic Normalization Coefficient}
\newacronym{COM}{COM}{center-of-mass}

\newacronym{NN}{NN}{nucleon-nucleon}
\newacronym{2NF}{2NF}{two-nucleon force}
\newacronym{3NF}{3NF}{three-nucleon force}
\newacronym{AV18}{AV18}{Argonne $v_{18}$}

\newacronym{QCD}{QCD}{Quantum Chromodynamics}
\newacronym{QED}{QED}{Quantum Electrodynamics}
\newacronym{QFT}{QFT}{Quantum Field Theory}

\newacronym{SVD}{SVD}{singular value decomposition}

\newacronym{EFT}{EFT}{effective field theory}
\newacronym{RG}{RG}{renormalization group}
\newacronym{SRG}{SRG}{Similarity Renormalization Group}

\newacronym{SVM}{SVM}{stochastic variational method}

\newacronym{HO}{HO}{Harmonic Oscillator}

\newacronym{HF}{HF}{Hartree-Fock}
\newacronym{HFB}{HFB}{Hartree-Fock-Bogoliubov}

\newacronym{QMC}{QMC}{Quantum Monte Carlo}
\newacronym{CC}{CC}{Coupled Cluster}
\newacronym{NCSM}{NCSM}{No Core Shell Model}
\newacronym{itNCSM}{IT-NCSM}{Importance Truncated No Core Shell Model}
\newacronym{ShM}{SM}{Shell Model}
\newacronym{DFT}{DFT}{density functional theory}
\newacronym{VMC}{VMC}{variational Monte Carlo}
\newacronym{GFMC}{GFMC}{Green's function Monte Carlo}
\newacronym{AFDMC}{AFDMC}{Auxiliary Field Diffusion Monte Carlo}
\newacronym{MFD}{MFD}{many fermion dynamics}
\newacronym{CI}{CI}{configuration interaction}
\newacronym{NCFC}{NCFC}{No Core Full Configuration}
\newacronym{FCI}{FCI}{Full Configuration Interaction}
\newacronym{MBPT}{MBPT}{many-body perturbation theory}

\newacronym{UV}{UV}{ultraviolet}
\newacronym{IR}{IR}{infrared}

\newacronym{NUCLEI}{NUCLEI}{NUclear Computational Low-Energy Initiative}
\newacronym{UNEDF}{UNEDF}{Universal Nuclear Energy Density Functional}

\newacronym{ChPT}{$\chi$-PT}{Chiral Perturbation Theory}
\newacronym{ChEFT}{$\chi$-EFT}{Chiral Effective Field Theory}

\newacronym{DCT}{DCT}{Discrete Cosine Transform}

\newacronym{DWBS}{DWBS}{Distorted Wave Born Series}
\newacronym{DWBA}{DWBA}{Distorted Wave Born Approximation}
\newacronym[longplural=ordinary differential equations]{ODE}{ODE}{ordinary differential equation}

\newcommand{\TmpCiteM}[1]{}

\renewcommand{\eqref}[1]{\textup{\ref{#1}}}
\title{
	\texorpdfstring{
		Improving Predictions with Reliable Extrapolation Schemes and
		Better Understanding of Factorization
	}{
		Improving Predictions with Reliable Extrapolation Schemes and
		Better Understanding of Factorization
	}
}
\author{Sushant N.\ More}
\advisorname{Prof.~Richard Furnstahl}
\degree{Doctor of Philosophy} 
\member{Prof.~Robert Perry}
\member{Prof.~Michael Lisa}
\member{Prof.~John Beacom}
\authordegrees{M.S.}
\graduationyear{2016}
\unit{Physics} 

\begin{document}

\maketitle
\disscopyright
\clearpage

\begin{abstract}

	New insights into the inter-nucleon interactions, developments in
	many-body technology, and the surge in computational capabilities
	has led to phenomenal progress in low-energy nuclear physics
	in the past few years.  Nonetheless, many calculations still lack
	a robust uncertainty quantification which is essential for making
	reliable predictions.  In this work we investigate two distinct sources
	of uncertainty and develop ways to account for them.

	Harmonic oscillator basis expansions are widely used in ab-initio nuclear
	structure calculations.
	Finite computational resources usually require that the basis be truncated
	before observables are fully converged, necessitating
	reliable extrapolation schemes.
	It has been demonstrated recently that errors introduced from basis
	truncation can be taken into account by focusing on the infrared and
	ultraviolet cutoffs induced by a truncated basis.
	We show that a finite oscillator basis effectively imposes a hard-wall
	boundary condition in coordinate space.
	We accurately determine the position of the hard-wall as a function of
	oscillator space parameters, derive infrared extrapolation formulas for
	the energy and other observables, and discuss the extension of this
	approach to higher angular momentum and to other localized bases.
	We exploit the duality of the harmonic oscillator to account for the errors
	introduced by a finite ultraviolet cutoff.

	Nucleon knockout reactions have been widely used to study and understand
	nuclear properties.  Such an analysis implicitly assumes that the
	effects of the probe can be separated from the physics of the target nucleus.
	This factorization between nuclear structure and reaction components
	depends on the renormalization scale and scheme, and has not been
	well understood.  But it is potentially critical for
	interpreting experiments and for extracting process-independent
	nuclear properties.
	We use a class of unitary transformations called the
	similarity renormalization group (SRG) transformations to systematically
	study the scale dependence of factorization for the simplest knockout process
	of deuteron electrodisintegration.
	We find that the extent of
	scale dependence depends strongly on kinematics, but in a \emph{systematic}
	way.
	We find a relatively weak scale dependence at the quasi-free kinematics
	that gets
	progressively stronger as one moves away from the quasi-free region.
	Based on examination of the relevant overlap matrix elements, we are able
	to qualitatively
	explain and even predict the nature of scale dependence based on the
	kinematics
	under consideration.

\end{abstract}

\dedication{To the memory of my late grandparents, who valued integrity and
 education above everything else.}
\begin{acknowledgments}

  It was an absolute pleasure to work with Prof.\ Richard Furnstahl.  Even after
  four years of working closely with Dick, I am still in awe of the time and
  effort he invests in his students.  His passion towards physics, dedication
  towards mentorship, and caring about the scientific community is exemplary.
  I certainly consider it a privilege to be his padawan and hope to have
  imbibed at least part of his scientific ethos.

  I am also thankful to my committee members, in particular Robert
  Perry.  Their constructive criticism and feedback has helped me develop
  professionally.  I was fortunate to learn from a few exceptional
  teachers at Ohio State.  I thank them for instilling in me a joy for learning
  advanced physics.  I would like to express my gratitude towards
  Scott Bogner and Thomas Papenbrock; they have been a constant source of
  support, and I look up to them as mentors.  I am grateful to
  Sabine Jeschonnek for helping with the deuteron disintegration theory.
  I have also benefited tremendously from the interactions with professors
  and peers in the wider nuclear physics community.

  My time at Ohio State coincided with some of the best
  post-docs in the group---Heiko, Kai, and Sebastian.  They were always ready
  to help and my PhD experience has
  been enriched immensely through interactions with them.  I have worked
  closest with Sebastian; I admire his efficiency and his attention to details.

  I was also lucky to overlap with excellent fellow graduate students in the group.
  It was fun growing up with my physics siblings---Kyle, Sarah, and Alex.
  Special mention goes to Kyle for doing a great job of an elder sibling.
  It was a pleasure to share office space with Brian, Russell, and Khal.
  I thank them for being a jolly company and for putting up with my
  idiosyncrasies.  The other physics graduate students that I would like
  to acknowledge are my gym buddies Hudson and Chuck and my former officemate
  Omar.  I have enjoyed discussion with them over a full gamut of topics and
  have always turned to them for friendly advice when needed.

  My friend Kaushik and I landed in the US together five years ago.  We have had
  a good run as flatmates since then.  From teaching me cooking to being my
  driving
  instructor, I am indebted to him for many things.  One of the best things
  that happened to me at OSU was joining the student group Sankalpa.  I thank
  Gayatri, Vinayak, and Keerthi for getting me started along that path.
  Friends formed through Sankalpa have been like a family to me and have
  helped me grow as a person.  I thank Mithila, Dhruv, and Garima for being
  my safety net; Janani, Pooja, Shubhendu, and Keshav for intellectually
  engaging discussions; and Vipul, Nandan, Subhasree, Bakul, and Madhura for
  being a lot of fun.

  I would like to take this opportunity to visit my roots and express gratitude
  towards my undergraduate teachers, friends, and relatives back in India.
  They trusted in me when I had little to show.  I value the decade-long
  friendship of my pals Sheetal Kumar, Ajit, Shishir, Vivek, and Lakshmi Priya.

  Finally, I must put on record the unremitting love of my doting yet
  disciplinarian parents and my headstrong but sweet younger sister.
  Their unwavering encouragement through thick and thin keeps me
  going and motivates me to scale new heights.

\end{acknowledgments}

\begin{vita}
\dateitem{May 2011}{Dual B.S. M.S., Indian Institute of Science Education and Research (IISER),
Pune, India}
\dateitem{December 2013}{M.S, The Ohio State University, Columbus, Ohio}
\begin{publist}
\raggedright
\pubitem{
  \noindent
  \textit{
   Deuteron electrodisintegration with unitarily evolved potentials 
  }\\
  S.~N.\ More,
S.~K\"{o}nig, R.~J.\ Furnstahl, and K.~Hebeler, Phys.\ Rev.\ C \textbf{92}, 064002 (2015)
}
\pubitem{
  \noindent
  \textit{
    Ultraviolet extrapolations in finite oscillator bases
  }\\
  S.~K\"{o}nig, S.~K.\ Bogner, R.~J.\ Furnstahl, S.~N.\ More, and T.~Papenbrock, Phys.\ Rev.\ C \textbf{90}, 064007 (2014)
}
\pubitem{
  \noindent
  \textit{
    Systematic expansion for infrared oscillator basis extrapolations
  }\\
  R.~J.\ Furnstahl, S.~N.\ More, and T.~Papenbrock, Phys.\ Rev.\ C \textbf{89}, 044301 (2014)
}
\pubitem{
  \noindent
  \textit{
    Universal properties of infrared oscillator basis extrapolations
  }\\
  S.~N.\ More, R.~J.\ Furnstahl, A.~Ekstr\"{o}m, G.~Hagen, and T.~Papenbrock, Phys.\ Rev.\ C {\textbf{87}}, 044326 (2013)
}
\pubitem{
  \noindent
  \textit{
    Non-local linear stability of ion beam eroded surfaces
  }\\
 S.~N.\ More, R.~Kree. Applied Surface Science 258 (2012) 4179-4185
}
\end{publist}
\begin{fieldsstudy}
\majorfield{Physics}
\end{fieldsstudy}

\end{vita}

\tableofcontents

\clearpage
\listoffigures

\clearpage
\listoftables



\startdoublespace

\cleardoublepage
\chapter{Introduction}
\label{chap:Intro}

	\section{Overview of Nuclear Physics}

\begin{figure}[htbp]
	 \centering
	 \includegraphics[width=0.9\textwidth]{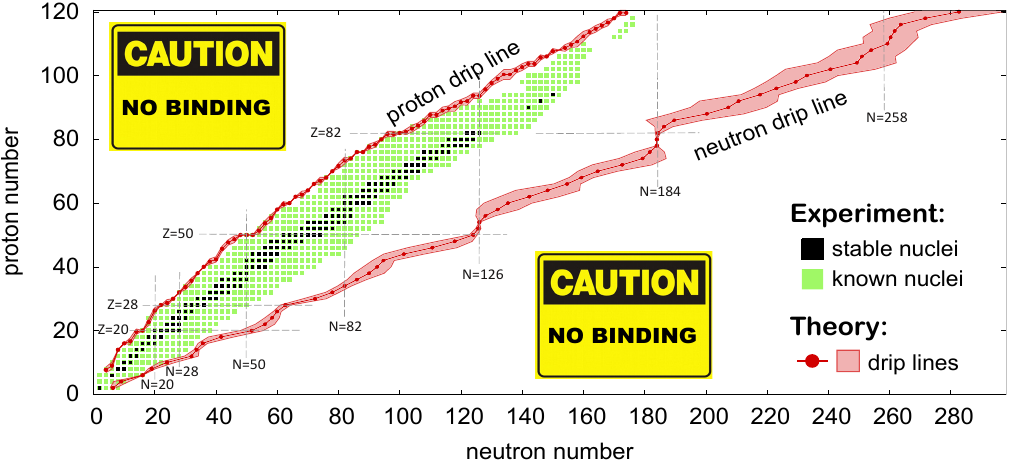}
	 \caption{Nuclear landscape.  A total of $288$ isotopes (black squares) are
	  	stable on the time scale of the solar system.  As more protons
			or neutrons are added to these stable nuclei, we enter the regime of
			short-lived radioactive nuclei (green squares).  `Drip lines' mark the
			limit of nuclear existence, where the last nucleon is no longer bound.
			The uncertainties around drip lines (in red)
	  	were obtained by averaging the results of different theoretical models.
			Figure from \cite{Long_range_plan}.}
	 \label{fig:Nuclear_landscape}
	\end{figure}

	Nuclear physics deals with the properties of atomic nuclei.
	The nuclear
	landscape shown in Fig.~\ref{fig:Nuclear_landscape} has been the
	traditional playground for nuclear physics.  The questions historically
	driving nuclear physics have been: how do protons and neutrons
	make stable nuclei and rare isotopes?  What are the limits of nuclear
	existence?
	What are the nuclear binding energies, excitation spectra, radii and so on?
	We would also like to describe nuclear reactions, make predictions about
	the shape of the nuclei and understand how the shape dictates the nuclear
	properties.

	By the mid-1970s, it was generally accepted that the nucleons
	(proton and neutrons)
	and other hadrons are composed of quarks, and that the quarks are held
	together through the exchange of gluons	\cite{Aubert:1974js}.  The
	following decades witnessed rapid
	development in the fundamental theory of strong interactions
	describing the interactions between quarks and gluons.  This theory goes
	by the name of Quantum Chromodynamics (QCD)
	\cite{Gross:2005kv}.  One of the active areas of
	investigation is obtaining the hadron structure from QCD.  This includes,
	for example, understanding the origin of proton spin, which is studied
	experimentally at Jefferson Laboratory
	\cite{Myhrer:2009uq}.  A related focus area
	is understanding the nature of the quark-gluon plasma (QGP)---the phase
	in which the universe is believed to exist for up to a few milliseconds after
	the Big Bang \cite{Martinez:2013xka}.  The energies involved in this
	subfield (few GeVs) are higher
	than the energies in `traditional' nuclear physics (few MeVs) introduced in
	the opening paragraph.  Therefore it is conventional to refer to the two
	subfields as high-energy nuclear physics and low-energy nuclear physics.
	The work in this thesis will mainly focus on questions in low-energy nuclear
	physics (LENP).

	Apart from the questions at the core of LENP, mentioned in connection to
	Fig.~\ref{fig:Nuclear_landscape}, inputs from LENP are immensely important
	in other areas as well.  One such broad area is that of nuclear astrophysics.
	The majority of the stable
	and known nuclei shown in Fig.~\ref{fig:Nuclear_landscape} were formed in
	big bang, stellar, or supernova nucleosynthesis
	\cite{Cyburt:2015mya, PAS:9305903}.  Inputs from LENP are
	critical in understanding the processes involved in nucleosynthesis and
	predicting the observed abundances of isotopes.
	Neutron stars are another fascinating astrophysical objects for low-energy
	nuclear physicists.  We would like to determine the equation of state for
	neutron stars and understand how and why stars explode
	\cite{Lattimer:2015eaa}.

	Finally, there are questions about the fundamental symmetries of the universe
	where
	nuclear physics hopes to make significant contributions.  For instance,
	why is there more matter than antimatter in the universe?  What is the nature
	of dark matter \cite{Feng:2010gw}?  What is the nature of the neutrinos
	(Majorana or Dirac
	fermions) and how have they shaped the evolution of the universe?
	In fact, as we will see later, accurate calculations of nuclear matrix
	elements are critical for the experiments undertaken to understand the
	nature of neutrinos \cite{Avignone:2007fu}.

	In addition to the broad scientific impact of nuclear physics that we have
	already mentioned, it also has many real-life applications.  Our knowledge of
	nuclei and ability to produce them has led to an increase in the quality of
	life
	for humankind.  Applications of nuclear physics encompass a diverse
	domain including but not limited to energy, security, medicine,
	radioisotope dating, and material sciences.

	\section{Checkered past; promising future}

	In 1935, Hideki Yukawa proposed the seminal idea of nuclear interactions
	being mediated by a massive boson \cite{Yukawa:1935xg}.   This could explain
	how protons and neutrons would stay bound in a nucleus, overcoming the Coulomb
	repulsion between protons.  The fact that such a model described scattering
	data well at low energies (few MeVs) and the eventual discovery of pions in
	1947 led to a wide acceptance of this model.  Very soon other heavy mesons
	($\rho$, $\omega$, $\sigma$) were discovered as well.  Scattering experiments
	also indicated that the strength of the nuclear potential depended on
	distance and at short distances the potential was repulsive.

\begin{figure}[htbp]
	 \centering
	 \includegraphics[width=0.6\textwidth]%
	 {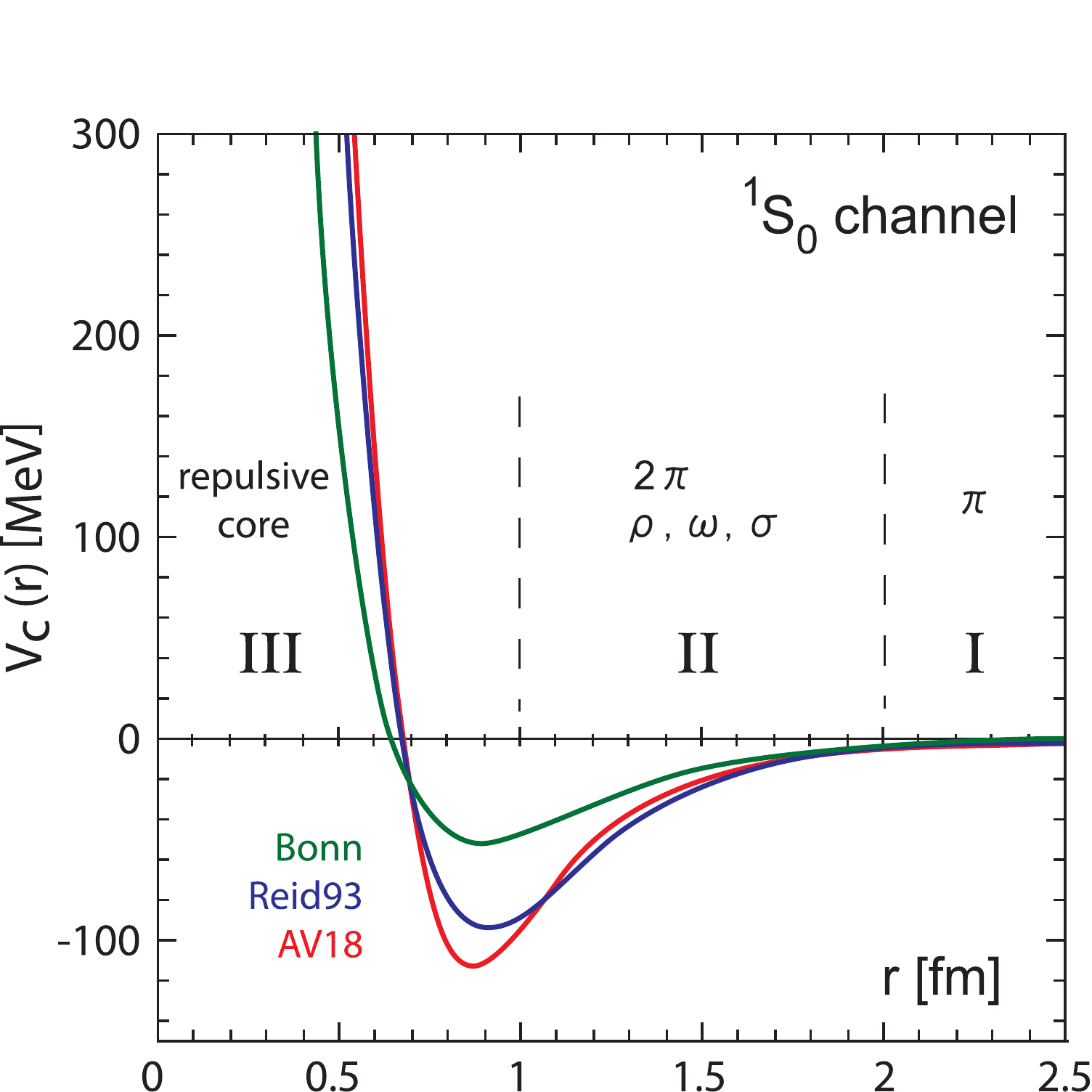}
	 \caption{AV18 \cite{Wiringa:1994wb}, Reid93 \cite{Stoks:1994wp}, and
	  	Bonn \cite{Machleidt:1989tm} potentials for $^1S_0$ channel as
			functions of internucleonic distance.  These potentials accurately
			describe neutron-proton scattering up to laboratory energies of
			300 MeV.  Regions I, II, and III correspond to long-range,
			intermediate-range, and short-range parts discussed in the text.
			Figure from \cite{Aoki:2008hh}. }
	 \label{fig:Nuclear_potentials}
	\end{figure}

	By 1950, there emerged an industry for coming up with better nuclear
	potentials.  These boson-exchange models shared some common features.
	The long-range part of the nucleonic interaction was given by pion exchange,
	the intermediate range was governed by multiple (mostly two) pion exchange,
	and short-range repulsion was thought to be because of overlap of nucleons.
	When heavy mesons were discovered, they were added to the intermediate range
	sector.  The pion, being the lightest meson, has the longest range.
	These general considerations form the basis for phenomenological potentials
	used even today as seen in Fig.~\ref{fig:Nuclear_potentials}.

	This intense effort is well summarized by Hans Bethe's quote in his essay
	`What Holds the Nucleus Together?' in Scientific American (1953):
	``In the past quarter century physicists have devoted a huge amount of
	experimentation and mental labor to this problem -- probably more man-hours
	than have been given to any other scientific question in the history of
	mankind.''
	The boson models did not have a smooth sailing though.  In particular
	the intermediate range multi-pion sector was beset with problems.  The
	pessimism this resulted in is palpable in Marvin Goldberger's comment
	in 1960: ``There are few problems in nuclear theoretical physics which
	have attracted more attention that that of trying to determine the
	fundamental interaction between two nucleons.  It is also true that
	scarcely ever has the world of physics owed so little to so many...It
	is hard to believe that many of the authors are talking about the same
	problem or, in fact, that they know what the problem is.'' A running joke
	was that nuclear physics is really `unclear' physics!

\begin{figure}[htbp]
		\centering
		\begin{subfigure}[c]{0.43\textwidth}
			\centering
			\includegraphics[width=\textwidth]
			{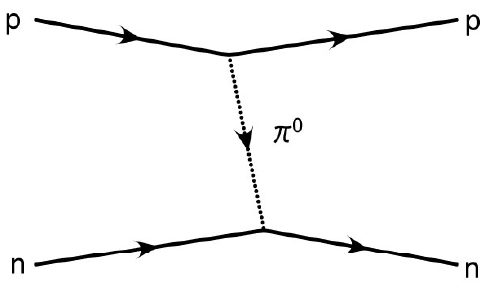}
			\caption{Inter-nucleon interaction in 1940}
			\label{fig:nucleon_pion_exchange}
		\end{subfigure}
		\hspace{0.07\textwidth}
		\begin{subfigure}[c]{0.42\textwidth}
			\centering
			\includegraphics[width=\textwidth]
			{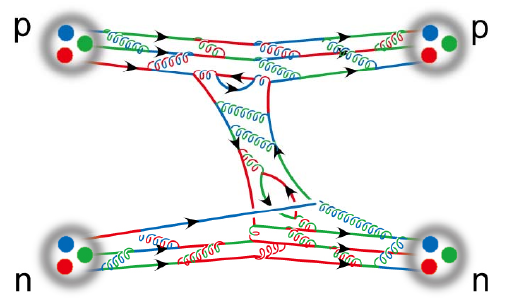}
			\caption{Inter-nucleon interaction in 1980}
			\label{fig:nucleon_gluon_exchange}
		\end{subfigure}
		\caption{Evolution of the inter-nucleon interaction picture over time.
		Figures from a talk by Witold Nazarewicz.}
		\label{fig:nucleon_interaction}
	\end{figure}

	There was relatively slow progress with regards to the development of
	internucleonic potentials in 1970's and 80's.  However, this period saw
	a rapid development of perturbative QCD.  It was realized that the nucleons
	and pions
	are composed of quarks which are held together by exchange of gluons
	(cf.~Fig.~\ref{fig:nucleon_interaction}).
	This pushed the effort to derive the nuclear
	potential from the `fundamental' theory of QCD.

	However, the effort to replace the hadronic descriptions at ordinary nuclear
	densities with a quark description as in Fig.~\ref{fig:nucleon_gluon_exchange}
	was not very fruitful.
	\begin{figure}[htbp]
	 \centering
	 \includegraphics[width=0.6\textwidth]%
	 {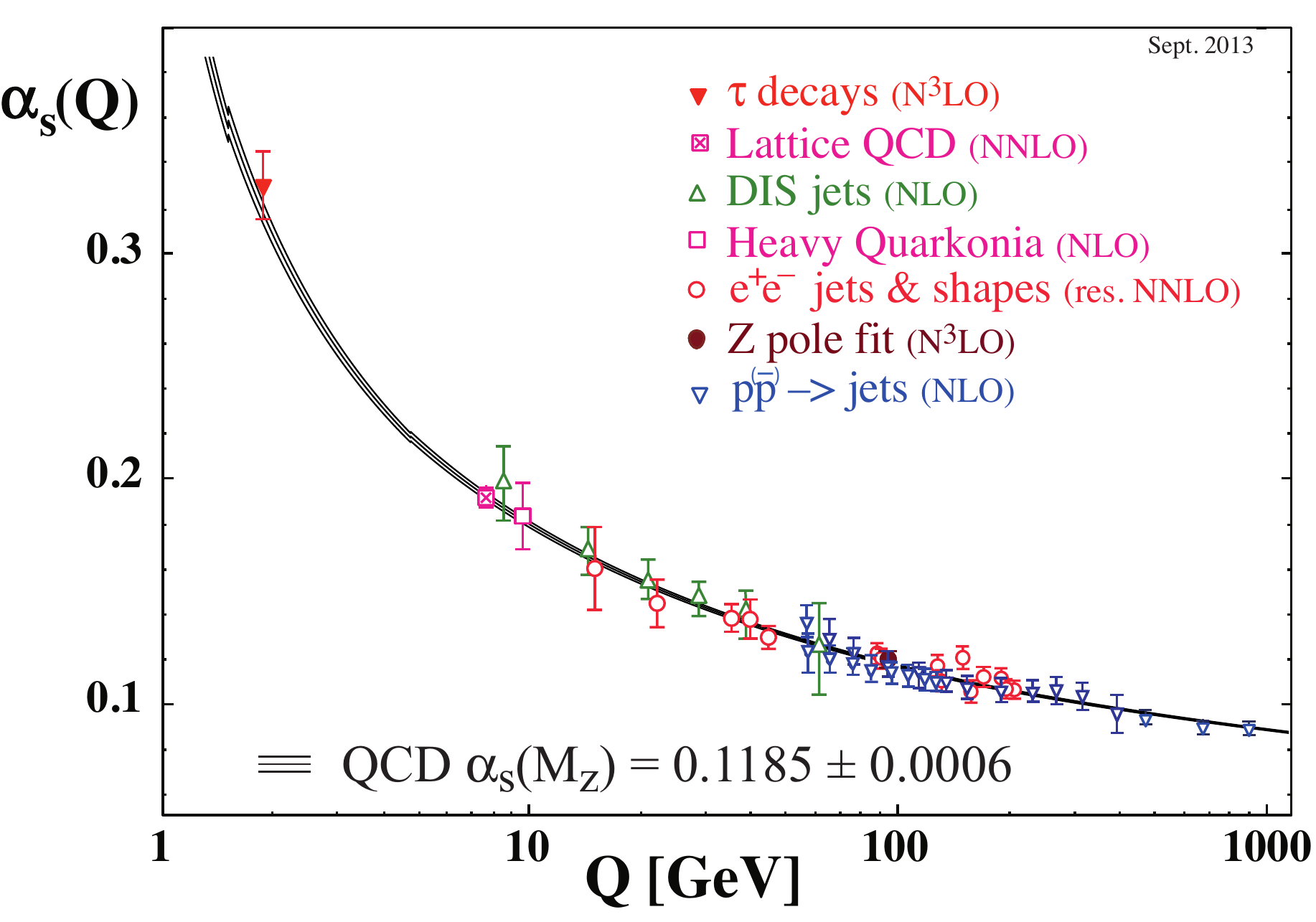}
	 \caption{Summary of measurements of the QCD coupling $\alpha_s$ as a
	 function of energy scale $Q$ \cite{Agashe:2014kda}. }
	 \label{fig:alpha_QCD_running}
	\end{figure}
	As seen in Fig.~\ref{fig:alpha_QCD_running}, the strength of the QCD
	coupling $\alpha_s$ increases with decreasing energies.  This makes
	QCD non-perturbative in the low-energy regime of nuclear physics, limiting
	the success of analytical calculations.

	Another aspect that makes low-energy nuclear physics difficult is that it is
	a many-body problem.  It exhibits some emergent phenomena that are difficult
	to capture in a reductionist approach.  This issue has been well-summarized by
	the famous article `More is different' by Phillip Anderson
	\cite{Anderson:1972pca} (albeit with a focus on many-body problem in
	condensed matter).

\begin{figure}[htbp]
	 \centering
	 \includegraphics[width=0.7\textwidth]%
	 {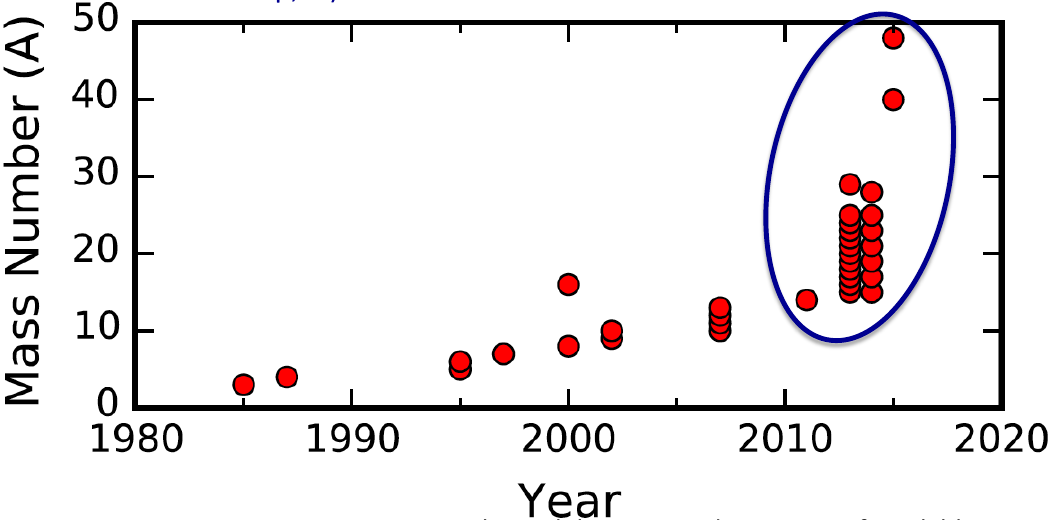}
	 \caption{LENP version of Moore's law and its violation.  $Y$ axis is the mass
	  	number of nuclei that can be calculated from ab-initio calculations.
			In past few years, it has been possible to push the ab-initio frontier to
			heavier nuclei.  Figure from a talk by Gaute Hagen. }
	 \label{fig:Moores_law_LENP}
	\end{figure}

	Despite all these challenges, great strides have been made in LENP
	in the last few decades.  As indicated in Fig.~\ref{fig:Moores_law_LENP}
	particularly the last few years have seen an explosion in the capabilities of
	low-energy nuclear theory.  This phenomenal progress has been possible due
	to the combination of a few factors---new insights about the nuclear force,
	developments in many-body technology, and a surge in computational
	capabilities.  In the following sections, we look briefly at each of these
	developments which will lead us to how the author's PhD work fits into the
	bigger picture.

	\section{Understanding the Force}
	\label{sec:recent_advances}

	We saw that non-pertubativeness of QCD at low energies
	(cf.~Fig.~\ref{fig:alpha_QCD_running}) motivated phenomenological
	descriptions of nuclear forces.  The two popular categories of
	phenomenological interactions are the meson exchange models (which we
	touched upon during discussion of Fig.~\ref{fig:Nuclear_potentials}) and
	local\footnote{The potential is local if
	$V(\bm{r}, \bm{r^\prime}) = V(\bm{r}) \delta(\bm{r} - \bm{r^\prime})$.}
	phenomenological potentials.
	Examples of these include
	the Bonn potentials \cite{Machleidt:1989tm, Machleidt:1987hj,
	Machleidt:2000ge} and the Argonne potential \cite{Wiringa:1994wb}.

	The Argonne potential is probably the most widely used phenemenological
	potential.  One of the reasons being that until recently, it was the only
	precision interaction usable for Quantum Monte Carlo calculations.  The
	Argonne interactions are built by writing down all the operators that
	satisfy the required symmetries---translational and Galilean invariance,
	rotational invariance in space and spin, rotational invariance in isospin,
	time reversal, and spatial reflection.  These operators are given below.
	\beq
	\wh{O}_i \in \{ \mathds{1}, \bm{\sigma_1} \cdot \bm{\sigma_2},
	S_{12}, \bm{L} \cdot \bm{S}, \bm{L}^2, \bm{L}^2 \bm{\sigma_1} \cdot
	\bm{\sigma_2}, (\bm{L} \cdot \bm{S})^2 \} \otimes
	\{\mathds{1}, \bm{\tau_1} \cdot \bm{\tau_2}\}
	\label{eq:AV18_14_operators}
	\eeq
	$S_{12} = 3 (\bm{\sigma_1} \cdot \wh{\bm{r}})
	(\bm{\sigma_2} \cdot \wh{\bm{r}}) - \bm{\sigma_1} \cdot \bm{\sigma_2}$
	is the tensor force.

	There are a total of 14 operators in Eq.~\eqref{eq:AV18_14_operators}.
	The AV18 potential has four more operators which are the charge-dependent
	and charge-symmetry breaking terms; they are small but needed to get
	$\chi^2/{\rm{dof}} \approx 1$ for np, nn, and pp scattering.  There are
	also AV8, AV6 potentials which use a limited set of operators.

	The AV18 potential is written as
	\beq
	\wh{V}_{18}(r) = \sum_{i = 1}^{18} V_i(r) \, \wh{O}_i \;,
	\eeq
	where $r$ is the inter-nucleon separation and
	$V_i (r) = V_{{\rm{EM}}} + V_{\pi}
	+ V_{{\rm{short~range}}}$.  The coefficients in the potential are fit to
	nucleon scattering up to $350$ MeV, and to deuteron bound state properties
	such as binding energy, radii, and quadrupole moment.
	A similar exercise has been done for 3N (3-body) interactions.  However,
	the large number of possible three-body operators makes it difficult to
	get rid of model dependence.

	Meson exchange models are formulated in terms of exchange of mesons taking
	into account the quantum nature of mesons (scalar, vector, pseudo-scalar,
	so on).
	The masses are those of real mesons, but couplings are fit parameters.
	In the simplest form, the interaction is a sum of Yukawa potentials,
	\beq
	V = \left(\frac{-g_s^2}{4 \pi}\right) \frac{\ee^{-m_s r}}{r} +
	 		\gamma_1^\mu \gamma_{2 \mu} \left(\frac{-g_\omega^2}{4 \pi}\right)
			\frac{\ee^{-m_\omega r}}{r} +
			\gamma_1^5 \gamma_{2}^5 \, \bm{\tau_1} \cdot \bm{\tau_2}
			\left(\frac{-g_{\pi}^2}{4 \pi}\right)
			\frac{\ee^{-m_{\pi} r}}{r}\;.
	\eeq

	Meson exchange potentials and the AV18 potential share the common shortcoming
	that there is no scope for systematic	improvements.  It is also unclear
	how to seek model independence and do robust uncertainty quantification.
	Finally, these models offer limited guidance on the strength and relevance of
	three- and higher-body forces.

	\subsection{Chiral EFT}

	\begin{figure}[htbp]
	 \centering
	 \includegraphics[width=0.6\textwidth]%
	 {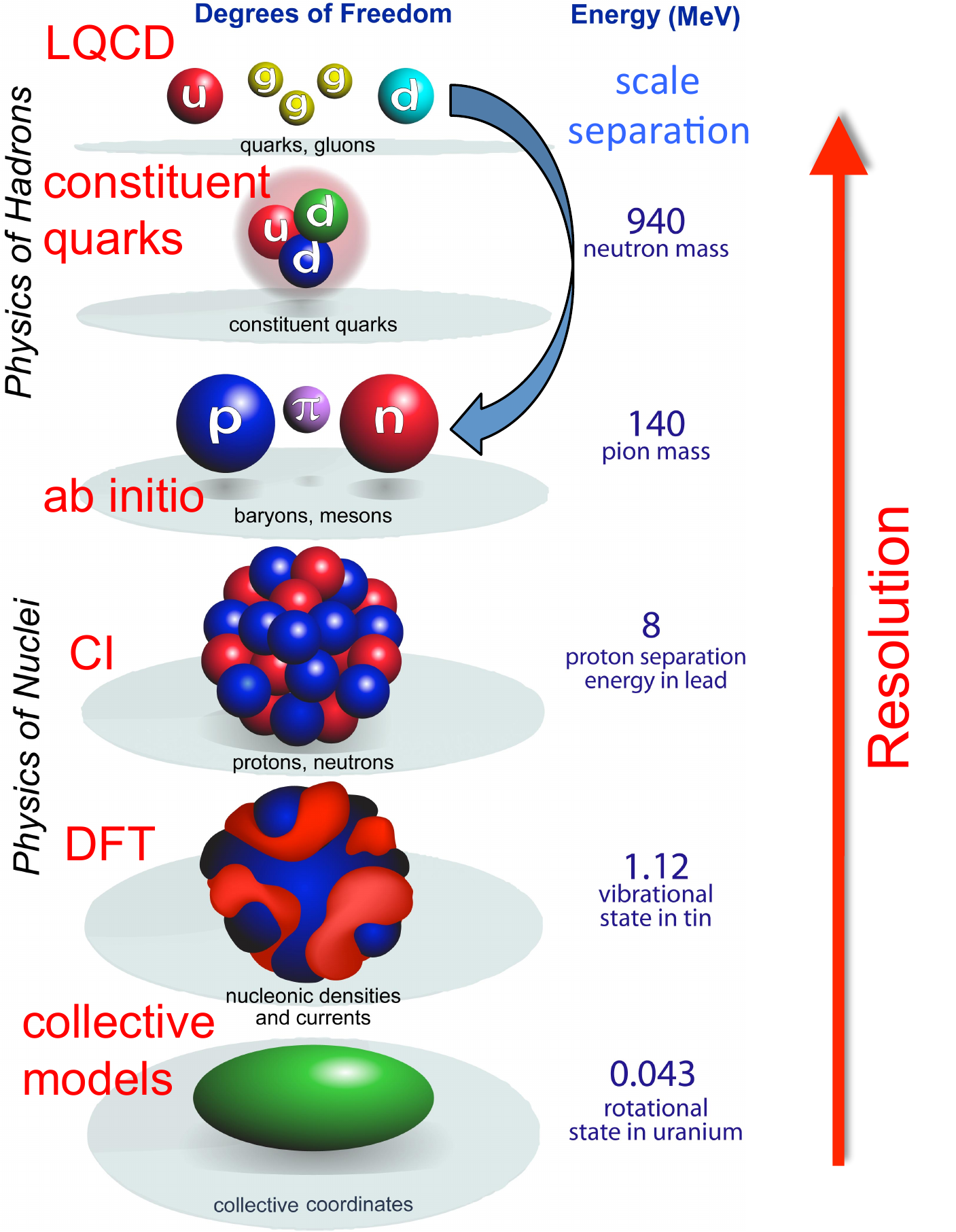}
	 \caption{Hierarchy of degrees of freedom and associated energy scales in
	  	nuclear physics \cite{LRP:2007}.}
	 \label{fig:degrees_of_freedom}
	\end{figure}
	An intriguing aspect of the world we live in, is that there are interesting
	phenomena at virtually all energy and length scales we can probe.  From TeV
	energies at the Large Hadron Collider (LHC) to the life-defining process of
	respiration which has the energy scale of only few meV, there are physical
	processes of interest at each step.  Nuclear physics spans a wide range of
	energy and length scales (cf.~Fig.~\ref{fig:degrees_of_freedom});
	a wider range than most subfields.  This hierarchy provides both
	challenges and opportunities.

	Figure~\ref{fig:degrees_of_freedom} indicates the relevant degrees of freedom
	for the given energy scales.  Even though degrees of freedom are a matter of
	choice, in practice, appropriate degrees of freedom often dictate the
	success of a theory.  To quote Steven Weinberg \cite{Guth:1984rq}:
	``You can use any degrees of freedom you want, but if you use the wrong ones,
	you'll be sorry.''
	Weinberg in his seminal paper \cite{Weinberg:1978kz} applied the concept
	of effective field theory (EFT) to low-energy QCD.  This effort
	proceeds by writing down the most general Lagrangian consistent with the
	(approximate) symmetries.

	QCD has the expected symmetries of translational, Galilean, and rotational
	invariance, and spatial reflection and time reversal.  Along with that,
	in the limit of vanishing quark masses, the QCD Lagrangian also possess an
	exact	chiral symmetry \cite{Peskin1995a}.  If the chiral symmetry holds,
	``left''- and
	``right''-handed fields do not mix.  As with any other continuous symmetry,
	spontaneous breaking of chiral symmetry leads to massless Goldstone boson(s).
	The masses of the up and down quark (quarks relevant in LENP) are both small
	($\sim 2 - 6$ MeV \cite{Agashe:2014kda}), but non-zero.  Therefore the
	chiral symmetry of the QCD Lagrangian is only approximate and the resulting
	Goldstone boson---pion---is light (compared to mass of nucleon), but not
	massless.

	Chiral EFT ($\chi$-EFT) uses nucleons and pions as degrees of freedom
	\footnote{Delta-full chiral EFTs also include the
	$\Delta$---a resonant state of the nucleon---as a degree of freedom. }.
	Heavy mesons are
	``integrated out''.  The crucial difference that distinguishes $\chi$-EFTs from
	meson theories of the 1950s is that they are constrained by chiral symmetry.
	Broken chiral symmetry serves as a connection with the underlying theory of
	QCD.
	A major advantage of $\chi$-EFT is that it permits systematic improvements
	and allows the possibility of having reliable uncertainty quantification.

	\begin{figure}[htbp]
	 \centering
	 \includegraphics[width=0.8\textwidth]%
	 {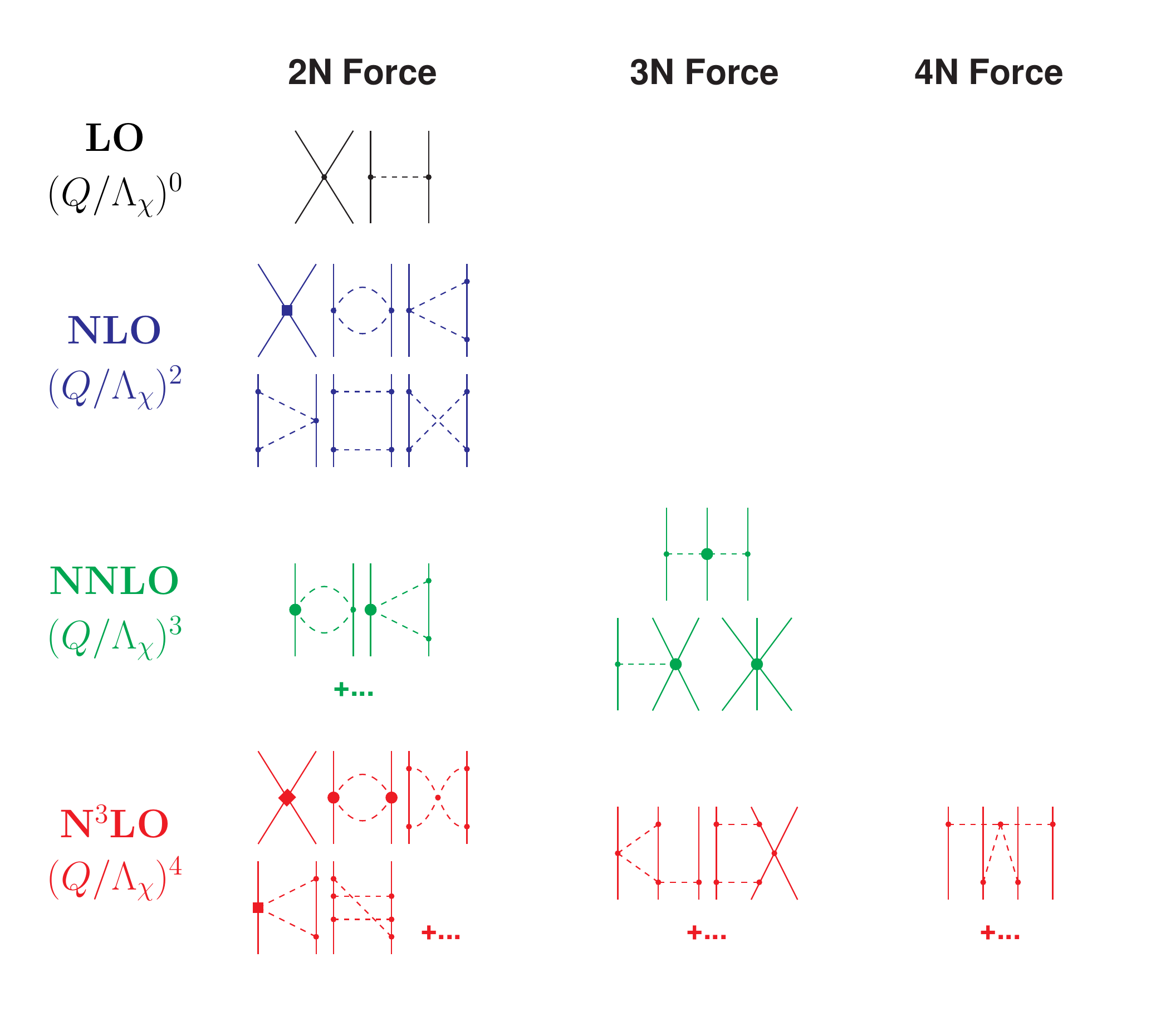}
	 \caption{Diagrams for a chiral Lagrangian at each order.  Solid lines are the
	 nucleons and dashed lines pions.  Figure from \cite{Machleidt:2011zz}.}
	 \label{fig:ch_EFT_diagrams}
	\end{figure}
	Diagrams in a $\chi$-EFT are shown in Fig.~\ref{fig:ch_EFT_diagrams}.  The
	three- and higher-body forces appear naturally in $\chi$-EFT with an
	expected hierarchy (higher-body forces are suppressed successively).  The
	coupling
	constants at the vertices in Fig.~\ref{fig:ch_EFT_diagrams} are called
	low-energy coupling constants (LECs); they encode the QCD physics and
	cannot be calculated in $\chi$-EFT.  LECs are fit to experimental data.
	Lattice QCD provides a promising method for extracting them in the near
	future (cf.~Subsec.~\ref{subsec:lattice_theories}).  Despite some concerns
	over
	power-counting (see Ref.~\cite{Griesshammer:2015osb} and references
	therein), it is
	fair to say that $\chi$-EFT has been a major breakthrough in low-energy
	nuclear theory.  For more details and applications of $\chi$-EFT, please
	see Refs.~\cite{Machleidt:2011zz} and \cite{Epelbaum:2005pn}.

	\section{Many-body methods}


	We learned in kindergarten quantum mechanics that a two-body problem
	can be solved by reducing it to a one-body problem.  The many-body
	problem is far more relevant to the nuclear physics, and is more
	challenging.
	For the nuclear properties such as the bound state energies, the
	equation we need to solve is the time-independent Schr\"{o}dinger equation
	\beq
	\wh{H} \ket{\Psi} = E \ket{\Psi} \;.
	\label{eq:Schrodinger_many_body}
	\eeq
	The Hamiltonian $\wh{H}$ is given by
	\beq
	\wh{H} = \sum_{\rm particles} \wh{T} + \sum_{\rm pairs}
	\wh{V}^{(2)}_{\rm pair} + \sum_{\rm triplets}
	\wh{V}^{(3)}_{\rm triplets} + \cdots \;,
	\eeq
	where $\wh{T}$ is the kinetic energy and $\wh{V}^{(2)}, \, \wh{V}^{(3)}$ are
	two- and three-body
	nuclear potentials (e.g., from $\chi$-EFT or phenomenological potentials).

	Over the years various methods have been developed to tackle
  this problem.  We will list some of the broad categories below.
	We only provide
  a brief explanation for each of them and refer the reader to the cited
  references for details.
	\bi
	\li
	\emph{Direct diagonalization:}
	This category involves expanding the many-body wave function $\ket{\Psi}$ in
	an appropriate complete basis.  Very often this basis is chosen to be the
	Slater determinant of harmonic oscillator (HO) wave functions.
	HO wave functions form a complete basis with discrete energy levels that
	depend on only one scale (the oscillator frequency).  Moreover, HO
	wave functions are simple analytic functions that go to zero at large
	distance just as nuclear bound state wave functions.

	Finite computational power forces us to truncate the infinite sum of
	Slater determinants at some point.  One of the common truncation scheme is
	the $\Nmax$ truncation.  It keeps all $A$ particle states
	$\ket{n_1 l_1 n_2 l_2 \ldots n_A l_A}$ such that
	\beq
	\sum_i 2 n_i + l_i \leq \Nmax \;.
	\eeq
	With $\Nmax$ truncation, solving the many-body Schr\"{o}dinger equation
	(Eq.~\eqref{eq:Schrodinger_many_body}) becomes a matrix diagonalization
	problem, and can be solved using standard algorithms. The $\Nmax$
	truncation in HO basis also allows
	separation of center-of-mass motion from relative motion, ensuring that only
	intrinsic properties are being calculated.

	\begin{figure}[htbp]
		\centering
		\begin{subfigure}[c]{0.42\textwidth}
			\centering
			\includegraphics[width=\textwidth]
			{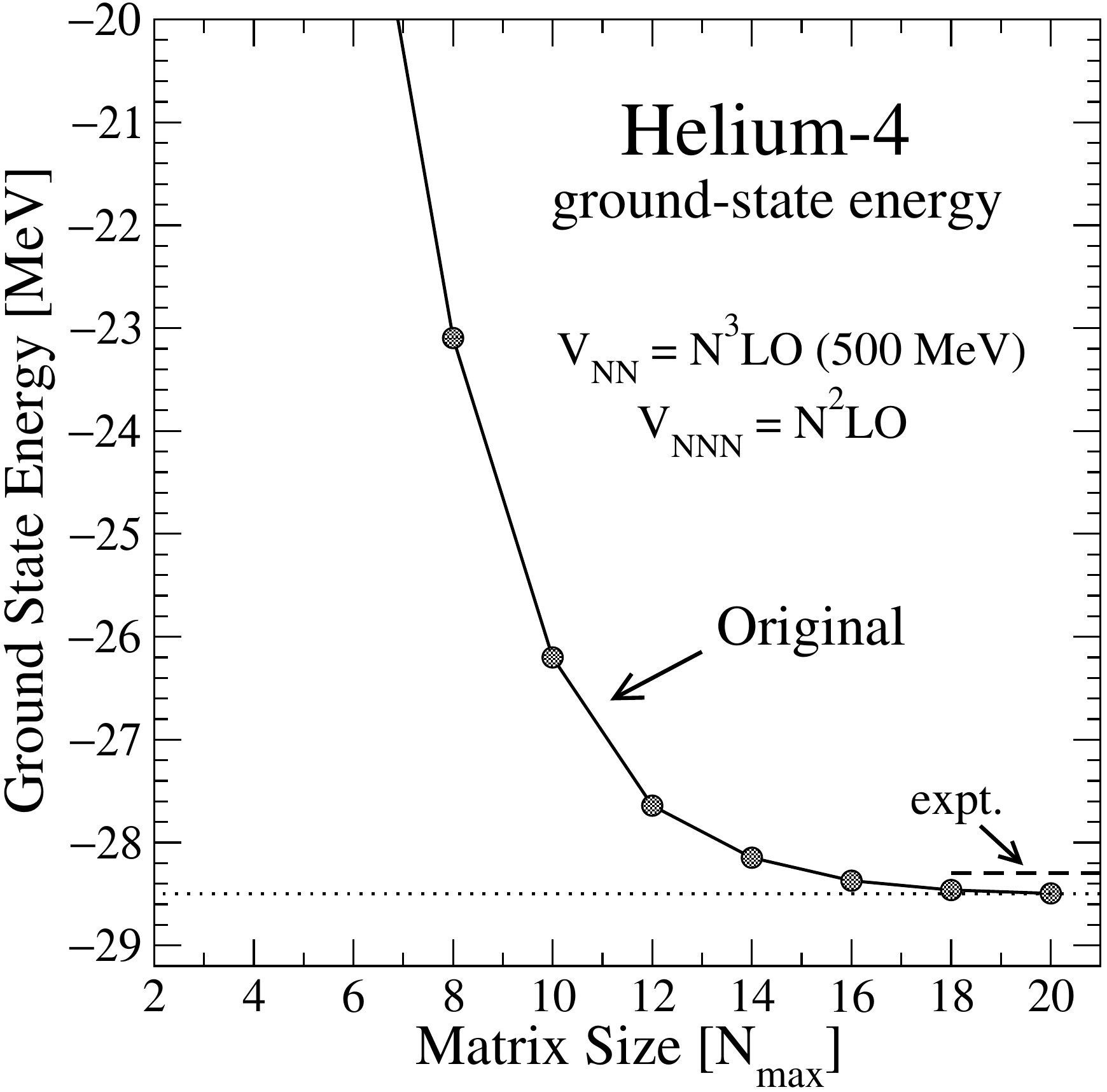}
			\caption{$^4$He energies as a function of $\Nmax$.}
			\label{fig:He4_vs_Nmax}
		\end{subfigure}
		\hspace{0.07\textwidth}
		\begin{subfigure}[c]{0.42\textwidth}
			\centering
			\includegraphics[width=\textwidth]
			{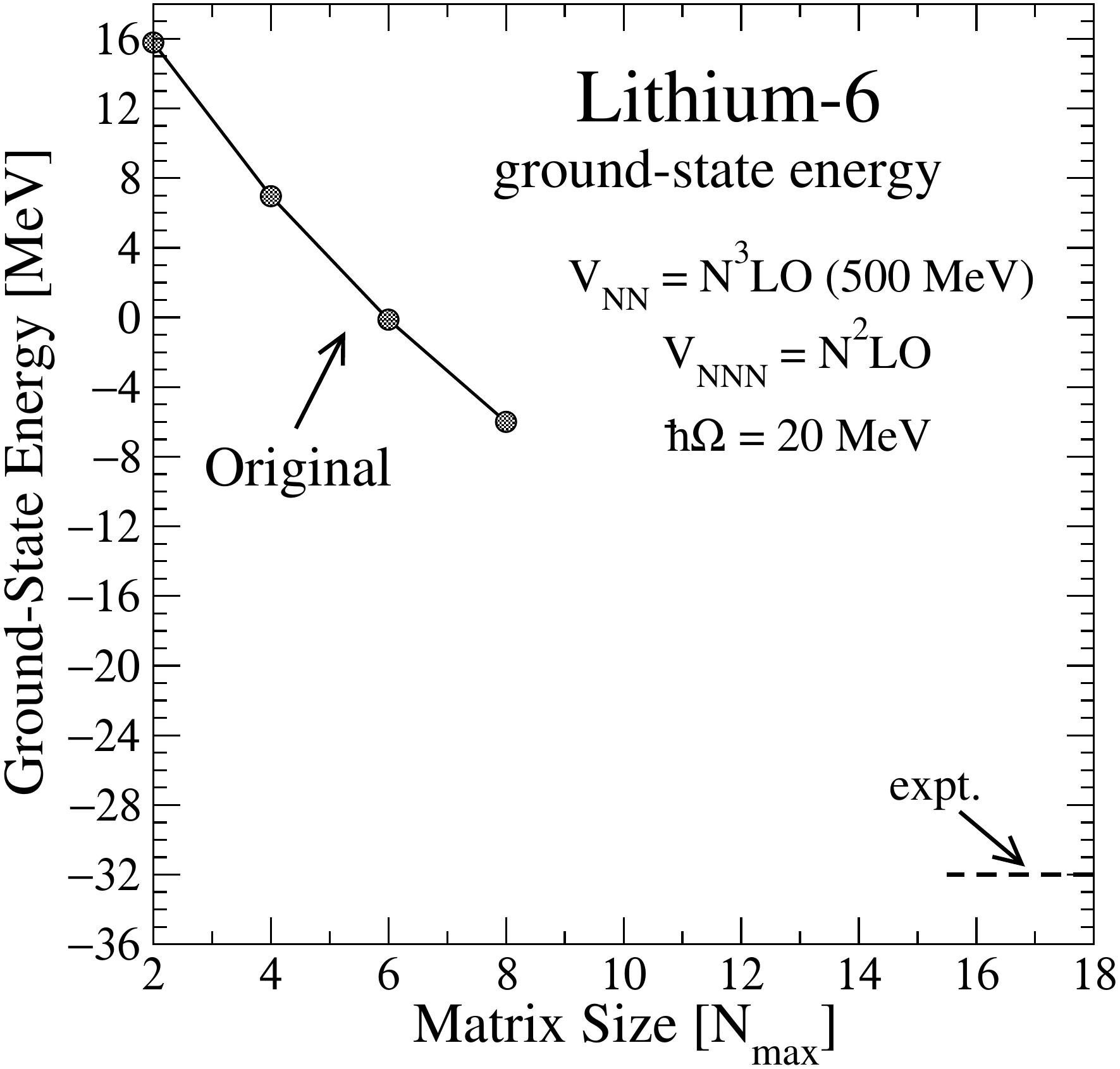}
			\caption{$^6$Li energies as a function of $\Nmax$.}
			\label{fig:Li6_vs_Nmax}
		\end{subfigure}
		\caption{Convergence of energies as a function of the truncation
			parameter $\Nmax$ \cite{Jurgenson:2010wy}.}
		\label{fig:Nmax_convergence}
	\end{figure}
	Figure~\ref{fig:Nmax_convergence} shows convergence plots for ground state
	energies for $^4$He and	$^6$Li.  As $\Nmax$ is increased the energies
	approach asymptotic values\footnote{When the asymptotic value doesn't
	match the experimental value, it points to the shortcomings in the nuclear
	Hamiltonian.}.
	\begin{figure}[htbp]
	 \centering
	 \includegraphics[width=0.45\textwidth]%
	 {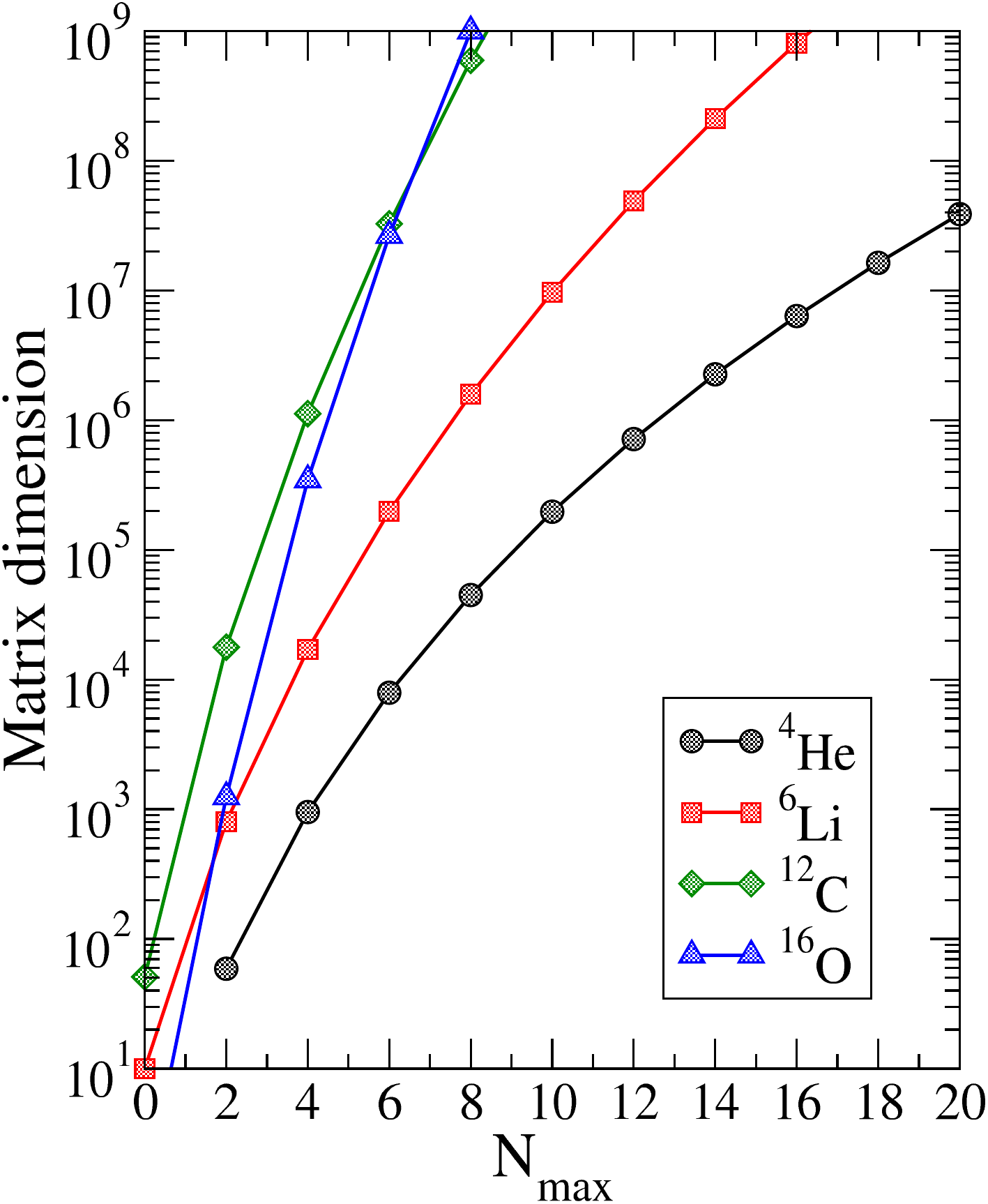}
	 \caption{Matrix dimension grows factorially with the number of nucleons.
	 	Figure courtesy of Pieter Maris.}
	 \label{fig:matrix_dimension_growth}
	\end{figure}
	A limitation of the diagonalization method is that the size of the Hamiltonian
	matrix that we need to diagonalize grows factorially as we go to higher $A$ and
	higher $\Nmax$ (cf.~Fig.~\ref{fig:matrix_dimension_growth}).  This often
	forces us to truncate the basis before convergence is reached
	(cf.~Fig.~\ref{fig:Li6_vs_Nmax}), and necessitates development of
	reliable extrapolation schemes.  Chapter~\ref{chap:Extrapolation}
	describes the author's original work devoted to the development of
	extrapolation schemes relevant to this problem.

	Many different methods fall under the broad umbrella of diagonalization
	methods.  This includes the traditional shell model~\cite{Brown:2001zz},
	the No Core Shell Model (NCSM)~\cite{Barrett:2013nh},
	the No Core Full Configuration (NCFC)~\cite{Maris:2008ax}, and the Importance
	Truncated No Core Shell Model (IT-NCSM)~\cite{Roth:2009cw}.
	The last three methods primarily differ in the nature of the truncation
	they employ.

	\li
	\emph{Monte Carlo methods:}
	These methods use the imaginary time evolution of the Schr\"{o}dinger
	equation
	\beq
	-\partial_\tau \ket{\Psi(\tau)} = \wh{H} \ket{\Psi(\tau)} \;.
	\label{eq:imag_time_SE}
	\eeq
	This equation can be solved by making an ansatz for
	$\ket{\Psi(\tau = 0)} \equiv \ket{\Psi_{\rm trial}}$.
	$\ket{\Psi_{\rm trial}}$ can be expanded in a complete basis of eigenvectors
	of $\wh{H}$ as
	\beq
	\ket{\Psi_{\rm trial}} = c_0 \ket{\Psi_0} +
	\sum_{i \neq 0} c_i \ket{\Psi_i} \;.
	\label{eq:Psi_trial_expansion}
	\eeq
	As long as the trial wave function is not orthogonal to the actual ground
	state, i.e, $c_0 \neq 0$ in Eq.~\eqref{eq:Psi_trial_expansion}, it can
	be shown that the imaginary time evolution (Eq.~\eqref{eq:imag_time_SE})
	projects out the ground state in the $\tau \to \infty$ limit.  There are many
	different ways to do the stochastic time evolution of
	Eq.~\eqref{eq:imag_time_SE} to project out the ground state.  The two
	most popular in nuclear physics are the Green's function Monte Carlo (GFMC)
	\cite{Pieper:2002ne, Pieper:2007ax}
	and the Auxiliary Field Diffusion Monte Carlo (AFDMC)
	\cite{Gandolfi:2007hs, Gezerlis:2014zia}.  Monte Carlo methods
	work best with a local potential.  The AV18 potential has been the
	potential	of choice for these methods.  Recently, it has been possible to
	derive the low-order $\chi$-EFT in local form, making its use possible
	for Monte Carlo methods \cite{Gezerlis:2014zia}.

	\li
	\emph{Coupled Cluster:} The energy of the many-body state $\Psi$ is given
	by
	\beq
	E = \mbraket{\Psi}{\wh{H}}{\Psi}\;.
	\label{eq:Schrodinger_CC}
	\eeq
	The coupled cluster (CC) method tries to build the state $\Psi$ from the
	reference state $\Phi$ (which for instance can be a Hartree-Fock state)
	using the transformation
	\beq
	\ket{\Psi} = \ee^T \ket{\Phi}\;.
	\label{eq:CC_defining_equation}
	\eeq
	Thus, Eq.~\eqref{eq:Schrodinger_CC} becomes
	\beq
	E = \mbraket{\Phi}{\ee^{-T}\wh{H} \ee^{T}}{\Phi}\;.
	\eeq
	The cluster operator $T$ in Eq.~\eqref{eq:CC_defining_equation}
	is defined with respect to the reference state.
	\beq
	T = T_1 + T_2 + \ldots + T_A \;.
	\label{eq:T_expansion}
	\eeq
	$T_n$ generates $n$-particles-$n$-holes excitations.  In practice,
	Eq.~\eqref{eq:T_expansion} is truncated at $T_2$
	(more recently corrections from $T_3$ are included as well).  CC
	scales much better than diagonalization or Monte Carlo methods discussed
	above and is therefore possible to use for medium-mass nuclei
	\cite{Hagen:2013nca}.

	\li
	\emph{Density Functional Theory:} Density Functional Theory (DFT) is based
	on the principle that the many-body ground state $\Psi(\bm{r_1}, \bm{r_2},
	\ldots, \bm{r_A})$ can be written as a functional of the density $\rho$,
	i.e, $\Psi = \Psi[\rho]$.
	Consequently, the energy (or any other observable) can be written
	as a functional of density
	\beq
	E[\rho] = \mbraket{\Psi[\rho]}{\wh{H}}{\Psi[\rho]} \;.
	\eeq

	DFT proceeds by writing down an energy density functional (EDF) guided by
	intuition and general theoretical arguments
	\cite{Drut:2009ce, Dobaczewski:2010gr}.  Inputs from experiments
	and exact calculations for simple few-body systems are also used for
	constructing the EDF.  The properties of the physical system are then found
	by the two step minimization of the EDF---first minimization is at a fixed
	density $\rho(\bm{r})$ and the second minimization is over $\rho(\bm{r})$.
	Once the EDF is decided upon, DFT does not scale prohibitively with
	the number of nucleons $A$, and is therefore the most popular for
	calculating properties of heavy-mass nuclei.

	\li
	\emph{IM-SRG:} In-Medium Similarity Renormalization Group (IM-SRG) is
	based on the SRG technique we will look at in Subsec.~\ref{subsec:SRG_intro}.
	It uses a series of continuous unitary transformations to decouple
	the ground state of the many-body Hamiltonian from the excitations.
	IM-SRG has made it possible to apply the ab-initio (starting with 2N and
	3N forces) methods to medium-mass nuclei and beyond.  Please see the
	Ref.~\cite{Hergert:2015awm} for a recent review on progress acheived by
	IM-SRG.
	\ei

	\subsection[``It from the bit"---lattice theories]%
	{``It from the bit"\footnote{The phrase ``It from the bit'' was originally
	used by John Wheeler while elucidating his ideas on digital physics.}
	---lattice theories}
	\label{subsec:lattice_theories}

	\medskip
	\subsubsection{Lattice QCD}

	We saw through Fig.~\ref{fig:alpha_QCD_running} that the largeness of the
	QCD coupling at low-energies makes it unamenable to analytical calculations.
	A well-established non-perturbative approach in this regime is lattice
	QCD \cite{Savage:2015eya}.  In lattice QCD, one discretizes space-time;
	fields representing quarks are defined at lattice sites and
	the gluon fields are defined on links connecting neighboring sites.

	Ideally we would like the lattice size to be as large as possible and the
	lattice spacing to be as
	small as possible.  However, lattice calculations are computationally
	extremely intensive, thereby severely
	constraining the lattice size and spacing.  Moreover, the computational
	cost of simulations
	scales with the quark mass roughly as $m_q ^{-4}$
	\cite{Detmold:2003rq}.  The simulations
	are therefore often done at quark masses larger than the physical quark
	masses.  From the Gell-Mann-Oakes-Renner relation $m_\pi ^2 \sim m_q$,
	and therefore the pion mass
	in the lattice calculations is larger than its physical value as well.
	The results
	are then extrapolated down to physical quark (or pion) mass.
	In the nuclear case, for accurate extrapolation one must also take into
	account the crucial
	non-analytic structure associated with chiral symmetry breaking
	\cite{Detmold:2001jb}.

	Recent calculations have been able to use the physical pion
	mass though only for very light hadrons \cite{Durr:2010aw}.  In near future,
	we hope to extract the low-energy constants in Fig.~\ref{fig:ch_EFT_diagrams}
	from lattice QCD.

	\medskip
	\subsubsection{Lattice EFT}

	Quarks and gluons have many degrees of freedom in terms of spin, color
	charge, and flavor.  This along with strong nonlinearity and
	non-perturbativeness of the problem makes getting nuclear physics from
	lattice QCD
	computationally difficult.  An alternative approach is to have nucleons
	on the lattice site rather than the quarks.
	\begin{figure}[htbp]
		\centering
		\begin{subfigure}[c]{0.42\textwidth}
			\centering
			\includegraphics[width=\textwidth]
			{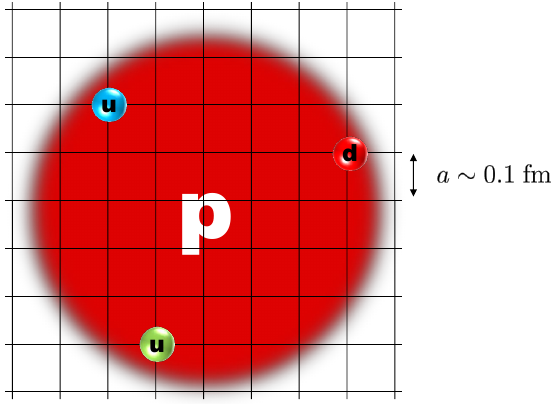}
			\caption{Lattice QCD~~~~~~~~~~~~}
			\label{fig:lattice_QCD}
		\end{subfigure}
		\hspace{0.1\textwidth}
		\begin{subfigure}[c]{0.42\textwidth}
			\centering
			\includegraphics[width=\textwidth]
			{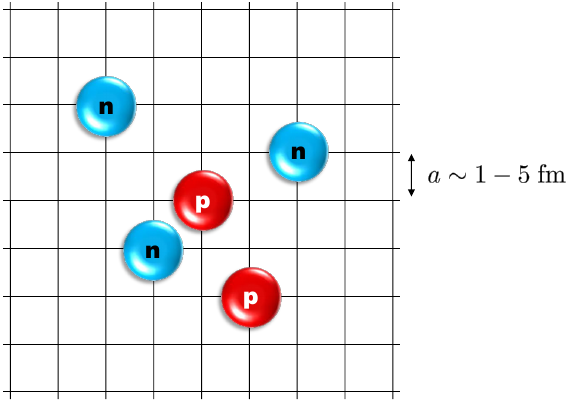}
			\caption{Lattice EFT~~~~~~~~~~~~}
			\label{fig:lattice_EFT}
		\end{subfigure}
		\caption{Lattice QCD vs.\ lattice EFT.  Figures from a talk by Dean Lee.}
		\label{fig:lattice_QCD_EFT}
	\end{figure}
	The difference between the two methods is illustrated in
	Fig.~\ref{fig:lattice_QCD_EFT}.  Lattice EFT combines the framework of
	effective field theory and computational lattice methods and is a
	promising tool for studying light nuclear systems.

	Lattice theories inherently come with the associated graininess and the
	finite size.  Thus, we have a cutoff for both the maximum length and the
	maximum momentum scale that we can have on a lattice.  This is qualitatively
	similar	to the harmonic oscillator basis truncation problem that we will
	look extensively at in Chapter~\ref{chap:Extrapolation}.  Various methods
	have been used to obtain the continuum limit from the lattice.  We will touch
	upon these in
	Subsec.~\ref{subsec:IR_front}, where we look at similarities and differences
	between the lattice methods and our work with oscillator basis truncation.

	\section{RG techniques}

	We saw in Fig.~\ref{fig:Nmax_convergence} that the convergence in many-body
	calculations is slow.  To understand why this is the case, recall from
	Fig.~\ref{fig:Nuclear_potentials} that the nuclear potentials have a strong
	short-range repulsion.  This hard core leads to high-momentum components in
	the potentials.
	\begin{figure}[htbp]
	 \centering
	 \includegraphics[width=0.45\textwidth]%
	 {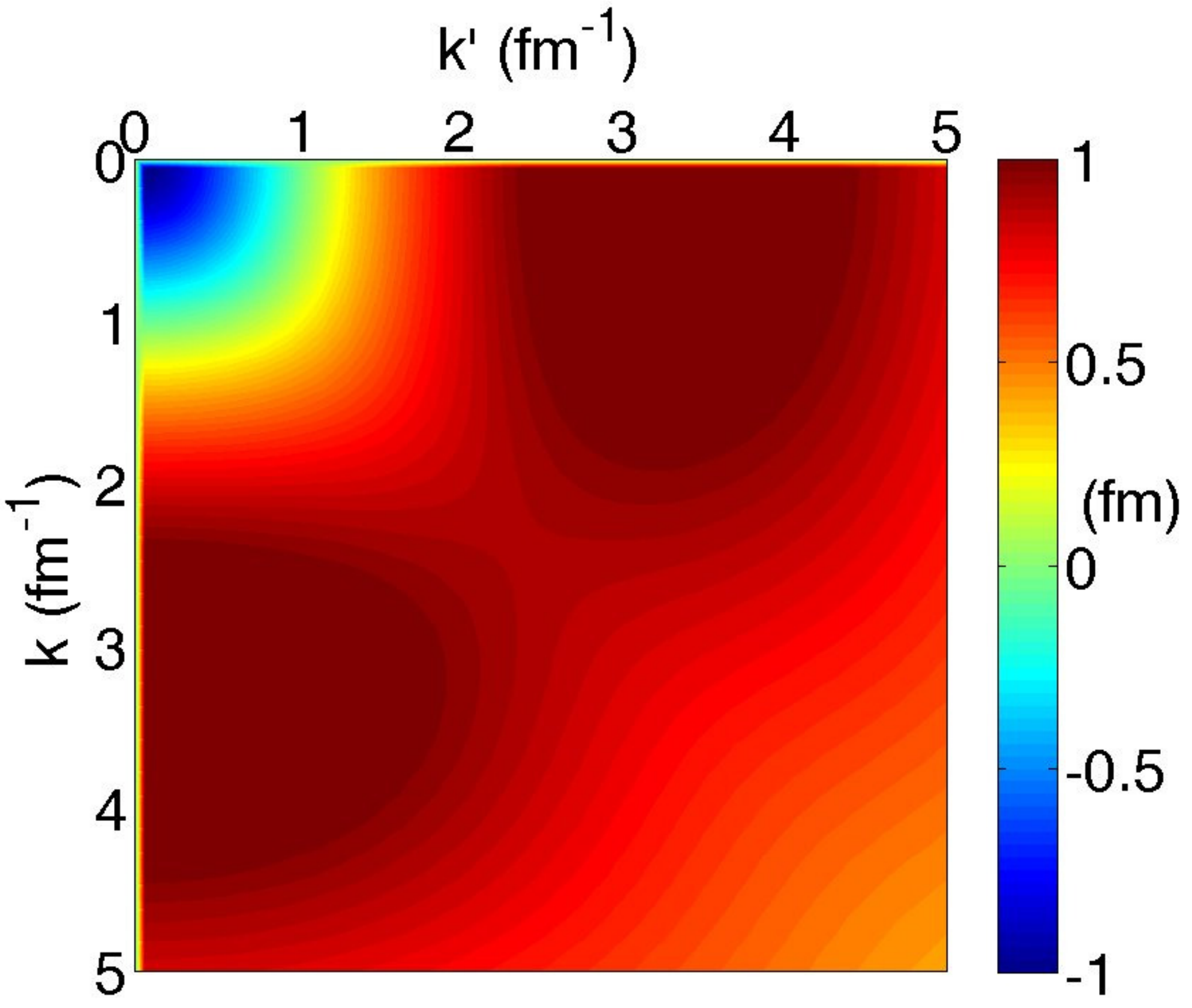}
	 \caption{The $^1 S_0$ AV18 potential in momentum space.  Figure from
	 \cite{Furnstahl:2013oba}.}
	 \label{fig:momentum_space_AV18}
	\end{figure}
	This can be seen in Fig.~\ref{fig:momentum_space_AV18} where we plot the
	AV18 potential in the $^1 S_0$ channel in momentum space.  We note that
	for the AV18 potential, $V(k, k^{\prime}) \to 0$ only for $k, \, k^\prime
	\gtrsim 25 {\rm{~fm}}^{-1}$.  However, the Fermi momentum of the nucleon
	in a heavy nucleus like $^{208}$Pb is only about $1.2 {\rm{~fm^{-1}}}$.
	The Fermi momentum sets the momentum scale of low-energy nuclear
	processes we wish to study.  Thus, we have a mismatch of resolution scales;
	the processes we wish to describe are $1 - 2 {\rm{~fm^{-1}}}$
	($200 - 400 {\rm{~MeV}}$), whereas the momentum scale in the potential
	is much higher.   To use the analogy due to Tom Banks, `it is like trying to
	understand the properties of waves in the ocean in terms of Feynmann
	diagrams'.  Though in principle this can be done, it makes
	calculations intractably complicated.

	This bring us back to Fig.~\ref{fig:degrees_of_freedom}.  The progression
	from top to bottom in Fig.~\ref{fig:degrees_of_freedom} can be thought of as
	reduction in resolution.  This can be established theoretically using
	renormalization group (RG) methods.  As mentioned before, the focus in LENP
	is the intermediate region, where nucleons are the degrees of freedom.
	But even within this limited region, the concept of changing resolution
	by RG methods has been extremely advantageous \cite{Furnstahl:2013oba}.

	\begin{figure}[htbp]
		\centering
		\begin{subfigure}[c]{0.55\textwidth}
			\centering
			\includegraphics[width=\textwidth]
			{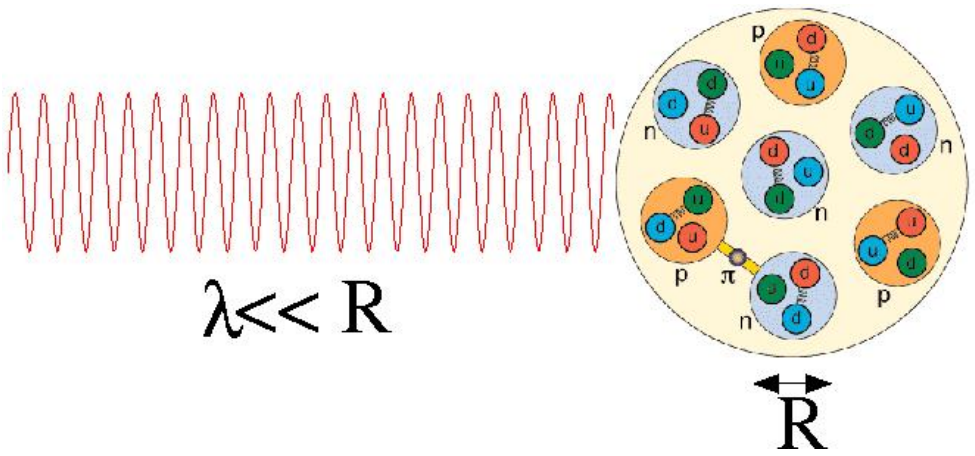}
			\caption{Nucleus under a high-resolution probe.}
			\label{fig:nucleus_high_resolution}
		\end{subfigure}
		\vskip 0.6cm
		\begin{subfigure}[c]{0.55\textwidth}
			\centering
			\includegraphics[width=\textwidth]
			{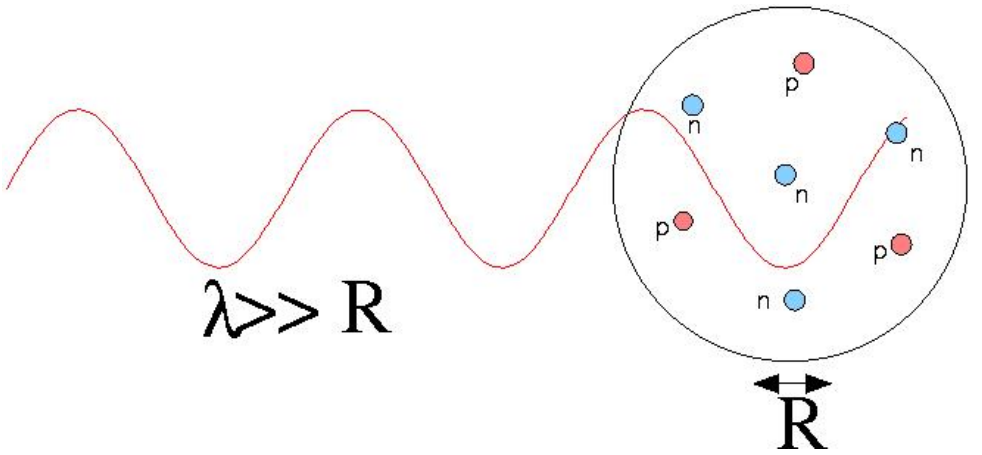}
			\caption{Fine details (nucleon substructure) not resolved when
			probed at low energies.}
			\label{fig:nucleus_low_resolution}
		\end{subfigure}
		\vskip 0.6cm
		\begin{subfigure}[c]{0.7\textwidth}
			\centering
			\includegraphics[width=\textwidth]
			{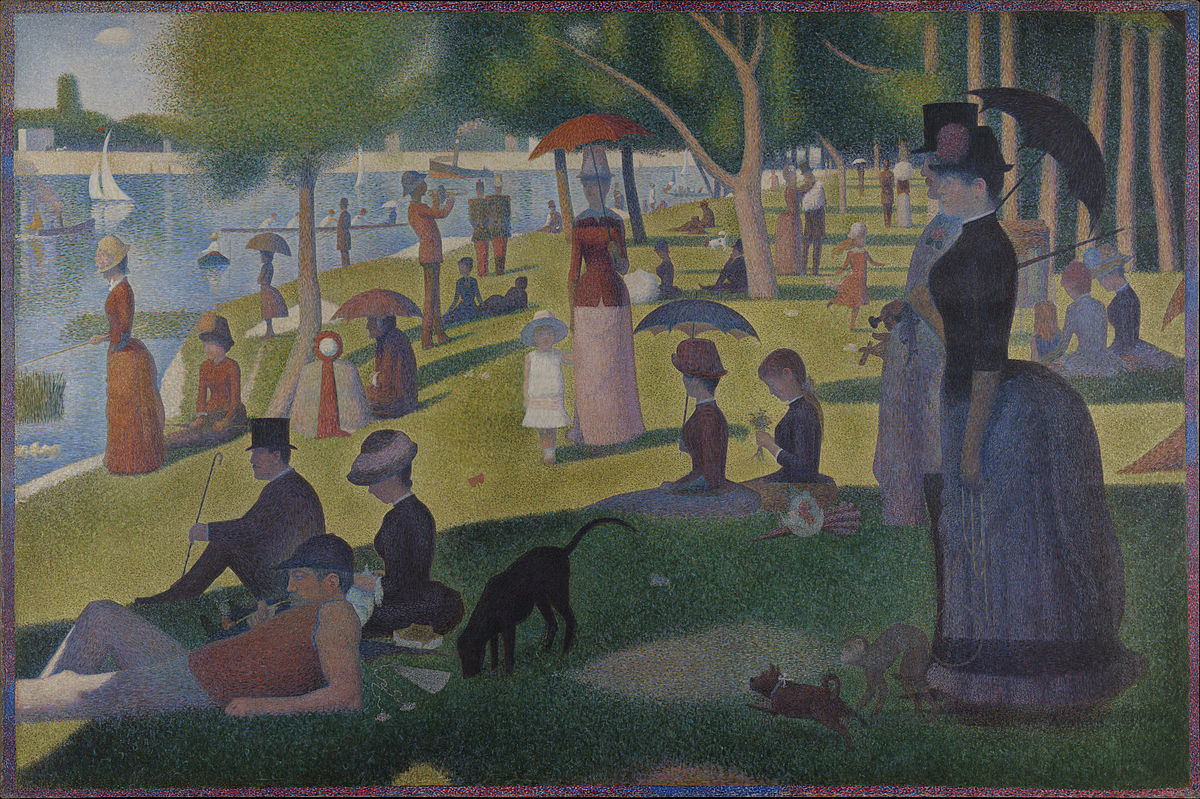}
			\caption{A painting by Georges Seurat, which is an example of
			pointillism.  Small, distinct dots of color are applied to form a
			pattern.  The pattern would be lost under a high-resolution probe.
			Image from Wikimedia Commons.}
			\label{fig:Georges_Seurat}
		\end{subfigure}
		\caption{Physics interpretation often changes with resolution.}
		\label{fig:physics_interpretation_resolution}
	\end{figure}
	We would like to stress that contrary to the popular notion, high resolution
	is not always the best thing, especially when the processes we are looking at
	are low-momentum.  Also, though the value of the calculated observable is
	independent of the resolution, the physical interpretation often changes
	with the resolution.
	This is illustrated in Fig.~\ref{fig:physics_interpretation_resolution}.

	\begin{figure}[htbp]
		\centering
		\begin{subfigure}[t]{0.45\textwidth}
			\centering
			\includegraphics[width=\textwidth]
			{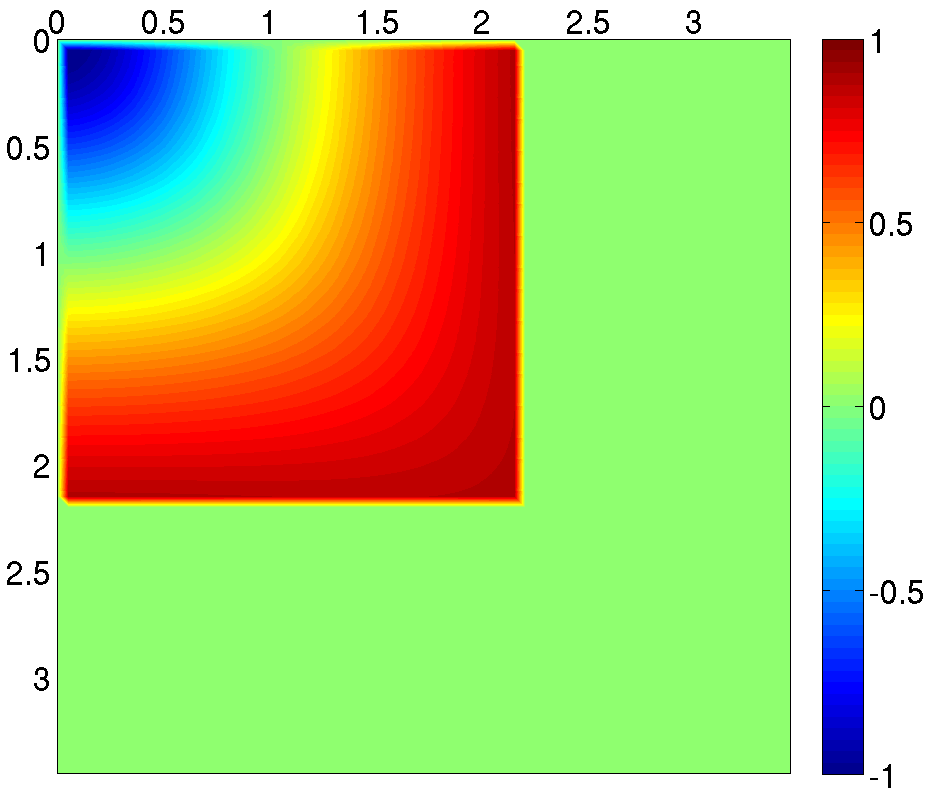}
			\caption{An illustration of a low-pass filter.}
			\label{fig:low_pass_filter_example}
		\end{subfigure}
		\hspace{0.07 \textwidth}
		\begin{subfigure}[t]{0.45\textwidth}
			\centering
			\includegraphics[width=\textwidth]
			{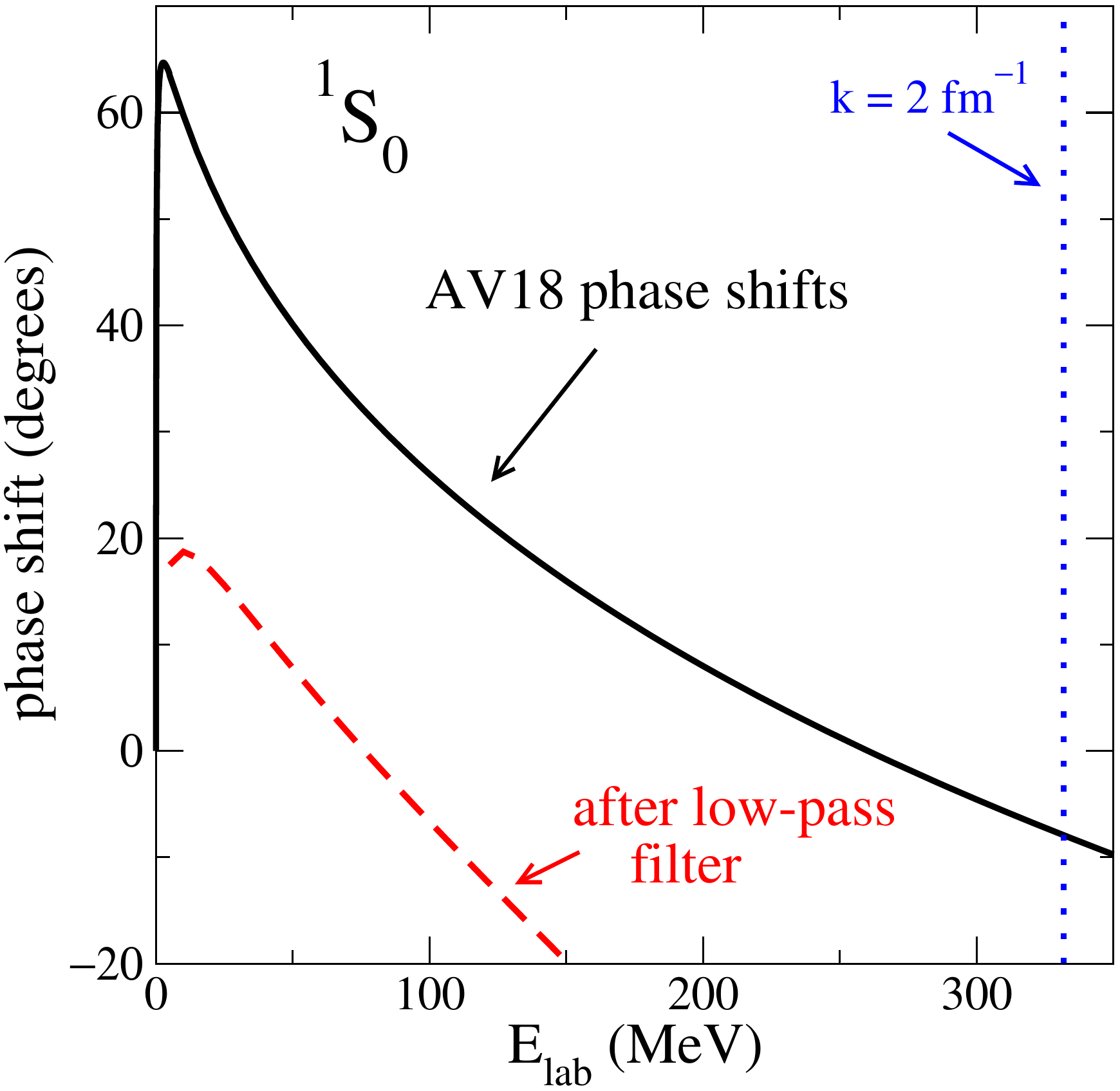}
			\caption{$^{1}S_0$ phase shifts for the AV18 potential and for the AV18
			potential after the low-pass filter which sets $V(k, k^\prime) = 0
			{\rm{~for~}} k, \, k^\prime > 2 {\rm{~fm^{-1}}}$ \cite{Furnstahl:2012fn}.}
			\label{fig:low_pass_filter_effect}
		\end{subfigure}
		\caption{Low-pass filter on nuclear potential---illustration and effect on
			phase shifts.}
		\label{fig:low_pass_filter_example_effect}
	\end{figure}
	One of the methods to get rid of the high-momentum components is to apply
	a `low-pass filter' on the potential
	(cf.~Fig.~\ref{fig:low_pass_filter_example}).  This is routinely done, for
	example, in image processing.  Compression of a digital photograph is
	achieved by Fourier transforming it, setting the high-momentum modes in
	the Fourier transform equal to zero, and then transforming back.
	However, as seen in Fig.~\ref{fig:low_pass_filter_example}, the
	straightforward application of a low-pass filter fails to reproduce
	nuclear phase shifts even at low energies.

	This failure of a low-pass filter can be understood by recalling from
	Fig.~\ref{fig:momentum_space_AV18} that the high and low-momentum
	modes are coupled.  For instance, consider (schematically) the expression
	for the tangent of phase shift in perturbation theory
	\beq
	\mbraket{k}{\wh{V}}{k} + \sum_{k^\prime}
	\dfrac{\mbraket{k}{\wh{V}}{k^\prime} \mbraket{k^\prime}{\wh{V}}{k}}
	{(k^2 - {k^\prime}^2)/m} + \cdots \;.
	\label{eq:phase_shift}
	\eeq
	The second term in Eq.~\eqref{eq:phase_shift} involves a sum over off-diagonal
	matrix elements of $\wh{V}$.  Therefore even phase shifts for small $k$
	will have significant contributions from high $k^\prime$ if the coupling
	matrix elements $\mbraket{k}{\wh{V}}{k^\prime}$ are large.

	A correct way to get rid of the high-momentum components is to use the
	RG evolution and lower the momentum cutoff in small steps.
	\begin{figure}[htbp]
		\centering
		\begin{subfigure}[t]{0.4\textwidth}
			\centering
			\includegraphics[width=\textwidth]
			{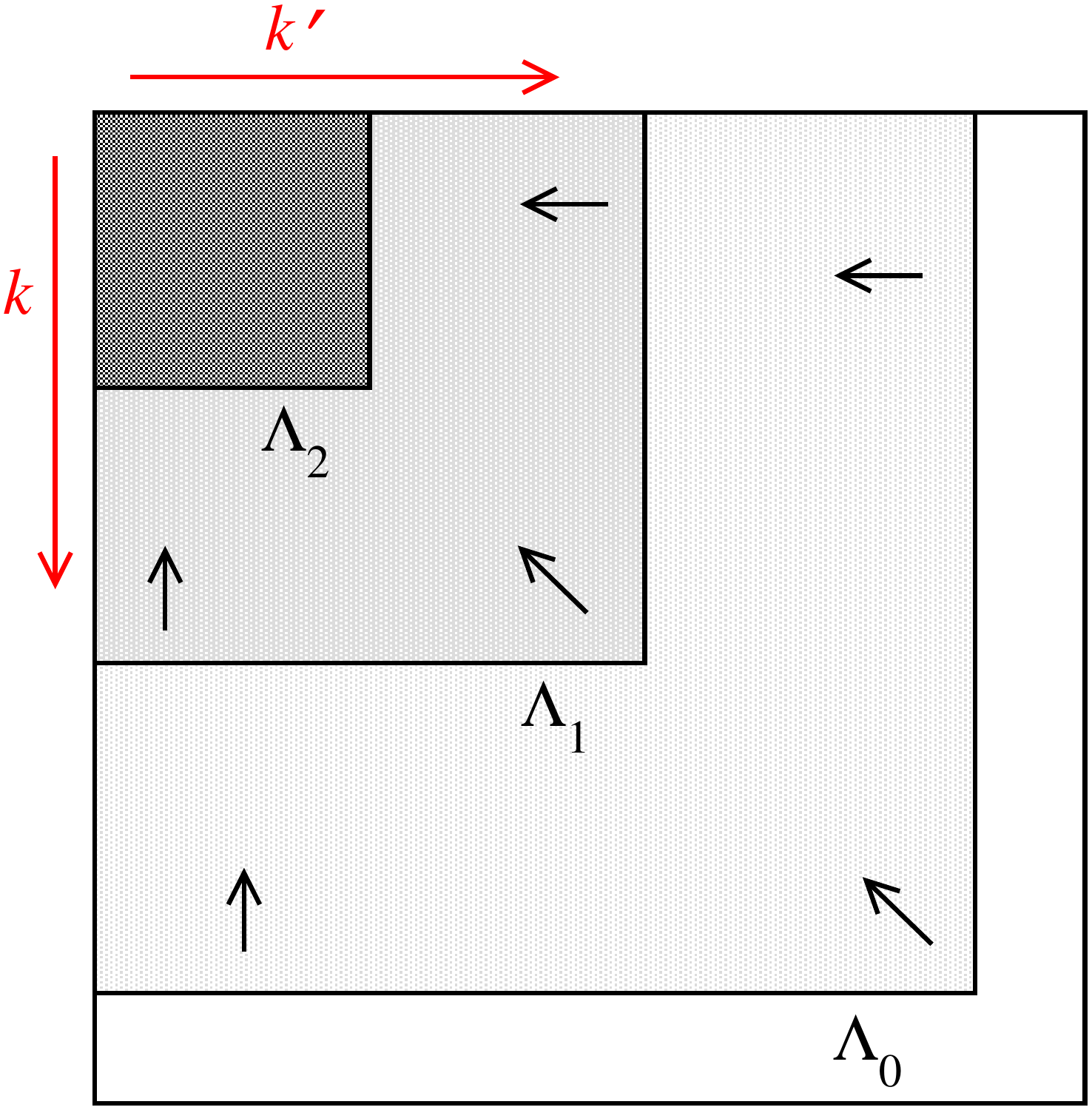}
			\caption{The $V_{{\rm{low~}}k}$ running in $\Lambda$.}
			\label{fig:V_lowk_running}
		\end{subfigure}
		\hspace{0.1 \textwidth}
		\begin{subfigure}[t]{0.4\textwidth}
			\centering
			\includegraphics[width=\textwidth]
			{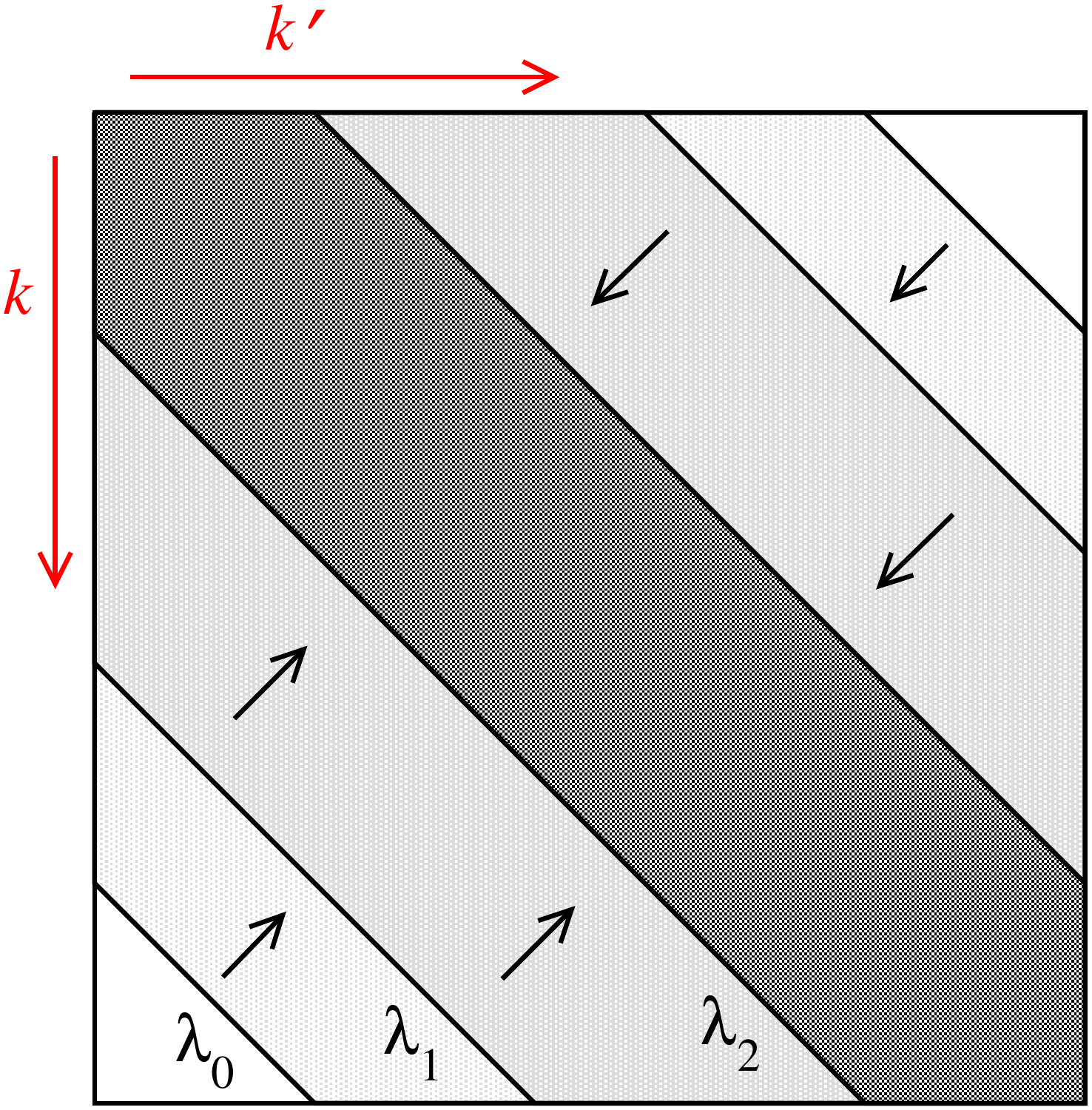}
			\caption{The SRG running in $\lambda$.}
			\label{fig:V_srg_running}
		\end{subfigure}
		\caption{Schematic illustration of two types of RG evolution.  For given
		$\Lambda_i$ or $\lambda_i$ the matrix elements outside the corresponding
		lines are negligible.  This decouples the high-momentum modes from the
		low-momentum ones.  Figure from \cite{Furnstahl:2013oba}.}
		\label{fig:V_lowk_V_srg_running}
	\end{figure}
	Two common choices for RG running are shown in
	Fig.~\ref{fig:V_lowk_V_srg_running}.  The RG running shown in
	Fig.~\ref{fig:V_lowk_running} is referred to as ``$V_{{\rm{low~}} k}$'' was
	historically developed first for LENP.  It attempts
	to get the potential in the low-pass-filter form through successive RG
	transformations \cite{Bogner:2009bt, Bogner:2003wn}.  Though successful
	for two-nucleon forces, it has been difficult to systematically treat
	many-body forces in the $V_{{\rm{low~}} k}$ framework.

 	A more recent approach through the similarity renormalization group (SRG) is
	illustrated in Fig.~\ref{fig:V_srg_running}.  SRG running drives the
	potential to a band-diagonal form.  It is possible to systematically account
	for the many-body forces in the SRG framework.  We will be using
	the SRG framework in our work presented in Chapter~\ref{chap:factorization}.
	Next we present a brief introduction to the SRG technique.

	\subsection{SRG}
	\label{subsec:SRG_intro}

	SRG for LENP was developed at Ohio State by S.~Bogner (then a post-doc),
	R.~Furnstahl, and R.~Perry \cite{Bogner:2006pc}.
	Their approach was inspired by the RG flows equations developed by
	Wegner for condensed matter applications \cite{Wegner:2000gi}. 
	The basic idea in SRG
	is to apply a series of unitary transformations $U_s$
	to transform the Hamiltonian $\wh{H}$ into a band-diagonal form shown in
	Fig.~\ref{fig:V_srg_running}:
	\beq
	\wh{H}_s = \wh{U}_s \,\wh{H}\, \wh{U}^\dag_s \;,
	\label{eq:SRG_H_transformation}
	\eeq
	where $s$ is the RG flow parameter.  $U_{s = 0} = \mathds{1}$.  In practice,
	instead of using Eq.~\eqref{eq:SRG_H_transformation}, the SRG evolution
	is done through the flow equation
	\beq
	\dfrac{d \wh{H}_s}{ds} = \left[[\wh{G}_s, \, \wh{H}_s], \, \wh{H}_s\right] \;.
	\label{eq:flow_eq_operator_form}
	\eeq
	$\wh{G}_s$ is the operator which generates the flow.  $\wh{T}_{\rm rel}$,
	the relative kinetic energy operator, is the most popular choice for
	$\wh{G}_s$, though other choices for $\wh{G}_s$ have been explored
	\cite{Li:2011sr}.

	Note that since we are just doing unitary transformations, eigenvalues such
	as energies are unchanged under SRG
	\beq
	E_n = \mbraket{\Psi_n}{\wh{H}}{\Psi_n}
	= {(} \langle \Psi_n | \wh{U}^\dagger_s {)} \wh{U}_s H \wh{U}^\dagger_s
     {(} \wh{U}_s | \Psi_n \rangle {)}\;.
	\eeq
	It is beneficial to change the flow parameter to $\lambda$, where
	$\lambda ^2 = 1/ \sqrt{s}$.  $\lambda$ has the units of momentum
	($\rm{fm^{-1}}$).  With this change of variable and setting
	$\wh{G}_s = \wh{T}_{\rm rel}$, for a given partial-wave channel,
	the Eq.~\eqref{eq:flow_eq_operator_form} in momentum basis becomes
	\beq
	\dfrac{dV_\lambda}{d\lambda}(k,k') \propto
      - (\epsilon_k - \epsilon_{k'})^2 V_\lambda(k,k')
      +  \sum_q (\epsilon_k + \epsilon_{k'} - 2\epsilon_q)
        V_\lambda(k,q) V_\lambda(q,k') \;.
	\label{eq:SRG_equation_mom_space}
	\eeq
	The first term on the right side of Eq.~\eqref{eq:SRG_equation_mom_space}
	drives the potential to the band diagonal form shown in
	Fig.~\ref{fig:V_srg_running} and the second term on right side of
	Eq.~\eqref{eq:SRG_equation_mom_space} makes sure that the unitarity
	is maintained.

	\begin{figure}[htbp]
	 \centering
	 \includegraphics[width=0.9\textwidth]%
	 {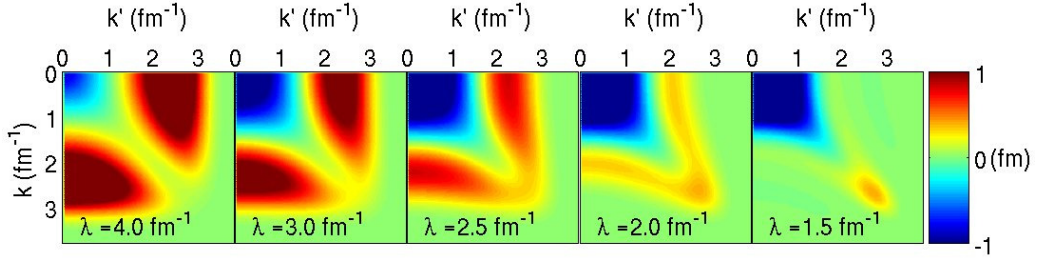}
	 \caption{SRG evolution of the chiral $\rm{N^3LO}$ potential
	 \cite{Epelbaum:2004fk} in the
	 $^3S_1$ channel.  Figure from \cite{Bogner:2009bt}.}
	 \label{fig:SRG_evolution_chiral_NN}
	\end{figure}
	Figure~\ref{fig:SRG_evolution_chiral_NN} shows
	Eq.~\eqref{eq:SRG_equation_mom_space} in action.  We see that as we
	go to lower $\lambda$, we achieve the decoupling between high- and
	low-momentum components.  Once this decoupling has been achieved, one can
	apply the low-pass filter and work with smaller matrices if desired.

	\medskip
	\subsubsection{Effects of SRG evolution}

	\begin{figure}[htbp]
	 \centering
	 \includegraphics[width=0.5\textwidth]%
	 {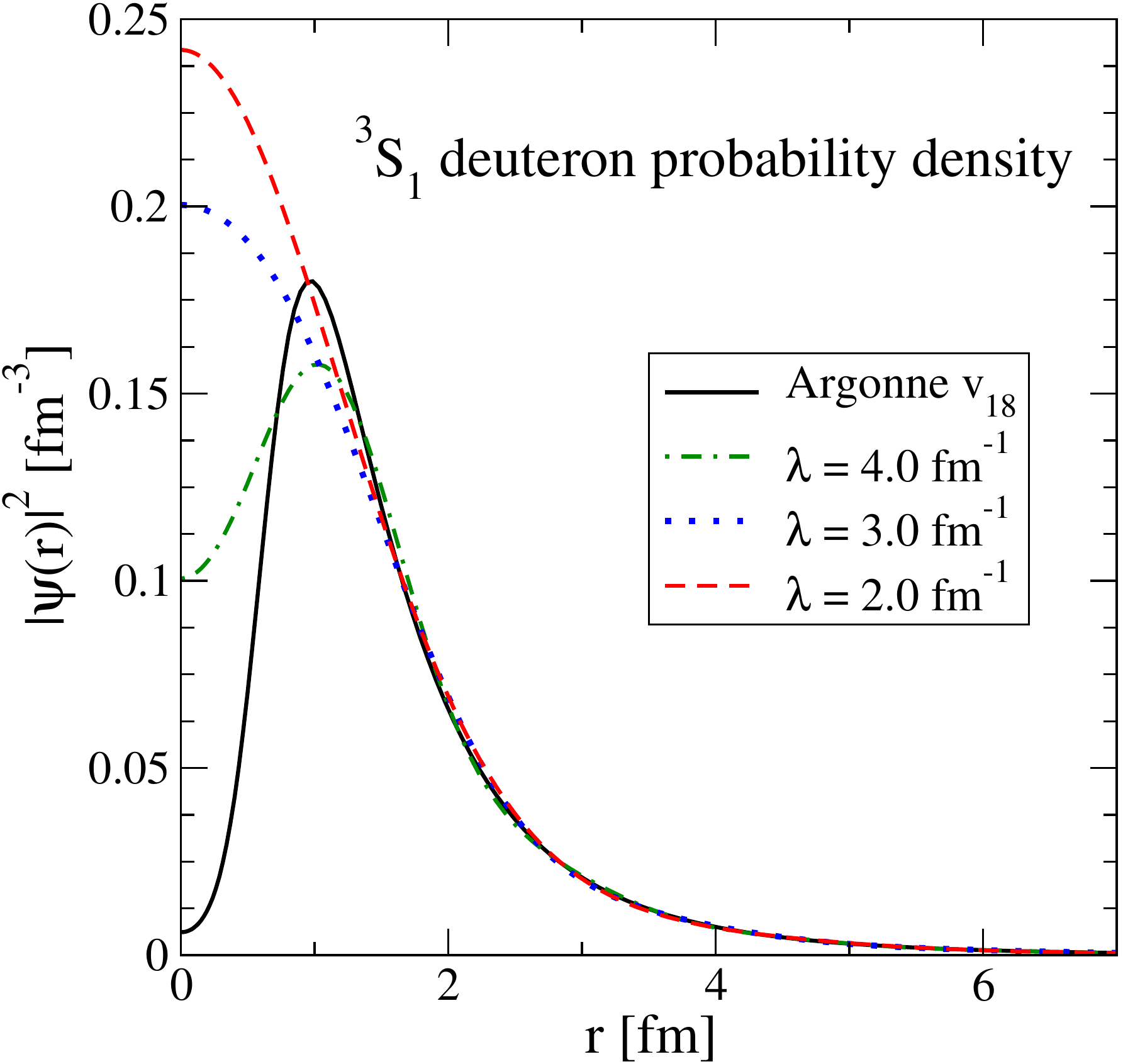}
	 \caption{$^3S_1$ probability density in deuteron for the AV18 potential and
	 the AV18 potential evolved to three SRG $\lambda$'s.  Figure from
	 \cite{Furnstahl:2012fn}.}
	 \label{fig:SRG_evolution_wf}
	\end{figure}
	The conventional nuclear potentials have a strong short-range repulsion.  As
	a result, the wave functions at short distance are suppressed.  This
	suppression is seen for the unevolved potential in
	Fig.~\ref{fig:SRG_evolution_wf}.  This suppression is called the
	short-range correlation (SRC).  The RG evolution gets rid of the
	short-range repulsion in the potential (it shifts the strength from
	high-momentum modes into the low-momentum modes).  The wave functions from
	the evolved potential therefore do not exhibit the SRC as seen in
	Fig.~\ref{fig:SRG_evolution_wf}.  It is also noteworthy that even though
	the deuteron wave functions change at short distances, the long-distance
	part is independent of the SRG $\lambda$.  The long-distance part of the
	wave function is related to the asymptotic normalization coefficient (ANC),
	which is an observable.  We will see later in
	Chapter~\ref{chap:Extrapolation} that the ANC being an observable plays an
	important role in establishing the universality of the extrapolation formulas
	derived in Chapter~\ref{chap:Extrapolation}.  The observables should be
	independent of the resolution scale, which at the end of the day is a
	theoretical choice.

	\begin{figure}[htbp]
		\centering
		\begin{subfigure}[t]{0.4\textwidth}
			\centering
			\includegraphics[width=\textwidth]
			{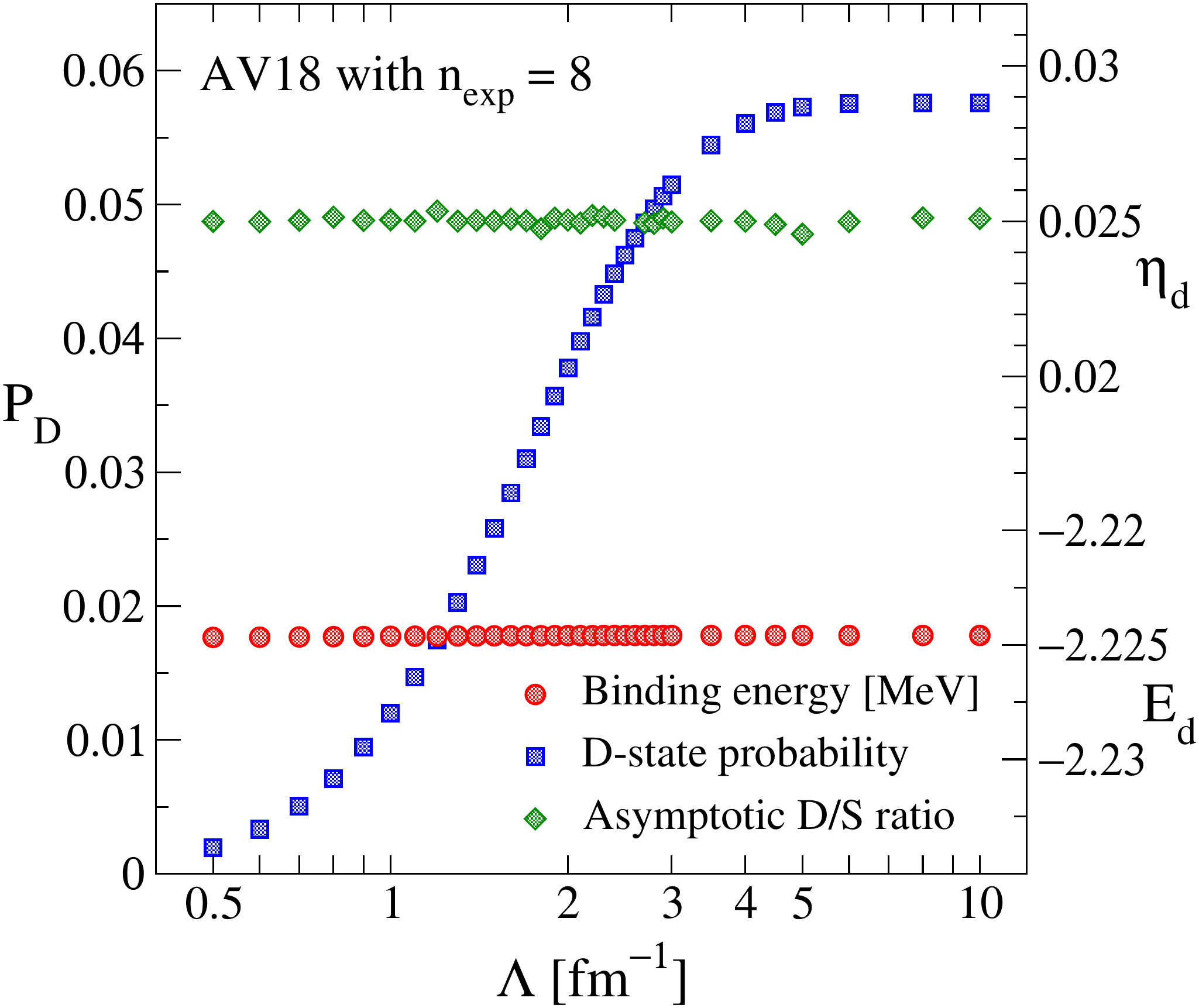}
			\caption{Effect of evolution on binding energy, the $D$-state
			probability, and the asymptotic $D/S$-state ratio $\eta_d$.  Figure from
			\cite{Bogner:2009bt}. }
			\label{fig:energy_prob_eta_evolv_effect}
		\end{subfigure}
		\hspace{0.1 \textwidth}
		\begin{subfigure}[t]{0.4\textwidth}
			\centering
			\includegraphics[width=0.9\textwidth]
			{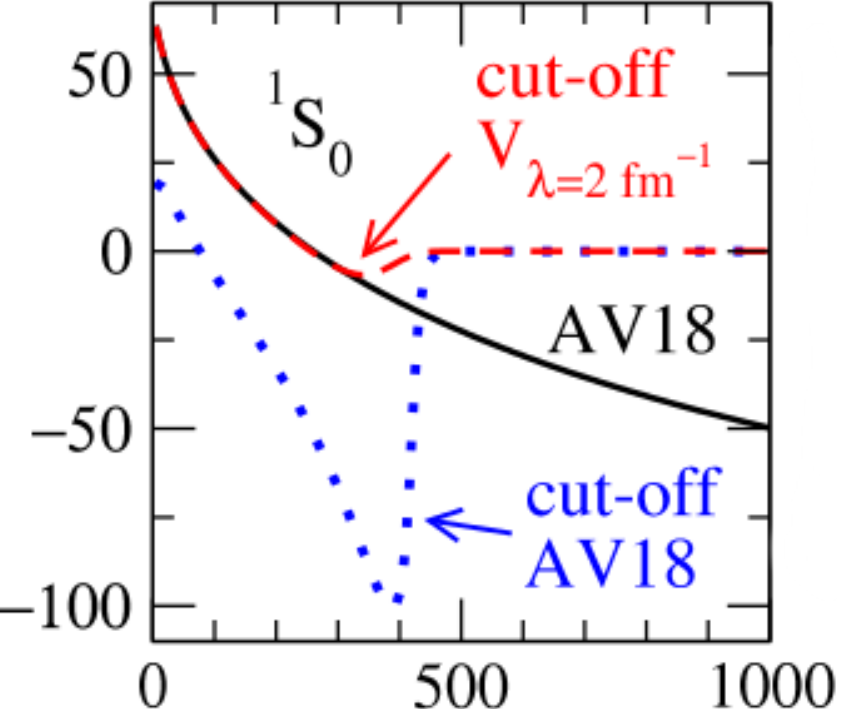}
			\caption{The solid
			black line indicate the phase shifts for the unevolved potential. The
			dotted blue line and the dashed red line are the phase shifts when the
			low-pass filter is applied
			to the unevolved potential and the evolved potential respectively.}
			\label{fig:phase_shift_SRG_evolv}
		\end{subfigure}
		\caption{Effect of RG evolution on (non-)observables.}
		\label{fig:SRG_effect}
	\end{figure}
	In Fig.~\ref{fig:SRG_effect}, we look at the effect of evolution on a few
	more (non-)observables.  We see in
	Fig.~\ref{fig:energy_prob_eta_evolv_effect} that the deuteron binding energy
	and the ratio of $D$-state ANC to the $S$-state ANC are independent of the
	resolution scale as expected.  However, the $D$-state probability depends
	on the resolution.  This indicates that the $D$-state probability of
	the deuteron is not an observable.  We return to the phase shifts with a
	low-pass filter problem in Fig.~\ref{fig:phase_shift_SRG_evolv}.
	We see that when the low-pass filter is applied to the SRG evolved potential,
	it reproduces the unevolved phase shifts up to the cut-off energy of the
	filter.  This discussion also makes it clear that the
	nuclear potential itself is not an observable and there is no ``correct''
	nuclear potential.  In fact, there is more than one ``correct'' potential.
	For instance, all the potentials in Fig.~\ref{fig:SRG_evolution_chiral_NN}
	give the same values for observables such as binding energies, phase shifts,
	etc.

	\begin{figure}[htbp]
		\centering
		\begin{subfigure}[t]{0.4\textwidth}
			\centering
			\includegraphics[width=\textwidth]
			{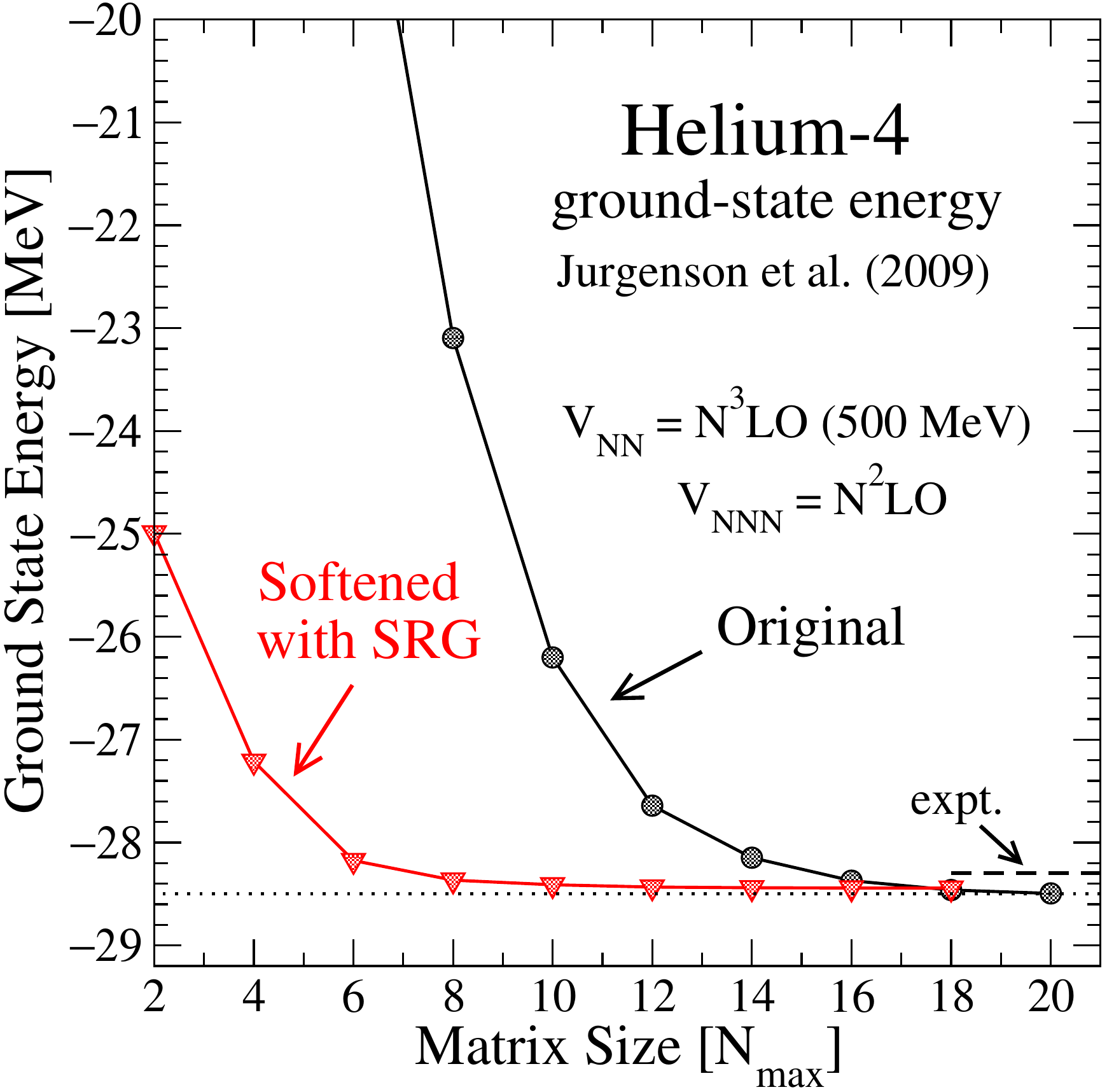}
		\end{subfigure}
		\hspace{0.1 \textwidth}
		\begin{subfigure}[t]{0.4\textwidth}
			\centering
			\includegraphics[width=\textwidth]
			{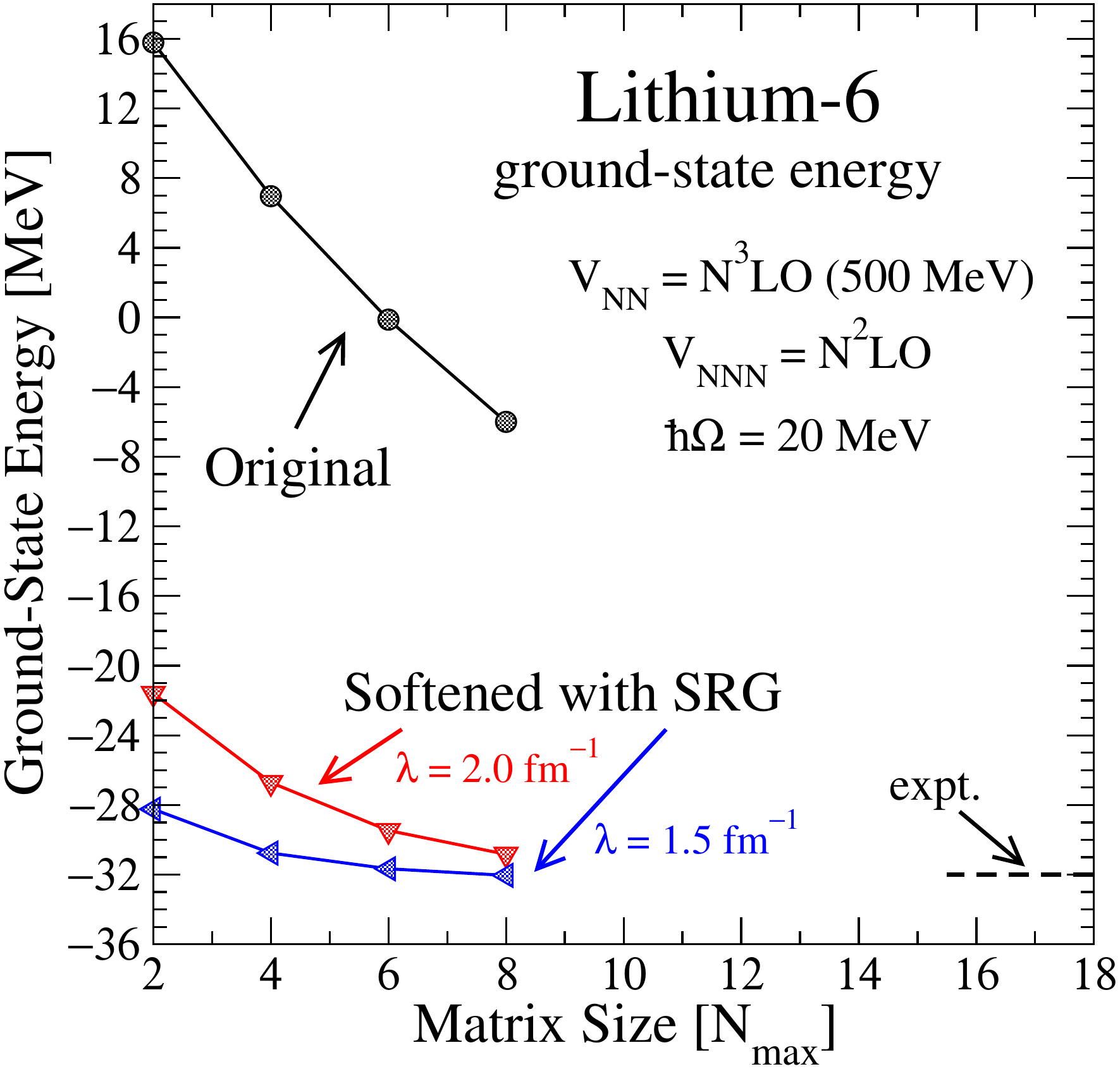}
		\end{subfigure}
		\caption{SRG evolution greatly accelerates the convergence in many-body
		calculations.  Figure from \cite{Furnstahl:2012fn}.}
		\label{fig:SRG_Nmax_convergence_He4_Li6}
	\end{figure}
	Since the SRG evolved potentials only have low-momentum modes, they are
	amenable to numerical calculations.  As seen in
	Fig.~\ref{fig:SRG_Nmax_convergence_He4_Li6}, SRG evolution greatly
	accelerates the convergence in many-body calculations.

	We note that even when we start with just two-body forces, SRG evolution
	introduces three- and higher-body forces.  This can be understood from the
	flow equation (Eq.~\eqref{eq:flow_eq_operator_form})
	\bea
	\dfrac{d \wh{H}_s}{ds} &=& \left[ [\sum \underbrace{a^\dag a}_{\wh{G}_s},
	\sum \underbrace{a^\dag a^\dag a a}_{{\rm 2-body}} ],
	\sum \underbrace{a^\dag a^\dag a a}_{{\rm 2-body}}  \right] \nonumber \\
	&=& \cdots + \sum \underbrace{a^\dag a^\dag a^\dag a a a}_{{\rm 3-body}}
	+ \cdots
	\label{eq:induced_forces_demonstration}
	\eea
	The second equality in Eq.~\eqref{eq:induced_forces_demonstration}
	demonstrates how the commutators give rise to three- and higher-body
	(up to $A$-body) forces.
	The initial
	potential in Fig.~\ref{fig:SRG_Nmax_convergence_He4_Li6}
	includes both two- and three-body forces; it has been demonstrated that
	3-body forces are crucial in getting the
	correct experimental values from theory \cite{Hammer:2012id}.
	To keep the invariance of energy with respect to the resolution scale
	(as in Fig.~\ref{fig:energy_prob_eta_evolv_effect}), it is important
	to keep also the induced 3-body forces \cite{Jurgenson:2010wy}.
	A major development
	in the SRG technology has been the ability to consistently evolve
	three-body forces \cite{Hebeler:2012pr, Wendt:2013bla}.

	A related important development is that of In-Medium SRG (IM-SRG)
	\cite{Hergert:2015awm}.  IM-SRG uses a reference state $\ket{\Phi}$ which is
	different from the particle vacuum $\ket{0}$ used in SRG.
	For example, $\Phi$ can be a Slater determinant that is fair
	approximation to nucleus' ground state.  Just like in SRG,
	IM-SRG then uses a series of unitary transformations to decouple the
	reference state from excitations.  IM-SRG also maintains the hierarchy of
	many-body forces, namely ${\rm 2N} \gg {\rm 3N} \gg {\rm 4 N} \cdots$.

	RG techniques have made possible calculations of medium-mass nuclei starting
	from inter-nucleonic interactions.  However, if we keep pushing the
	calculations to higher-mass nuclei, we run into the same problem as
	indicated in Fig.~\ref{fig:Li6_vs_Nmax}, i.e, we run out of
	computational power before we reach convergence.
	So, along with RG techniques we also need reliable extrapolation
	techniques that will allow us to extrapolate the results from
	finite $\Nmax$ to $\Nmax = \infty$.  This problem will
	form the basis of Chapter~\ref{chap:Extrapolation}.

	\section{Path forward for LENP}

	We have seen that nuclear theory has come a long way from the pion theories
	of the $1940$'s.  The focus these days is on doing precision calculations and
	making reliable predictions.
	The tool box of a nuclear theorist includes a wide variety of
	techniques.  In most cases, different techniques have complimentary
	strengths.  In other cases, alternative methods provide a means to
	cross-check answers.
	Having made great strides in the evaluation
	of the bound state properties, the push recently has been on
	calculating resonant and scattering states, and using the improved
	understanding of nuclear structure to study nuclear reactions.

	To pick a particular example, let's focus on the neutrinoless double
	beta decay ($0 \nu \beta \beta$) example.  $0 \nu \beta \beta$ is of
	wide interest because of its potential to shed light on the nature of
	neutrinos (i.e, if neutrino is a Majorana or a Dirac fermion).
	$0 \nu \beta \beta$ has not been observed yet and there are experiments
	around the world looking for this decay \cite{Garfagnini:2014nla}.
	The experiments need guidance from theory to help design the
	experiment and to interpret the measurements.  The theoretical
	calculation of $0 \nu \beta \beta$
	cross section involves computing a nuclear matrix element for the
	transition.
	\begin{figure}[htbp]
	 \centering
	 \includegraphics[width=0.7\textwidth]%
	 {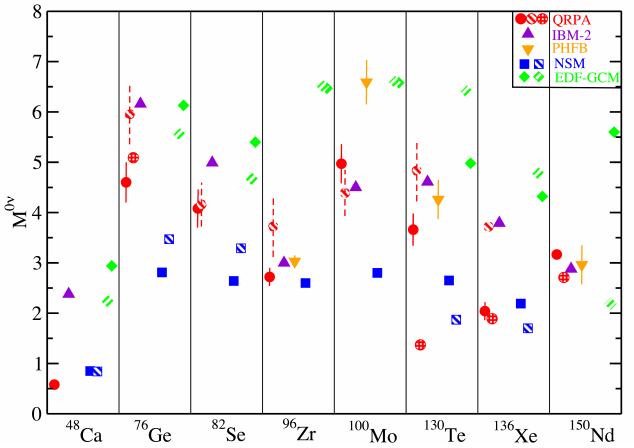}
	 \caption{Nuclear matrix element for $0 \nu \beta \beta$ for various nuclei
	 calculated using different methods.  The legends indicate the different
	 methods used for calculations.  Figure from a talk by Petr Vogel.}
	 \label{fig:0nu_double_beta_matrix_elements}
	\end{figure}
	Figure~\ref{fig:0nu_double_beta_matrix_elements} indicates the current level
	of agreement for calculations from different methods.  We see that these
	results differ by a factor of two.  The reaction cross section is proportional
	to the matrix element squared, and therefore the current level of
	agreement translates to an uncertainty of factor four in the cost of the
	experiment.

	We thus see that it is important that the theoretical calculations are
	accompanied by a reliable uncertainty estimate.  The sources of uncertainty in
	a theoretical calculation can broadly be classified into two
	categories---uncertainty in the input (e.g., shortcomings in the
	potential, limitations of the assumptions made, etc.) and
	uncertainty arising from the method (e.g., errors due to various
	approximations made while solving).  As presented in the next section,
	the work in this thesis focuses on investigating two common sources
	of uncertainties.

	\section{Thesis organization}

	The next two chapters in this thesis will present the author's
	original work as a PhD student.
	As demonstrated in Fig.~\ref{fig:Li6_vs_Nmax}, we often run out of
	computational power before the convergence is reached, necessitating a need
	for trustworthy extrapolation schemes.
	Chapter~\ref{chap:Extrapolation} describes our published work in this
	regard \cite{More:2013rma,Furnstahl:2013vda,Konig:2014hma}.
	The extrapolation schemes we developed were physically motivated as opposed
	to phenomenological forms previously in use.
	The author of this thesis was a lead author on Ref.~\cite{More:2013rma}.
	All authors of Ref.~\cite{Furnstahl:2013vda} contributed equally.
	We will therefore look in detail at the results presented in
	Refs.~\cite{More:2013rma, Furnstahl:2013vda} in this thesis.
	The author's contribution to work presented in Ref.~\cite{Konig:2014hma}
	was secondary, and hence only the summary from that work will be
	presented.

	Theoretical calculations of nuclear cross sections involve evaluating
	nuclear structure (which involves description of the initial and final state)
	and	nuclear reaction (which involves a description of the probe).  To make
	accurate predictions, it is important to understand the uncertainty
	stemming from the renormalization scale and scheme dependence of
	nuclear structure and reaction components.  We address this problem in
	Chapter~\ref{chap:factorization} by using the SRG to look at the scale
	dependence of deuteron electrodisintegration.  This effort was led by the
	author and has been published in Ref.~\cite{More:2015tpa}.

	Both these projects were the first of its kind, and are fertile grounds for
	further	development.  In fact, the work presented in
	Chapter~\ref{chap:Extrapolation} has already had a lot of impact as testified
	by the number of articles citing our publications.  We believe that the
	work in Chapter~\ref{chap:factorization} will also receive wide
	attention soon.  We present the details of our calculations in Appendix, which
	would enable anyone interested to reproduce our results.

\cleardoublepage
\chapter{Extrapolation}
\label{chap:Extrapolation}

	As we have seen in the introduction, the harmonic oscillator (HO) basis is
	routinely used in low-energy nuclear physics (LENP) calculations.  We also saw
	that
	the size of Hamiltonian matrix that we need to diagonalize grows factorially
	with the number of nucleons (cf.~Fig.~\ref{fig:matrix_dimension_growth}),
	severely restricting the number of terms that
	can be kept in the basis expansion.
	The single particle nuclear wave
	function with the $\Nmax$ truncation introduced in
	Chapter~\ref{chap:Intro} is given by
	\beq
	\psi_{N_{\rm max}}^{\Omega}(r) = \sum_{\alpha = 0}^{N_{\rm max}} c_{\alpha}
	\varphi_{\alpha}^{\Omega}(r) \;.
	\label{eq:single_particle_truncation}
	\eeq
	$\varphi_{\alpha}^{\Omega}(r)$ in Eq.~\eqref{eq:single_particle_truncation}
	are
	the HO wave functions;
	$\Omega$ is the frequency of the HO \footnote{
	In LENP, the oscillator frequency is often denoted by $\Omega$ rather
	than $\omega$.}.
	For reference, the $S$-wave HO wave function is given by
	\beq
	\varphi_{\alpha}^{\Omega}(r) =  \mathcal{N} \ee^{
		\frac{- \mu \Omega}{2 \hbar} r^2} L_{\alpha}^{1/2}
		(\frac{\mu \Omega}{\hbar} r^2) \;,
	\label{eq:HO_S_wave_written_out}
	\eeq
	where $\mathcal{N}$ is the normalization constant, $\mu$ is the
	reduced mass, and $L_{\alpha}^{1/2}$ denotes the generalized Laguerre
	polynomial.

\begin{figure}[h]
		\centering
		\includegraphics[width=0.55\textwidth]
		{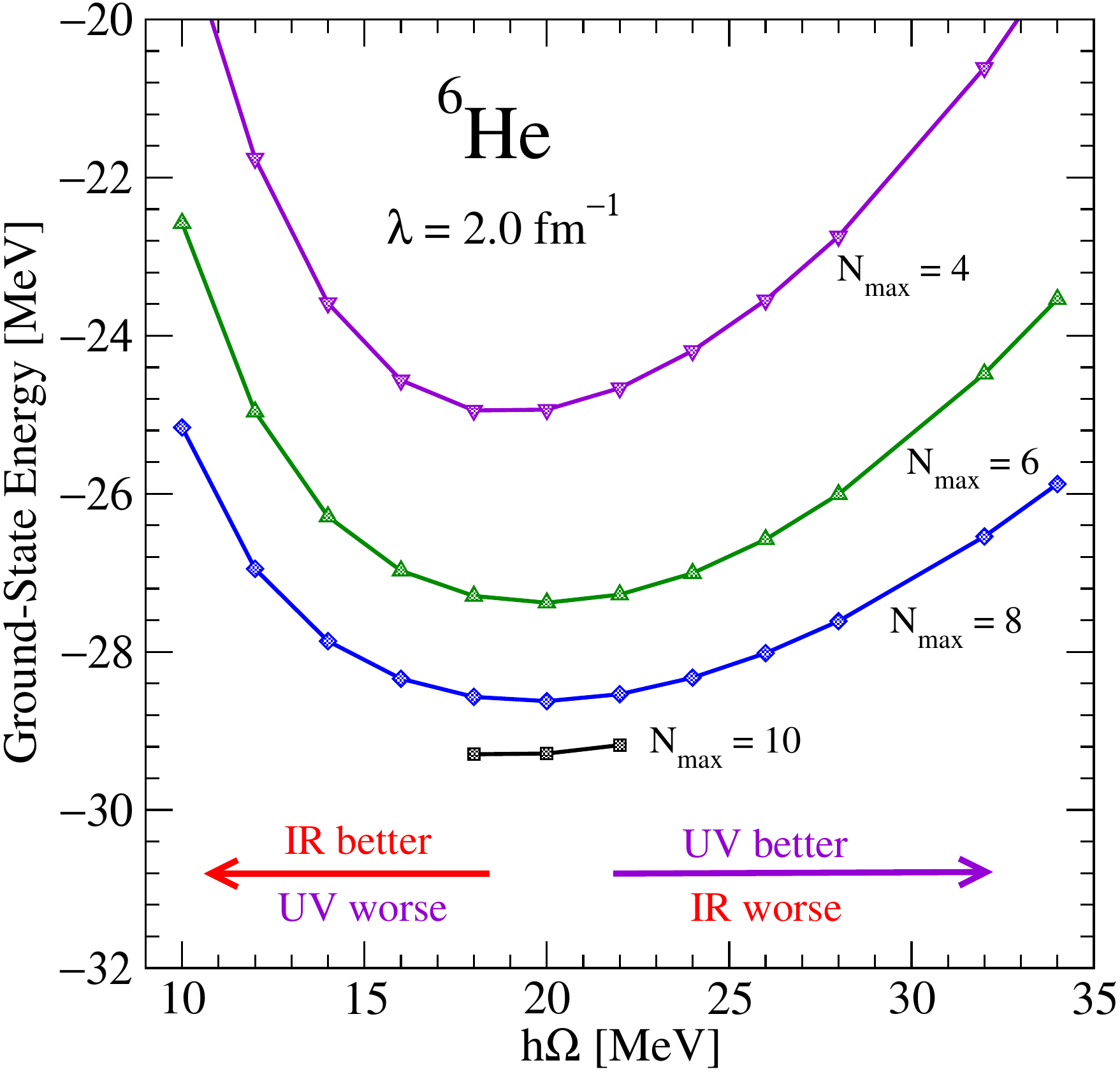}
		\caption{Ground state energy for $^6$He as a function of $\Nmax$ and
		  $\Omega$.  Figure taken from \cite{Furnstahl2012}. }
		\label{fig:H6_function_Omega}
	\end{figure}

	The energy obtained in the HO basis---$E(\Nmax, \Omega)$---is a
	function of $\Nmax$ and $\Omega$.  This is illustrated in
	Fig.~\ref{fig:H6_function_Omega}.
	We see that as we go to higher $\Nmax$, the curves get flatter with respect
	to $\Omega$, or in other words the dependence on $\Omega$ drops out.

\begin{figure}[h]
	\centering
	\includegraphics[width=0.55\textwidth]
	{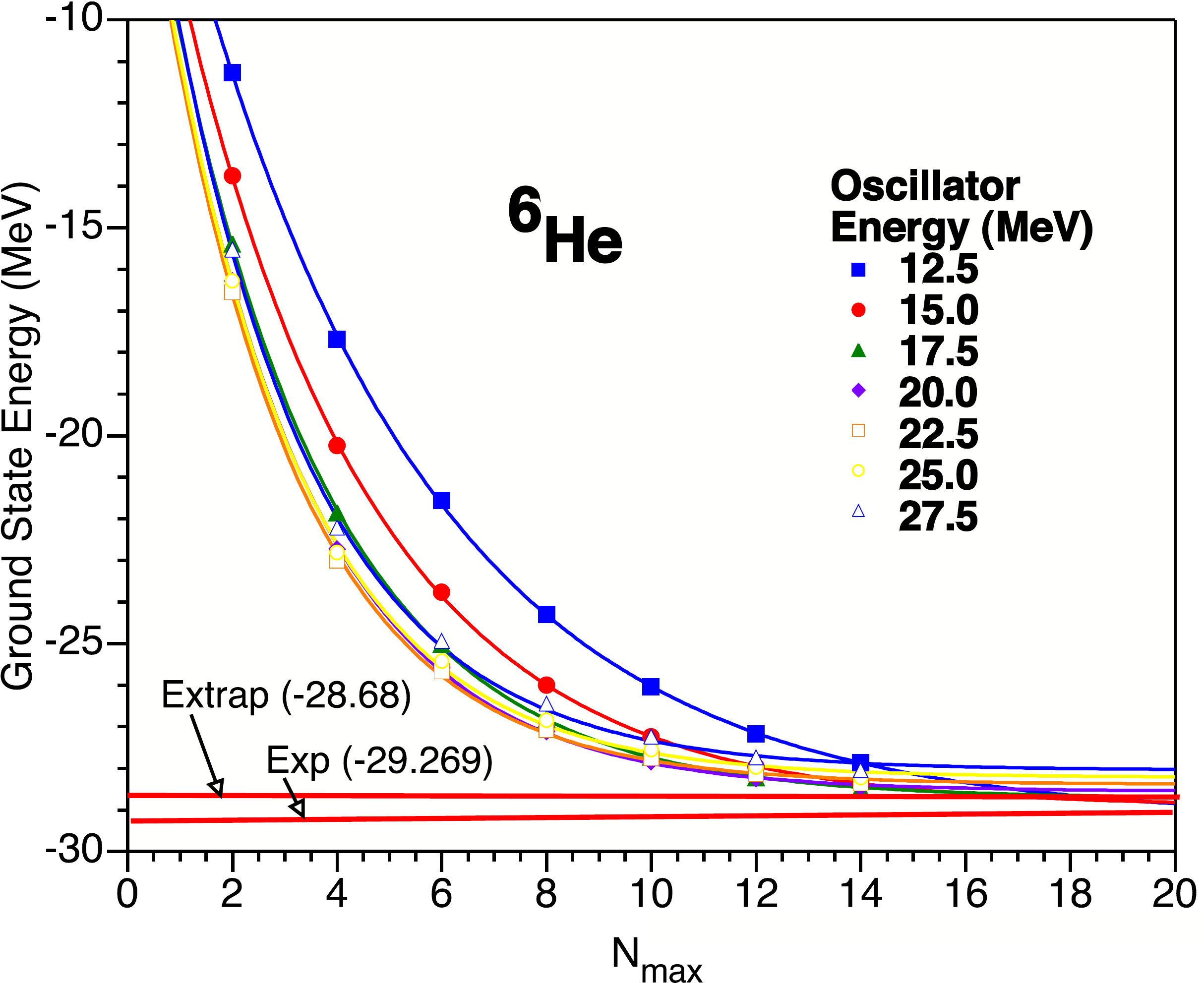}
	\caption{The estimate for exact $^6$He ground state energy using
	  Eq.~\eqref{eq:exp_Nmax_extrapolation}.  Extrapolated answer from the
		constrained fit and the experimental binding energy are indicated by
		horizontal lines.  Figure from \cite{Maris2009}.}
	\label{fig:exp_Nmax_extrapolation_6He}
	\end{figure}

	The goal is to extrapolate to $\Nmax = \infty$ from a finite $\Nmax$.
	The most widely used extrapolation scheme employs an exponential in $\Nmax$
	form
	\beq
	E(\Nmax) = \Einf + a e^{-c \Nmax}\;,
	\label{eq:exp_Nmax_extrapolation}
	\eeq
	where $a$ and $c$ are determined separately for each $\hw$ (with the
	option of constraining the fit to get the same asymptotic $\Einf$ value).
	Figure~\ref{fig:exp_Nmax_extrapolation_6He} shows estimate for the ground
	state energy for $^6$He obtained using the extrapolation form of
	Eq.~\eqref{eq:exp_Nmax_extrapolation}.

	The exponential in $\Nmax$ extrapolation is widely used in literature and
	seems to work quite well
	\cite{Hagen:2007hi,Bogner:2007rx,Forssen:2008qp,Maris:2008ax,Roth:2009cw}.
	There are however many open questions about this extrapolation scheme such as
	the answer for $\Einf$ depends on the oscillator frequency $\Omega$ and it is
	not clear which is the best choice for $\Omega$.  The terms $a$ and $c$ in
	Eq.~\eqref{eq:exp_Nmax_extrapolation} are fit to data.  There is no way to
	extract these terms for one nucleus and use it to predict something else.
	Moreover, the physical motivation for an exponential in $\Nmax$ extrapolation
	is slim at best.  It has been claimed that for larger nuclei $\Nmax$ is a
	logarithmic measure of the number of states \cite{Bogner:2007rx}.
	This would account for the exponential behavior, but there is no
	demonstration that it follows in general or with a specific
	logarithmic dependence.

	An alternative approach to extrapolations is motivated by effective field
	theory (EFT) and based instead on explicitly considering the infrared (IR)
	and ultraviolet (UV) cutoffs imposed by a finite oscillator
	basis~\cite{Coon:2012ab}.  The truncation in the oscillator basis introduces
	a maximum length scale (or an IR cutoff) and also a maximum momentum scale
	(or an UV cutoff).  These length and momentum scales can be motivated by
	the classical turning points denoted by $L_0$ and $\Lambda_0$ respectively.
	We have
	\begin{align}
	L_0 = \sqrt{2 (\Nmax + 3/2)} b \;,  \nonumber\\
	\Lambda_0 = \sqrt{2 (\Nmax + 3/2)} \hbar/b \;.
	\label{eq:L0_Lam0_cutoff}
	\end{align}
	$b$ is the oscillator length given by
	$\displaystyle b = \sqrt{\hbar/ m \Omega}$.
	The errors due the finite IR (UV) cutoff are called the IR (UV) errors.
	To draw a lattice analogy, IR errors stem from the finite box size,
	and the UV errors are a result of the finite lattice spacing (or the
	graininess of the lattice).  Ideally, we would like the box size to be as
	large as possible and the lattice spacing to be as small as possible.
	Because of finite computational power, this is not always possible though
	and therefore we need reliable extrapolation schemes in both IR as well as UV.

	This approach of thinking of the HO truncation in terms of IR and UV cutoffs
	has led to lot of development in the past three years.
	Note that $b$ appears in the numerator for $L_0$ and in the denominator
	for $\Lambda_0$, so it is not possible to make both cutoffs large
	simultaneously.
	We can choose the oscillator parameters such that one of the cutoffs in
	Eqs.~\eqref{eq:L0_Lam0_cutoff} is large, making the errors due to that cutoff
	small and focus on the errors due to the other cutoff.
	The first attempt and
	test for a theoretically motivated IR correction was made in
	\cite{Furnstahl2012}.  These corrections were made theoretically sound
	in \cite{More:2013rma, Furnstahl:2013vda}.  In \cite{Konig:2014hma} we looked
	at the UV correction for the deuteron.
	Our papers \cite{More:2013rma, Furnstahl:2013vda, Konig:2014hma} have led to
	physically motivated extrapolation schemes and will form the basis of the next
	two sections.  Insights from our work have also led to development of
	extrapolation schemes (both in IR and UV) for the many-body case by other
	groups.  We will touch upon these developments in
	Subsec.~\ref{subsec:IR_front}.

	\section[Infrared story]{Infrared story
	\footnote{Based on \cite{More:2013rma} and \cite{Furnstahl:2013vda}}}
	\label{sec:IR_story}

	As mentioned in the introduction, there was a lack of well motivated
	extrapolation schemes in LENP and this is where our work comes in.
	We started with the two-body case because it is more tractable
	mathematically.  Note that extrapolation is usually not necessary for the
	two-body problem, because convergence is reached before we run out of
	computational power.  This allows us to test our extrapolation formulas.
	Once we establish that the approach works for the two-body case, we can hope
	to extend the approach to few- and many-body case.

	As mentioned earlier, IR cutoff effectively puts system in a finite box.
	We need to find appropriate box length such that
	\beq
	E(\Nmax) = E(L_{\rm box})\;.
	\label{eq:Nmax_Lbox_correspondence}
	\eeq
	Note that our original problem was to find $\Einf \equiv E(\Nmax = \infty)$
	given $E$ at a finite $\Nmax$.  Once we make the correspondence in
	Eq.~\eqref{eq:Nmax_Lbox_correspondence} and express energy as a function of
	box length, we can use various techniques (discussed later) to estimate
	$E (L_{\rm box} = \infty)$ which equals $E(\Nmax = \infty)$.

	\subsection{Tale of tails}
	\label{subsec:tale_of_tails}

	The box size is usually bigger than the range of the potential.  Thus
	imposing the IR cutoff modifies the asymptotic part (or tail) of the
	bound-state wave function.  Our early work focused on trying to estimate the
	appropriate box size by matching the tails of wave functions in the truncated
	HO basis to the tails of wave functions in boxes of different lengths.

	Our strategy was to use a range of model potentials for which the
	Schr\"odinger equation can be solved
	analytically or to any desired precision numerically to broadly test
	and illustrate various features, and then turn to the deuteron for a
	real-world example.  In particular we considered:
	\bea
	V_{\rm sw}(r) &=& -V_0\, \theta(R-r)   \qquad \mbox{[square well]}
	\;,
	\label{eq:Vsw}
	\\
	V_{\rm exp}(r) &=& -V_0\, e^{-(r/R)}  \qquad\ \ \mbox{[exponential]}
	\;,
	\\
	V_{\rm g}(r) &=& -V_0\, e^{-(r/R)^2}  \qquad\ \mbox{[Gaussian]}
	\;,
	\label{eq:Vg}
	\\
	V_{\rm q}(r) &=& -V_0\, e^{-(r/R)^4} \qquad\ \mbox{[quartic]}
	\;,
	\label{eq:Vq}
	\eea
	where for each of the models we work in units with $\hbar = 1$, reduced mass
	$\mu=1$, and $R=1$, but consider different values for $V_0$.  For the
	realistic potential we use the Entem-Machleidt 500\,MeV chiral EFT
	N$^3$LO potential~\cite{Entem:2003ft} and unitarily evolve it with the
	similarity renormalization group (SRG).  These potentials provide a
	diverse set of tests for universal properties.  Because we can go to
	very high \hw\ and $\Nmax$ for the two-particle bound states (and
	therefore large $\LamUV$), it is possible to always ensure that UV
	corrections are negligible.

\begin{figure}[h]
		\centering
		\includegraphics[width=0.6 \textwidth]
		{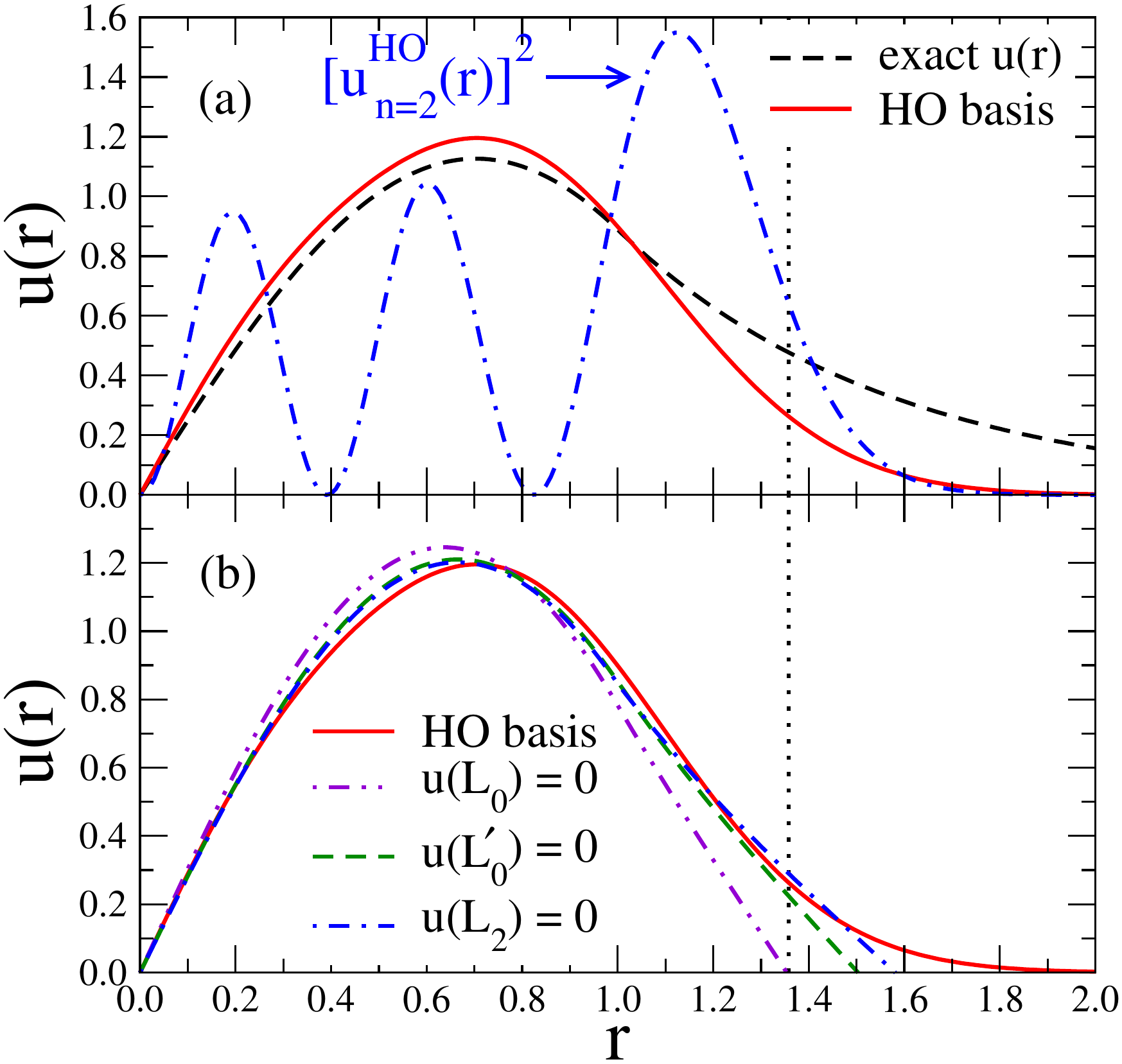}
		\caption{(a) The exact radial wave
    function (dashed) for a square well Eq.~\eqref{eq:Vsw} with depth
	  $V_0=4$ (and $\hbar = \mu = R = 1$) is compared to the wave function
		obtained from an HO basis truncated at $\Nmax = 4$ with $\hw=6$ (solid).
		The spatial extent of the wave function obtained from the HO basis
		truncation is dictated by the square of HO wave function for the highest
		radial quantum number (dot-dashed).
		(b) The wave functions obtained from imposing a Dirichlet
	  boundary condition (bc) at $L_0$, $\LA$ and $L_2$ are compared to the wave
		function in truncated HO basis. }
		\label{fig:sq_well_tail_matching}
	\end{figure}

	We start with empirical considerations before presenting an
  analytical understanding.  An example of how this correspondence between
	the HO truncation and a hard wall at specific length plays out is
	presented in Fig.~\ref{fig:sq_well_tail_matching}.
	In the top panel, the exact
	ground-state radial wave function (dashed) for the square well in
	Eq.~\eqref{eq:Vsw} is compared to the solution in an oscillator basis
	truncated at $\Nmax = 4$ determined by
	diagonalization (solid).  The truncated basis cuts off the tail of the
	exact wave function because the individual basis wave functions have a
	radial extent that depends on \hw\ (from the Gaussian part;
	cf.~Eq.~\eqref{eq:HO_S_wave_written_out}) and on the
	largest power of $r$ (from the polynomial part).  The latter is given
	by $\Nmax = 2n + l$.  With $\Nmax = 4$ and $l=0$, this means that $n=2$ gives
	the largest power.

	The cutoff will then be determined by the $n=2$ oscillator wave
	function, $u_{n=2}^{\rm HO}(r)$, whose square (which is the relevant
	quantity) is also plotted in the top panel (dot-dashed).  It is
	evident that the tail of the wave function in the truncated basis is
	fixed by this squared wave function.  Our premise is that the HO truncation
	is well modeled by a hard-wall (Dirichlet) bc at $r=L$.
	If so, the question remains how best to \emph{quantitatively} determine $L$
	given	$\Nmax$ and $\hw$.  In the bottom panel of
	Fig.~\ref{fig:sq_well_tail_matching} we show the wave functions	for several
	possible choices for $L$.
	$L_0$ corresponds to choosing the classical
	turning point (i.e. the half-height point of the tail of
	$[u^{HO}_{n=2}(r)]^2$); it is manifestly too small.  The authors of
	\cite{Furnstahl2012} advocated an improved choice for $L$ given by
	\beq
	\LA = L_0 + 0.54437\, b\, (L_0/b)^{-1/3} \;.
  \label{eq:LA}
	\eeq
	The length $\LA$ in Eq.~\eqref{eq:LA} is obtained by linear extrapolation
	from the slope at the half-height point.
	However, choosing
	\beq
	L = L_2 \equiv \sqrt{2 (\Nmax + 3/2 + 2)}b
	\label{eq:L2_def}
	\eeq
	was found to work the best in almost all examples.

	The most direct illustration of this conclusion comes from the
	bound-state energies.  In the example in Fig.~\ref{fig:sq_well_tail_matching},
	the exact energy (in dimensionless units) is $-1.51$ while the
	result for the basis truncated at $\Nmax=4$ is $-1.33$, which is therefore
	what we	hope to reproduce.  With $L_0$, the energy is $-0.97$, with $\LA$ it
	is $-1.21$, and with $L_2$ it is $-1.29$.  While this is only one
	example of a model problem, we have found that $L_2$ always gives a
	better energy estimate than $\LA$ (and something like
	$L_3 \equiv \sqrt{2 (\Nmax + 3/2 + 3)}b$ is almost always worse).

\begin{figure}[h]
	\centering
	\includegraphics[width=0.6\textwidth]
	  {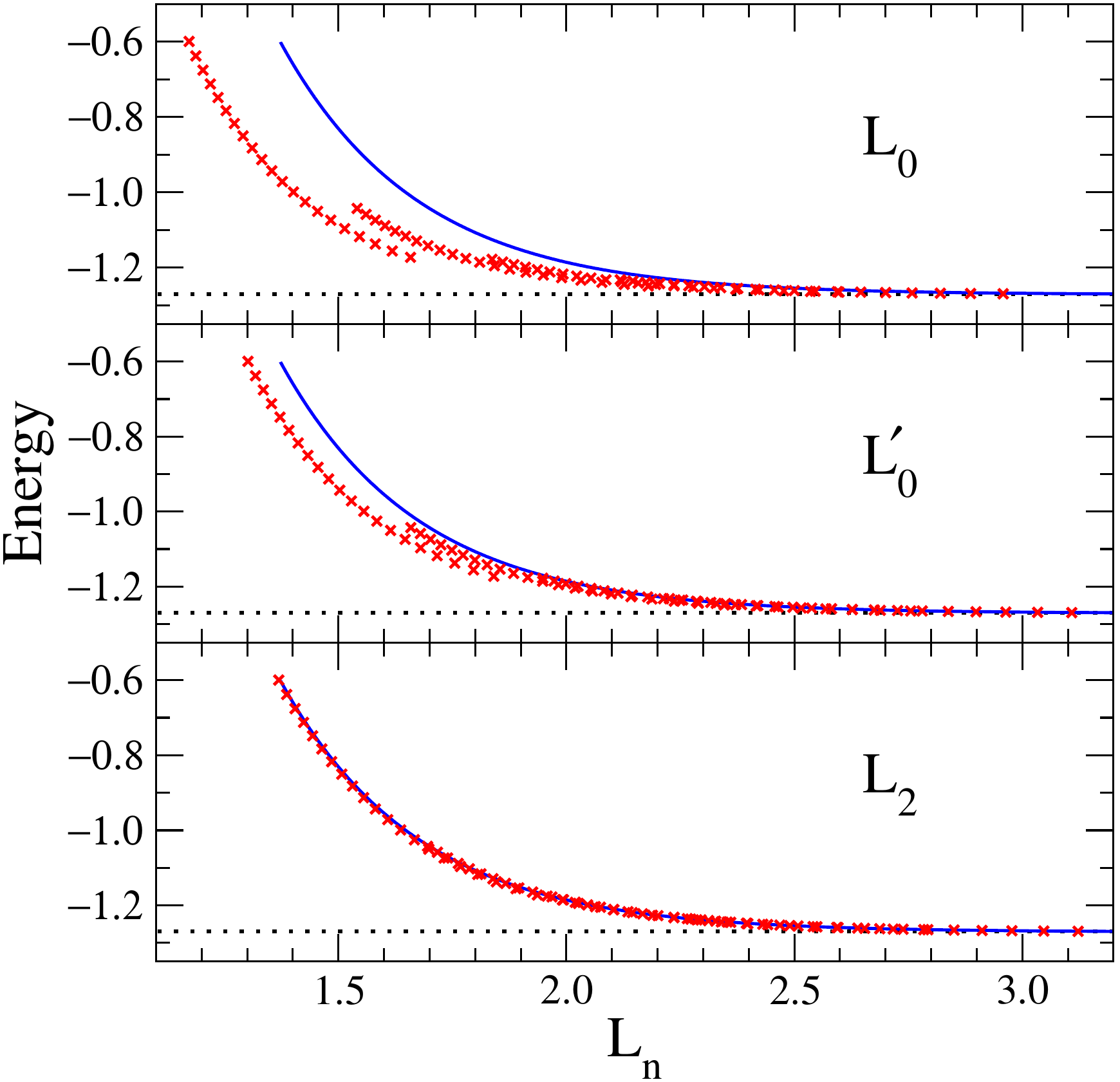}
	\caption{Ground-state energies versus $L_0$ (top),
		$\LA$ (middle), and $L_2$ (bottom) for a Gaussian potential well
		Eq.~\eqref{eq:Vg} with $V_0=5$ and $R=1$.
		The crosses are the
		energies from HO basis truncation.
		The energies obtained by
		numerically solving the Schr{\"o}dinger equation with a Dirichlet
		bc at $L$ lie on the solid line.
		The horizontal dotted lines mark
		the exact energy $\Einf=-1.27$.}
  \label{fig:spatial_cutoff_vs_HO_gauss}
  \end{figure}

\begin{figure}[h]
	\centering
	\includegraphics[width=0.6\textwidth]
	  {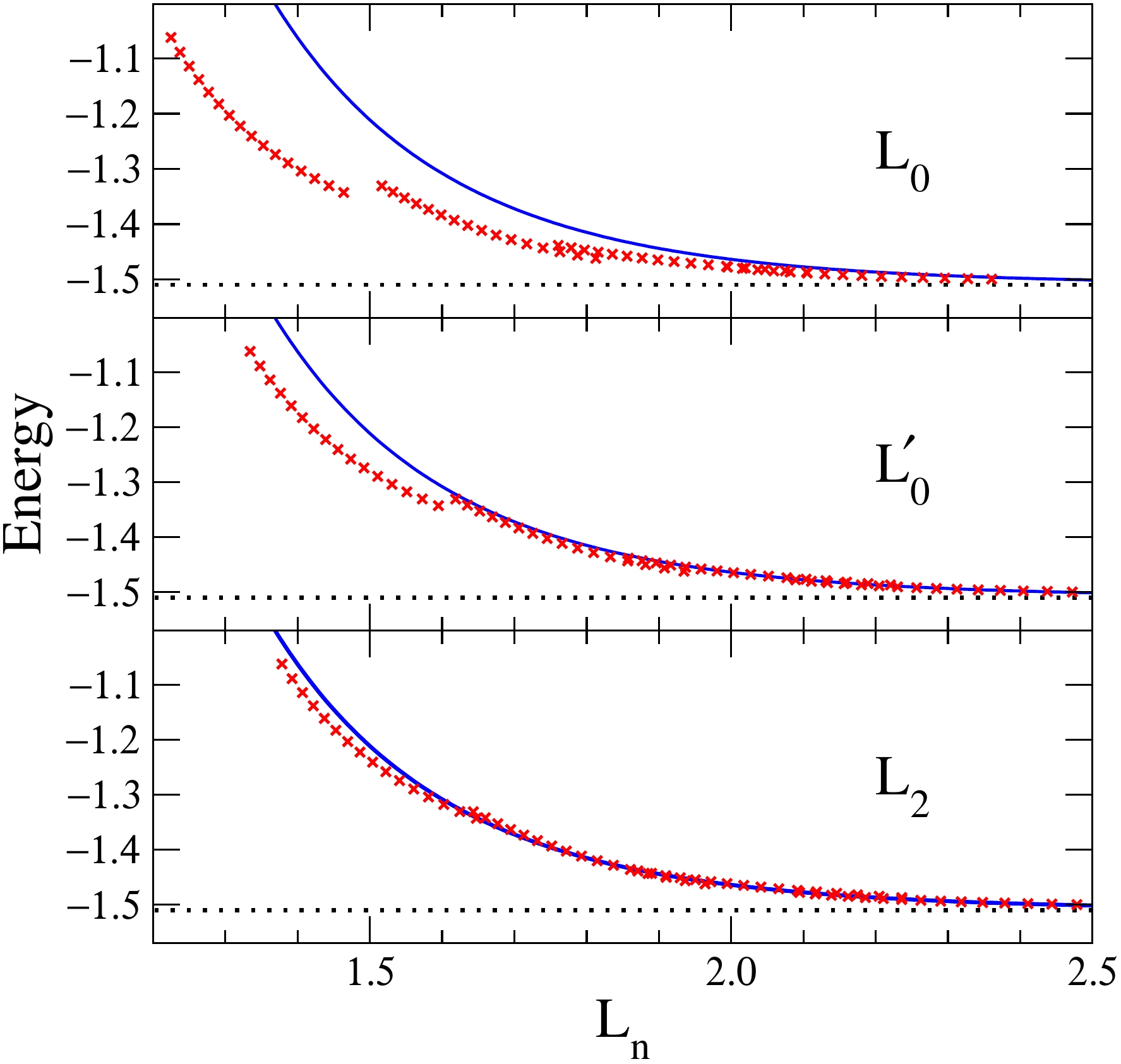}
	\caption{Ground-state energies versus $L_0$ (top),
	  $\LA$ (middle), and $L_2$ (bottom) for a square well potential well
	  Eq.~\eqref{eq:Vsw} with $V_0=4$ and $R=1$.
	  The crosses are the
	  energies from HO basis truncation.
	  The energies obtained by
	  numerically solving the Schr{\"o}dinger equation with a Dirichlet
	  bc at $L$ lie on the solid line.
	  The horizontal dotted lines mark
	  the exact energy $\Einf=-1.51$.}
  \label{fig:spatial_cutoff_vs_HO_square_well}
  \end{figure}

	Another signature that demonstrates the suitability of $L_2$ is that
	points from many different $\hw$ and $\Nmax$ values all lie on the same
	curve.  Figures~\ref{fig:spatial_cutoff_vs_HO_gauss} and
	\ref{fig:spatial_cutoff_vs_HO_square_well} show the energies from a
	wide range of HO truncations for $L_0$, $\LA$ and $L_2$ for the
	Gaussian well and the square well potential, respectively.  The
	energies for different $\hbar\Omega$ and $\Nmax$ lie on the same smooth
	and unbroken curve if we use $L_2$ but not with the other choices.  For
	$L=L_0$ and $L=\LA$, one finds that sets of points with different
	$\hbar\Omega$ but same $\Nmax$ fall on smooth, $\Nmax$-dependent curves.  For
	the square well, there are small discontinuities visible even for
	$L=L_2$.  At the square well radius, the wave function's second
	derivative is not smooth, and this is difficult to approximate with a
	finite set of oscillator functions.  This lack of UV convergence is
	likely the origin of the very small discontinuities.
	As a further test, we solve the Schr{\"o}dinger equation with a
	vanishing Dirichlet bc (solid lines in
	Figs.~\ref{fig:spatial_cutoff_vs_HO_gauss} and
	\ref{fig:spatial_cutoff_vs_HO_square_well}), and compare to the
	energies obtained from the HO truncations (crosses).  The finite
	oscillator basis energies are well approximated by a Dirichlet
	bc with a mapping from the oscillator $\hbar\Omega$
	and $\Nmax$ to an equivalent length given by $L_2$.  Note that for large
	$\Nmax$, the differences between $L_0$, $\LA$ and $L_2$ may be smaller
	than other uncertainties involved in nuclear calculations, but for
	practical calculations one will want to use small $\Nmax$ results, where
	these considerations are very relevant.

	These results from model calculations are consistent with those from
	realistic potentials applied to the deuteron.  To illustrate this, we
	use the N$^3$LO 500\,MeV potential of Entem and
	Machleidt~\cite{Entem:2003ft}.  We generate results in an HO basis
	with \hw\ ranging from $1$ to $100\,\mbox{MeV}$ and $\Nmax$ from $4$ to
	$100$ (in steps of 4 to avoid HO artifacts for the
	deuteron~\cite{Bogner:2007rx}).  We then restrict the data to where UV
	corrections are negligible (see Section~\ref{sec:UV_story}).
	\begin{figure}[h]
	\centering
  \includegraphics[width=0.6\textwidth]
	{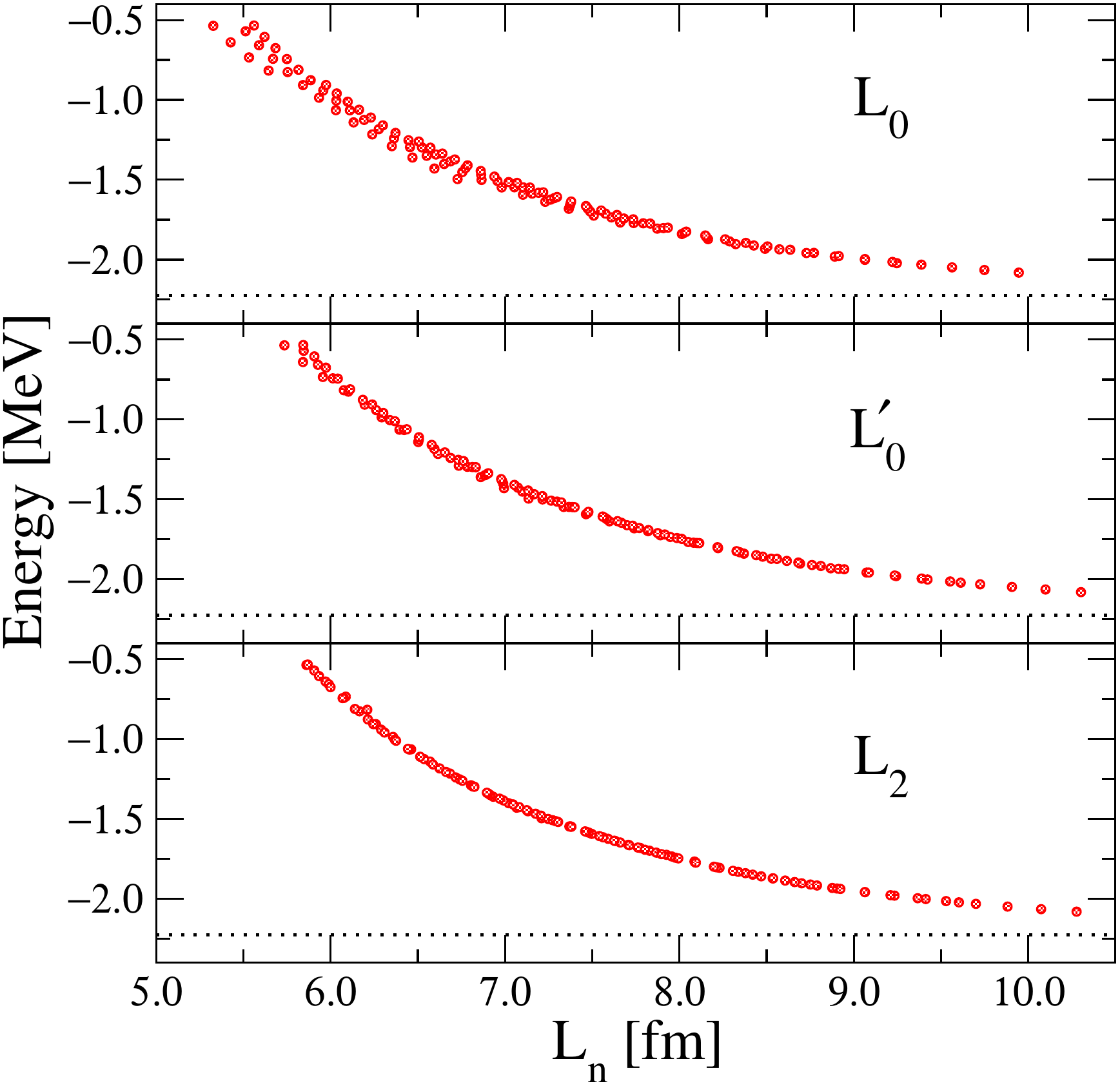}
  \caption{Ground-state energies versus $L_0$ (top),
  	$\LA$ (middle), and $L_2$ (bottom) for the Entem-Machleidt 500\,MeV
  	N$^3$LO potential~\cite{Entem:2003ft}.  The horizontal dotted lines
  	mark the exact energy $\Einf = -2.2246\,\mbox{MeV}$. }
  \label{fig:spatial_cutoff_vs_HO_deuteron}
  \end{figure}
	Figure~\ref{fig:spatial_cutoff_vs_HO_deuteron} shows that the
	criterion of a continuous curve with the smallest spread of points
	clearly favors $L_2$.

	\medskip
	\subsubsection{Analytical derivation of $L_2$}

	In the asymptotic region (this is the region where IR cutoff is imposed), the
	potential is negligible and the only relevant
	part of the Hamiltonian is the kinetic energy or the $p^2$ operator.
	In what follows, we analytically compute the smallest eigenvalue
	$\kappa^2_{\rm min}$ of $p^2$ in a finite oscillator basis and will
	see that $\kappa_{\rm min} = \pi/L_2$. In the remainder of this
	Subsection, we set the oscillator length to one. We focus on $s$-waves
	and thus consider wave functions that are regular at the origin, i.e.
	the radial wave functions are identical to the odd wave functions of
	the one-dimensional harmonic oscillator.

	The localized eigenfunction of the operator $p^2$ with smallest
	eigenvalue $\kappa^2$ is
	\bea
	\label{eq:eigen_l0}
	\psi_{\kappa}(r) = \left\{\begin{array}{ll}
	\sin{\kappa r}\ , & 0 \le r \le {\pi\over\kappa}\\
	0 \ , & r > {\pi\over\kappa}
	\end{array}\right. \; .
	\eea
	We employ the $s$-wave oscillator functions
	\bea
	\varphi_{2n+1}(r) &=&(-1)^n \sqrt{2 n!\over\Gamma(n+3/2)} r
	L^{1\over 2}_n\left(r^2\right)
	e^{-{r^2\over 2}} \nonumber\\
	&=&\left(\pi^{1\over 2} 2^{2n} (2n+1)!\right)^{-1/2}H_{2n+1}(r)
	e^{-{r^2\over 2}} \; ,
	\eea
	with energy $E=(2n +3/2)\hbar\Omega$. Here, $L_n^{1/2}$ denotes the
	Laguerre polynomial, and it is convenient to rewrite this function in
	terms of the Hermite polynomial $H_n$. We expand the
	eigenfunction in Eq.~\eqref{eq:eigen_l0} as
	\beq
	\label{expand}
	\psi_{\kappa}(r) = \sum_{n=0}^\infty c_{2n+1}(\kappa)\varphi_{2n+1}(r) \; .
	\eeq
	Before we turn to the computation of the expansion coefficients
	$c_{2n+1}(\kappa)$, we consider the eigenvalue problem for the
	operator $p^2$.  We have
	\beq
	p^2 = a^\dagger a +{1\over 2} -{1\over 2}
	\left(a^2 +\left(a^\dagger\right)^2\right) \;,
	\eeq
	where $a$ and $a^\dagger$ denote the annihilation and creation
	operator for the one-dimensional harmonic oscillator, respectively.
	The matrix of $p^2$ is tridiagonal in the oscillator basis.
	For the matrix representation, we order the basis states as
	$(\varphi_1, \varphi_3, \varphi_5, \ldots)$. Thus, the eigenvalue
	problem $p^2-\kappa^2=0$ becomes a set of rows of coupled linear
	equations. In an infinite basis, the eigenvector $(c_1(\kappa),
	c_3(\kappa), c_5(\kappa), \ldots )$ identically satisfies every row of
	the eigenvalue problem for any value of $\kappa$. In a finite basis
	$(\varphi_1, \varphi_3, \varphi_5,\ldots \varphi_{2n+1})$, however,
	the last row of the eigenvalue problem
	\beq
	\label{quant}
	\left(2n+3/2 -\kappa^2\right) c_{2n+1}(\kappa) =
	 {1\over
	  2}\sqrt{2n}\sqrt{2 n+1} \, c_{2n-1}(\kappa) \; ,
	\eeq
	can only be fulfilled for certain values of $\kappa$, and this is the
	quantization condition. To solve this eigenvalue problem we need
	expressions for the expansion coefficients $c_{2n+1}(\kappa)$
	for $n\gg 1$. Those can be derived analytically as follows.

	We rewrite the eigenfunction in Eq.~\eqref{eq:eigen_l0} as a Fourier transform
	\beq
	\psi_{\kappa} (r) = \sqrt{2\over \pi} \int\limits_0^\infty dk
	\tilde{\psi}_{\kappa}(k) \sin kr \; ,
	\eeq
	and expand the sine function in terms of oscillator functions as
	\beq
	\sin kr = \sqrt{\pi\over 2} \sum_{n=0}^\infty (-1)^n \varphi_{2n+1}(r)
	\varphi_{2n+1}(k) \; .
	\eeq
	Thus, the expansion coefficients in Eq.~\eqref{expand} are given in
	terms of the Fourier transform $\tilde{\psi}_\kappa(k)$ as
	\beq
	\label{integ}
	c_{2n+1}(\kappa) = (-1)^n\int\limits_0^\infty dk\, \tilde{\psi}_{\kappa}(k)
	\varphi_{2n+1}(k) \; .
	\eeq
	So far, all manipulations have been exact.  We need an expression for
	$c_{2n+1}(\kappa)$ for $n\gg 1$ and use the asymptotic expansion
	\beq
	\label{approxwf}
	\varphi_{2n+1}(k) \approx {(-1)^n\sqrt{2}\over \pi^{1/4}}
	{(2n-1)!!\over \sqrt{(2n)!}}
	\sin (\sqrt{4n+3}k) \; ,
	\eeq
	which is valid for $|k|\ll \sqrt{2n}$, see~\cite{gradshteyn}.
	Using this approximation, one finds (making use of Fourier transforms)
	\bea
	\label{phi}
	c_{2n+1}(\kappa) &\approx& \pi^{1/4} {(2n-1)!!\over \sqrt{(2n)!}}
	\psi_{\kappa}(\sqrt{4n+3}) \nonumber\\
	&=&\pi^{1/4}{(2n-1)!!\over \sqrt{(2n)!}}
	\sin (\sqrt{4n+3}\kappa) \; ,
	\eea
	with $\kappa \le \pi/\sqrt{4n+3}$ due to Eq.~\eqref{eq:eigen_l0}.

	Let us return to the solution of the quantization
	condition in Eq.~\eqref{quant}.  We make the ansatz
	\beq
	\kappa = {\pi\over\sqrt{4n+3+2\Delta}} \; ,
	\eeq
	and must assume that $\Delta > 0$.  This ansatz is well motivated, since the
	naive semiclassical estimate $\kappa = \pi/L_0$ yields $\Delta=0$.  We
	insert the expansion coefficients of Eq.~\eqref{phi} into Eq.~\eqref{quant}
	and consider its leading-order approximation
	for $n\gg 1$ and $n\gg \Delta$. This yields
	\beq
	\Delta = 2
	\eeq
	as the solution. Recalling that a truncation of the basis at
	$\varphi_{2n+1}$ corresponds to the maximum energy
	$E=(2n+3/2)\hbar\Omega$, we see that we must identify
	$\Nmax \equiv N = 2n$. Thus,
	$\kappa_{\rm min} = \pi/L_2$ is the lowest momentum in a finite
	oscillator basis with $n \gg 1$ basis states (and not $1/b$ as stated in
	Ref.~\cite{Coon:2012ab}).  It is clear from its
	very definition that $\pi/L_2$ is also (a very precise approximation of)
	the infrared cutoff in a finite	oscillator basis, and that $L_2$ (and not $b$
	as stated in Refs.~\cite{Stetcu:2006ey,Stetcu:2007ms}) is the radial extent
	of the oscillator basis and the analog to the extent of the lattice in the
	lattice computations \cite{Luscher:1985dn}.

	The derivation of our key result $\kappa_{\rm min}=\pi/L_2$ is based
	on the assumption that the number of shells $N$ fulfills $N\gg 1$.
	Table~\ref{tab:L0_L2_k_min_comparison} shows a comparison of
	numerical results for
	$\kappa_{\rm min}$ in different model spaces. We see that
	$\pi/L_2$ is a very good approximation already for $N=2$, with a
	deviation of about 1\%.

	\begin{table}[ht]
	\centering
	\begin{tabular}{|c|c|c|c|}\hline
	$N$ & $\kappa_{\rm min}$ & $\pi/L_2$ & $\pi/L_0$ \\\hline
	   0 & 1.2247 & 1.1874 & 1.8138\\
	   2 & 0.9586 & 0.9472 & 1.1874\\
	   4 & 0.8163 & 0.8112 & 0.9472\\
	   6 & 0.7236 & 0.7207 & 0.8112\\
	   8 & 0.6568 & 0.6551 & 0.7207\\
	  10 & 0.6058 & 0.6046 & 0.6551\\
	  12 & 0.5651 & 0.5642 & 0.6046\\
	  14 & 0.5316 & 0.5310 & 0.5642\\
	  16 & 0.5035 & 0.5031 & 0.5310\\
	  18 & 0.4795 & 0.4791 & 0.5031\\
	  20 & 0.4585 & 0.4582 & 0.4791\\\hline
	\end{tabular}
	\caption{Comparison between the lowest momentum $\kappa_{\rm min}$, $\pi/L_2$,
	 and $\pi/L_0$ for model spaces with up to $N$ oscillator quanta.}
	\label{tab:L0_L2_k_min_comparison}
	\end{table}

	Note that this approach can be generalized to other localized
	bases.  The (numerical) computation of the lowest eigenvalue of the momentum
	operator $p^2$ yields the box size $L$ corresponding to the employed Hilbert
	space.

	\medskip
	\subsubsection{EFT-like approach}

	We mentioned that the relevant operator for IR truncation is $p^2$.  To
	get a better understanding of the correspondence between the HO truncation
	and a hard wall at $L_2$, let's compare the spectrum of $p^2$ in the two
	cases.
	\begin{figure}[h]
	\centering
	\includegraphics[width=0.6\textwidth]
	{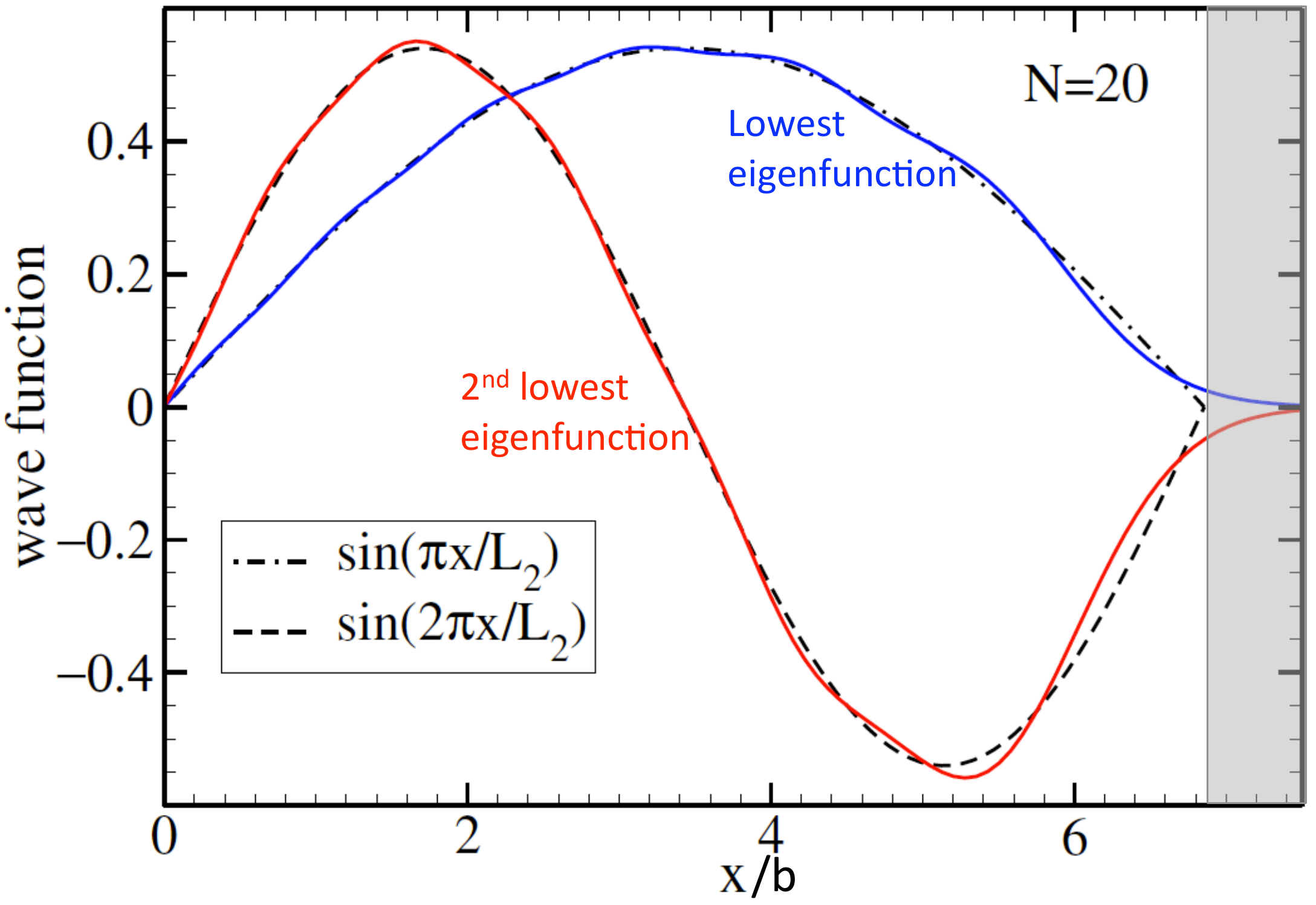}
	\caption{Eigenfunctions of $p^2$ in the truncated HO basis compared to those
	  in a box of size $L_2$. }
	\label{fig:mom_eigen_fns}
	\end{figure}
	In Fig.~\ref{fig:mom_eigen_fns}, we compare the low-lying eigenfunctions of
	$p^2$ in the truncated HO basis to the eigenfunctions in a box of size $L_2$.
	As we will see later, the asymptotic or near the wall difference between the
	two eigenfunctions are high-momentum effects irrelevant for the
	long-wavelength physics of the bound states.

\begin{figure}[h]
	\centering
	\includegraphics[width=0.6\textwidth]{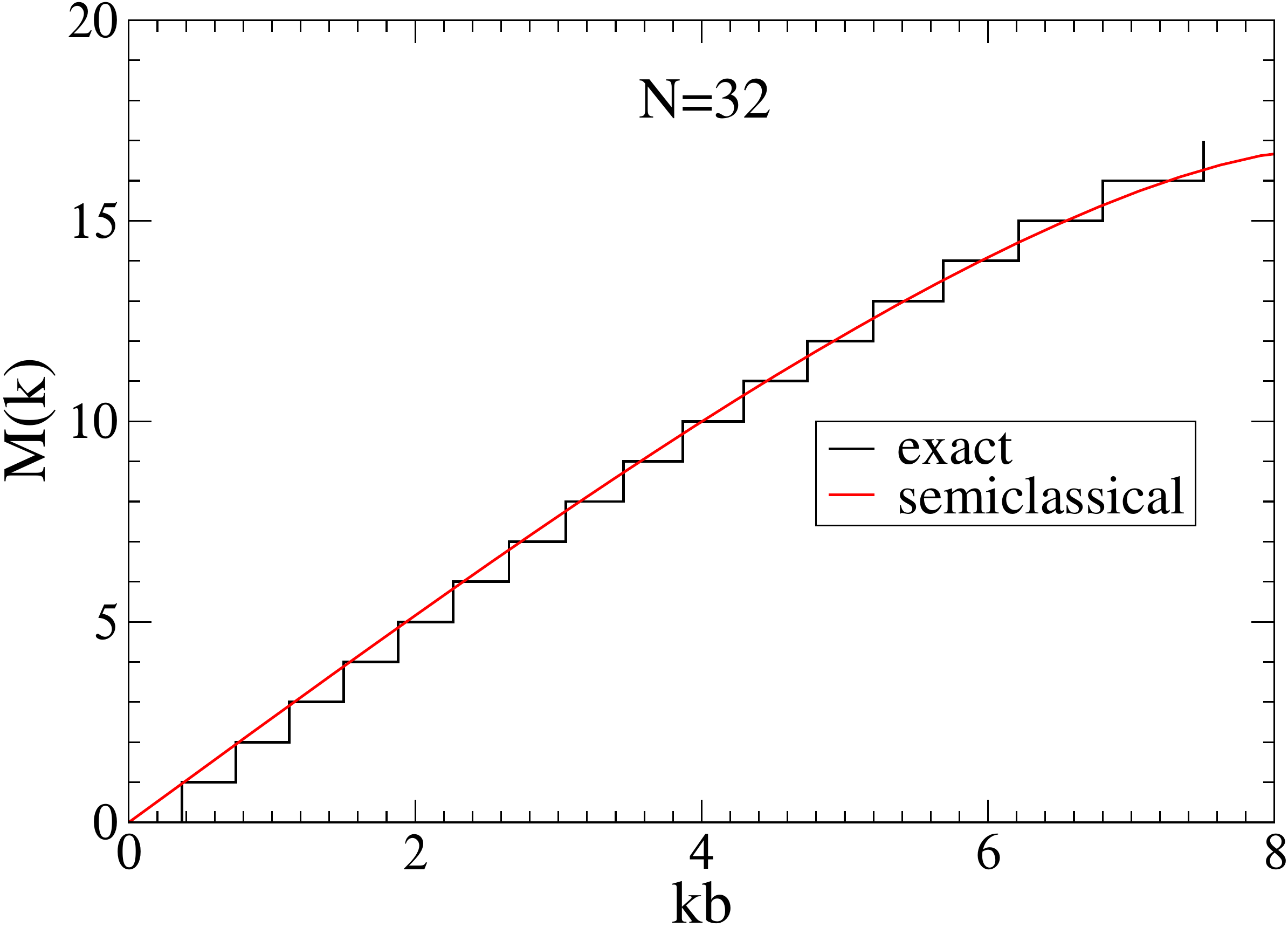}
	\caption{The staircase function of the $s$ states of
	  the operator $p^2$ in a finite oscillator basis with $N=32$ (black)
	  compared to its semiclassical estimate (smooth red curve). $M(k)$
	  denotes the number of states of the operator $p^2$ with eigenvalues
	  $p^2\le\hbar^2 k^2$.}
	\label{fig:staircase}
	\end{figure}

	Another way to look at this is to compute the number $M(k)$ of ($s$-wave)
	states up to a momentum $k$.  We find
	\bea
	  M(k)&=& {\rm Tr} \left[\Theta\left(\hbar^2k^2-p^2\right)
	   \Theta\left(E-{p^2\over 2m}-{m\over 2}\Omega^2 r^2\right) \right]
	  \nonumber\\
	  &\approx&{1\over 2\pi\hbar} \int\limits_{-\hbar k}^{\hbar k}\! dp
	  \int\limits_0^\infty\! dr \,
	   \Theta\left(\hbar^2k^2-p^2\right)
	   \Theta\left(E-{p^2\over 2m}-{m\over 2}\Omega^2 r^2\right)
	   \; .
	\eea
	Here, we apply the semiclassical approximation and write the trace
	as a phase-space integral.  We assume $\hbar^2k^2/(2m)\le
	E$, perform the integrations and use $E/(\hbar\Omega)= N+3/2$
	\footnote{For the sake of brevity we replace $\Nmax$ by $N$}. This
	yields
	\bea
	  M(k) = {bk\over 2\pi}\sqrt{2N+3-b^2k^2}
	   +{N+3/2\over \pi}\arcsin{bk\over\sqrt{2N+3}}\;,
	  \label{Mstair}
	\eea
	where $b$ is the oscillator length.
	Figure~\ref{fig:staircase} shows a comparison between the quantum
	mechanical staircase function and the semiclassical
	estimate of Eq.~\eqref{Mstair} for $N=32$.  For sufficiently small values of
	$kb\ll\sqrt{2N}$, the number of $s$-wave momentum eigenstates grows
	linearly, and inspection of Eq.~\eqref{Mstair} shows that the slope at
	the origin is $L_0/\pi$ semiclassically.  The linear growth of the
	number of eigenstates of $p^2$ with $k$ clearly demonstrate that --- at
	not too large values of $kb$ --- the spectrum of $p^2$ in the oscillator
	basis is similar to the spectrum of $p^2$ in a spherical
	box.

\begin{figure}[h]
	\centering
	\includegraphics[width=0.65\textwidth]
	{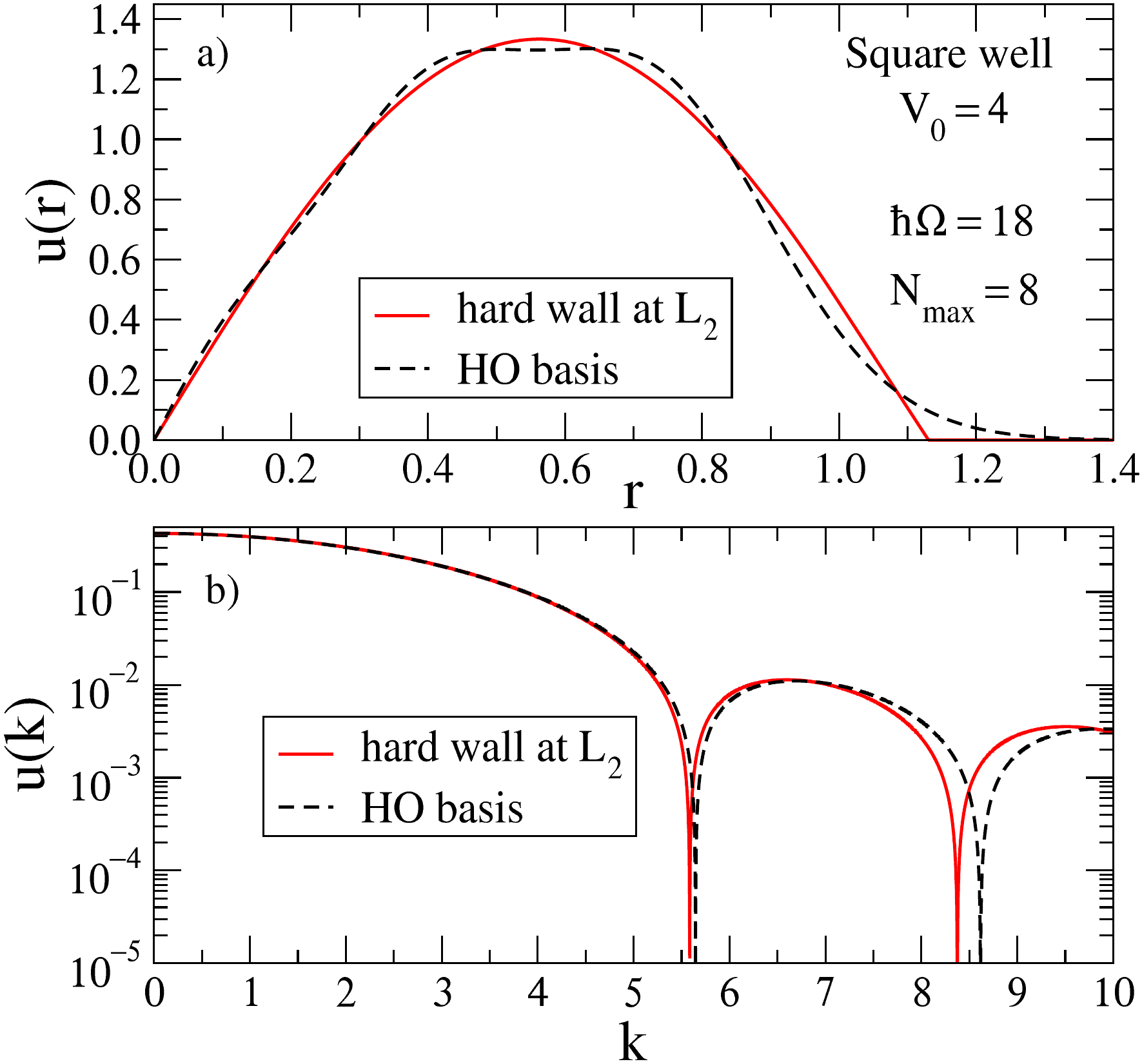}
	\caption{Ground-state wave functions for a square well potential of depth
	  $V_0=4$ (see Eq.~\eqref{eq:Vsw}; lengths are in units of $R$ and energies
		in units of $1/R^2$ with $\hbar^2/\mu = 1$) from solving the Schr\"odinger
	  equation with a truncated harmonic oscillator basis with $\hbar\Omega = 18$
	  and $N = 8$ (dashed) and with a
	  Dirichlet bc at $r=L_2$ given from Eq.~\eqref{eq:L2_def}
		(solid).  The coordinate-space radial wave functions in a) exhibit a
		difference at $r$ near 1.5, but the Fourier-transformed wave functions in
		b) are in close agreement at low $k$, showing that the differences are
		high-momentum modes.}
	\label{fig:HO_q_and_rspace_sqwell}
	\end{figure}

	As a final example of the correspondence between the HO truncation and the
	hard wall at $L_2$, we look at the ground state wave functions of a square
	well in the two bases in Fig.~\ref{fig:HO_q_and_rspace_sqwell}.
	The binding momentum in this case is $1.7$ (in units of $1/R$).  The
	Fourier-transformed wave functions differ at much larger momentum and this
	difference is irrelevant for the long-wavelength physics of bound states.
	Thus the use of Dirichlet bc to take into account the HO
	truncation is similar in spirit to the use of contact interactions to
	describe the effect of unknown short-ranged forces on long-wavelength probes.

	\subsection{Cashing in on the hard wall correspondence}
	\label{subsec:make_cash}

	We have so far focused on establishing how HO truncation is analogous to
	putting the system in a spherical box of a specified radius.  Now let's see
	how this correspondence helps us in getting the exact energy $\Einf$.

	\medskip
	\subsubsection{Linear energy method}

	Our first approximation to the IR correction
	is based on what is known in quantum chemistry
	as the linear energy method~\cite{Djajaputra:2000aa}.  Given a
	hard-wall bc at $r=L$ beyond the range of the
	potential, we write the energy compared to that for $L=\infty$ as
	\beq
	 E_L = E_{\infty}+\Delta E_L
	 \;.
	\eeq
	We seek an estimate for $\Delta E_L$, which is assumed to be small,
	based on an expansion of the wave function in $\Delta E_L$.  Let
	$u_E(r)$ be a radial solution with regular bc at the
	origin and energy $E$.  For convenience in using standard quantum
	scattering formalism below, we choose the normalization corresponding
	to what is called the ``regular solution'' in
	Ref.~\cite{taylor2006scattering}, which means that $u_E(0) = 0$ and
	the slope at the origin is unity for all $E$.  We denote the
	particular solutions $u_{E_L}(r)\equiv u_L(r)$ and $u_{\Einf}(r)
	\equiv u_\infty (r)$. Then there is a smooth expansion of $u_E$ about
	$E=\Einf$ at fixed $r$, so we approximate~\cite{Djajaputra:2000aa}
	\beq
	  u_L(r)\approx u_\infty (r) + \Delta E_L
	  \left.\frac{du_E(r)}{dE}\right|_{E_{\infty}}
	  + \mathcal{O}(\Delta E_L^2) \;,
	  \label{eq:linear_energy_approx}
	\eeq
	for $r\leq L$.
	By evaluating Eq.~\eqref{eq:linear_energy_approx} at $r=L$ with the
	bc $u_L(L)=0$, we find
	\beq
	  \Delta E_L \approx -u_\infty(L) \left(\left.\frac{d u_E(L)}
	  {dE}\right|_{\Einf}\right)^{-1}
	  \;,
	  \label{eq:Delta_EL}
	\eeq
	which is the estimate for the IR correction.

	\begin{figure}[h]
	\centering
	\includegraphics[width=0.6\textwidth]{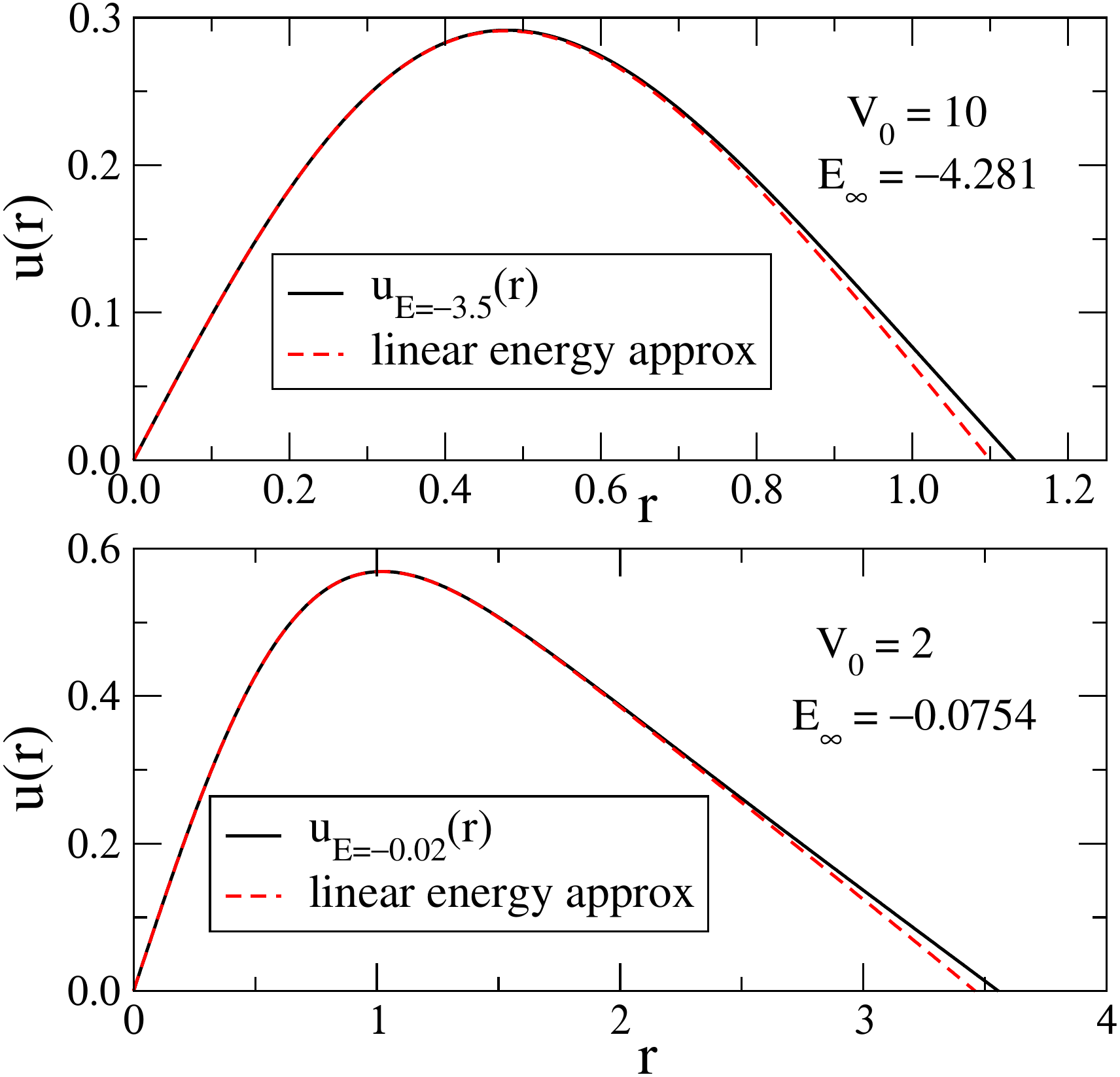}
	\caption{Testing the linear energy approximation
	  Eq.~\eqref{eq:linear_energy_approx} for (a) deep ($V_0=10$) and (b)
	  shallow ($V_0=2$) Gaussian potential well Eq.~\eqref{eq:Vg}
		($\hbar = \mu = R=1$).
	  The solid lines are the exact solutions $u_L(r)$ for energies $-3.5$
	  and $-0.020$, respectively, whose zero crossings determine the
	  corresponding values for $L$.}
	\label{fig:linear_energy_approx}
	\end{figure}

	We can check the accuracy of the linear energy
	approximation~(Eq.~\eqref{eq:linear_energy_approx}) by numerically solving
	the Schr\"odinger equation with a specified energy.  This determines
	$L$ as the radius at which the resulting wave function vanishes. Then
	we compare this wave function for $r \leq L$ to the right side of
	Eq.~\eqref{eq:linear_energy_approx}, with the derivative calculated
	numerically.  Figure~\ref{fig:linear_energy_approx} shows
	representative examples for a deep and shallow Gaussian potential.  In
	these examples and other cases, the approximation to the wave function
	is good, particularly in the interior.  The estimates for $\Delta E_L$
	using the right side of Eq.~\eqref{eq:Delta_EL} are within a few to
	ten percent: 0.68 versus 0.70 and 0.050 versus 0.055 for the two
	cases.

	The good approximation to the wave function suggests that for the
	calculation of other observables the linear energy approximation will
	be useful.  For observables most sensitive to the long distance
	(outer) part of the wave function, such as the radius, this has
	already been shown to be true~\cite{Furnstahl:2012qg}.  But the good
	approximation to the wave function at small $r$ means that corrections
	for short-range observables should also be controlled, with the
	dominant contribution in an extrapolation formula coming from the
	normalization.

	Next we derive an expression for the derivative in
	Eq.~\eqref{eq:Delta_EL}.  To start with we assume we have a single
	partial-wave channel.  For general $E < 0$, the asymptotic form of the
	radial wave function for $r$ greater than the range of the potential is
	\beq
	  u_E(r)\overset{r \gg R}\longrightarrow A_E(e^{-k_E r}+\alpha_E e^{+k_E r})
	  \;,
	  \label{eq:asymptotic_form_uE}
	\eeq
	with $u_\infty(r)\overset{r \gg R}\longrightarrow \Ainf e^{-\kinf r}$
	for $E=\Einf$.  We take the derivative of
	Eq.~\eqref{eq:asymptotic_form_uE} with respect to energy, evaluate at
	$E=\Einf$ using $\alpha_{\Einf}=0$ and $dk_E/dE = -\mu/(\hbar^2 k_E)$,
	to find
	\bea
	    \left.\frac{d u_E(r)}{d E}\right|_{\Einf} &=&
	    \Ainf \left.\frac{d \alpha_E}{d E}\right|_{\Einf} e^{+\kinf r}
	    +
	    \Ainf \frac{\mu}{\hbar^2}\frac{r}{\kinf} e^{-\kinf r}
	    +
	    \left.\frac{d A_E}{d E}\right|_{\Einf} e^{-\kinf r}
	    \;.
	  \label{eq:entire_correction}
	\eea
	We now evaluate at $r=L$ and anticipate that the $e^{+\kinf L}$ term
	dominates:
	\beq
	  \left.\frac{d u_E(L)}{d E}\right|_{\Einf}
	  \approx \Ainf \left.\frac{d\alpha_E}{d E}\right|_{\Einf} e^{+\kinf L}
	   + \mathcal{O}(e^{-\kinf L})
	   \;.
	  \label{eq:substitute_this}
	\eeq
	Substituting Eq.~\eqref{eq:substitute_this} into
	Eq.~\eqref{eq:Delta_EL}, we obtain
	\beq \Delta E_L \approx -\left[ \left.\frac{d \alpha_E}{d
	      E}\right|_{\Einf}\right]^{-1} e^{-2 \kinf L} + \mathcal{O}(e^{-4
	  \kinf L}) \;.
	\eeq
	Note that this result is independent of	the normalization of the
	wave function.

	To calculate the derivative explicitly, we turn to scattering theory,
	following the notation and discussion in
	Ref.~\cite{taylor2006scattering}.  In particular, the asymptotic form
	of the regular scattering wave function $\phi_{l,k}$ for orbital
	angular momentum $l$ and for positive energy $E \equiv \hbar^2 k^2/2\mu$
	is given in terms of the Jost function
	$\Jost_l(k)$~\cite{taylor2006scattering},
	\beq
	  \phi_{l,k}(r) \longrightarrow
	    \frac{i}{2}[\Jost_l(k)\hat{h}_l ^{-}(k r)-\Jost_l(-k)\hat{h}_l^{+}(k r)]
	    \;,
	    \label{eq:phi_asymp}
	\eeq
	where the $\hat{h}_l^{\pm}$ functions (related to Hankel functions)
	behave asymptotically as
	\beq
	\hat{h}_l ^{\pm}(k r)
	    \overset{r\rightarrow\infty}\longrightarrow e^{\pm i (k r - l \pi/2)}
	    \;.
	\eeq
	The ratio of the Jost functions appearing in Eq.~\eqref{eq:phi_asymp}
	gives the partial wave $S$-matrix $s_l(k)$:
	\beq
	   s_l(k) = \frac{\Jost_l(-k)}{\Jost_l(+k)}
	   \;,
	   \label{eq:s_l}
	\eeq
	which is in turn related to the partial-wave scattering amplitude
	$f_l(k)$ by
	\beq
	  f_l(k) = \frac{s_l(k) - 1}{2i k}
	  \;.
	  \label{eq:f_l}
	\eeq
	We will restrict ourselves to $l=0$ for simplicity; the generalization
	to higher $l$ is straightforward and will be considered later.

	To apply Eq.~\eqref{eq:phi_asymp} to negative energies, we
	analytically continue from real to (positive) imaginary $k$.  So,
	\bea
	   \phi_{0,ik_E}(r)
	       &\overset{r \gg R}\longrightarrow&
	         \frac{i}{2}\bigl(
		    \Jost_0(ik_E)e^{k_E r}-\Jost_0(-ik_E) e^{-k_E r} \bigr)
		 \nonumber \\
	       &=&
	       -\frac{i}{2}\Jost_0(-ik_E) \bigl(
	           e^{-k_E r} - \frac{\Jost_0(-ik_E)}{\Jost_0(ik_E)} e^{k_E r}
	         \bigr)
	       	\;,
	  \label{eq:asymptotic_form_phiE}
	\eea
	where $R$ is the range of the potential.  Upon comparing to
	Eq.~\eqref{eq:asymptotic_form_uE} we conclude that
	\beq
	  \alpha_E = -\frac{\Jost_0(ik_E)}{\Jost_0(-ik_E)}
	   = -\frac{1}{s_0(ik_E)} \;.
	  \label{eq:alphaE_and_S}
	\eeq
	Note that Eq.~\eqref{eq:alphaE_and_S} is consistent with the
	bound-state limit of Eq.~\eqref{eq:asymptotic_form_uE}: at a bound
	state where $\Einf = -\hbar^2 \kinf^2/2\mu$ there is a simple pole in the $S$
	matrix, which means $\alpha_E = 0$ as expected (no exponentially
	rising piece).

	From Ref.~\cite{taylor2006scattering} we learn that the residue as a
	function of $E$ of the partial wave amplitude $f_l(E)$ at the
	bound-state pole is $\displaystyle (-1)^{l+1} \ANC^2 \hbar^2/2 \mu$,
	where $\ANC$ is	the asymptotic normalization coefficient (ANC).  The ANC is
	defined by the large-$r$ behavior of the
	\emph{normalized} bound-state wave function:
	\beq
	  u_{\rm norm}(r)\overset{r\gg R}\longrightarrow \ANC e^{-\kinf r}
	  \;.
	  \label{eq:definition_of_ANC}
	\eeq
	Thus, near the bound-state pole (with $E = \hbar^2 k^2/2\mu$),
	\beq
	  f_0(k)  \approx  \frac{- \hbar^2\ANC^2}{2\mu (E-\Einf)}
	      =  \frac{-\ANC^2}{k^2 + \kinf^2} \;.
	\eeq
	or, using Eqs.~\eqref{eq:f_l} and \eqref{eq:alphaE_and_S},
	\beq
	  \alpha_E(k) \approx -\frac{k^2+\kinf^2}{k^2+\kinf^2-2 i k \ANC^2}
	  \;.
	  \label{eq:alphaE_k}
	\eeq
	Now,
	\beq
	  \left.\frac{d \alpha_E}{d E}
	  \right|_{\Einf}=\frac{d\alpha_E/dk\vert_{k=i\kinf}}
	  {dE/dk|_{k=i\kinf}}
	  \;,
	\eeq
	so using Eq.~\eqref{eq:alphaE_k} we find
	\beq
	  \left.\frac{d\alpha_E}{d k}\right|_{k=i\kinf}=\frac{-i}{\ANC^2}
	  \;,
	\eeq
	and therefore
	\beq
	  \left.\frac{d\alpha_E}{dE}\right|_{\Einf}=\frac{-\mu}{\hbar^2 \kinf \ANC^2}
	  \;.
	\eeq
	Putting it all together, we have
	\beq
	  \Delta E_L = \frac{\hbar^2 \kinf \ANC^2}{\mu} e^{-2 \kinf L}
	    + \mathcal{O}(e^{-4 \kinf L})
	    \;.
	    \label{eq:complete_IR_scaling}
	\eeq
	Equation~\eqref{eq:complete_IR_scaling} matches the result
	\beq
	E(L) = \Einf + A e^{-2 \kinf L} + \mathcal{O}(e^{-4 \kinf L})
	\label{eq:exp_L_extrapolation}
	\eeq
	in \cite{Furnstahl2012}, but now we have identified
	$\displaystyle A = \hbar^2\kinf\ANC^2/\mu$.

	In \cite{More:2013rma}, we advocated including second term in
	Eq.~\eqref{eq:entire_correction} for weakly bound states (small $\kinf$
	makes the term $\displaystyle \Ainf \frac{\mu}{\hbar^2}\frac{r}{\kinf}
	e^{-\kinf r}$
	non-negligible).  Including this term was also seen to give better prediction
	for weakly bound states like deuteron.  As we pointed out in
	\cite{Furnstahl:2013vda}, a better way to arrange an expression for
	$\Delta E$ is to have a systematic expansion in powers of $e^{-2 \kinf L}$.
 	Keeping the second term in Eq.~\eqref{eq:entire_correction}, generates terms
	in higher powers of $\mathcal{O}(e^{-2 \kinf L})$.  However, these higher
	order terms also arise from the $\mathcal{O}(\Delta E_L ^2)$ term in
	Eq.~\eqref{eq:linear_energy_approx} and to be consistent
	we need to take in account contributions up to a given order from both the
	sources (i.e., Eqs.~\eqref{eq:linear_energy_approx} and
	\eqref{eq:entire_correction}).  As we will see below, relating $k_L$
	($k_L$ is the binding momentum when we have a hard wall at length $L$)
	directly to the $S$-matrix allows us to transparently obtain systematic
	expansion for $\Delta E$ in powers of $e^{-2 \kinf L}$.

	\medskip
	\subsubsection{The $S$-matrix way}

	In \cite{Furnstahl:2013vda}, we returned to Eq.~\eqref{eq:asymptotic_form_uE}
	and noted that the bc uniquely fixed the coefficient
	$\alpha_E$.  We need $u_E(r = L) = 0$ which fixes
	\beq
	\alpha_E = -e^{-2 k_E L}\;.
	\label{eq:alpha_def}
	\eeq
	To make the $L$ dependence explicit, we modify the notation and let
	$k_L \equiv k_E$.  Comparing Eqs.~\eqref{eq:alpha_def} and
	\eqref{eq:alphaE_and_S}, we have
	\beq
	e^{-2 k_L L} = \left[s_0(i k_L)\right]^{-1} \;.
	\label{eq:basiceq}
	\eeq
	We then use appropriate parametrization for $s_0$ valid in the complex $k$
	region and solve the transcendental
	equation~\eqref{eq:basiceq} for $k_L$ and thereby find $E_L$.

	If the potential has no long-range part that introduces a singularity
	in the complex $k$ plane nearer to the origin than the bound-state
	pole (which is the case, for example, for the deuteron when we assume that
	the longest-ranged interaction is from pion exchange), then the
	continuation of the positive-energy partial-wave S-matrix (i.e., the
	phase shifts) to the pole should be unique.  Because $|k_L| <
	|\kinf|$, $s_0(i k_L)$ and therefore $k_L$ and the energy shift $E_L$
	should be determined solely by observables.

	The leading term in an expansion of $k_L - \kinf$ using
	Eq.~\eqref{eq:basiceq} comes from the bound-state pole, at which $s_0$
	behaves like~\cite{newton2002scattering}
	\beq
	  s_0(k) \approx \frac{-i\ANC^2}{k-i\kinf}
	  \;.
	  \label{eq:purepole}
	\eeq
	Note that $\ANC$ here is
	the asymptotic normalization coefficient (ANC) defined in
	Eq.~\eqref{eq:definition_of_ANC}.
	Substituting Eq.~\eqref{eq:purepole} into Eq.~\eqref{eq:basiceq} yields
	\beq
	  k_L - \kinf \approx -\ANC^2 e^{-2k_L L} \approx -\ANC^2 e^{-2\kinf L}
	  \;.
	  \label{eq:kLatLO}
	\eeq
	This is the leading-order (LO) result for $k_L$ obtained in
	Eq.~\eqref{eq:complete_IR_scaling}.  Note that in
	Eq.~\eqref{eq:complete_IR_scaling}, $\displaystyle \Delta E_L \equiv
	E_L - \Einf = \kinf^2/2 - k_L^2/2$.  We set $\displaystyle \hbar^2/ \mu = 1$.
	The notation $\kinf$ for the exact binding momentum make sense in this
	context, because in the exact case, the hard wall is at $L = \infty$.

	Iterations of the intermediate equation  in \eqref{eq:kLatLO} motivate the
	NLO parameterization of $k_L$ as
	\beq
	  k_L = \kinf + A e^{-2 \kinf L} + (B L + C) e^{-4 \kinf L}  +
		\mathcal{O}(e^{-6 \kinf L})
	  \;,
	  \label{eq:kLexpansion}
	\eeq
	with $A = -\ANC^2$.  In general we can substitute this expansion into
	Eq.~\eqref{eq:basiceq} using an parametrized form of the S-matrix,
	then expand in powers of $e^{-2 \kinf L}$ and equate $e^{-2 \kinf L}$,
	$L e^{-4 \kinf L}$, and $e^{-4 \kinf L}$ terms on both sides of the
	equation.  However, while both $A$ and $B$ are uniquely determined by
	the pole in $s_0(k)$ at $k=i\kinf$, $C$ is only determined unambiguously if
	$s_0(k)$ is consistently parameterized away from the pole.  For
	example, the two parametrizations
	\beq
	\label{eq:S0_1}
	    s_0(i k_L) \approx \frac{\kinf^2 - k_L^2 + 2 k_L \ANC^2}{\kinf^2 - k_L^2}
	\eeq
	and
	\beq
	  s_0(k) \approx \frac{-\ANC^2}{2\kinf}\,\frac{k+i\kinf}{k- i\kinf}
	  \label{eq:s0_Newton}
	\eeq
	yield different results for $C$. The first
	parametrization~(Eq.~\eqref{eq:S0_1}) is based on a particular form for the
	partial-wave scattering amplitude near the
	pole~\cite{taylor2006scattering}, and was employed in
	Ref.~\cite{More:2013rma}.  The second paramerization
	(Eq.~\eqref{eq:s0_Newton}) correctly incorporates that the S-matrix also has
	a zero at $-i\kinf$~\cite{newton2002scattering}.  In neither case,
	however, do we have a sufficiently general parametrization that allows
	us to unambiguously determine $C$.

	For the complete NLO energy correction, we start from the
	general expression for the S-matrix
	\beq
	  s_0(k) = \frac{k\,\cot\delta_0(k) + ik}{k\,\cot\delta_0(k) -ik}
	  \;,
	  \label{eq:s0_exact}
	\eeq
	and use an effective range expansion to substitute for $k\,\cot
	\delta_0(k)$.  In particular, we use an expansion
	around the bound-state pole rather than about zero energy,
	namely~\cite{wu2011scattering,Phillips:1999hh},
	 \beq
	  k\,\cot\delta_0(k) = -\kinf + \frac12 \rho_d(k^2 + \kinf^2)
	     + w_2(k^2+\kinf^2)^2 +  \cdots
	     \;.
	     \label{eq:eff_range_kinf}
	\eeq
	To match the residue at the S-matrix pole as in Eq.~\eqref{eq:purepole},
	we identify
	\beq
	 \rho_d = \frac{1}{\kinf} - \frac{2}{\ANC^2}
	  \;.
	  \label{eq:rhoD_gamma_rel}
	\eeq
	$w_2$ is a low-energy observable like $\ANC$ and $\kinf$.
	Now we substitute Eq.~\eqref{eq:eff_range_kinf} into Eq.~\eqref{eq:s0_exact}
	and use Eq.~\eqref{eq:kLexpansion} to expand both sides of
	Eq.~\eqref{eq:basiceq}, equating terms with equal powers of
	$e^{-2\kinf L}$ and $L$.  The resulting expansion for the binding
	momentum to NLO is
	\begin{align}
    [k_L]_{\rm NLO} &=  \kinf - \ANC^2 e^{-2\kinf L}  - 2  L \ANC^4
		e^{-4\kinf L}
    \nonumber  \\
    &\null - \ANC^2 \left(1 - \frac{\ANC^2}{2\kinf} - \frac{\ANC^4}{4\kinf^2}
		+ 2 \kinf w_2 \ANC^4 \right) e^{-4\kinf L} \;.
    \label{eq:complete_k_correction_NLO}
  \end{align}
	Using $\Delta E_L \equiv E_L - \Einf = \kinf^2/2 - k_L^2/2$,
	the correction for the energy due to finite $L$ is
	\begin{align}
  	[\Delta E_L]_{\rm NLO} &=  \kinf \ANC^2 e^{-2\kinf L}
  	+ 2\kinf L\ANC^4 e^{-4\kinf L}
    \nonumber \\
  	& \null\quad +
    \kinf\ANC^2 \Bigl( 1-\frac{\ANC^2}{\kinf}-\frac{\ANC^4}{4\kinf^2}
    + 2\kinf w_2 \ANC^4  \Bigr)  e^{-4\kinf L} \;.
    \label{eq:complete_E_correction_NLO}
	\end{align}
	In what follows we use LO to refer to the first term in this expansion and
	L-NLO to
	refer to the first two terms (the second term should dominate the full
	NLO expression when $\kinf L$ is large).  We also note that
	higher-order terms in Eq.~\eqref{eq:eff_range_kinf} (e.g., terms
	proportional to $(k^2+\kinf^2)^3$ and higher powers) do not affect the
	binding momentum or energy predictions
	Eqs.~\eqref{eq:complete_k_correction_NLO} and
	\eqref{eq:complete_E_correction_NLO} at NLO.

	As a special case, let us consider the zero-range limit of a
	potential. In this case $\rho_d = w_2 = 0$, $\ANC^2 = 2\kinf$, and
	\beq
	   [s_0(ik_L)]^{-1} = \frac{\kinf-k_L}{\kinf+k_L}\;.
	\eeq
	The expansion for $k_L$ in a form similar to Eq.~\eqref{eq:kLexpansion}
	can be extended to arbitrary order using Eq.~\eqref{eq:basiceq}.

	We note finally that the leading corrections beyond NLO scale as $L^2
	e^{-6\kinf L}$. While we do not pursue a derivation of such high-order
	corrections here, the knowledge of the leading form is useful in some of
	the error analysis we present in Subsec.~\ref{subsec:pudding_proof}.

	\medskip
	\subsubsection{Differential method}

	Because we seek the change in energy with respect to a cutoff, it is
	natural to formulate the problem in the spirit of renormalization
	group methods by seeking a flow equation for the bound-state energy as
	a function of $L$.  Such an approach is already documented in the
	literature, for example in Refs.~\cite{Arteca1984} and
	\cite{Fernandez1981}, and it provides us with an alternative method that
	does not directly reference the S-matrix.  The basic equation is
	\beq
	  \frac{\partial E_L}{\partial L} = -\frac12 \frac{|u'_L(L)|^2}
		{\int_0^L |u_L(r)|^2\, dr}
	  \;.
	  \label{eq:dEdL}
	\eeq
	Here the prime denotes a derivative with respect to $r$.  Given an
	expression for the right-hand side in terms of observables ($\kinf$,
	$\ANC$, and so on) and $L$, we can simply integrate to find the energy
	correction for a bc at $L$
	\beq
	  \Delta E_L \equiv E_L - \Einf = \int_{\Einf}^{E_L}\! dE\,
	     = \int_\infty^L\! \frac{\partial E_L}{\partial L} dL
	     \;.
	\eeq
	To derive Eq.~\eqref{eq:dEdL}, we start with
	\beq
	  \frac{\partial}{\partial L}\left [
	   \int_0^L u_L(r) H u_L(r)\, dr = E_L \int_0^L\! dr\, u_L(r)^2
	  \right]
	  \;,
	\eeq
	which yields (after some cancellations)
	\beq
	    \frac12 \left.\left( \frac{\partial u_L(r)}{\partial r}
	         \frac{\partial u_L(r)}{\partial L}
	         \right)\right|_0^L
	   =
	  \frac{\partial E_L}{\partial L} \int_0^L\! dr\, u_L(r)^2
	         \;.
	         \label{eq:DeltaEL}
	\eeq
	The left-hand side is a surface term from partially integrating the
	kinetic energy in $H$.  The lower limit vanishes because $u_L(0) = 0$
	for any $L$.  Finally, we replace the partial derivative with respect
	to $L$ at the upper limit using
	\beq
	   \frac{\partial u_L(L)}{\partial L} = - \frac{\partial u_L(L)}{\partial r}
	   \;,
	\eeq
	which follows from expanding $u_{L'}(L') = 0$ about $u_{L}(L) = 0$ for
	$L' = L + \Delta L$.

	To apply Eq.~\eqref{eq:dEdL}, we start with $u_L(r)$ in the asymptotic
	region, as given by
	\beq
  	u_L(r) \overset{r \gg R}{\longrightarrow}  \left(e^{-k_L r} - e^{-2k_L L}
		e^{k_L r}\right)
   	\;.
   	\label{eq:uLasymp2}
  \eeq
	The normalization
	constant $\gamma_L$ is chosen so that the integral of $u_L(r)^2$ from
	0 to $L$ is unity; it becomes the ANC $\ANC$ as
	$L\rightarrow\infty$.  Thus
	\beq
	   u_L'(L) = -2 \gamma_L k_L e^{-k_L L}
	   \;.
	   \label{eq:uLprime}
	\eeq
	Now we need to expand $k_L$ and $\gamma_L$ about $\kinf$ and $\ANC$,
	respectively.  The leading term is trivial: $k_L \rightarrow \kinf$
	and $\gamma_L \rightarrow \ANC$, so the only $L$ dependence in
	$u_L'(L)^2$ is in $e^{-2 \kinf L}$ and the integration in
	\eqref{eq:dEdL} is immediate:
	\begin{align}
	\Delta E_L
	     = \int_\infty^L\! \frac{\partial E_L}{\partial L} dL
	     =  -2 \ANC^2 \kinf^2 \int_\infty^L\! e^{-2\kinf L} \, dL
	     =  \kinf \ANC^2 e^{-2 \kinf L} +  \mathcal{O}(e^{-4 \kinf L})
	     \;.
	\end{align}
	This is the same LO result for $\Delta E_L$ found by other methods.

	To go to NLO we need an expression for $\gamma_L$.  In the zero-range
	(zr) limit, $\gamma_L$ is given completely in terms of $k_L$ using the
	normalization condition (because the asymptotic form in
	Eq.~\eqref{eq:uLasymp2} holds
	over the entire range of the integral)
	\begin{align}
	  \gamma_L^2 &= \left[\int_0^L\! dr\, (e^{-k_L r} - e^{-2k_L L}e^{k_L r})^2
		\right]^{-1}
	  =
	  2 k_L(1 + 4 k_L L e^{-2 k_L L}) + \mathcal{O}(e^{-4 k_L L}) \;.
	\end{align}
	We expand $k_L$ everywhere in Eq.~\eqref{eq:dEdL} using
	Eq.~\eqref{eq:uLprime} and our LO result
	\beq
	   k_L = \kinf (1 - 2 e^{-2 \kinf L}) \;.
	\eeq
	Here, we neglected terms that are $\mathcal{O}(e^{-6 \kinf L})$ or smaller.
	We need to expand $e^{-2 k_L L}$ in $u_L'(L)$ to get
	\beq
	   e^{-2 k_L L} = e^{-2 \kinf L} (1 + 4 \kinf L e^{-2 \kinf L})
	    + \mathcal{O}(e^{-6 \kinf L}) \;.
	\eeq
	(Elsewhere it suffices to replace $e^{-2 k_L L}$ by $e^{-2 \kinf L}$ to NLO.)
	So we find that
	\begin{align}
	  \frac{\partial E_L}{\partial L} &= -\frac12 (4 \gamma_L^2 k_L^2
		e^{-2 k_L L})
	  \nonumber \\
	  &\approx -2  [2\kinf (1 - 2 e^{-2 \kinf L})(1 + 4 \kinf L
		e^{-2 \kinf L})]
	  \nonumber \\
	  & \  \null \times [\kinf^2 (1 - 4 e^{-2\kinf L})][e^{-2 \kinf L}
		(1 + 4 \kinf L e^{-2 \kinf L})]
	  \nonumber \\
	  &\approx
	  -4 \kinf^3 e^{-2 \kinf L} -  8 \kinf^3 (4\kinf L - 3)e^{-4 \kinf L}
	  + \mathcal{O}(e^{-6 \kinf L}) \;,
	\end{align}
	and then finally
	\begin{align}
	[\Delta E_L]_{\rm zr, NLO}
	     &= \int_\infty^L\! \frac{\partial E_L}{\partial L} dL
	     \nonumber \\
	     &=
	     2 \kinf^2 e^{-2 \kinf L} +  4 \kinf^2 (2\kinf L - 1)e^{-4 \kinf L}
	     + \mathcal{O}(e^{-6 \kinf L})
	     \;,
	\end{align}
	in agreement with Eq.~\eqref{eq:complete_E_correction_NLO}
	with $\ANC^2 = 2\kinf$ and $w_2 = 0$.
	We can take this procedure to higher order by
	using a more general expansion for $k_L$.

	To extend the differential method
	to higher order for nonzero range, we must parametrize
	$\gamma_L$ to account for the part of the integration within the
	range of the potential; e.g., in terms of the effective range.  However,
	we have not found a clear advantage in doing this
	compared to the straightforward S-matrix method.

	\subsection{The proof is in the pudding}
	\label{subsec:pudding_proof}

	In this subsection we will test the Eq.~\eqref{eq:complete_E_correction_NLO}
	for various test models and for the deuteron.  Note that for the cases
	that we test Eq.~\eqref{eq:complete_E_correction_NLO}, the exact answer
	$\Einf$ is already known (either by exact analytical calculation or by using
	large number of basis states).  So we can compare how good the prediction from
	Eq.~\eqref{eq:complete_E_correction_NLO} is by comparing to exact answer.
	In cases where the exact values for $\Einf$ and $\ANC$ are not known, our
	approach suggests that we can use the exponential in $L$ fit of
	Eq.~\eqref{eq:exp_L_extrapolation} to extract $\Einf$ and $\kinf$.

	Based on the results
	presented in Subsec.~\ref{subsec:tale_of_tails}, we use $L_2$ in all our
	further analyses.
	It is important that we isolate the IR corrections in making these
	tests.  The truncation in the HO basis also introduces an
	ultraviolet error inversely proportional to the ultraviolet cutoff
	$\Lambda_{\rm UV} \approx \sqrt{2 \mu \hbar \Omega (N+3/2)}$.  In the
	results here we use combinations of $\hbar \Omega$ and $N$ values such
	that the UV error in each case can be neglected compared to the IR
	error.

\begin{figure}[h]
	\centering
	\includegraphics[width=0.6\textwidth]
	{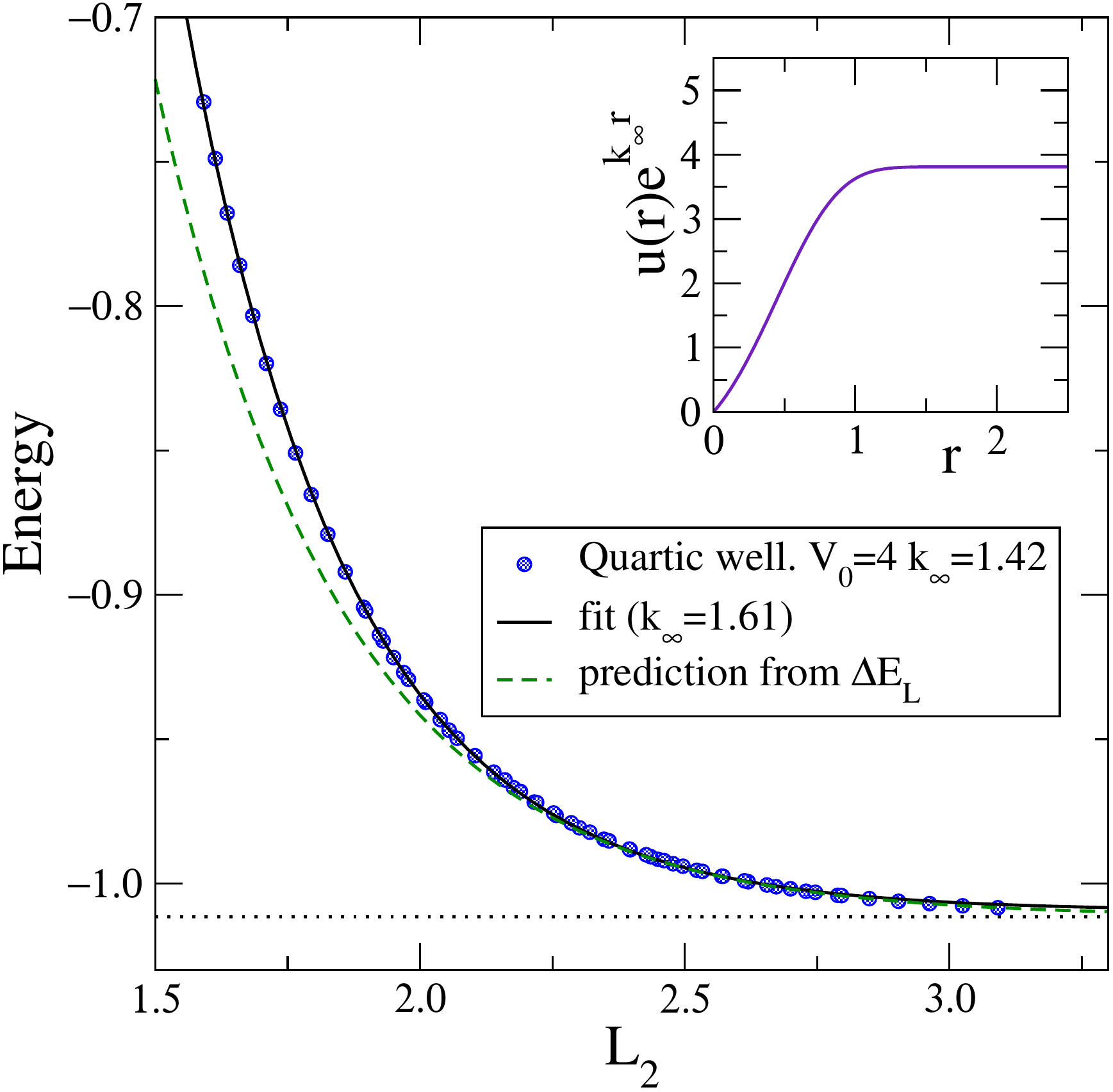}
  \caption{Energy versus $L_2$ for a quartic potential
    well Eq.~\eqref{eq:Vq} for a wide range of $N$ and $\hw$
    (circles) ($\hbar = \mu = R=1$).  The solid line is a fit to
		Eq.~\eqref{eq:exp_L_extrapolation}
    with $A$, $\kinf$ and $\Einf$ as fit parameters while the dashed line is
		the prediction from Eq.~\eqref{eq:complete_IR_scaling}.
    The horizontal line is the exact energy, $\Einf=-1.0115$.
    The inset illustrates the calculation of the asymptotic
    normalization coefficient (ANC) from the (normalized) wave
    function.}
		\label{fig:IR_quartic_inset_ANC}
	\end{figure}

	For each of the model potentials, the radial Schr\"odinger equation is
	accurately solved numerically in coordinate space for the energy,
	which yields $\kinf$, and the wave functions.  The asymptotic
	normalization coefficient $\ANC$ is found by multiplying the wave
	function by $e^{\kinf r}$ and reading off its asymptotic value.  This
	is illustrated in the inset of Fig.~\ref{fig:IR_quartic_inset_ANC},
	which also shows the onset of the plateau that defines the asymptotic
	region in $L_2$ where we expect our correction formulas to hold.  For
	the deuteron, the Hamiltonian is diagonalized in
	momentum space to find $\kinf$, and then an extrapolation to the pole
	is used to find the $s$-wave and $d$-wave ANCs~\cite{Amado:1979zz}.
	In the present subsection we use only the $s$-wave ANC for the deuteron.

	The derivations in Subsec.~\ref{subsec:make_cash} imply that the
	energy corrections should have the same exponential form and
	functional dependence on the radius $L$ at which the wave
	function is zero, independent of the potential and for
	any bound state.
	Here we make some representative tests of a direct fit of
	Eq.~\eqref{eq:exp_L_extrapolation} in comparison to applying
	Eq.~\eqref{eq:complete_IR_scaling}.

	Figure~\ref{fig:IR_quartic_inset_ANC} shows results for a quartic
	potential with a moderate depth.  The fit to
	Eq.~\eqref{eq:exp_L_extrapolation} is very good over a large range in $L_2$
	for which the energy changes by 30\%, and the prediction for $\Einf$
	is accurate to 0.2\%.  However, the fit value of $\kinf$ is 1.61
	compared to the exact value of 1.42.  The dashed curve shows the
	prediction from Eq.~\eqref{eq:complete_IR_scaling} using the exact
	$\kinf$ and $\ANC$.  It is evident that the approximation is very good
	above $L_2 > 2$ but increasingly deviates at smaller $L_2$.

	\begin{figure}[h]
	\centering
	\includegraphics[width=0.6\textwidth]
	{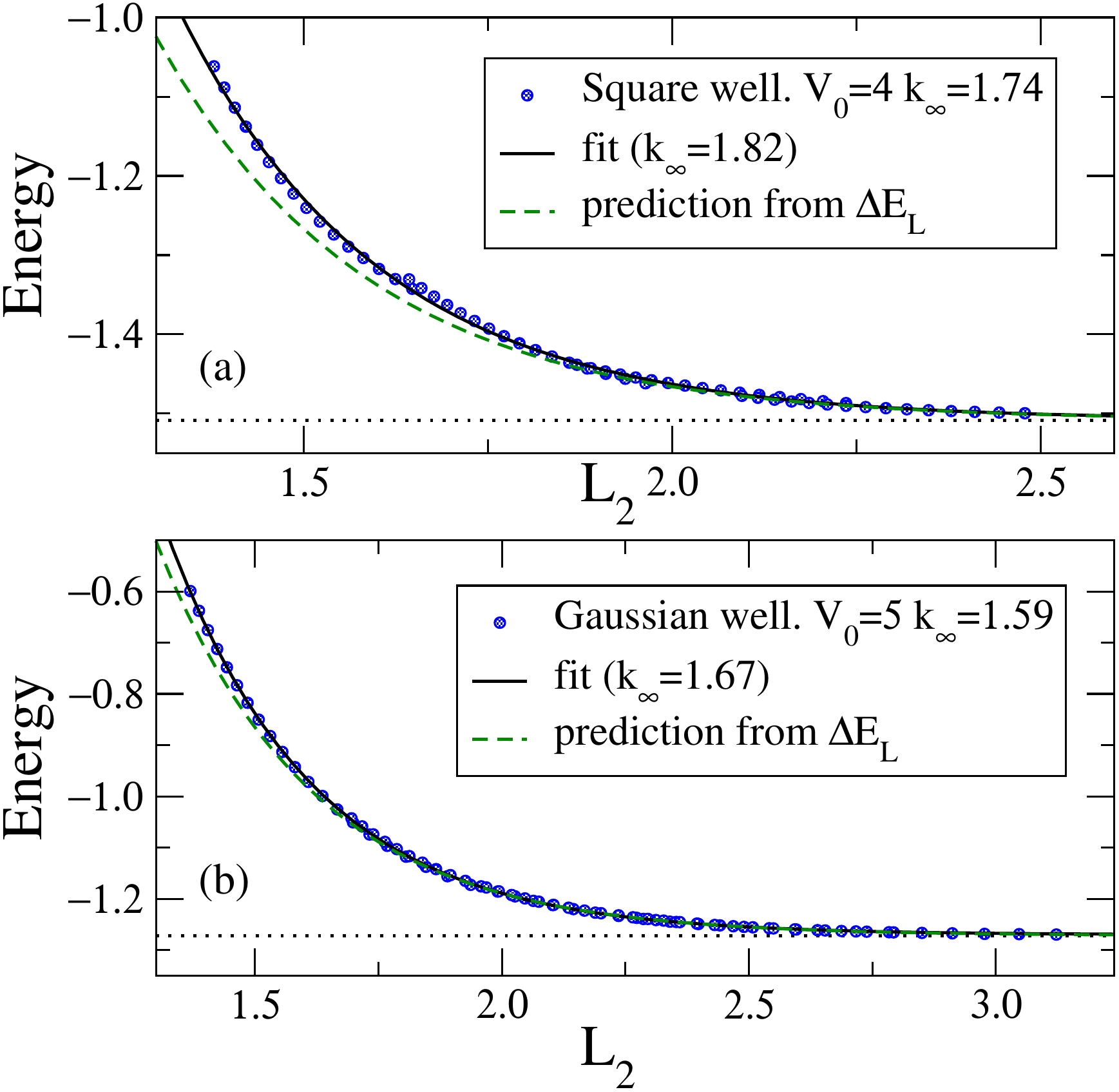}
	\caption{Energy versus $L_2$ for moderate-depth (a) square
  	well Eq.~\eqref{eq:Vsw} and for (b) Gaussian potential
  	well Eq.~\eqref{eq:Vg} ($\hbar = \mu = R=1$) for a wide range of $N$ and
		$\hw$ (circles).  The solid line is a fit to
		Eq.~\eqref{eq:exp_L_extrapolation} with $A$, $\kinf$ and $\Einf$ as
  	fit parameters while the dashed line is the
  	prediction from Eq.~\eqref{eq:complete_IR_scaling}.  The horizontal dotted
		lines are the exact energies;
   	square well: $\Einf = -1.5088$, Gaussian well: $\Einf = -1.2717$}
	\label{fig:universal_sq_gauss_wells}
	\end{figure}

	In Fig.~\ref{fig:universal_sq_gauss_wells}, examples are shown for
	square well and Gaussian potentials with a moderate depth.  Again we
	find a good fit to an exponential fall-off in $L_2$, but in these
	cases not only are the energies well predicted (again to better than
	0.2\%) but the fit values of $\kinf$ are within 5\% of the exact results.
	\begin{figure}[h]
	\centering
	\includegraphics[width=0.6\textwidth]
	{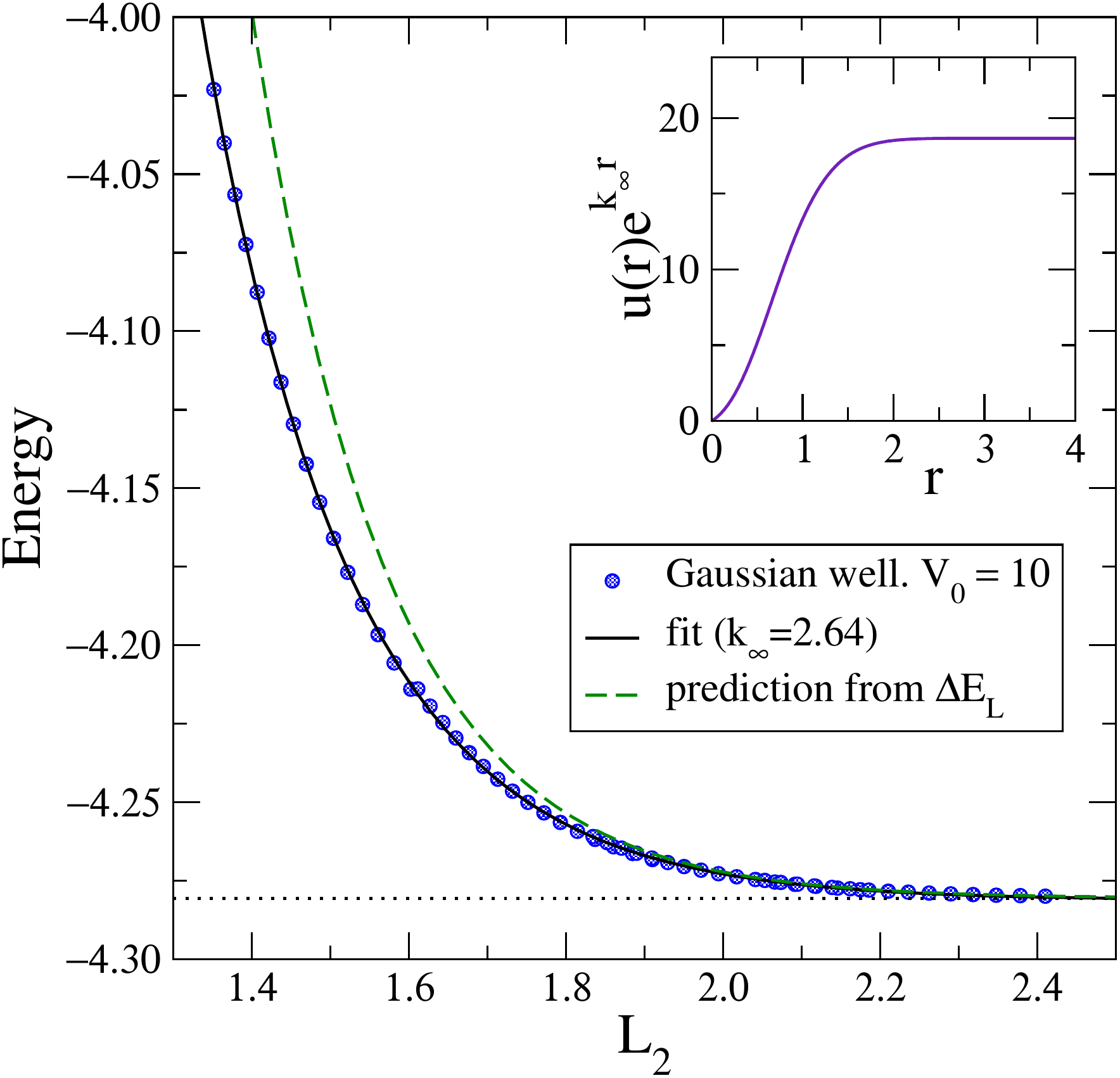}
	\caption{Energy versus $L_2$ for the deeply bound
  	ground state of a Gaussian potential for a wide range of $N$ and
  	$\hw$ (circles) ($\hbar = \mu = R = 1$).  These are compared to the
		prediction of
	  Eq.~\eqref{eq:complete_IR_scaling} (dashed).  The solid line
	  is a fit to Eq.~\eqref{eq:exp_L_extrapolation} with $A$, $\kinf$ and
		$\Einf$ as fit parameters.  The horizontal dotted line
	  is the exact energy, $\Einf=-4.2806$.}
	\label{fig:deep_gauss_ANC_inset}
	\end{figure}
	\begin{figure}[h]
	\centering
	\includegraphics[width=0.6\textwidth]
	{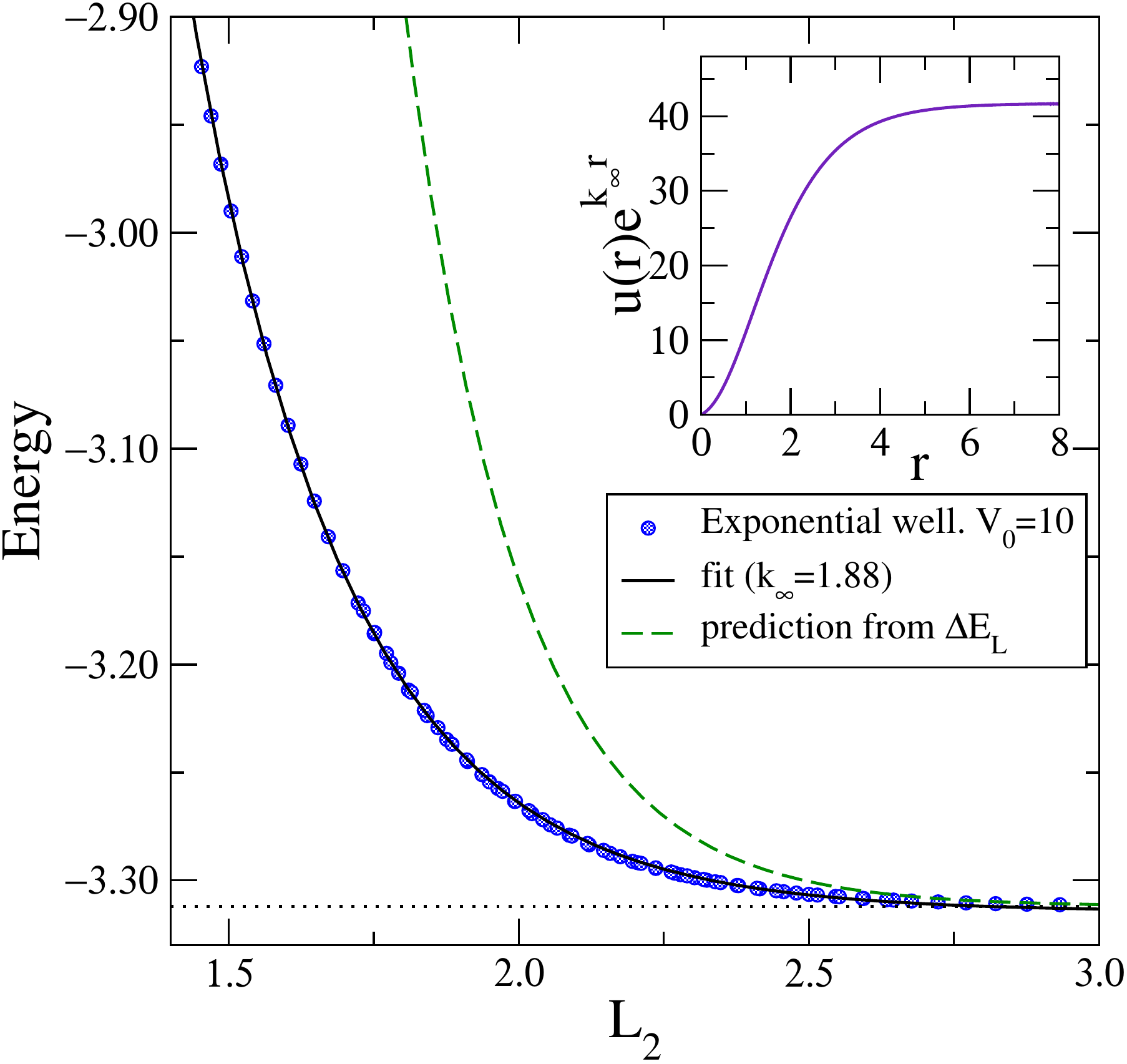}
	\caption{Energy versus $L_2$ for the deeply bound
	  ground state of an exponential potential well for a wide range of
	  $N$ and $\hw$ (circles) ($\hbar = \mu = R = 1$).  These are compared to the
		predictions of Eq.~\eqref{eq:complete_IR_scaling} (dashed).  The solid line
	  is a fit to Eq.~\eqref{eq:exp_L_extrapolation} with $A$, $\kinf$ and
		$\Einf$ as fit parameters.  The horizontal dotted line
	  is the exact energy, $\Einf=-3.3121$.}
	\label{fig:deep_exp_ANC_inset}
	\end{figure}
	For deeply bound states, Eq.~\eqref{eq:complete_IR_scaling} fails
	for a different reason.  The error in
	Eq.~\eqref{eq:complete_IR_scaling} is proportional to $e^{-4 \kinf L}$,
	so one might expect that the prediction to become increasingly
	accurate as the state becomes more bound.  However, as seen in
	Figs.~\ref{fig:deep_gauss_ANC_inset} and \ref{fig:deep_exp_ANC_inset},
	results for deep Gaussian and exponential potential wells do not match
	this expectation.  In deriving the energy corrections we used
	the asymptotic form of the wave functions.  This is valid only in the
	region $r \gg R$, where $R$ is the range of the potential.  The
	potentials at the smaller values of $L_2$ shown in the figures are not
	negligible.  Indeed, it is evident from the insets in
	Figs.~\ref{fig:deep_gauss_ANC_inset} and \ref{fig:deep_exp_ANC_inset}
	that we are not in the asymptotic region for those values of $L$.
	The lesson is that when applying the IR extrapolation schemes
	discussed in the present paper we need to make sure that the two
	conditions for its applicability are fulfilled.  First, we need $N$
	sufficiently large
	for $L_2$ to be the correct box size
	(see Table~\ref{tab:L0_L2_k_min_comparison}).
	Second we need $L_2$ to be the
	largest length scale in the problem under consideration.

	\begin{figure}[h]
	\centering
	\includegraphics[width=0.6\textwidth]
	{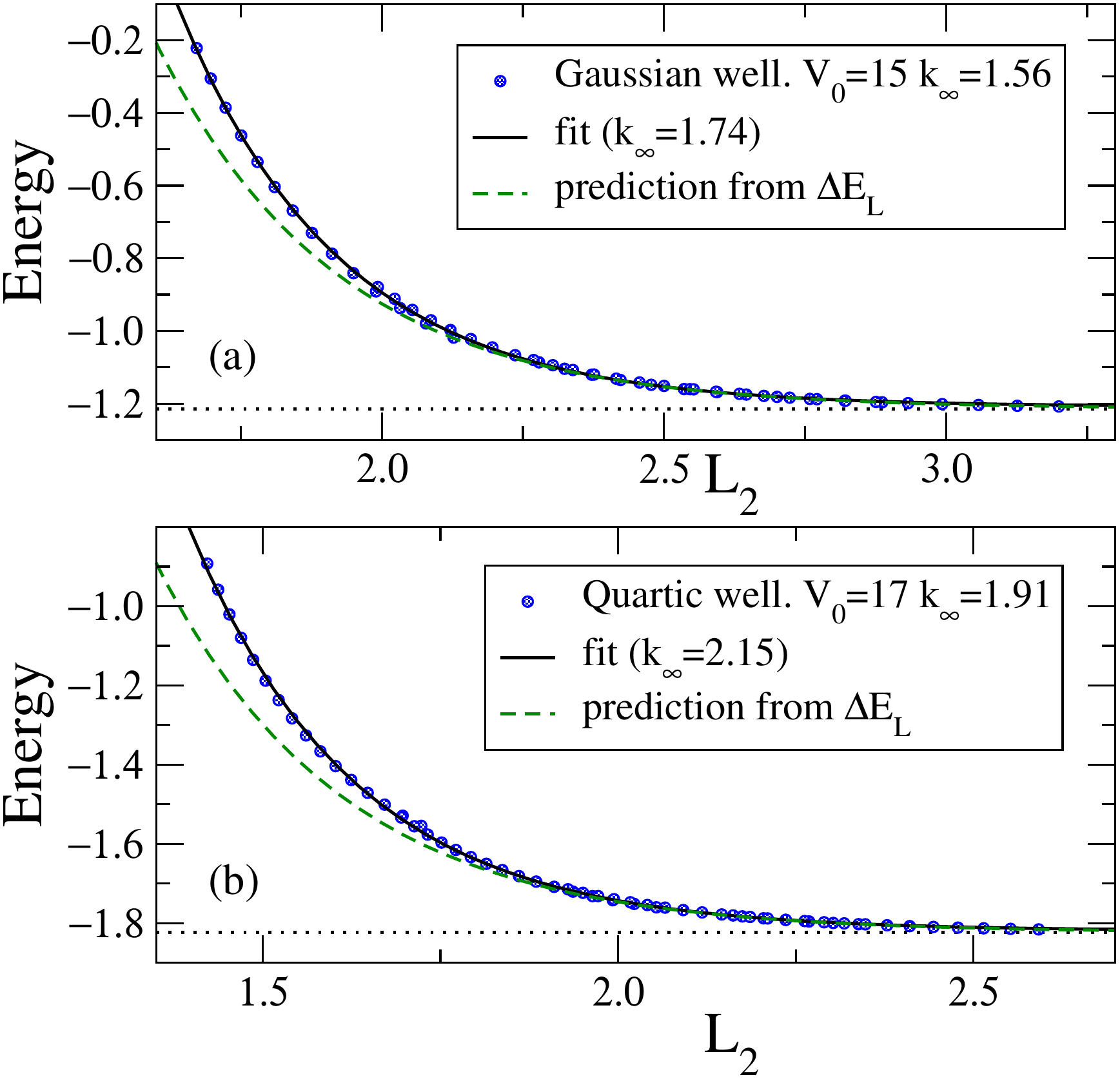}
	\caption{Energy versus $L_2$ for the first excited
	  states of deep (a) Gaussian Eq.~\eqref{eq:Vg} and (b) quartic
	  Eq.~\eqref{eq:Vq} potential wells for a wide range of $N$ and $\hw$
	  (circles) ($\hbar = \mu = R=1$).  The solid line is a fit to
		Eq.~\eqref{eq:exp_L_extrapolation}
	  with $A$, $\kinf$ and $\Einf$ as fit parameters while the dashed line is
		the prediction from Eqs.~\eqref{eq:complete_IR_scaling}.
	  The horizontal dotted lines are the exact energies for the first excited
		states; Gaussian well: $\Einf = -1.2147$, quartic well: $\Einf = -1.8236$}
	  \label{fig:IR_excited_gauss_quartic_wells}
	\end{figure}
	The results so far are for the ground state of the potential.  However,
	the derivations in the Subsec.~\ref{subsec:make_cash} should
	also hold for excited states.  This is so because the generalization of
	the results in Subsec.~\ref{subsec:tale_of_tails} shows that
	$(j\pi/L_2)^2$ is a very good approximation to the $j^{\rm th}$
	eigenvalue of the operator $p^2$ for $j \ll N$.  In
	Fig.~\ref{fig:IR_excited_gauss_quartic_wells} representative results
	for excited states from two model potentials are shown.  We find the
	same systematics as with the ground-state results: the exponential fit
	works very well but the extracted $\kinf$ is only correct at about the 10\%
	level.  In assessing the success of	Eq.~\eqref{eq:complete_IR_scaling}, we
	note that these excited states in deep potentials are comparable to the
	ground states in moderate-depth potentials shown in
	Fig.~\ref{fig:universal_sq_gauss_wells}.  The discussion there applies
	here as well, namely that the prediction from
	Eq.~\eqref{eq:complete_IR_scaling} is very good at large $L_2$, but
	increasingly deviates at smaller $L_2$.

	\begin{figure}[h]
	\centering
	\includegraphics[width=0.6\textwidth]
	{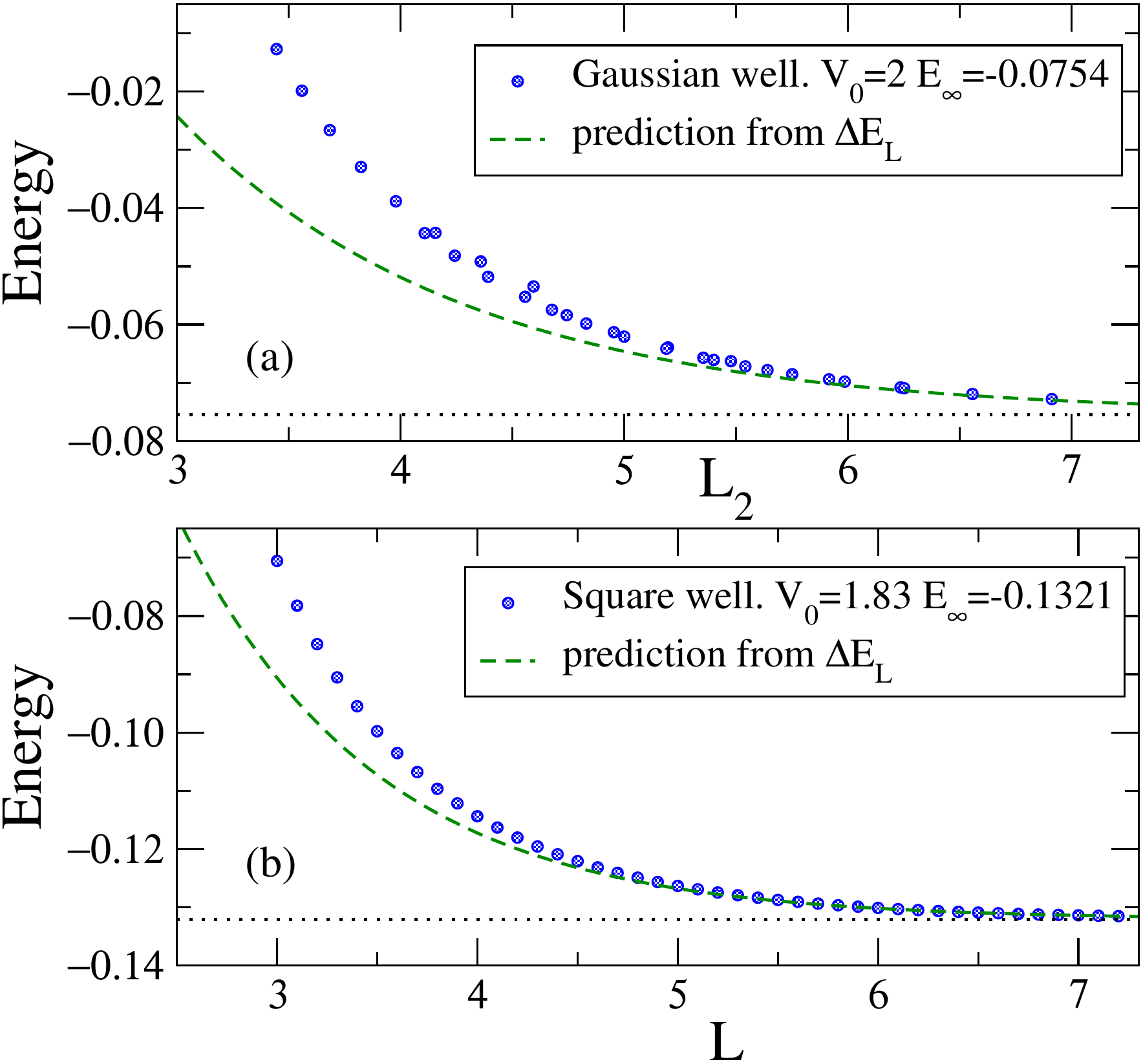}
	\caption{ (a) Ground-state energy versus $L_2$ for
    model Gaussian potential. (b) Energy versus $L$
    for the square well.  The energies for the square well are from solving
		the	Schr\"odinger equation
    exactly with a Dirichlet bc on wave functions at
    $r=L$. The dashed line is the prediction from
    Eqs.~\eqref{eq:complete_IR_scaling}.  The depths of these model potentials
    are chosen so that the scaled
    energies (with $\hbar = \mu = R=1$) are the same as the deuteron binding
		energy.}
	\label{fig:prediction_deuteron_equivalent}
	\end{figure}
	The case of weakly bound states is of special interest because of the
	correspondence to deuteron which is also a weakly bound shallow state.
	Figure~\ref{fig:prediction_deuteron_equivalent} (a) shows ground-state
	energies for many different $N$ and $\hw$ versus $L_2$ using Gaussian model
	potentials whose parameters are chosen so that the energies are the
	same as the deuteron binding energy (scaled to units with $\hbar=1$,
	$\mu=1$, $R=1$).
	In Fig.~\ref{fig:prediction_deuteron_equivalent} (b)
	we have a `deuteron-like' square well. The energies in this case are
	obtained by	solving the Schr\"odinger equation exactly with a Dirichlet
	bc on wave
	functions at $r=L$.  The prediction from Eq.~\eqref{eq:complete_IR_scaling}
	fails to reproduce the data except at the highest values of $L_2$.
	Eq.~\eqref{eq:complete_IR_scaling} keeps only the leading-order (LO)
	corrections	to $\Delta E$.  For the weakly bound states, the higher-order
	corrections become important.  The corrections to the energy up to the next
	to leading order (NLO) were noted in Eq.~\eqref{eq:complete_E_correction_NLO}.
	Note that the NLO correction involves the low-energy constant $w_2$.  In
	certain cases, this constant $w_2$ can be calculated and $\Delta E$ up to NLO
	can be obtained.  Now we turn to the test of
	Eq.~\eqref{eq:complete_E_correction_NLO}.

	\medskip
	\subsubsection{NLO and systematics of the correction}

	For the square-well potential of Eq.~\eqref{eq:Vsw}, the parameters in
	Eq.~\eqref{eq:complete_E_correction_NLO} can be calculated easily.
	The $S$-wave scattering phase shift for	the square well is
	\beq
	 \delta_0 (k) =
	  \tan^{-1}\left[\sqrt{{k^2 \over k^2 + \eta^2}} \tan(\sqrt{k^2+\eta^2}
	  R)\right] - k R\;,
	  \label{eq:delta0}
	\eeq
	with $\eta = \sqrt{2 V_0}$. 	Analytically continuing the effective
	range expansion by taking $k \rightarrow i k_L$
	in Eqs.~\eqref{eq:eff_range_kinf} and \eqref{eq:delta0}, we obtain
  \begin{multline}
	{i k_L \sqrt{\eta^2 - k_L^2} -k_L^2 \tan( \sqrt{\eta^2 - k_L^2} R)
	\tan(i k_L R) \over i k_L \tan(\sqrt{\eta^2 - k_L^2} R) -
	\sqrt{\eta^2 - k_L^2} \tan(i k_L R) }
	=  \\
	-\kinf
	+ \frac12{\rho_d} (\kinf^2-k_L^2) + w_2 (\kinf^2 - k_L^2)^2
	+ \mathcal{O}\left((\kinf^2-k_L^2)^3\right)	\;.
	\label{eq:get_w2_sq_well}
	\end{multline}
	The branch for the square-root is fixed by the requirement that
	$\tan \delta (i\kinf) = -i$.  Note from Eq.~\eqref{eq:s0_exact} that
	this builds in the requirement that the $S$-matrix has a pole at
	$i \kinf$.  To get $\rho_d$ ($w_2$) we differentiate once (twice) each
	side of Eq.~\eqref{eq:get_w2_sq_well} with
	respect to $k_L$ and then set $k_L = \kinf$.  The $\rho_d$ obtained in
	this way is consistent with Eq.~\eqref{eq:rhoD_gamma_rel} when $\ANC$
	is obtained by the large $r$ behavior of the bound-state wave function
	as defined in Eq.~\eqref{eq:definition_of_ANC}.

	The square well with a Dirichlet bc at $L > R$ can be solved analytically.
	The wave functions inside and outside the square well are
	\beq
  u^<_L(r) = C \sin\kappa_L r \,,
  \qquad
  u^>_L(r) = D(e^{-k_L r} - e^{-2 k_L L} e^{+k_L r})
  \;,
	\eeq
	which builds in the boundary condition $u^>_L(L) = 0$.
	The interior wave number $ \kappa_L = \sqrt{\eta^2 - k_L^2}$ and
	$k_L = \sqrt{2 |E_L|}$.
	Matching the logarithmic derivatives at $r=R$ for $E=\Einf$ yields
	\beq
	  \kappa_\infty \cot{\kappa_\infty R} = -\kinf
	\eeq
	and with the boundary condition at $L$ we get:
	\beq
	 \kappa_L \cot \kappa_L R = - k_L \frac{1 + e^{-2k_L (L-R)}}
	 {1-e^{-2k_L (L-R)}} \;.
	 \label{eq:sq_well_matchingL}
	\eeq
	We expand both sides of Eq.~\eqref{eq:sq_well_matchingL} in powers of
	\beq
	  \Delta k \equiv k_L - \kinf\;.
	  \label{eq:Delta_k_def}
	\eeq
	We write the left-hand side of Eq.~\eqref{eq:sq_well_matchingL} as
	\beq
	\kappa_L \cot \kappa_L R = \kappainf \cot \kappainf R + \mathcal{A}
	(\Delta k)  + \mathcal{B} (\Delta k)^2+\cdots \;,
	\label{eq:LHS_matchingL_expansion}
	\eeq
	and obtain the coefficients $\mathcal{A}$, $\mathcal{B}$ by Taylor
	expanding $\kappa_L	\cot (\kappa_L R)$ around $\kinf$.
	We write $\Delta k$ as
	\beq
	\label{eq:Delta_k_def2} \Delta k = k_{(1)} + k_{(2)} + \cdots \;.
	\eeq
	Here $k_{(1)} \sim e^{-2 \kinf L}$ is the LO correction, $k_{(2)}
	\sim e^{-4\kinf L}$ is the NLO correction and so on, and we truncate the
	expressions consistently to obtain the energy correction for the
	square well to the desired order.  The results of the general S-matrix
	and square-well-only Taylor expansion methods of calculating energy
	corrections are found to match explicitly at LO, \LNLO, and NLO.
	We remind the reader that we use \LNLO~to denote terms proportional
	to $L e^{-4 \kinf L}$.

	\begin{figure}[h]
	\centering
	\includegraphics[width=0.6\textwidth]
	{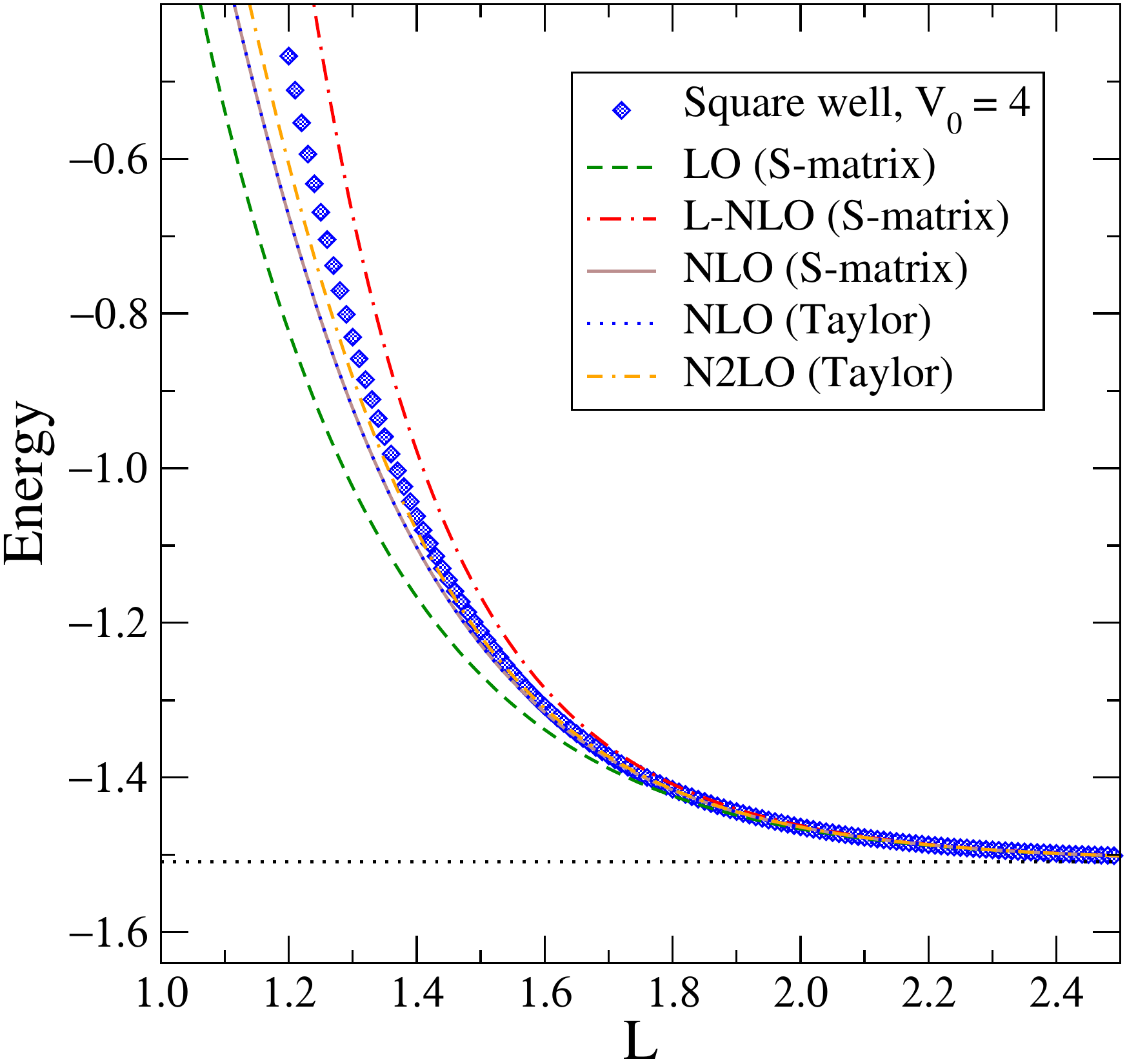}
	\caption{Bound-state energy for a square well of depth
	  $V_0=4$ (lengths are in units of $R$ and energies in units of
	  $1/R^2$ with $\hbar^2/\mu = 1$) from solving the Schr\"odinger
	  equation with a Dirichlet bc at $r=L$.  The diamonds
	  are exact results for each $L$ while the horizontal dotted line is
	  the energy for $L\rightarrow\infty$, $\Einf = -1.5088$.  The dashed,
	  dot-dashed and solid lines are predictions for the energy using the
	  systematic correction formula
	  Eq.~\eqref{eq:complete_E_correction_NLO} at LO (first term only),
	  \LNLO\ (first two terms), and full NLO (all terms), respectively.
	  The dotted curve on top of the solid line and the dot-double-dashed
	  lines are respectively the NLO and N2LO predictions for the square
	  well from the Taylor expansion using Eqs.~\eqref{eq:sq_well_matchingL}
		and \eqref{eq:LHS_matchingL_expansion}.}
	\label{fig:sqwell_curvesV4}
	\end{figure}
	\begin{figure}[h]
	\centering
	\includegraphics[width=0.6\textwidth]
	{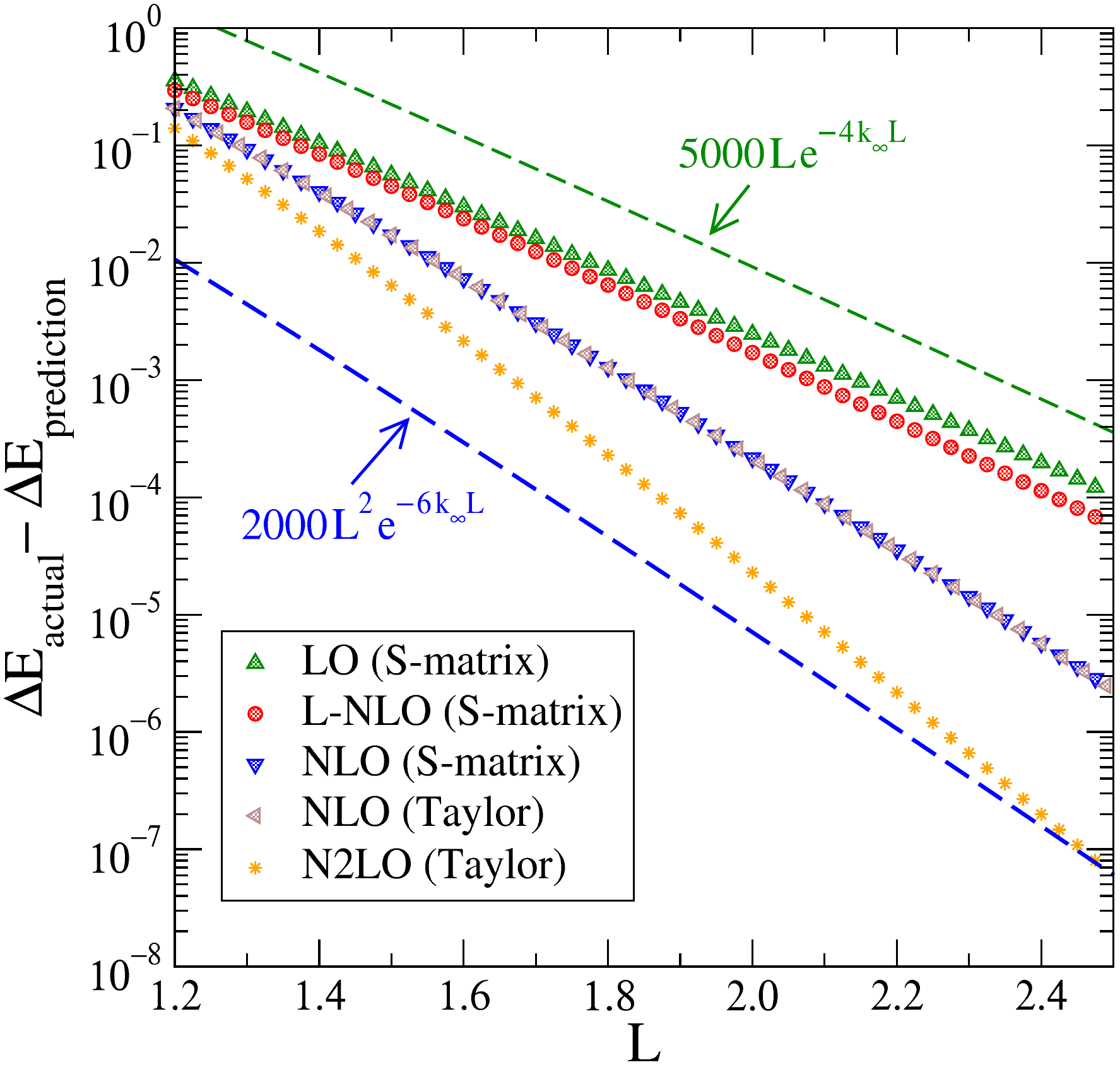}
	\caption{Error plots of the energy correction at each
	  $L$ for the square well of Fig.~\ref{fig:sqwell_curvesV4} ($V_0 =
	  4$) predicted at different orders by
	  Eq.~\eqref{eq:complete_E_correction_NLO} and by the Taylor expansion
	  method, each compared to the exact energy.
	  Lines proportional to $L e^{-4\kinf L}$ (dashes) and $L^2 e^{-6\kinf L}$
		(with arbitrary normalization) are plotted for comparison to
	  anticipated error slopes. }
	\label{fig:sq_well_error_log_plotsV4}
	\end{figure}
	Figure~\ref{fig:sqwell_curvesV4} compares the energy corrections
	for the general S-matrix method at LO and NLO for a representative
	square-well potential with one bound state to the exact energies.  The
	Taylor expansion results for the square well at NLO and N2LO (which is
	proportional to $e^{-6\kinf L}$) are also plotted.  We note that the
	predictions are systematically improved as higher-order terms are
	included and that keeping terms only up to \LNLO~overestimates the
	energy correction.  Also as seen in Fig.~\ref{fig:sqwell_curvesV4},
	the full NLO energy correction predicted by
	Eq.~\eqref{eq:complete_E_correction_NLO}, with $w_2$ determined by
	Eq.~\eqref{eq:get_w2_sq_well}, matches the `exact' NLO result obtained
	by Taylor expansion.  This confirms that
	Eq.~\eqref{eq:complete_E_correction_NLO} is indeed the complete energy
	correction at NLO.

	To see if the errors decrease with the implied systematics, we plot
	the difference of actual energy corrections and the energy corrections
	predicted at different orders on a log-linear scale in
	Fig.~\ref{fig:sq_well_error_log_plotsV4}.  We observe that the errors
	successively decrease at each fixed $L$ as we go from LO to NLO to
	N2LO.  The up triangles in Fig.~\ref{fig:sq_well_error_log_plotsV4} are
	$\Delta E_{\rm actual} - \Delta E_{\rm LO}$.  From
	Eq.~\eqref{eq:complete_E_correction_NLO} the dominant omitted
	correction in $\Delta E_{\rm LO}$ is proportional to $L e^{-4 \kinf L}$.
	As seen in Fig.~\ref{fig:sq_well_error_log_plotsV4}, the slope
	of $\Delta E_{\rm actual} - \Delta E_{\rm LO}$ is roughly $L
	e^{-4 \kinf L}$, as expected.  We also note that $\Delta E_{\rm \LNLO}$
	is only a marginal improvement over $\Delta E_{\rm LO}$ and that
	$\Delta E_{\rm actual} - \Delta E_{\rm NLO}$ has the expected slope of
	$L^2 e^{-6 \kinf L}$.  We again see a perfect agreement between the
	results obtained from the S-matrix method
	(Eq.~\eqref{eq:complete_E_correction_NLO}) and those obtained from the
	Taylor expansion of Eq.~\eqref{eq:sq_well_matchingL}.  We have also
	studied deeper square wells with more than one bound state
	and verified that our results apply even in the presence of excited
	states.

	\begin{figure}[h]
	\centering
	\includegraphics[width=0.55\textwidth]
	{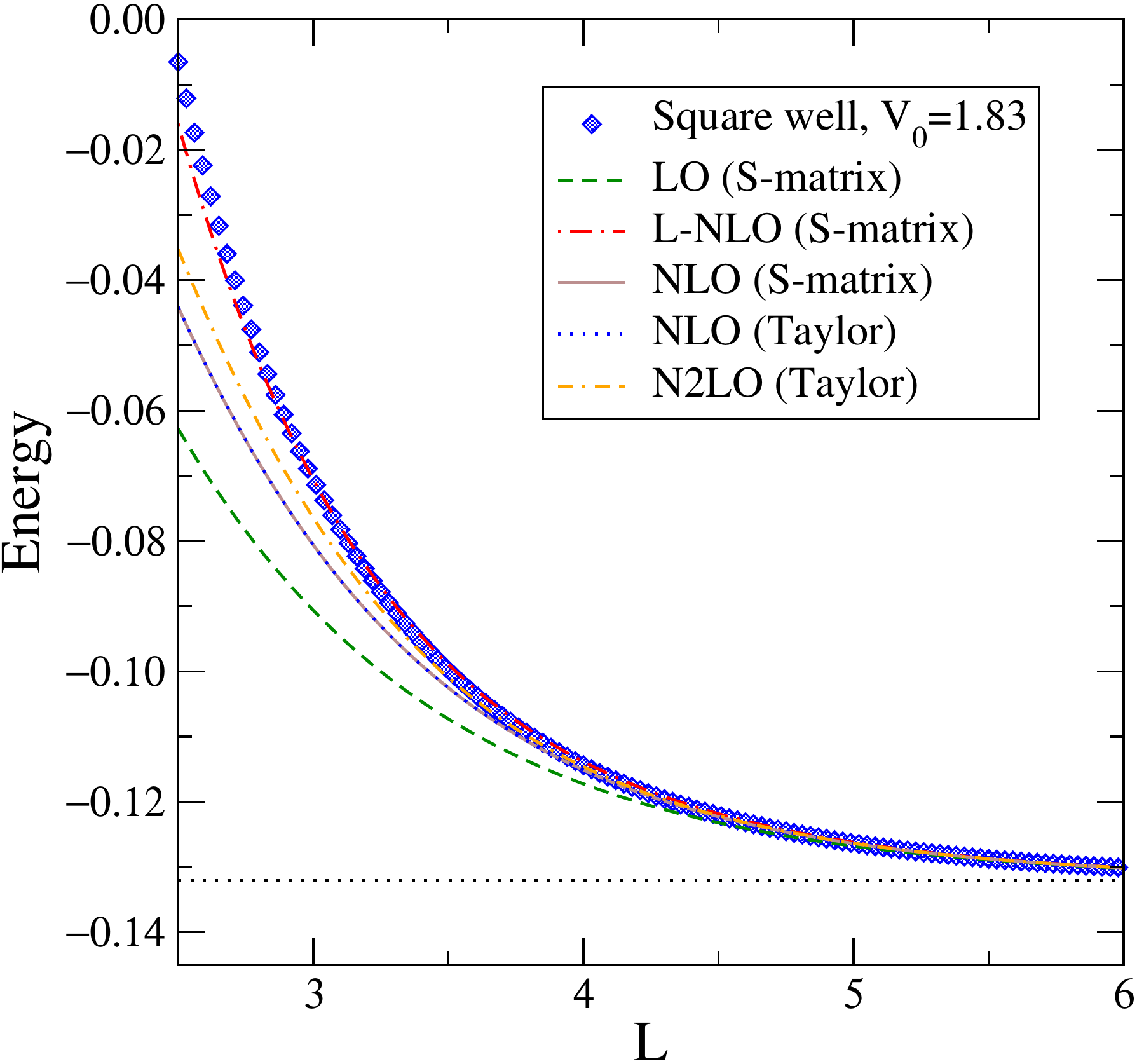}
	\caption{Bound-state energy for a square well of depth
	  $V_0=1.83$ (units with $R=1$), which simulates a deuteron, from
	  solving the Schr\"odinger equation with a Dirichlet bc at $r=L$.
		The horizontal dotted line is the exact energy,
	  $\Einf = -0.1321$ and the other curves are as the same as in
	  Fig.~\ref{fig:sqwell_curvesV4}.}
	\label{fig:sqwell_curves}
	\end{figure}
	\begin{figure}[h]
	\centering
	\includegraphics[width=0.55\textwidth]
	{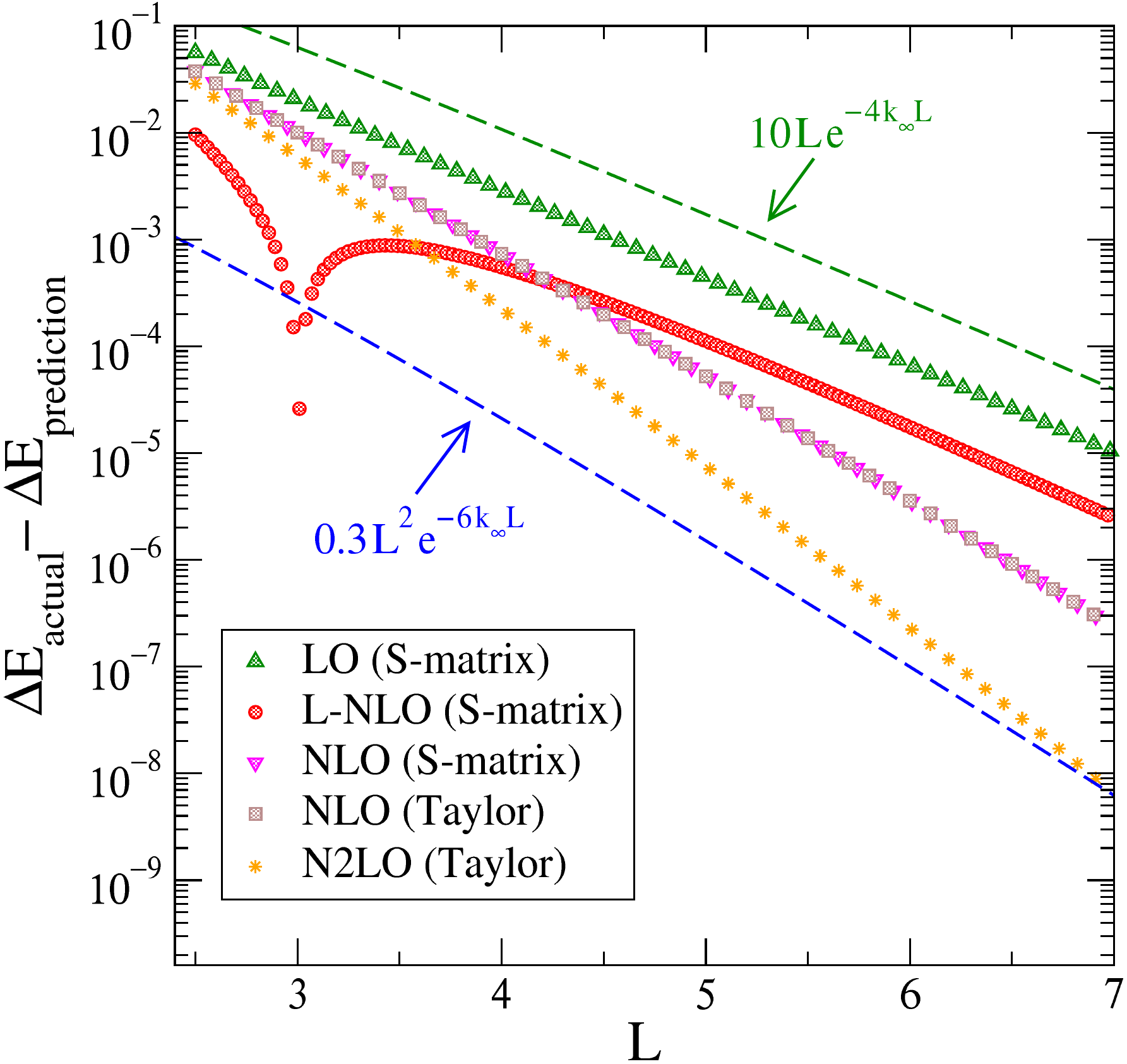}
	\caption{Comparison of the actual energy correction due
	  to truncation to the energy correction predicted to different orders
	  by Eq.~\eqref{eq:complete_E_correction_NLO} for a square well
	  (Eq.~\eqref{eq:Vsw}) with $V_0 = 1.83$ and $R=1$.}
	\label{fig:sq_well_error_log_plots}
	\end{figure}
	In Figs.~\ref{fig:sqwell_curves} and
	\ref{fig:sq_well_error_log_plots}, the same analysis is done but now
	with the depth of the square well adjusted so that the exact binding
	energy is the same as the deuteron binding energy scaled to the units
	$\hbar=1$, $\mu=1$ and $R=1$. An important difference in this case
	compared to the deeper square well is that the \LNLO~prediction gives
	a very close estimate for the truncated energies.  However as seen in
	Fig.~\ref{fig:sq_well_error_log_plots}, the improvement achieved by
	the \LNLO~prediction is not systematic with $L$.  At large $L$,
	$\Delta E-\Delta E_{\rm \LNLO}$ has the same slope as $\Delta E -
	\Delta E_{\rm LO}$ and is not the dominant NLO correction.  In this
	regard, the proximity of \LNLO~ prediction to the actual data in
	Fig.~\ref{fig:sqwell_curves} should not be over-emphasized.
	\begin{figure}[h]
	\centering
	\includegraphics[width=0.6\textwidth]
	{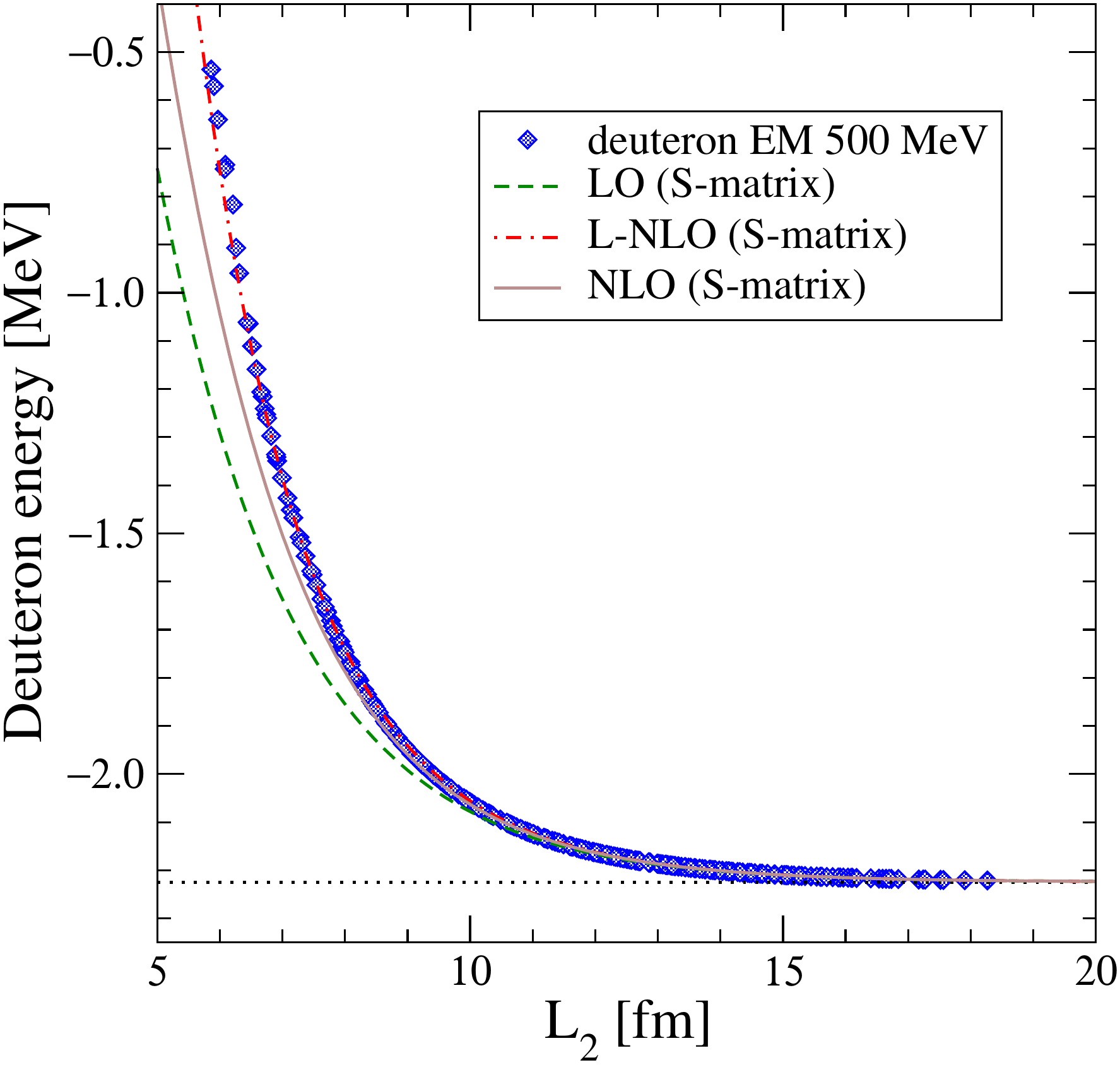}
	\caption{Deuteron energy versus $L_2$ (see
	  Eq.~\eqref{eq:L2_def}) for the chiral N$^3$LO (500\,MeV) potential
	  of Ref.~\cite{Entem:2003ft}.  To eliminate the UV contamination we
	  only plot results for $\hw > 49$~MeV.  The dashed, dot-dashed and solid
	  lines are respectively the LO (first term in
	  Eq.~\eqref{eq:complete_E_correction_NLO}), \LNLO~(first two terms in
	  Eq.~\eqref{eq:complete_E_correction_NLO}) and the full NLO (all the
	  terms in Eq.~\eqref{eq:complete_E_correction_NLO}) predictions for
	  the energy correction.  The horizontal dotted line is the deuteron
	  energy.}
	\label{fig:deuteron_curves}
	\end{figure}
	\begin{figure}[h]
	\centering
	\includegraphics[width=0.6\textwidth]
	{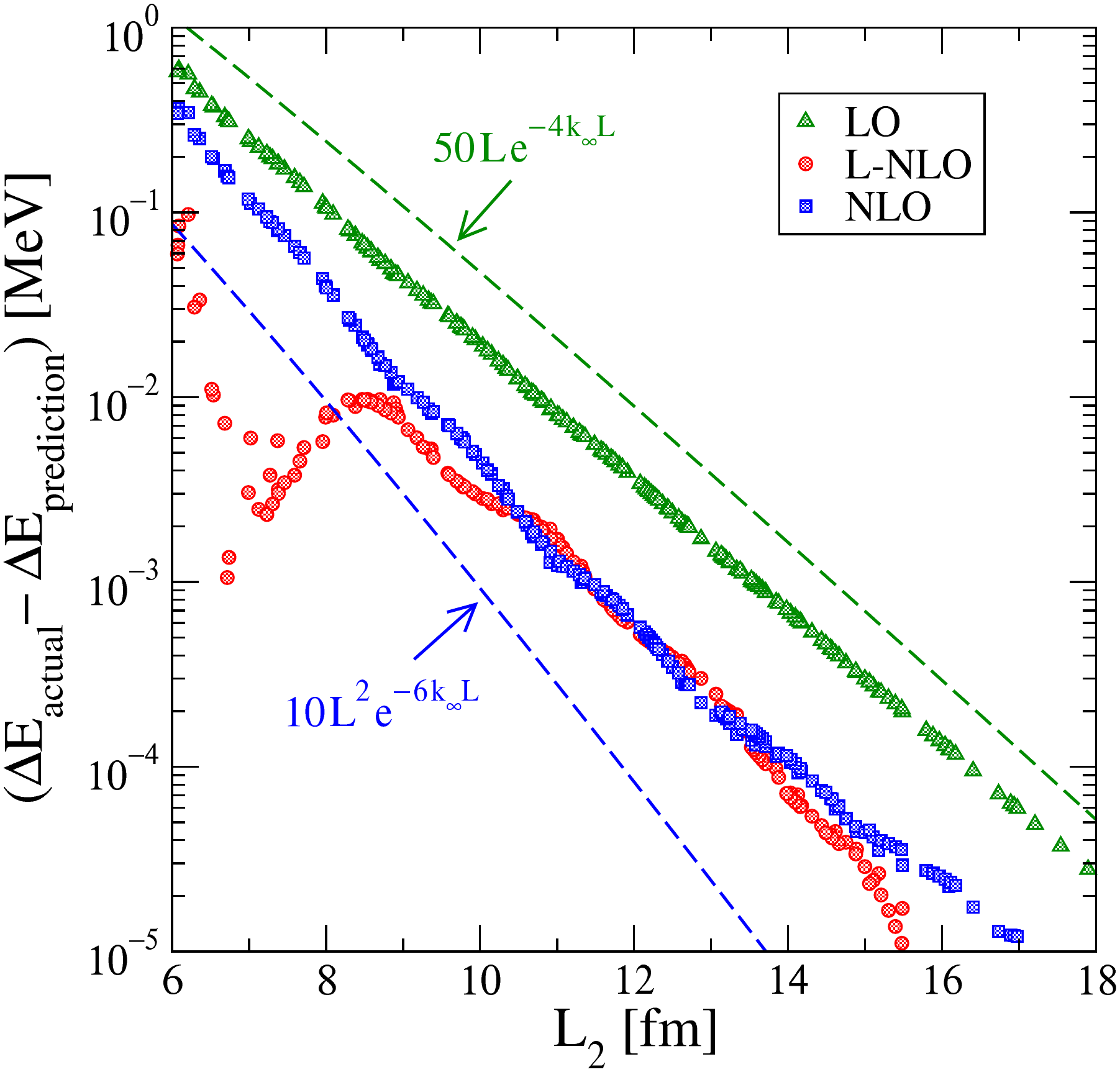}
	\caption{Comparison of the actual energy correction due
	  to HO basis truncation ($\hw$ restricted to be greater than $49$~MeV to
	  eliminate UV contamination) for the deuteron to the energy
	  correction predicted to different orders from
	  Eq.~\eqref{eq:complete_E_correction_NLO}.  For the parameter $w_2$ in
	  Eq.~\eqref{eq:complete_E_correction_NLO} we use the value reported
	  in \cite{Phillips:1999hh}.}
	\label{fig:deuteron_error_log_plots}
	\end{figure}
	Figures~\ref{fig:deuteron_curves} and
	\ref{fig:deuteron_error_log_plots} show analogous results for the
	deuteron calculated with the chiral EFT potential of
	Ref.~\cite{Entem:2003ft}.  We use the HO basis and predict the ($l=0$) energy
	correction from Eq.~\eqref{eq:complete_E_correction_NLO} assuming a
	Dirichlet bc at $L_2$ given by Eq.~\eqref{eq:L2_def}.  We only include
	energies for which $\hw > 49$~MeV, which is sufficient to render UV
	corrections negligible.  For the parameter $w_2$ in
	Eq.~\eqref{eq:complete_E_correction_NLO} we use $w_2
	= 0.389$ as reported in \cite{Phillips:1999hh}.  We also note that the
	$\rho_d$ value reported in \cite{Phillips:1999hh} satisfies
	Eq.~\eqref{eq:rhoD_gamma_rel}, where $\ANC$ now is the $s$-wave ANC.
	The $y$-axis minimum is dictated by the limited precision of
	the ANC and $w_2$ values.
	We notice again that the close agreement of the \LNLO~prediction to
	the deuteron data is not systematic while the full corrections to the
	LO and NLO predictions have the anticipated slopes except at large
	$L_2$.  In the next Subsec.~\ref{subsec:higher_angular_momenta} we
	extend our formulas to $l>0$, which
	enables us to include contributions from the $d$-wave at LO.  This
	becomes noticable on the error plot for large $L_2$ (see
	Fig.~\ref{fig:deuteron_error_plot_l2}).
	\begin{figure}
	\centering
 	\includegraphics[width=0.6\columnwidth]
	{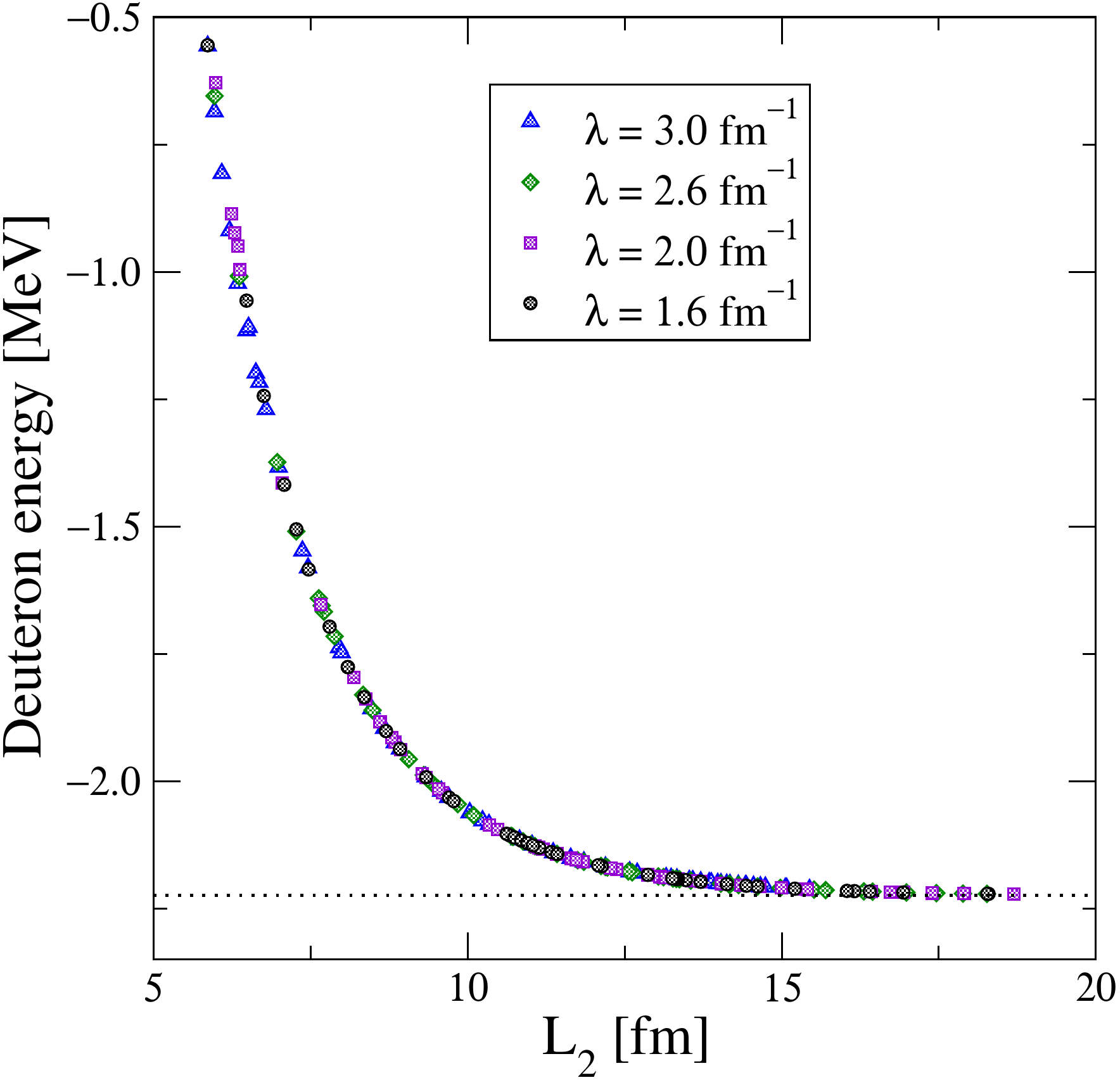}
 	\caption{Deuteron energy versus $L_2$ for the
  	potential of Ref.~\cite{Entem:2003ft} evolved by the SRG to four
   	different resolutions (specified by $\lambda$).  To eliminate the
   	UV contamination we only keep points for which $\hw >40$.  The
   	horizontal dotted line is the deuteron binding energy.}
	\label{fig:SRG_data_collapse}
	\end{figure}

	As a final test of the universal applicability of the correction
	formula Eq.~\eqref{eq:complete_E_correction_NLO}, we consider a sequence
	of unitarily equivalent	potentials for the deuteron.  In particular,
	we use the similarity renormalization group (SRG)~\cite{Bogner:2009bt}
	to evolve the initial Entem-Machleidt
	potential to four values of the SRG evolution parameter $\lambda$.
	Because the transformation is exactly unitary (up to very small
	numerical errors) at the two-body level, the measurable quantities
	such as phase shifts, bound-state energies, and ANCs are unchanged.
	From Eq.~\eqref{eq:complete_E_correction_NLO} or more generally from
	Eq.~\eqref{eq:basiceq}, we see that the IR energy correction can written
	in terms of observables ($S$-matrix near the bound state can be
	parametrized in terms of low-energy observables) and therefore should be
	independent of the SRG scale.  This is verified in
	Fig.~\ref{fig:SRG_data_collapse}.

	\begin{figure}
		\centering
	 \includegraphics[width=0.6\textwidth]
	 {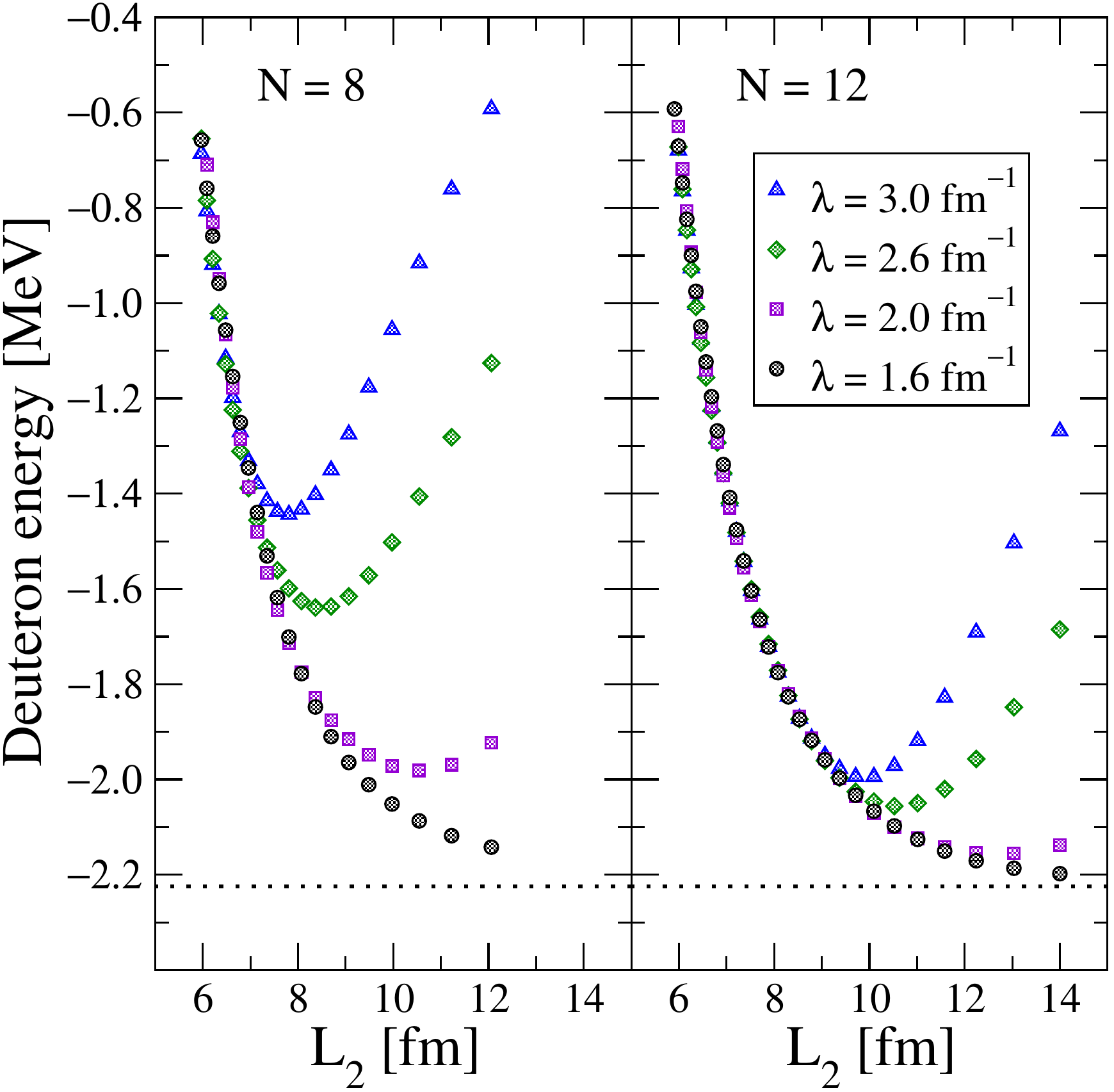}
	 \caption{The same SRG-evolved potentials as in
	   Fig.~\ref{fig:SRG_data_collapse} are used to generate energies, but
	   with $N$ fixed at (a) 8 and (b) 12 and no restriction on
	   $\hw$.  Thus UV corrections are not negligible everywhere.  The
	   horizontal dotted line is the deuteron binding energy.}
	 \label{fig:SRG_data_collapse_contamination}
	\end{figure}
	As $\lambda$ decreases, the SRG systematically reduces the coupling
	between high-momentum and low-momentum potential matrix elements, thereby
	lowering the effective UV cutoff.  Thus these potentials are useful
	tools to assess the role of UV corrections.  This is exploited in
	Fig.~\ref{fig:SRG_data_collapse_contamination} where we relax the condition
	that the UV corrections are small compared to IR corrections.  In
	particular, we fix $N$ at 8 and 12 and scan through the full range of
	$\hw$.  We observe that with increasing $L_2$, each of the curves with
	a given $\lambda$ eventually deviates from the universal curve,
	first with $\lambda = 3.0\fmi$ and then later with decreasing
	$\lambda$ or with higher $N$.  We can understand this in terms of the
	behavior of the induced UV cutoff.  For fixed $N$, Eq.~\eqref{eq:L2_def}
	tells us that increasing $L_2$ means increasing $b$ (or decreasing
	$\hw$).  But at fixed $N$, $\LamUV \propto 1/b$, so the UV cutoff will
	be decreasing and the corresponding UV energy correction increasing.
	Thus the curves at fixed $\lambda$ correspond to the curves
	seen in conventional plots of energy versus $\hw$ (e.g., see
	Ref.~\cite{Bogner:2007rx}).  The softer potentials (lower $\lambda$)
	will have lower intrinsic UV cutoffs and therefore they are only
	affected for larger $L_2$. The minima for each $\lambda$ are when IR
	and UV corrections are roughly equal.

	\medskip
	\subsubsection{Comparison to conventional extrapolation schemes}

	The extrapolation formulas in Eqs.~\eqref{eq:complete_IR_scaling} and
	\eqref{eq:complete_E_correction_NLO} with $L=L_2$ are theoretically founded.
	Thus the functional form for the energy extrapolation that we have is an
	exponential in $L$.  As mentioned in the introduction to this chapter,
	a popular	phenomenological choice is an exponential in $N$ extrapolation
	\big(Eq.~\eqref{eq:exp_Nmax_extrapolation}\big).  From
	Eq.~\eqref{eq:L2_def} we see that exponential in $N$ extrapolation
	corresponds to gaussian in $L$.  Authors of
	Ref.~\cite{Tolle:2012cx} investigated the convergence properties of
	genuine and smeared contact interactions in
	an effective theory of trapped bosons and found that the smearing
	changed a power law dependence of the convergence to an exponential
	dependence.  Here we will consider all three functional dependences on
	$L$: exponential, Gaussian, and power law.
	\begin{figure}[h]
	\centering
	\includegraphics[width=0.6\textwidth]
	{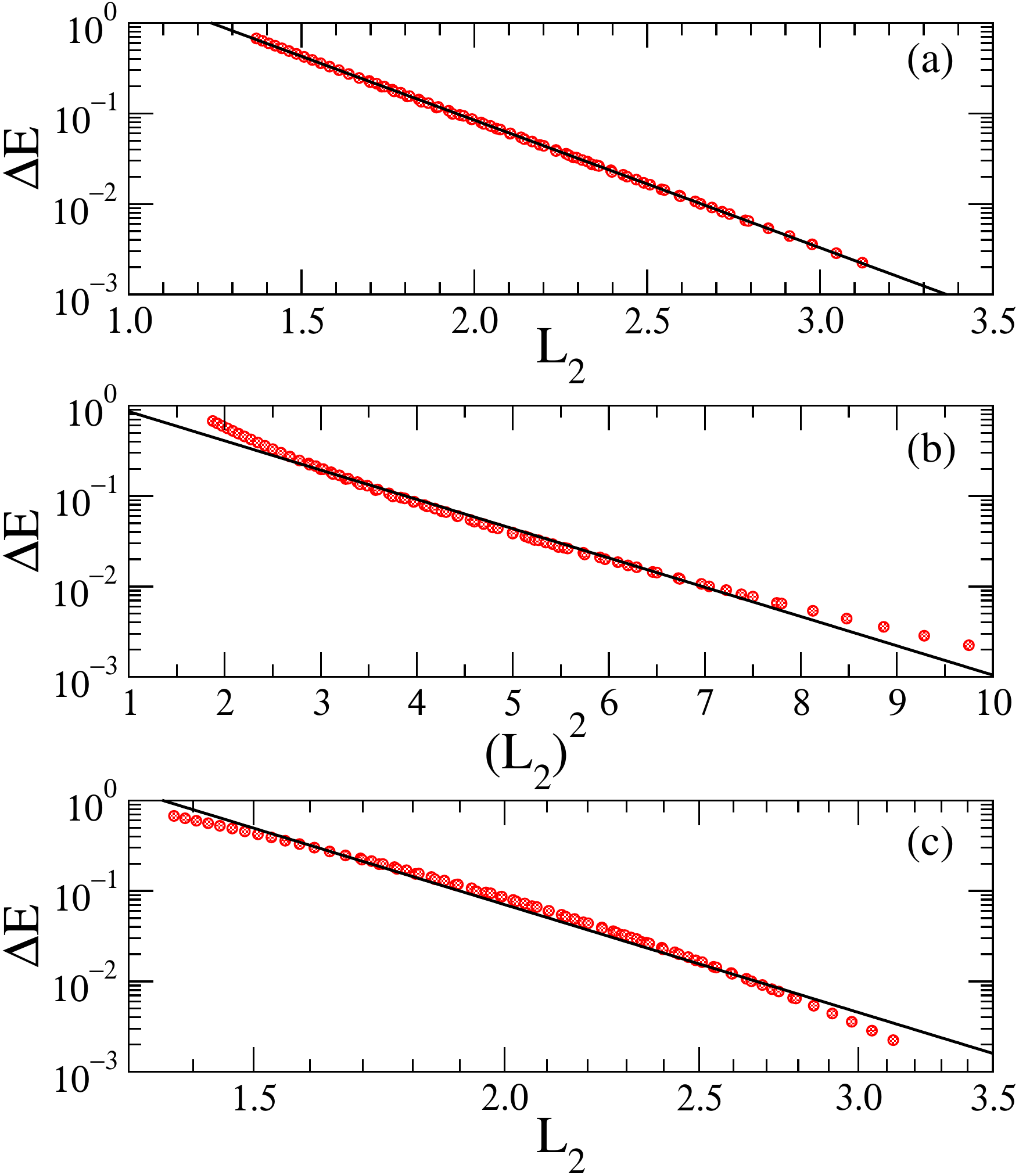}
	\caption{The IR energy correction $\Delta E_L$ versus $L_2$
		for a Gaussian potential well Eq.~\eqref{eq:Vg} with
		$V_0 = 5$ (and $\hbar = \mu = R=1$) using a wide range of $N$ and $\hw$.
		The energies are fitted with (a) exponential, (b) Gaussian,
		and (c) power law dependence on $L_2$.}
	\label{fig:Vgauss_other_forms}
	\end{figure}
	\begin{figure}[h]
	\centering
	\includegraphics[width=0.6\textwidth]
	{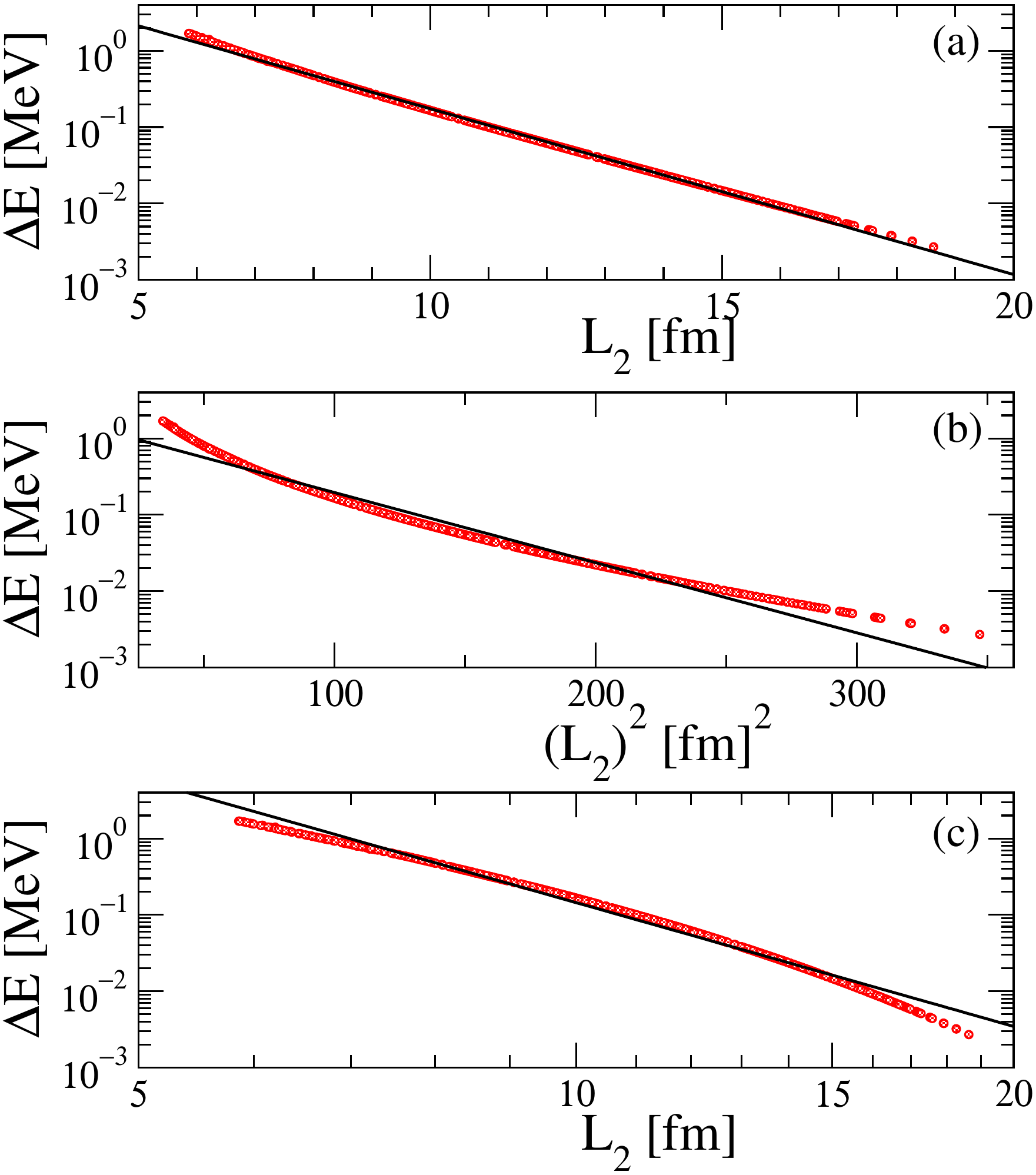}
	\caption{The IR energy correction $\Delta E_L$
		versus $L_2$ for the deuteron calculated with the chiral EFT potential from
		Ref.~\cite{Entem:2003ft} using a wide range of $N$ and $\hw$.
		The energies are fitted with (a) exponential, (b) Gaussian,
		and (c) power law dependence on $L_2$.}
	\label{fig:deuteron_other_forms}
	\end{figure}
	A purely empirical test can be made for our models and the deuteron
	because we can calculate the exact $\Einf$, plot $\Delta E(L_2) \equiv
	E(L_2) - \Einf$ against $L_2$, and then attempt to fit each of the
	three choices of $\Delta E(L_2)$.  Figure~\ref{fig:Vgauss_other_forms}
	shows the results for a representative model potential (a Gaussian)
	with moderate depth while Fig.~\ref{fig:deuteron_other_forms} shows
	the results for the deuteron.  The plots are made so that the
	candidate form would yield a straight line if followed precisely.  We
	see that the exponential form is an excellent fit for the model
	throughout the range of $L_2$ and a reasonable but not perfect fit for
	the deuteron.  The not perfect fit for the deuteron can be explained
	by noting that it is a weakly bound state.  We see from
	Figs.~\ref{fig:deuteron_curves} and \ref{fig:deuteron_error_log_plots}
	that for deuteron, the correction will be a sum of exponential terms.
	In contrast to the exponential extrapolation, Gaussian and power law fits
	fail over the full range of $L_2$.  This is consistent with
	Tolle {\it et al.}~\cite{Tolle:2012cx}.  For limited ranges of $L_2$ a
	Gaussian does	provide a reasonable fit (and should give a good extrapolation
	for	$\Einf$ if close enough to convergence).  This is consistent with the
	apparent success of exponential in $N$ extrapolation observed in the
	literature.  However, we see that globally exponential in $L$ is clearly
	superior.  Moreover, the exponent and coefficient (see
	Eq.~\eqref{eq:complete_IR_scaling}) are physically motivated whereas in the
	exponential in $N$ case (Eq.~\eqref{eq:exp_Nmax_extrapolation}) these
	parameters are fit to data.

	\subsection{Higher angular momenta}
	\label{subsec:higher_angular_momenta}

	The deuteron ground state is a mixture of an $s$ and a $d$ state, and
	the $s$ and $d$ asymptotic normalization coefficients (as well as the
	$d$-to-$s$ state ratio of about 2.5\%) are observables.  The
	extrapolation formulas so far were derived for $s$
	states, and it is of interest to extend these to nonzero angular
	momenta $l$.  We do so in two steps.  First, we show that $L_2$ is also
	the relevant effective hard-wall radius for oscillator wave functions
	with nonzero angular momenta.  Second, we derive the energy correction
	for nonzero angular momenta.

	\medskip
	\subsubsection{$L$ for nonzero angular momenta}

	For the derivation of the relevant IR length scale at $l>0$ we closely
	follow the derivation for $l=0$ presented in
	Subsec.~\ref{subsec:tale_of_tails}.  We compute the smallest eigenvalue
	$\kappa^2$ of the squared momentum operator $\hat{p}^2$ in a finite
	oscillator basis and identify $\kappa =x_l/L$ (with $x_l$ being the
	smallest positive zero of the spherical Bessel function $j_l$). This
	identification, and the form of the corresponding eigenfunctions are,
	of course, guided by the Dirichlet bc at $r=L$.  We set the oscillator
	length $b=1$.  Because this is the only length scale here, the results
	are general and can be extended to any $b$ with simple rescaling.
	The normalized radial	oscillator wave function of energy
	\beq
	  E = 2n + l + 3/2
	\eeq
	is $\psi_{nl}(r)=u_{nl}(r)/r$ with
	\beq
	u_{nl}(r)=\sqrt{2n!\over\Gamma(n+l+3/2)} r^{l+1} e^{-r^2/2}
	L_{n}^{l+1/2}(r^2) \;.
	\eeq
	Here, $L_n^{l+1/2}$ denotes the generalized Laguerre polynomial.

	In this basis, the operator $\hat{p}^2$ of the momentum squared is
	tridiagonal with matrix elements
	\begin{align}
	\langle u_{ml}|\hat{p}^2|u_{nl}\rangle &= (2n+l+3/2)\delta_m^n \nonumber\\
	   & \qquad \null +\sqrt{n+1}
	   \sqrt{n+l+3/2}\,\delta_m^{n+1} \nonumber \\
	  & \qquad \null +\sqrt{n}\sqrt{n+l+1/2}\,\delta_m^{n-1} \;.
	\end{align}
	For the eigenfunction of $\hat{p}^2$ with smallest eigenvalue
	$\kappa^2$ at angular momentum $l$, we make the ansatz
	$\psi_{\kappa l}(r)/r$ with
	\bea
	\label{eq:eigen_l_gen}
	\psi_{\kappa l}(r) =\left\{\begin{array}{cc}
	\kappa r j_l(\kappa r) \;, & 0\le \kappa r \le x_l \;, \\
	0 \;, & \kappa r > x_l \;.\end{array}\right.
	\eea
	Here, $j_l$ is the regular spherical Bessel function and $x_l$ is its
	smallest positive zero.  Clearly, these eigenfunctions are those of a
	particle in a spherical cavity with a Dirichlet bc at $x_l/\kappa$. In
	an infinite basis, the wave function $\psi_{\kappa l}(r)/r$ is an
	eigenfunction of $\hat{p}^2$ for any non-negative value of $\kappa$.
	In a finite oscillator basis, only discrete momenta $\kappa$ are
	allowed.  For their computation we expand the eigenfunction as
	\beq
	\psi_{\kappa l}(r) = \sum_{m=0}^{n} c_m(\kappa) u_{ml}(r) \;,
	\eeq
	where we supress the dependence of the admixture coefficients
	$c_m(\kappa)$ on $l$, which is kept fixed throughout this derivation.

	The last row of the matrix eigenvalue problem for $\hat{p}^2$ is
	\beq
	\label{quantize}
	(2n+l+3/2 -\kappa^2)c_n(\kappa) = -\sqrt{n}\sqrt{n+l+1/2}\, c_{n-1} \;,
	\eeq
	and this becomes the quantization condition for $\kappa$.  The direct
	computation of the coefficients $c_n(\kappa)$ seems difficult.
	Instead, we make a Fourier-Bessel expansion
	\beq
	\label{fourierbessel}
	\psi_{\kappa l}(r)= \sqrt{2\over \pi}\int\limits_0^\infty\! dk\,
	   \tilde{\psi}_{\kappa l}(k)\, kr j_l(kr) \;,
	\eeq
	and use
	\beq
	kr j_l(kr) =\sqrt{\pi\over 2}\sum_{n=0}^\infty (-1)^n u_{nl}(k) u_{nl}(r) \;.
	\eeq
	Thus,
	\beq
	\psi_{\kappa l}(r)= \sum_{n=0}^\infty (-1)^n u_{nl}(r)
	\int\limits_0^\infty\! dk\, \tilde{\psi}_{\kappa l}(k) u_{nl}(k) \;,
	\eeq
	and the admixture coefficients are therefore
	\beq
	c_n(\kappa)= (-1)^n \int\limits_0^\infty dk\, \tilde{\psi}_{\kappa l}(k)
	u_{nl}(k) \;.
	\eeq

	So far, our formal manipulations have been exact.  We now employ an
	asymptotic approximation of the generalized Laguerre polynomials
	(which enters the $u_{nl}(k)$) in terms of Bessel functions, valid for
	$n\gg 1$, see Eq.~15 of Ref.~\cite{Deano2013}.  This yields
	\begin{align}
	u_{nl}(k) \approx {2^{1-n}\over \pi^{1/4}} \sqrt{(2n+2l+1)!\over (n+l)! n!}
	   \,(4n+2l+3)^{-{l+1\over 2}}
	   \sqrt{4n+2l+3} k\, j_l(\sqrt{4n+2l+3}k) \;,
	\end{align}
	and
	\begin{align}
	c_n(\kappa) \approx C_{nl} \sqrt{2\over \pi}\int\limits_0^\infty dk \,
	\tilde{\psi}_{\kappa l}(k)	\sqrt{4n+2l+3} k
	 j_l(\sqrt{4n+2l+3}k) \;.
	\end{align}
	Here, $C_{nl}$ is a constant that does not depend on $\kappa$.  The key
	point is that the asymptotic expansion in terms of Bessel functions
	allows us now to employ the definition in Eq.~\eqref{fourierbessel} to
	evaluate the integral
	\begin{align}
	 \sqrt{2\over \pi}\int\limits_0^\infty dk\, & \tilde{\psi}_{\kappa l}(k)
	\sqrt{4n+2l+3} k\, j_l(\sqrt{4n+2l+3}k) \nonumber \\
	&=  \psi_{\kappa l}(\sqrt{4n+2l+3}) \nonumber \\
	&= \sqrt{4n+2l+3}\, \kappa\, j_l(\sqrt{4n+2l+3}\kappa) \;.
	\end{align}
	Putting it all together, we find
	\begin{align}
	c_n(\kappa) = {2^{1/2-n}(-1)^n \pi^{1/4} \over (4n+2l+3)^{l/2}}
	\sqrt{(2n+2l+1)!\over (n+l)!n!}
	\kappa\, j_l(\sqrt{4n+2l+3}\kappa) \;.
	\end{align}
	We insert this expression for $c_n(\kappa)$ into the quantization
	condition of Eq.~\eqref{quantize} and make the ansatz
	\beq
	\kappa ={x_l\over\sqrt{4n+2l+3+2\Delta}} \ .
	\eeq
	Assuming the limit $n\gg 1$ and $n\gg l$ in the quantization condition
	then yields
	\beq
	\Delta=2 \ .
	\eeq
	Thus, $\Delta$ does not depend on $l$ in this limit, and the result
	is consistent with the $l=0$ result of Ref.~\cite{More:2013rma}. In
	other words, the extent of the position space in finite oscillator
	basis with maximum radial quantum number $n$ and angular momentum $l$ is
	\bea
	\label{L2}
	  L_2 &=& \sqrt{2(2n+l+3/2+2)}b \nonumber\\
	&=& \sqrt{2(N+3/2+2)}b \;,
	\eea
	in accord with Eq.~\eqref{eq:L2_def}.

	Table~\ref{tab:lowest_kappa} shows numerical comparisons for $l=0,1,2$
	and a range of $n$ of the exact minimum momentum $\kappa$ and the
	estimate $x_l/L_2$ (with $x_0=\pi$, $x_1\approx 4.49341$, $x_2\approx
	5.76346$).  The estimates are accurate approximations of the exact
	results even for small $N=2n+l$, but the accuracy decreases somewhat
	with increasing orbital angular momentum. In some practical
	calculations it might thus be of advantage to directly employ the
	numerical results for $L_2$ instead of the approximate analytical
	expression~(Eq.~\eqref{L2}).

	\begin{table}[h]
	\begin{tabular}{c|c|c|c||c|c|c|c||c|c|c|c}
	$l$& $n$ & $\kappa$  &${x_l/ L_2}$ &
	$l$& $n$ & $\kappa$  &${x_l/ L_2}$ &
	$l$& $n$ & $\kappa$  &${x_l/ L_2}$ \\
	\hline
	0 &  0  & 1.2247  & 1.1874   &
	1 &  0  & 1.5811  & 1.4978   &
	2 &  0  & 1.8708  & 1.7378   \\
	0 &  1  & 0.9586  & 0.9472   &
	1 &  1  & 1.2764  & 1.2463   &
	2 &  1  & 1.5423  & 1.4881   \\
	0 &  2  & 0.8163  & 0.8112   &
	1 &  2  & 1.1047  & 1.0898   &
	2 &  2  & 1.3509  & 1.3222   \\
	0  &  3  &  0.7236  &  0.7207  &
	1  &  3  &  0.9892  &  0.9805  &
	2 &    3  &    1.2191  &     1.2018    \\
	0   &  4   &   0.6568   &    0.6551    &
	1 &    4 &     0.9042   &    0.8987  &
	2 &    4  &    1.1207  &     1.1092   \\
	0  &   5   &   0.6058   &    0.6046    &
	1 &    5 &     0.8382   &    0.8344  &
	2 &    5   &   1.0432   &    1.0352    \\
	0  &   6   &   0.5651   &    0.5642    &
	1 &    6 &     0.7850   &    0.7822  &
	2  &   6  &    0.9801  &     0.9742  \\
	0   &  7 &     0.5316    &   0.5310    &
	1  &   7 &     0.7408   &    0.7387  &
	2 &    7  &    0.9274  &     0.9229   \\
	0  &   8    &  0.5035    &   0.5031    &
	1  &   8 &     0.7033   &    0.7018  &
	2 &    8  &    0.8824  &     0.8789   \\
	0  &  9  &    0.4795   &    0.4791     &
	1 &    9 &     0.6711   &    0.6698  &
	2  &   9   &   0.8435  &     0.8407    \\
	0  &   10 &   0.4585  &     0.4582    &
	1 &    10 &   0.6429  &     0.6419   &
	2  &   10   & 0.8093 &      0.8070  \\
	\end{tabular}
	\caption{Comparison of the exact lowest momentum $\kappa$ with the analytical
		estimate $x_l/L_2$ for $l=0, 1, 2$ and $0 \leq n \leq 10$.}
	\label{tab:lowest_kappa}
	\end{table}

	\medskip
	\subsubsection{Energy correction for finite angular momentum}

	Let us extend our $l=0$ result for $[\Delta E]_{\rm LO}$ to $l>0$ following
	the	$S$-matrix method in Subsec.~\ref{subsec:make_cash}.
	For orbital angular momentum $l$, the asymptotic wave function is
	\beq
  u_L(r) \overset{r \gg R}{\longrightarrow}  k_L r\Bigl(h_l^{(1)}(ik_L r) -
	{h_l^{(1)}(ik_L L)\over h_l^{(1)}(-ik_L L)}
	h_l^{(1)}(-ik_Lr)\Bigr) \;.
  \label{eq:uLasympl}
	\eeq
	Here, $h_l^{(1)}$ denotes the spherical Hankel function of the first
	kind (or the spherical Bessel function of the third
	kind)~\cite{abramowitz1964}.  By definition $u_L(L)=0$.

\begin{figure}[h]
	\centering
	\includegraphics[width=0.6\textwidth]
	{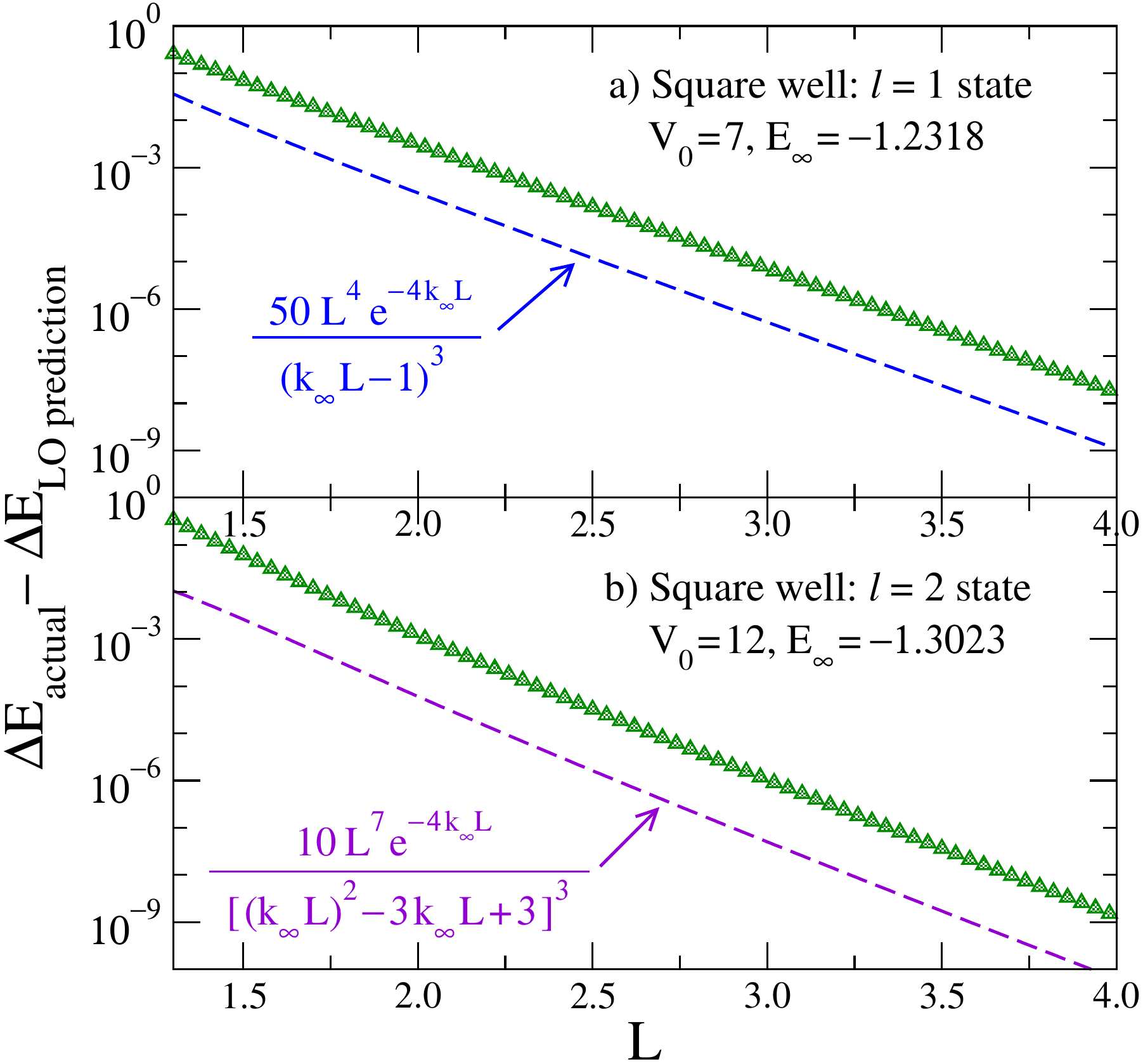}
	\caption{Error plots of the energy correction at each $L$ for (a) $l = 1$ and
	(b) $l = 2$ square-well states predicted at leading order by
	Eqs.~\eqref{eq:DeltaE_l1} and \eqref{eq:DeltaE_l2} compared to the exact
	energy.  Lines proportional to the expected \LNLO~residual errors are plotted
	for comparison.}
	\label{fig:sq_well_error_plot_higher_l}
	\end{figure}

	In complete analogy to the case of $s$ waves (e.g., using
	Eqs.~\eqref{eq:basiceq} and \eqref{eq:purepole} for general $l$), the
	correction $\Delta E$ of the energy at leading order is
	\beq
	[\Delta E]_{\rm LO} = - \kinf \left(\gamma_\infty^{(l)}\right)^2
	{h_l^{(1)}(ik_L L)\over h_l^{(1)}(-ik_L L)} \ .
	\label{eq:ELO_general_l}
	\eeq
	We note that
	\beq
	{h_l^{(1)}(ix)\over h_l^{(1)}(-ix)} \approx - e^{-2x}
	\eeq
	for $x\gg 1$.  In particular, for $l=1$
	\beq
	[\Delta E]_{\rm LO} =\kinf \left(\gamma_\infty^{(1)}\right)^2
	{\kinf L+1\over \kinf L-1} \, e^{-2\kinf L}  \;,
	\label{eq:DeltaE_l1}
	\eeq
	and for $l=2$
	\beq
	[\Delta E]_{\rm LO} = \kinf \left(\gamma_\infty^{(2)}\right)^2
	{(\kinf L)^2 +3 \kinf L +3\over (\kinf L)^2-3\kinf L +3}\, e^{-2\kinf L} \;.
	\label{eq:DeltaE_l2}
	\eeq
	These corrections are tested in Fig.~\ref{fig:sq_well_error_plot_higher_l}.
	For coupled channels, the leading energy correction will be the sum
	of the LO corrections for the individual angular momenta.
	We note that lattices with periodic bc lead to energy shifts that
	depend on the angular momentum~\cite{Konig:2011nz}.
	In contrast, the basis truncations we consider in this work are
	variational and thus always yield a positive energy correction.

	\begin{figure}[h]
	\centering
	\includegraphics[width=0.6\textwidth]
	{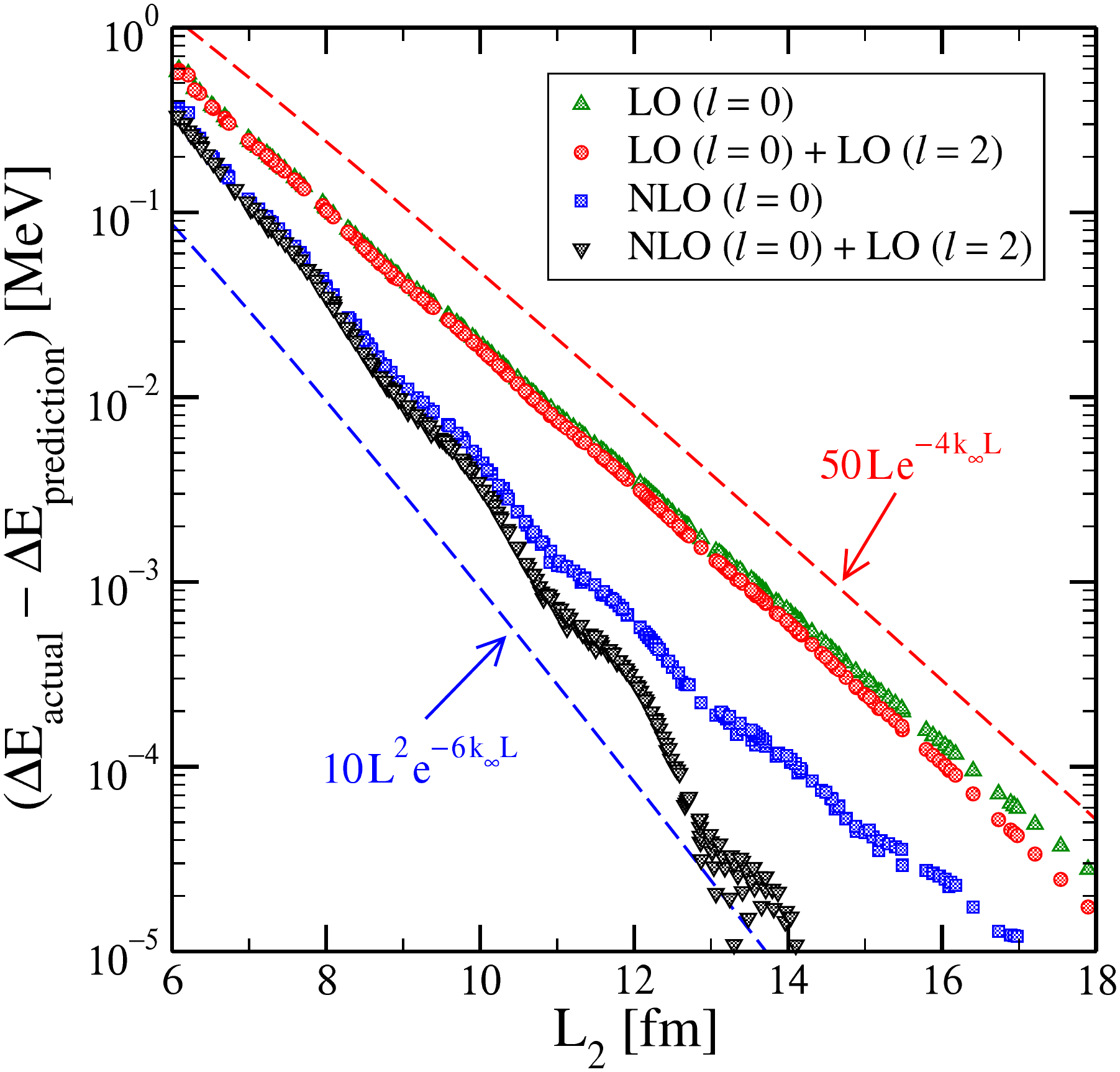}
	\caption{Residual error for the deuteron energy due to
	  HO basis truncation as a function of $L = L_2$ (with $\hw > 49$~MeV
	  to eliminate UV contamination) after subtracting $l=0$ energy
	  corrections at different orders from
	  Eq.~\eqref{eq:complete_E_correction_NLO} and the $l=2$ correction
	  from Eq.~\eqref{eq:DeltaE_l2}.  For the parameter $w_2$ in
	  Eq.~\eqref{eq:complete_E_correction_NLO} we use the value reported
	  in \cite{Phillips:1999hh}.}
	\label{fig:deuteron_error_plot_l2}
	\end{figure}
	We return to the deuteron and take
	$|\gamma_\infty^{(2)}/\gamma_\infty^{(0)}| \approx 0.0226/0.8843$ from
	Ref.~\cite{Machleidt:2011zz}.  Then
	\beq
	[\Delta E]_{\rm LO} =\kinf \left(\gamma_\infty^{(0)}\right)^2
	e^{-2\kinf L}
	  \left[ 1 +
	\left|\frac{\gamma_\infty^{(2)}}{\gamma_\infty^{(0)}}\right|^2
	\frac{(\kinf L)^2 +3 \kinf L +3}{(\kinf L)^2-3\kinf L +3}\right]
	\;.
	\eeq
	This formula is tested in Fig.~\ref{fig:deuteron_error_plot_l2} with
	the same deuteron calculations as in
	Fig.~\ref{fig:deuteron_error_log_plots}.  We note that the deviation
	after subtraction of the NLO ($l=0$) result does not exhibit the
	$\exp(-6k_\infty L)$ falloff but is rather consistent with an
	$\exp(-4\kinf L)$ falloff at large $L$.  We attribute this to the
	missing LO $d$-state correction.  Due to the small value of the
	$d$-to-$s$ state ratio, the $d$-wave correction is small, but it makes
	a perceptible shift of the $s$-wave LO result.  When added to the NLO
	$l=0$ correction, the large $L_2$ behavior of the error is brought
	somewhat closer in line with the predicted dependence of $L^2
	e^{-6\kinf L}$.  We note, however, that the NLO correction is not complete
	due to the missing $l=2$ correction.  Calculating NLO corrections for $l>0$
	remains an open question.  Particularly, because that would entail taking
	into account the admixture between different channels.

	\subsection{Radii and phase shifts}
	\label{subsec:radii_phase_shifts}

	\medskip
	\subsubsection{Radii}

\begin{figure}[h]
	\centering
	\includegraphics[width=0.6\textwidth]
	{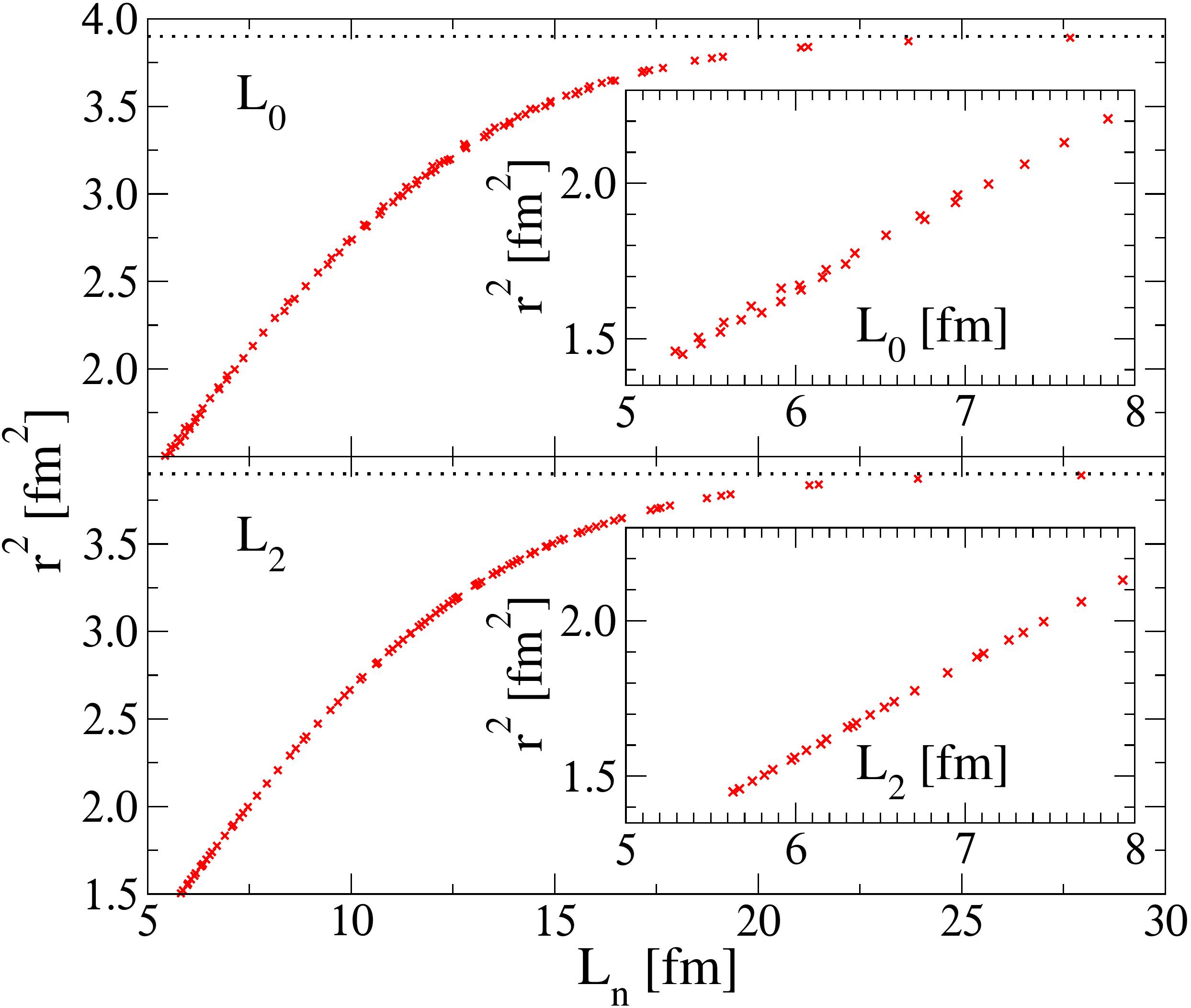}
	\caption{Deuteron radius squared versus $L_0$ (top) and
  	$L_2$ (bottom) for the Entem-Machleidt 500\,MeV N$^3$LO
  	potential~\cite{Entem:2003ft}.  The horizontal dotted lines mark the
  	exact radius squared $r^2_{\infty} = 3.9006~{\rm fm}^2$. The insets
  	show a magnification of data at smaller lengths $L_n$.}
	\label{fig:deuteron_radii}
	\end{figure}

	Along with binding energies, another nuclear observable that we looked at
	was the radius squared.  Figure~\ref{fig:deuteron_radii} shows the
	numerical results for the squared radius for the deuteron calculated in the
	HO basis.  Analogous to Figs.~\ref{fig:spatial_cutoff_vs_HO_square_well} and
	\ref{fig:spatial_cutoff_vs_HO_deuteron}, we see that the results for
	the squared radius fall on a continuous curve with minimal spread when
	plotted as a function of $L_2$ (but not as a function of $L_0$).

	Though the squared radius is a long-ranged
	operator, its matrix elements will still be modified at short distances
	by renormalizations or  similarity transformations of the Hamiltonian,
	see, e.g., Ref.~\cite{Stetcu:2004wh}.  Thus we cannot expect an extrapolation
	law for the radius that depends entirely on observables.
	Instead, we seek a formula that identifies the $L$ dependence but leaves
	parameters to be fit.  We define
	\beq
	  \rsqav_L = \rsqinf + \Delta\rsqav_L \;,
	\eeq
	where
	\beq
	  \Delta\rsqav_L =
	  \frac{\int_0^L |u_L(r)|^2\, r^2\,dr}{\int_0^L |u_L(r)|^2\, dr}
	  - \frac{\int_0^\infty |u_\infty(r)|^2\, r^2\,dr}{\int_0^\infty
		|u_\infty(r)|^2\, dr} \;.
	  \label{eq:Delta-rsq}
	\eeq
	The strategy is to isolate the polynomial $L$ dependence by splitting the
	necessary integrals into an interior part and an exterior part:
	\beq
  \int_0^L\! r^n |u_L(r)|^2\, dr =
  \int_0^R\! r^n |u_L(r)|^2\, dr + \int_R^L\! r^n |u_L(r)|^2\, dr \;,
  \label{eq:rnintegral}
	\eeq
	where $R$ is sufficiently large so that the asymptotic form of $u_L(r)$ from
	Eq.~\eqref{eq:uLasymp2} can be used in the second integral.
	Our expression for $\Delta\rsqav_L$ is independent of the normalization
	of $u_L(r)$, so we are free to choose it so that the large $r$ form is
	exactly given by Eq.~\eqref{eq:uLasymp2}.

	The first integral in Eq.~\eqref{eq:rnintegral} will depend on the details
	of the interior wave function	and therefore on the potential, but
	the linear energy method shows us that to $\mathcal{O}(e^{-2\kinf L})$
	the $L$ dependence is isolated.  In particular, the dependence on $L$ of
	$u_L(r)$ in	Eq.~\eqref{eq:linear_energy_approx} is confined to
	$\Delta E_L = \kinf\ANC^2 e^{-2\kinf L}$ because $du_E(r)/dE|_{\Einf}$ for
	$r < R$ is independent of $L$ with our choice of normalization.  Thus the
	integral over $r$ cannot introduce polynomial $L$ dependence and we can
	conclude that
	\beq
  \int_0^R\! r^n |u_L(r)|^2\, dr = \mathcal{O}(L^0) e^{-2\kinf L}
	     + \mathcal{O}(e^{-4\kinf L}) \;.
	\eeq
	The $\mathcal{O}(L^0)$ coefficient will depend on the potential, so we
	will treat it as a parameter to be fit.

	The second integral can be directly evaluated to $\mathcal{O}(e^{-2\kinf L})$
	using Eq.~\eqref{eq:uLasymp2} and $[k_L]_{LO} = \kinf - \ANC^2 e^{-2\kinf L}$
	to expand $|u_L(r)|^2$.  For $n=0$ we find
	\begin{align}
	  \int_R^L\! &|u_L(r)|^2\, dr =  \frac{1}{2\kinf}e^{-2\kinf R} + \Bigl[
	  \frac{\ANC^2}{\kinf} \Bigl(R + \frac{1}{2\kinf}\Bigr)e^{-2\kinf R} + 2R -
		2L \Bigr]e^{-2\kinf L} + \mathcal{O}(e^{-4\kinf L}) \;,
	  \label{eq:asympnorm}
	\end{align}
	and for $n=2$ we find
	\begin{align}
  	\int_R^L\! r^2 |u_L(r)|^2\, dr =
  	\frac{1}{2\kinf^3}\Bigl[ \frac12 + \kinf R + (\kinf R)^2 \Bigr]
		e^{-2\kinf R}
    \nonumber \\
		\null +  \Bigl[ \frac{\ANC^2}{\kinf^4} \Bigl( \frac34 + \frac32\kinf R +
		\frac32(\kinf R)^2 + (\kinf R)^3 \Bigr)e^{-2\kinf R}  \nonumber \\
    \null + \frac{1}{\kinf^3} \Bigl( \frac23 (\kinf R)^3 - \kinf L - \frac23
		(\kinf L)^3 \Bigr) \Bigr]e^{-2\kinf L}
		+ \mathcal{O}(e^{-4\kinf L})  \;.
    \label{eq:asymprsq}
	\end{align}
	Note that it is necessary to keep the expansion of $|u_L(r)|^2$ up to
	$e^{-4\kinf L}$	until after doing the integrals because terms proportional
	to $e^{-4\kinf L} e^{2\kinf r}$ will be leading order.

	When we use Eqs.~\eqref{eq:asympnorm} and \eqref{eq:asymprsq} and our previous
	result for the interior integrals in Eq.~\eqref{eq:Delta-rsq}, expanding
	consistently to $\mathcal{O}(e^{-2\kinf L})$, we will mix $R$-dependent
	terms with the $L$ dependence.  However, we can immediately conclude that
	the general form to this order is (with $\beta \equiv 2\kinf L$)
	\bea
	 	\rsqav_L \approx {\rsqinf}[ 1 - (c_0 \beta^3 +c_1 \beta + c_2) e^{-\beta}]
	  \;.
		\label{rad}
	\eea
	Here, $\rsqinf$, $c_0$, $c_1$, and $c_2$ are fit parameters while $\kinf$
	should be determined from fitting the energy. 	This form has been verified
	explicitly for finite-range model potentials (e.g., square well
	and delta shell). 	The approximation in Eq.~\eqref{rad} should be valid
	in the asymptotic regime $\beta\gg 1$.  In practice, one needs
	$\beta\gtrsim 3$ so that the dominant $\beta^3$ correction is
	approximately an order of magnitude larger than the subleading terms
	(with $c_1$ and $c_2$ expected to be roughly the same size as
	$c_0$ or smaller).

	If we take the zero-range limit $R\to 0$ of the potential, we
	arrive at the simple expression
	\beq
	 \frac{\Delta\rsqav_L}{\rsqinf}
	 \approx -\left(\frac{(2\kinf L)^3}{3} -4\right) e^{-2\kinf L} \;.
	 \label{eq:rsqzerorange}
	\eeq
	Note that in this limit the correction becomes independent of the potential.
	Equation~\eqref{eq:rsqzerorange} suggests that for a short-range potential,
	the $c_1$ and $c_2$ terms will give comparable contributions for moderate
	$\beta$, and therefore will be difficult to determine reliably.

	\begin{figure}[h]
		\centering
		\begin{subfigure}[c]{0.45\textwidth}
			\centering
			\includegraphics[width=\textwidth]
			{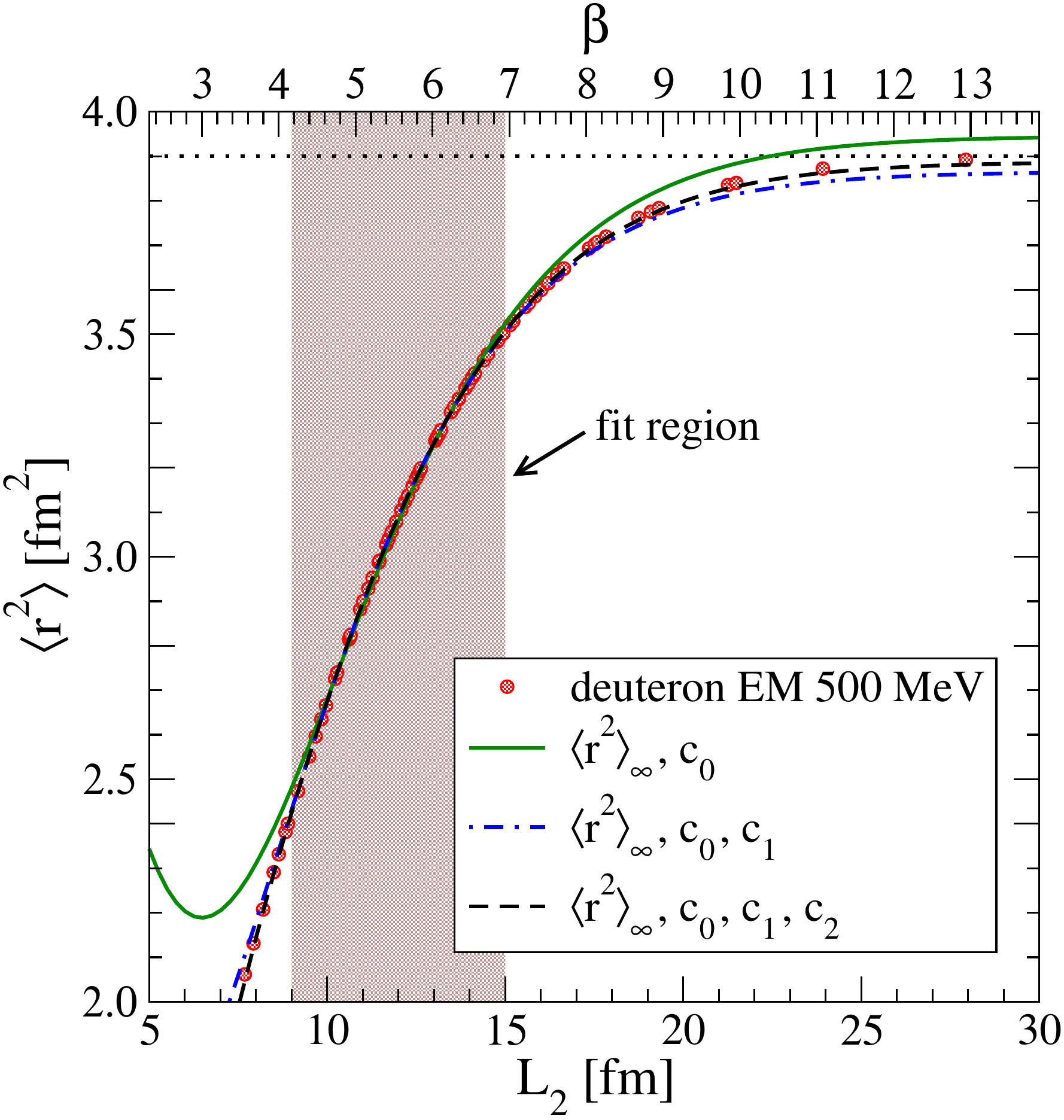}
			\caption{Fit region: $L$ from $9$ to $15$ fm.}
			\label{fig:deuteron_radius_fit_9to15}
		\end{subfigure}
		\hspace{0.05\textwidth}
		\begin{subfigure}[c]{0.45\textwidth}
			\centering
			\includegraphics[width=\textwidth]
			{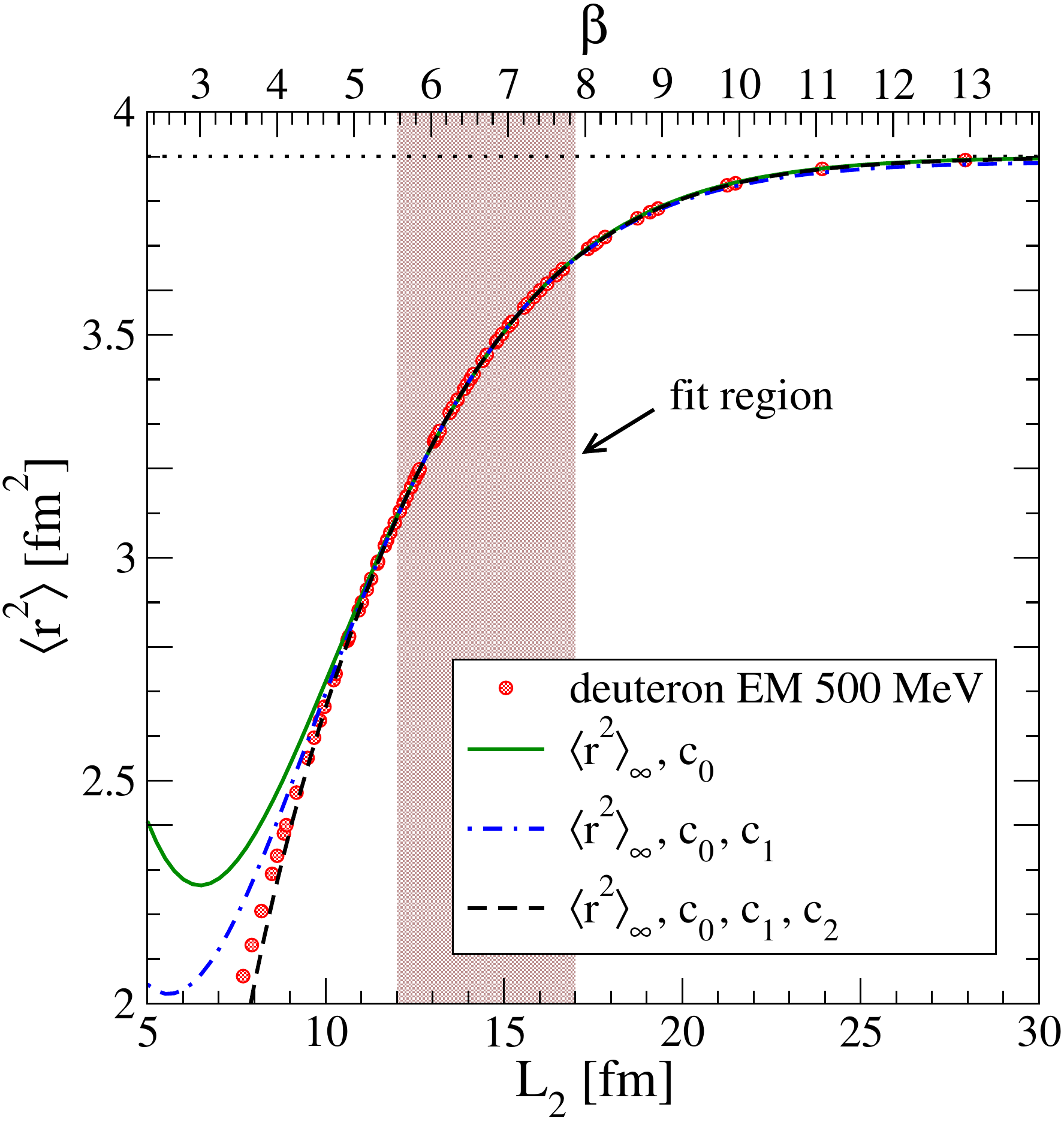}
			\caption{Fit region: $L$ from $12$ to $17$ fm.}
			\label{fig:deuteron_radius_fit_12to17}
		\end{subfigure}
		\caption{Deuteron radius squared versus $L_2$ for the chiral
		  N$^3$LO (500\,MeV) potential of Ref.~\cite{Entem:2003ft}.  To
			eliminate the UV contamination we only plot results for $\hw > 49$~MeV.
			The solid, dot-dashed, and dashed lines are results from fitting
			Eq.~\eqref{rad} in the shaded region to find $\rsqinf$ and one,
			two, or three of the $c_i$ constants, respectively.
  		The horizontal dotted line is the deuteron radius squared.}
		\label{fig:deuteron_radius_fit}
	\end{figure}
	Sample fits of Eq.~\eqref{rad} for the deuteron are shown in
	Fig.~\ref{fig:deuteron_radius_fit}.  Results are obtained for fitting
	one, two, and all three $c_i$ constants to radii calculated
	with the same truncated oscillator basis parameters used for
	Fig.~\ref{fig:deuteron_curves}.  The fit region is for $L_2$
	between 9 and 15\,fm, where the calculations only show a small
	amount of curvature and between 12 and 17\,fm.  All points are equally
	weighted.
	\begin{table}[h]
		\begin{tabular}{ |c|c|c|c|c| }
			\hline
	 		\multicolumn{5}{|c|}{fit region: $L$ from $9$ to $15$ fm. Exact:
	 		$\rsqinf = 3.901$ fm$^2$, $\kinf = 0.232$ fm$^{-1}$} \\ \hline
			& $\rsqinf$ & $c_0$ & $c_1$ & $c_2$ \\ \hline
			$1$ term & $3.945 \pm 0.006$ & $0.331 \pm 0.002$ & -- & -- \\ 
	 		$2$ terms & $3.865 \pm 0.001$ & $0.266 \pm 0.001$ & $1.131 \pm 0.018$ & --
			\\ 
	 		$3$ terms & $3.887 \pm 0.001$ & $0.308 \pm 0.002$ & $-1.239 \pm 0.135$ &
			$7.167 \pm 0.408$ \\ \hline
	 \end{tabular}
	 \caption{Coefficients from fitting Eq.~\eqref{rad} to the deuteron data
	    using one, two, or three $c_i$ constants as shown in
			Fig.~\ref{fig:deuteron_radius_fit_9to15}}
	 \label{tab:coeff_deuteron_radius_fit_9to15}
  \end{table}
	\begin{table}
		\begin{tabular}{ |c|c|c|c|c| }
	  \hline
	  \multicolumn{5}{|c|}{fit region: $L$ from $12$ to $17$ fm. Exact:
		$\rsqinf = 3.901$ fm$^2$, $\kinf = 0.232$ fm$^{-1}$} \\ \hline
	  & $\rsqinf$ & $c_0$ & $c_1$ & $c_2$ \\ \hline
	  $1$ term & $3.899 \pm 0.001$ & $0.312 \pm 0.000$ & -- & -- \\ 
	  $2$ terms & $3.888 \pm 0.001$ & $0.293 \pm 0.001$ & $0.503 \pm 0.026$ & --
		\\ 
	  $3$ terms & $3.898 \pm 0.001$ & $0.339 \pm 0.006$ & $-3.577 \pm 0.508$ &
		$15.339 \pm 1.908$ \\ \hline
	  \end{tabular}
		\caption{Coefficients from fitting Eq.~\eqref{rad} to the deuteron data
 	    using one, two, or three $c_i$ constants as shown in
 			Fig.~\ref{fig:deuteron_radius_fit_12to17}}
		\label{tab:coeff_deuteron_radius_fit_12to17}
	\end{table}
	The values for $\rsqinf$ and the coefficients $c_i$'s from the fits in
	Fig.~\ref{fig:deuteron_radius_fit} are reported in
	Tables~\ref{tab:coeff_deuteron_radius_fit_9to15} and
	\ref{tab:coeff_deuteron_radius_fit_12to17}.
	For all of these fits, the value of $c_0$ is
	fairly stable, ranging from 0.27 to 0.33 (note that
	$c_0 = 1/3$ in the zero-range limit).  In contrast, $c_1$
	and $c_2$ are not well determined (even the sign of $c_1$
	varies).  This is consistent with
	fits using the square-well potential, where analytic expressions
	for the $c_i$s can be found.  We find that $\rsqinf$
	and $c_0$ are well determined by fits in analogous
	regions but that $c_1$ and $c_2$ are not.
	If we push the analysis by taking the fit region between 7 and 13\,fm,
	the $\rsqinf$ prediction using only $c_0$ breaks down, giving
	4.21\,fm$^2$.  However, the fit with all three $c_i$s is still
	reasonable, giving 3.86\,fm$^2$.  This indicates the importance of using
	the correct functional form for extrapolation.
	Further studies are needed to test how these trends might carry
	over to $A>2$ nuclei.  It is worthwhile to note that this approach can
	be generalized to any coordinate space operator.

	\medskip
	\subsubsection{Phase shifts}

	The argument for computing scattering phase shifts is as
	follows: The oscillator basis appears as a spherical box of size
	$L$.  For low momenta we have $L=L_2$, but at higher momentum $L$
	deviates slightly from $L_2$, and can be determined from the
	eigenvalues of the operator $p^2$.  Thus, the positive-energy states
	computed in the oscillator basis can be used to extract phase shifts.

	In a fixed harmonic oscillator basis ($N,\hbar\Omega$), the
	computation of the phase shifts for a given partial wave $^{2S+1}l_J$
	with orbital angular momentum $l$ proceeds as follows: First, one
	computes the discrete eigenvalues $p_i^2$ of the operator $p^2$ for
	orbital angular momentum $l$.  Second, we need to determine the
	momentum dependent box size $L_i=L(p_i)$.  Assuming that the $i^{\rm th}$
	momentum eigenstate is the $i^{\rm th}$ eigenstate of a
	spherical box, we must determine the $i^{\rm th}$ zero of the
	spherical Bessel function.  Thus $j_l(p_iL_i/\hbar) = 0 $ determines
	$L(p_i)$.  We evaluate the smooth function $ L(p)$ for arbitrary
	momentum $p$ by interpolating between the discrete momenta
	$p_i$.  Third, we compute the discrete positive energies $E_i =
	\hbar^2k_i^2/(2m)$ of the neutron-proton system in relative
	coordinates for the partial wave $^{2S+1}l_J$, and compute the phase
	shifts from the Dirichlet boundary condition at $r=L$, i.e.
	\beq
	\tan\delta_l(k_i) = { j_l(k_iL(\hbar k_i)) \over \eta_l(k_iL(\hbar k_i))}\;.
	\eeq
	Here $\eta_l$ is the spherical Neumann function.  In practice one
	repeats this procedure for several values of $\hbar\Omega$ in order to
	get sufficiently many datapoints that fall onto a smooth curve.
	Note that for $E_i$ or $k_i$ are obtained from diagonalizing the nuclear
	Hamiltonian in HO basis whereas $p_i$ are obtained from diagonalizing the
	momentum squared operator.

	\begin{figure}[h]
	\centering
	\includegraphics[width=0.6\textwidth]
	{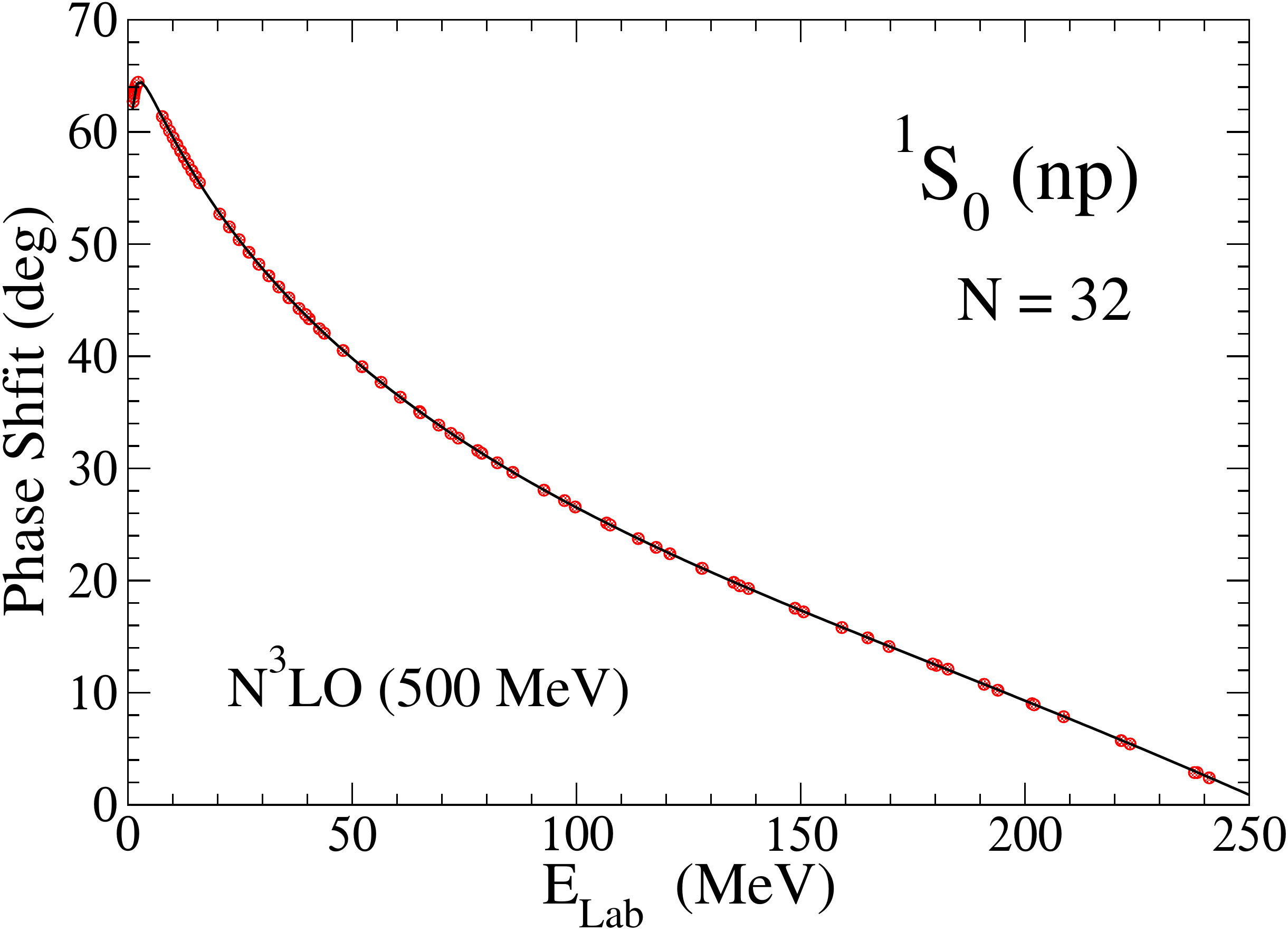}
	\caption{The $^1$S$_0$ phase shifts (in degrees) of the N$^3$LO
	  chiral interaction (solid line) compared to the phase shifts
	  computed directly in the harmonic oscillator basis (circles).}
	\label{fig:1s0phase}
	\end{figure}
	\begin{figure}[h]
	\centering
	\includegraphics[width=0.6\textwidth]
	{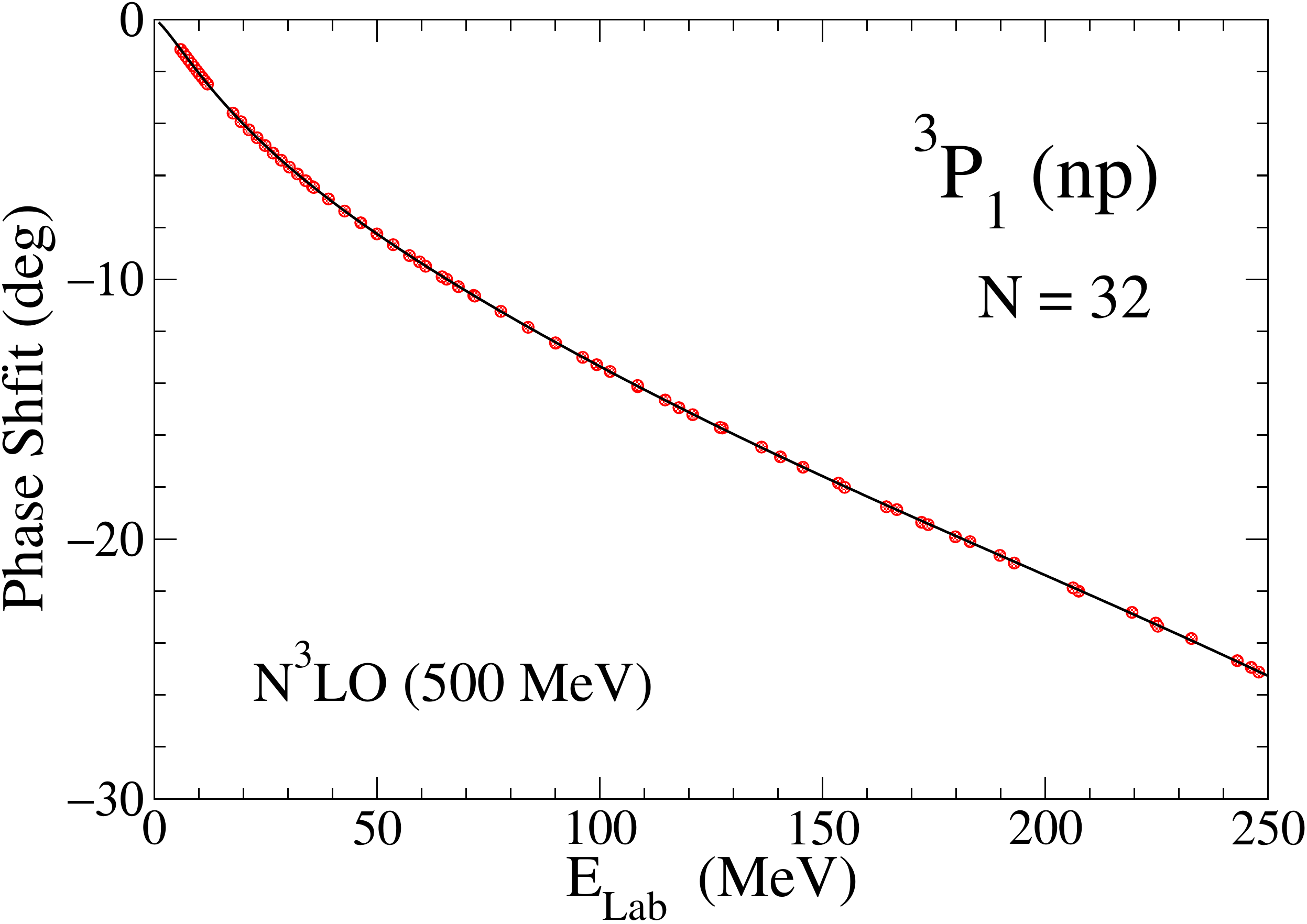}
	\caption{The $^3$P$_1$ phase shifts (in degrees) of the N$^3$LO
	  chiral interaction (solid line) compared to the phase shifts
	  computed directly in the harmonic oscillator basis (circles).}
	\label{fig:3p1phase}
	\end{figure}
	As examples we compute the scattering phase shifts for the $^1$S$_0$ and
	$^3$P$_1$	partial waves in model spaces with $N=32$ and
	$\hbar\Omega=20,22,\ldots,40$~MeV.  Our calculations are based on the
	Entem-Machleidt 500\,MeV chiral EFT N$^3$LO
	potential~\cite{Entem:2003ft}.  Figures~\ref{fig:1s0phase} and
	\ref{fig:3p1phase} show the results and compares them to the numerically
	exact phase	shifts.  For smaller $N$ than our current choice, the
	computed phase shifts start to deviate from exact phase shifts at
	higher energies.  However, if one is interested only in low-energy
	phase shifts and observables such as the scattering length and the
	effective range, a smaller harmonic oscillator basis is sufficient.

	There are other methods to compute scattering phase shifts in the
	harmonic oscillator basis.  Bang \emph{et al.} \cite{bang2000} used the
	method of harmonic oscillator representation of scattering equations
	(HORSE) for this purpose, and more recent
	works~\cite{Luu:2010hw,Stetcu:2009ic} computed phase shifts to develop an
	EFT for nuclear interactions directly in the oscillator
	basis~\cite{Stetcu:2006ey}.  References~\cite{Luu:2010hw,Stetcu:2009ic}
	build on the results by Busch {\it et al.}~\cite{busch1998} and their
	generalization~\cite{Bhattacharyya:2006fg} to finite range corrections,
	and extract scattering information from the energy shifts of bound
	states in a harmonic oscillator potential.  The resulting EFTs are
	quite efficient for contact interactions and systems such as ultracold
	trapped fermions, but nuclear potentials with a finite range require
	an extrapolation of $\Omega\to 0$~\cite{Luu:2010hw}.  The approach
	presented in this Subsection is more direct, as no external oscillator
	potential is employed.  We note that the phase shift analysis presented
	here can be extended to coupled channels as well.

	Finally, we note again that the approach of this Section can be
	utilized in other localized basis sets.  All that is required is the
	diagonalization of the operator $p^2$ in the employed basis set, which
	yields the (momentum dependent) box size.

	\section[Ultraviolet story]
	{Ultraviolet story \footnote{Based on \cite{Konig:2014hma}}}
	\label{sec:UV_story}

  In Sec.~\ref{sec:IR_story} we worked in the region where the UV errors were
	small and focused on the IR errors.  However, for many methods full
	suppression of the UV errors is not feasible and the need to understand
	UV corrections remains.  In all cases the UV effect is a systematic error
	that must be quantified.  In addition, this error worsens for harder
	nucleon--nucleon potentials	that may still be of interest.
	The understanding of the UV errors formed the focus of our work in
	Ref.~\cite{Konig:2014hma}.  The author of this thesis contributed to the work
	presented in \cite{Konig:2014hma}.  However, the UV effort was mainly
	spearheaded by Sebastian K\"{o}nig, the post-doc in the group.  To avoid
	misappropriation of credit, only a summary of results from
	\cite{Konig:2014hma} will be presented in this section.
	The summary presented here hopes to elucidate how the UV results tie into
	our broader agenda of developing reliable extrapolations schemes for nuclear
	calculations.  We refer the reader to \cite{Konig:2014hma} for the additional
	UV extrapolation details.

	We follow the strategy of Sec.~\ref{sec:IR_story} by focusing on the
	two-body problem and exactly solvable examples to establish the true
	UV behavior for the	simple systems and then make correspondence to the
	nuclear systems.

	\subsection{Duality and momentum-space boxes}
	\label{subsec:UV_cutoff_duality}

	To briefly recap results in Subsec.~\ref{subsec:tale_of_tails}---we
	demonstrated there that a
	truncated oscillator basis with highest excitation energy
	$N\Omega$ effectively imposes a spherical hard-wall boundary condition
	at a radius depending on $N$ and $b$.  The optimal effective radius
	$\Leff$ can be determined by matching the smallest eigenvalue
	$\kappa^2$ of the squared momentum operator $p^2$ in the finite
	basis to the corresponding eigenvalue of the spherical box, namely
	$\kappa=\pi/L$ (for $\ell=0$).  The value can be established numerically,
	but an accurate approximation for the two-body system is
	\begin{equation}
	\label{eq:L2_def}
	  \Leff = L_2\equiv\sqrt{2(N+3/2+2)}b \,.
	\end{equation}
	Note that $L_2$ differs by $\mathcal{O}(1/N)$ from the naive estimate
	$L_0\equiv\sqrt{2(N+3/2)}b$.  In localized bases that differ from the
	harmonic oscillator, $L$ can also be determined from a numerical
	diagonalization of the operator $p^2$.

\begin{figure}[thbp]
	\centering
	\includegraphics[width=0.6\textwidth]
	{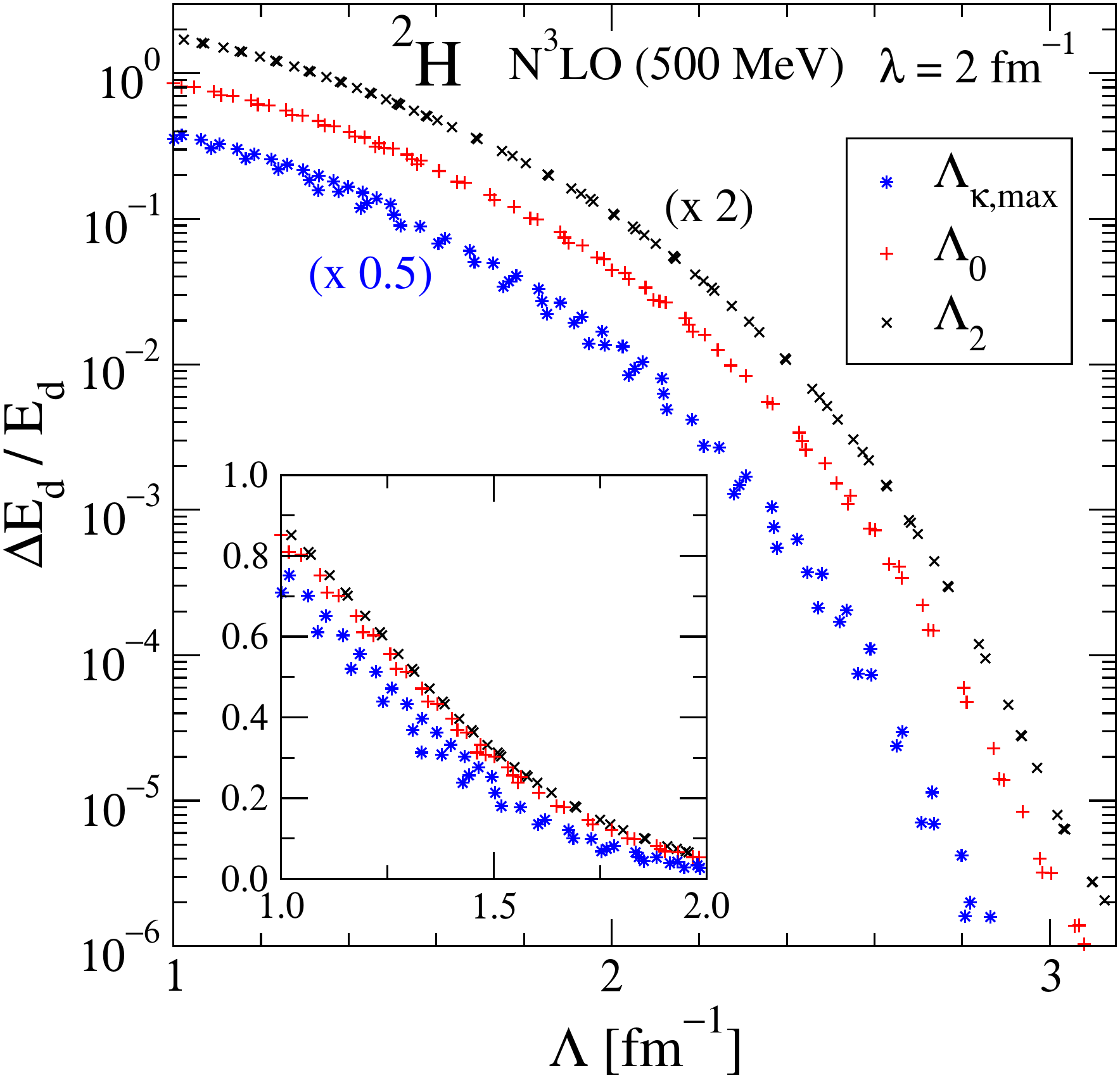}
	\caption{Relative error of deuteron binding energy
	  plotted vs. lengths $\Lambda_2$, $\Lambda_0$, and $\Lambda_{\kappa,
	  {\rm max}}$ (multiplied by factors 2, 1, and $1/2$, respectively,
	  to separate the curves.  Inset: The same values on a linear
	  scale and without the separation factors.}
	\label{fig:deuteron_lambdas}
	\end{figure}

	The dual nature of the harmonic oscillator Hamiltonian
	\begin{equation}
	 H_\mathrm{HO} = \frac{p^2}{2\mu} + \frac{\mu\Omega^2r^2}{2}
	\label{eq:H-HO}
	\end{equation}
	(\ie, under $p \leftrightarrow \mu\Omega r$) implies that the
	truncation of the basis will effectively impose a sharp cutoff at a
	momentum $\Lameff$ depending only on $N$ and $b$.  The analog matching
	condition leads us to consider the smallest eigenvalue (denoted $\rho$) of
	the operator $r^2$ evaluated in that truncated basis.  This eigenvalue is
	identical to the smallest (squared) distance that can be realized in the
	oscillator basis.  Thus it corresponds to a lattice spacing on a grid
	and therefore sets the highest momentum available.  As we see
	in Fig.~\ref{fig:deuteron_lambdas}, the square root of the largest
	eigenvalue of the squared momentum operator, which might be a natural guess
	for the effective UV cutoff, is not an accurate estimate for $\Lameff$.
	From steps completely analogous (dual) to those given in
	Subsec.~\ref{subsec:tale_of_tails} for the IR case, we find that
	the solution (in a subspace with fixed angular momentum $\ell$) is
	\begin{equation}
	\rho = \frac{x_\ell b}{\sqrt2}\left(\Nmax+\frac32+\Delta\right)^{\!-1/2}
	\label{eq:rho-Nmax-Delta}
	\end{equation}
	with $\Delta=2$ to leading order.  The constant $x_\ell$ in the prefactor is
	the first positive zero of the spherical Bessel function $j_\ell$.  Since the
	UV cutoff is given by $x_\ell/\rho$, it drops out again in our final
	result:
	\begin{equation}
	\Lambda_2 \equiv \sqrt{2(\Nmax+3/2+2)}/b \,.
	\label{eq:Lambda-2-simple}
	\end{equation}
	Hence, we have shown that the proper effective UV cutoff imposed by
	the basis truncation is given by $\Lambda_2$, which differs by a correction
	term from the naive estimate
	\begin{equation}
	 \Lambda_0 \equiv \sqrt{2(\Nmax+3/2)}/b
	\label{eq:Lambda-0}
	\end{equation}
	that one obtains by simply considering the maximum single-particle energy
	level represented by the truncated basis.  We note that subleading
	corrections	to $\Delta=2$, which by duality apply equally to the IR and UV
	cutoff, are	derived in Appendix of Ref.~\cite{Konig:2014hma}.

	Fig.~\ref{fig:deuteron_lambdas} shows the relative error when plotted
	against three cutoff variables, $\Lambda_2$, $\Lambda_0$, and
	$\Lambda_{\kappa_{\rm max}}$.
	The	calculations use the 500\,MeV N$^3$LO nucleon-nucleon $NN$ potential
	of Ref.~\cite{Entem:2003ft}, evolved by the	SRG~\cite{Bogner:2006pc} to
	$\lambda = 2\,\fmi$.
	$\Lambda_{\kappa_{\rm max}}$ is defined as the square root
	of the largest eigenvalue of the squared momentum operator in the finite
	oscillator basis, which one might naively expect to be a natural choice.
	However, of the cases considered this actually gives the largest scatter
	in data.  From the fact that we get an essentially
	smooth curve  only for $\Lambda_2$, we conclude that this identification of
	the	relevant UV cutoff is correct.
	\begin{figure}[thbp]
	\centering
	\includegraphics[width=0.6\textwidth]%
	{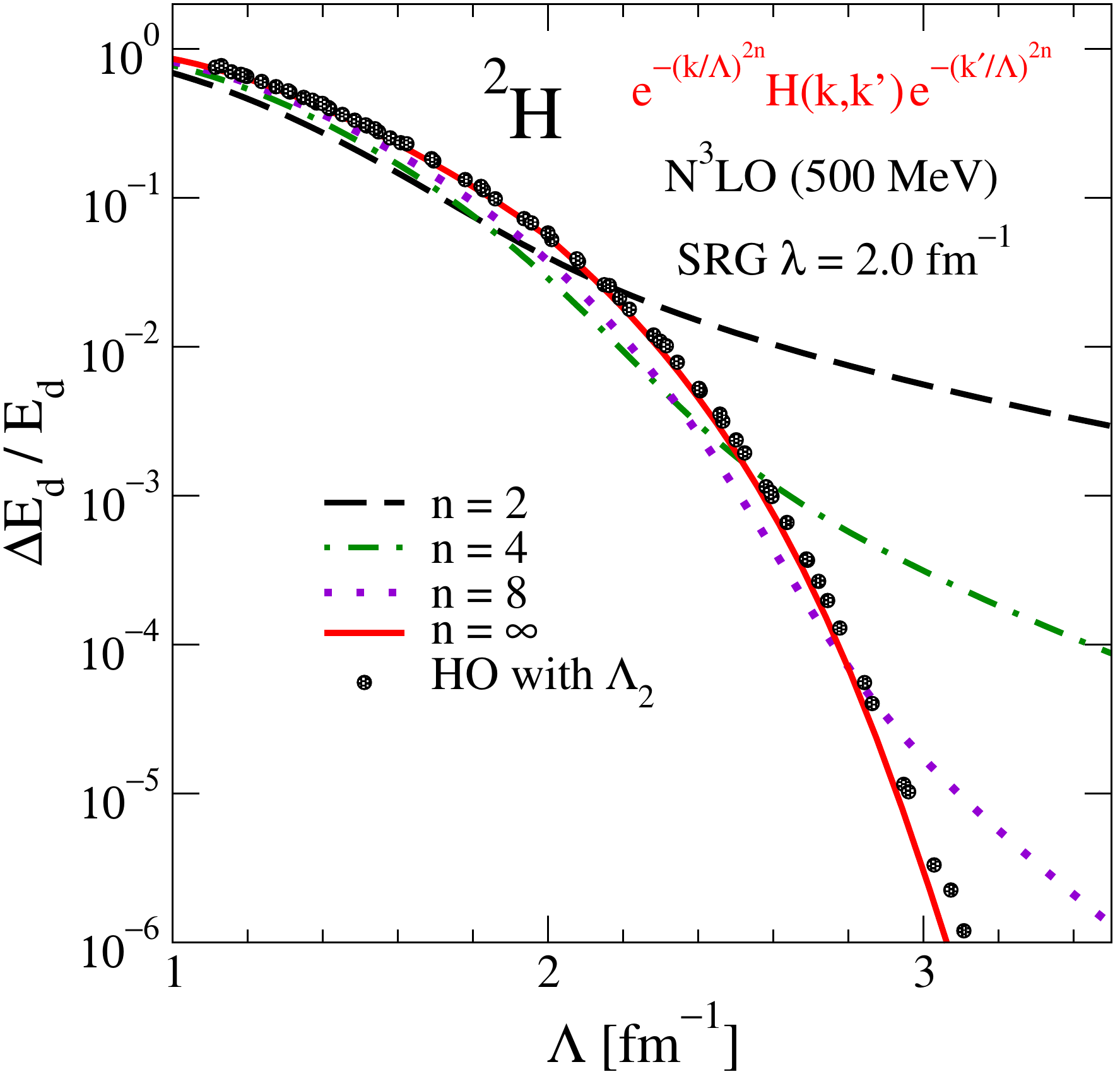}
	\caption{Calculations of the relative error in the
	   deuteron energy as a function of $\Lambda_2(\Nmax,\hw)$.  Circles
	   represent a wide range of oscillator parameters $\Nmax$ and
	   $\hw$ that are IR converged.  The series of lines shows energies for
	   which the Hamiltonian has been smoothly cutoff with exponent $n$.
	   The solid line corresponds to a sharp cutoff.}
	\label{fig:deuteron_vs_cutLam}
	\end{figure}
	In Fig.~\ref{fig:deuteron_vs_cutLam}, deuteron calculations are plotted as
	a function of
	$\Lambda = \Lambda_2(\Nmax,\Omega)$ along with several other functions
	of $\Lambda$ given by the relative error from the same Hamiltonian,
	but now smoothly cut off as
	\begin{equation}
	 H_{\rm cut}(k,k') = \ee^{-(k^2/\Lambda^2)^n} H(k,k')
	 \ee^{-(k'{}^2/\Lambda^2)^n} \,,
	\end{equation}
	for $n=2,4,8$ and $\infty$.  The latter corresponds to a sharp cutoff.
	We find that the curve from a sharp cutoff tracks the
	truncated-oscillator points through many orders of magnitude.
	This validates the claim that the error from oscillator basis
	truncation is well reproduced by applying instead a sharp
	cutoff in momentum at $\Lambda_2$.

	\subsection{Separable approximations}
	\label{subsec:Separable_approximation}

	We showed in Ref.~\cite{Konig:2014hma} that for a separable interaction of
	the form
	\begin{equation}
	V(k',k) = g\,\eta(k') \eta(k) \,,
	\label{eq:pot-RegContact}
	\end{equation}
	UV energy correction formula can be exactly derived.  For
	potential in Eq.~\eqref{eq:pot-RegContact}, the cutoff dependent
	binding momentum $\kappa_\lambda$ is given by the quantization condition
	\begin{equation}
	 {-}1 = 4\pi a \int_0^\Lambda\!\dd k\,
	 \frac{k^2\,\eta^2_\lambda(k)}{\kappa^2_\Lambda+k^2} \,,
	\label{eq:quant-sep}
	\end{equation}
	which is straightforward to solve numerically.

	However, most	interactions used in practical calculations do not have this
	convenient simple form (at least not in nuclear physics).  Still,
	as shown in Ref.~\cite{Konig:2014hma},	Eq.~\eqref{eq:quant-sep} can be put
	to some use using separable approximations.
	Methods to obtain separable approximations for a given potential have
	been known and used for quite a while (see, \eg,
	Refs.~\cite{Harms:1970hd,Ernst:1973zzb,elgaroy1998}).
	The technique we use is called \emph{unitary pole approximation
	(UPA)}~\cite{Ernst:1973zzb,Lovelace:1964aa}.  Assuming that for an
	arbitrary potential $\hat{V}$ we know a (bound) eigenstate $\ket{\psi}$, we
	can	construct a rank-1 separable approximation in momentum space by setting
	\begin{equation}
	 \hat{V}_\text{sep} =
	 \frac{\ket{\eta}\bra{\eta}}
	 {\mbraket{\psi}{\hat{V}}{\psi}} =
	 \frac{\hat{V}\ket{\psi}\bra{\psi}\hat{V}}
	 {\mbraket{\psi}{\hat{V}}{\psi}} \,.
	\label{eq:UPA-op}
	\end{equation}
	In other words, we have
	\begin{equation}
	 \eta(k) = \mbraket{k}{\hat{V}}{\psi}
	 \label{eq:form_factor_def}
	\end{equation}
	for the momentum-space ``form factor,'' and the coupling strength
	$g=\mbraket{\psi}{\hat{V}}{\psi}$ is, of course, independent of any
	particular representation.  From Eq.~\eqref{eq:UPA-op} one immediately
	sees that
	\begin{equation}
	 \hat{V}_\text{sep}\ket{\psi} = \hat{V}\ket{\psi} \,.
	\label{eq:Vsep-psi-V-psi}
	\end{equation}
	This means that the separable approximation is constructed in such a
	way that it exactly reproduces the state $\ket{\psi}$ used for its
	construction.  The potential from Eq.~\eqref{eq:UPA-op} reproduces the exact
	half off-shell T-matrix at the energy corresponding to the state
	$\psi$, and more sophisticated approximations (separable potential of
	rank $>1$) can be constructed by using more than a single
	state~\cite{Ernst:1973zzb}.  Since we are only interested in
	performing the UV extrapolation for a single state, however, the
	rank-1 approximation is sufficient.  To assess
	to what extent it actually reflects the UV behavior of a calculation
	based on the \emph{original} potential, we first considered some
	examples where the separable approximation can be constructed
	analytically such as the square well (Eq.~\eqref{eq:Vsw}) and the
	P\"{o}schl-Teller potential of the form
	\begin{equation}
	\VPT(r) = -\frac{\alpha^2 \beta(\beta-1)}{\cosh^2(\alpha r)}\;.
	\label{eq:V-PT}
	\end{equation}
	For given values of $\alpha$ and $\beta$, this potential has an
	analytically known bound-state spectrum.  Motivated by the success of the
	separable approximation (Eq.~\eqref{eq:UPA-op}) for the toy models,
	we moved on to the deuteron.  Here we will just look at a few representative
	results for the deuteron.

	A difficulty in applying the separable approximation directly to the deuteron
	is that the form factor $n(k)$ in Eq.~\eqref{eq:form_factor_def} depends on
	the deuteron wave function.  The exact wave function of course can not be
	calculated due to the truncation in the HO basis.  We use the best wave
	function available from the largest oscillator space and set
	\beq
	\eta(k) = \mbraket{k}{\hat{V}}{\psi}_{\rm HO,~best}\;.
	\label{eq:eta_best}
	\eeq
	As we know, deuteron has both $S$- and $D$-wave components.  This is taken
	into account by letting $\eta \to \eta_S ^2 + \eta_D ^2$.

\begin{figure}[thbp]
	\centering
	\includegraphics[width=0.6\textwidth]%
	{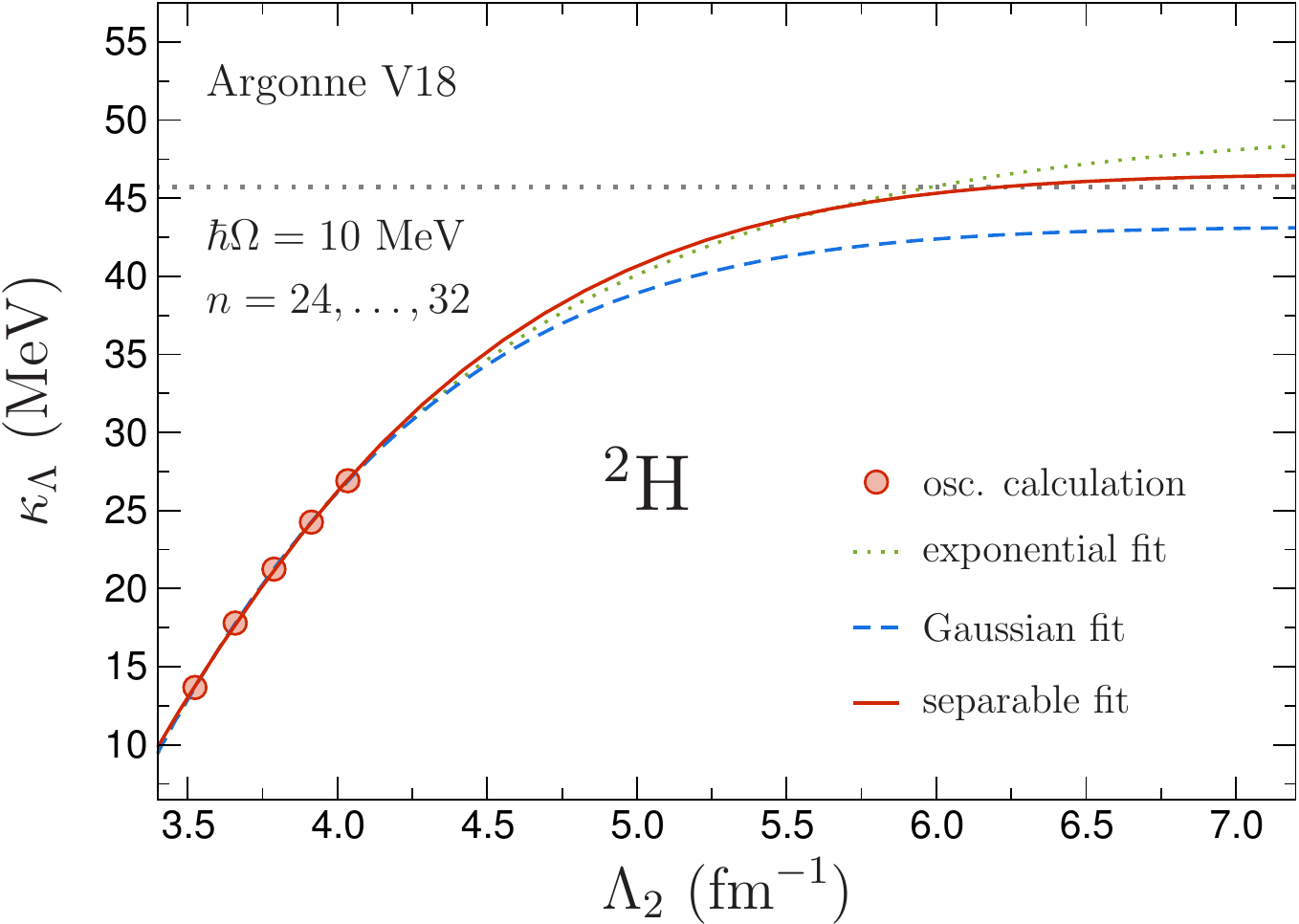}
	\caption{Comparison of UV extrapolations for a deuteron state
   calculated with the AV18 potential of Ref.~\cite{Wiringa:1994wb}.
   Circles: oscillator results.
   Dotted line: exponential extrapolation (Eq.~\eqref{eq:exp_fit_mom_UV}).
   Dashed line: Gaussian extrapolation (Eq.~\eqref{eq:gauss_fit_mom_UV}).
   Solid line: simplest separable extrapolation (Eq.~\eqref{eq:fit-eta-simple}).
   Dotted horizontal lines indicate the exact result for the binding
   momentum.}
	\label{fig:deuteron_sep_fit_av18}
	\end{figure}

\begin{figure}[thbp]
	\centering
	\includegraphics[width=0.6\textwidth]%
	{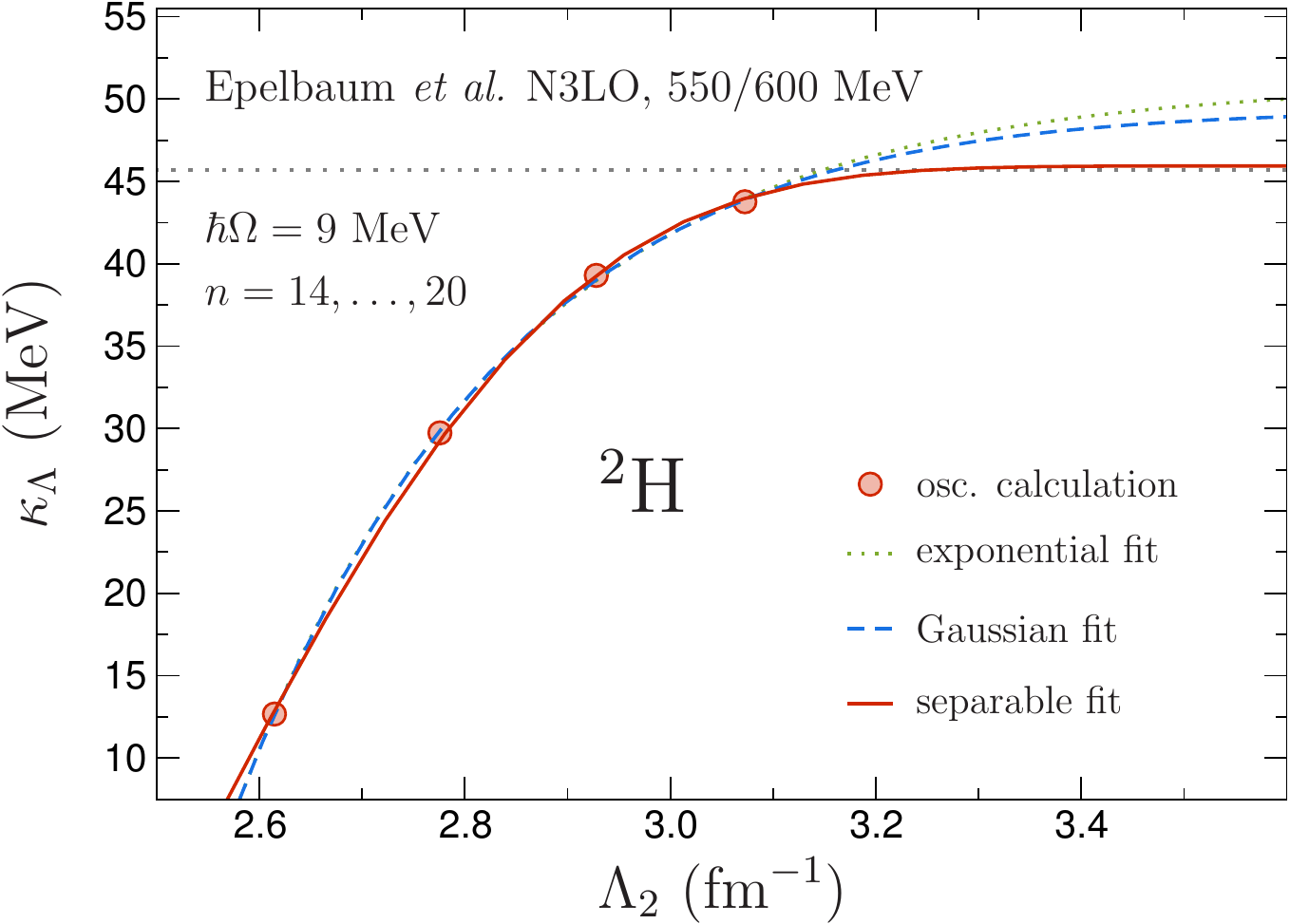}
	\caption{Calculations of UV extrapolations for a deuteron state
	   calculated with the Epelbaum~\etal N3LO ($550$/$600~\MeV$ cutoff)
	   potential of Ref.~\cite{Epelbaum:2004fk}.  The legend description is the
	   same as in Fig.~\ref{fig:deuteron_sep_fit_av18}.}
	\label{fig:deuteron_sep_fit_Epelbaum}
	\end{figure}

\begin{figure}[thbp]
	\centering
	\includegraphics[width=0.6\textwidth]%
	{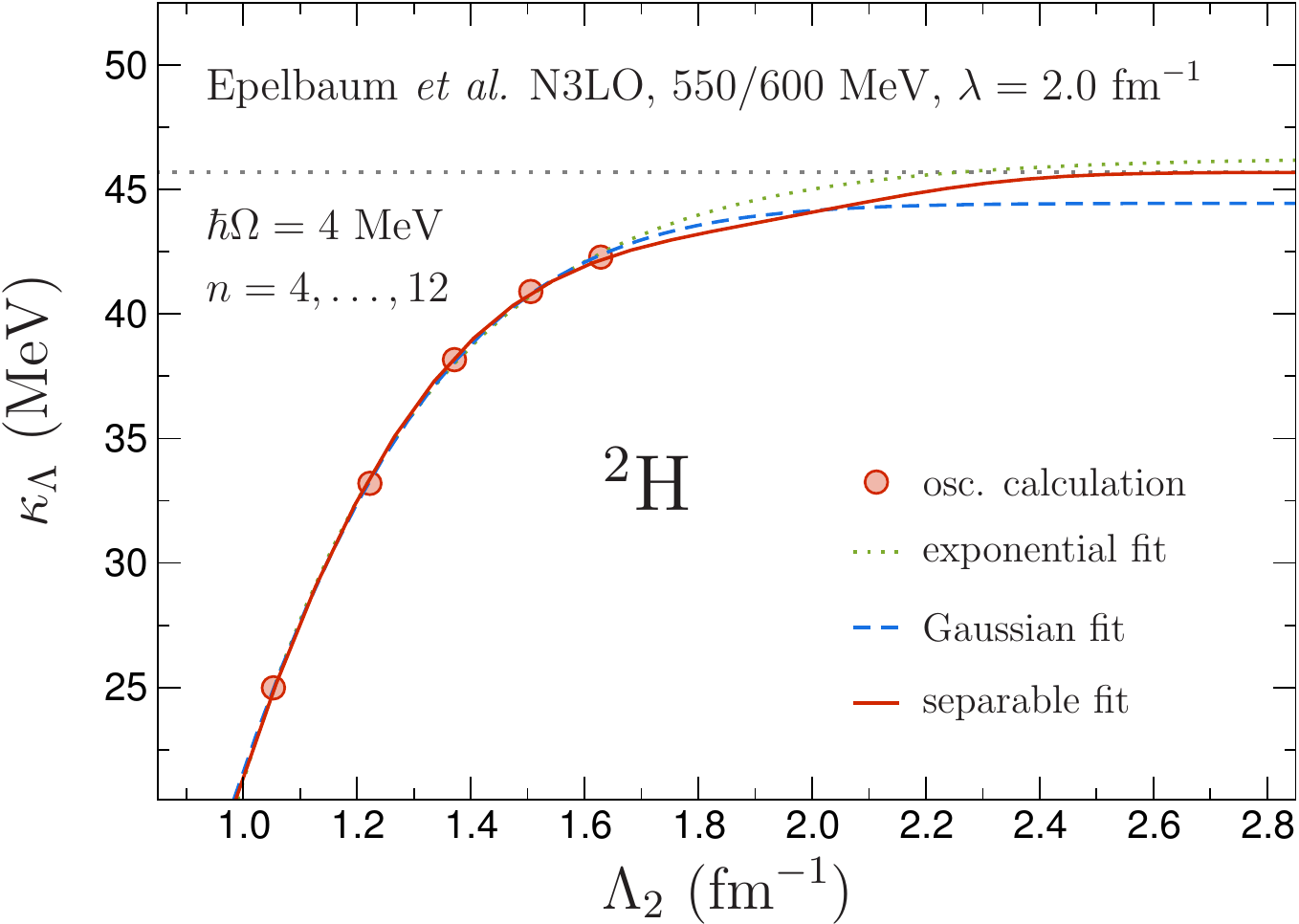}
	\caption{Calculations of UV extrapolations for a deuteron state
		 calculated with the Epelbaum~\etal N3LO ($550$/$600~\MeV$ cutoff)
		 potential of Ref.~\cite{Epelbaum:2004fk}.  The legend description is the
		 same as in Fig.~\ref{fig:deuteron_sep_fit_av18}.}
	\label{fig:deuteron_sep_fit_Epelbaum_evolved}
	\end{figure}

	As derived in Ref.~\cite{Konig:2014hma}, the simplest fit formula inspired
	by separable approximation is
	\begin{equation}
	 \kappa_\Lambda = \kappa_\infty
	  - A \int\nolimits_\Lambda^\infty \dd k\,\eta(k)^2 \,,
	\label{eq:fit-eta-simple}
	\end{equation}
	In Figs.~\ref{fig:deuteron_sep_fit_av18}, \ref{fig:deuteron_sep_fit_Epelbaum},
	and \ref{fig:deuteron_sep_fit_Epelbaum_evolved} we show the results
	obtained for the deuteron from fitting to Eq.~\eqref{eq:fit-eta-simple}.
	We compare the result for separable fit to two phenomenological choices.
	The exponential fit
	\beq
	\kappa_\Lambda = \kappa_\infty - a \, \ee^{-b \Lambda} \;,
	\label{eq:exp_fit_mom_UV}
	\eeq
	and the gaussian fit
	\beq
	\kappa_\Lambda = \kappa_\infty - a \, \ee^{-b \Lambda^2} \;.
	\label{eq:gauss_fit_mom_UV}
	\eeq
	We see from Figs.~\ref{fig:deuteron_sep_fit_av18},
	\ref{fig:deuteron_sep_fit_Epelbaum}, and
	\ref{fig:deuteron_sep_fit_Epelbaum_evolved} that the fit from the
	separable approximation is superior to phenomenological fits.
	The separable approximation allows extrapolation even when we are far from
	convergence (this is especially evident in
	Fig.~\ref{fig:deuteron_sep_fit_av18}).  It is also worthwhile to note
	that the separable fit (Eq.~\eqref{eq:fit-eta-simple}) has just
	two fit parameters $\kappainf$ and $A$, whereas the phenomenological
	fits have three free parameters $\kappainf$, $a$, and $b$.  The reason
	that the separable fit does well is that it puts in the information
	we already know about the state through Eq.~\eqref{eq:eta_best}.

	The IR and UV corrections exhibit a complementary mix of universal and
	non-universal characteristics.  The IR corrections are dictated by
	asymptotic behavior and are consequently determined by observables,
	independent of the details of the interaction.  So unitarily
	equivalent potentials---such as those	generated by renormalization-group
	running---will have the same corrections.
	In contrast, because they probe short-range features, UV corrections
	depend on	the details of the interaction (and the state under
	consideration).  However, the IR correction depends on the number of
	nucleons, whereas as we will see in Subsec.~\ref{subsec:UV_front}
	the UV correction is expected to scale simply with the number of particles.

	\section{Moving forward and related developments}

  In this section, we will list the open questions with respect to both IR
	and UV extrapolations.  As mentioned before, this approach of mapping the
	HO truncation to IR and UV cutoffs and using them to obtain physically
	motivated extrapolation formulas was rigorously developed for the first
	time by us \cite{More:2013rma,Furnstahl:2013vda,Konig:2014hma}.  This
	pioneering work has sprouted many new developments by extending our work.
	We will briefly touch upon some of these related developments.

	\subsection{IR front}
	\label{subsec:IR_front}

  \medskip
	\subsubsection{Open questions}

	As discussed towards the end of Subsec.~\ref{subsec:higher_angular_momenta},
	the NLO IR correction is incomplete due to the missing $l=2$ correction.
	It might be challenging to derive NLO corrections to the binding
	energies for nuclei with $A> 2$, particular for nuclei with nonzero
	ground-state spin.  Here, many different orbital anglar momenta can
	contribute to the ground-state wave function, and one would presumably
	need to know the admixture of the different channels quite accurately.
	Our results show that nonzero orbital angular momenta yield
	corrections in inverse powers of $\kinf L$ to the LO energy
	extrapolation.  On the other hand, the leading contributions to
	bound-state energies in finite model spaces fall off as $\exp{(-2\kinf
	L)}$ for all orbital angular momenta.  This makes extrapolations
	feasible in practice.

	The formulation in terms of S-matrix analytic structure is closely
	related to methods used to analyze break-up reactions, which provides a
	link to $A>2$ extrapolations.  Indeed, in Ref.~\cite{Furnstahl:2012qg}
	the basic form of the LO extrapolation proportional to $e^{-2\kinf L}$
	was based on interpreting $\kinf$ in terms of the one-particle
	separation energy.  More generally, the asymptotic many-body wave
	function is dominated by configurations corresponding to the break-up
	channels with the lowest separation energies and it is
	their modification by the hard wall at $L$ that will be associated
	with the energy shift $\Delta E_L$.  This is in turn dominantly
	described by the S-matrix near poles at the corresponding separation
	binding momenta.  Future work will seek to clarify the precise nature
	of the more general expansion (including the effects of the Coulomb
	interaction) and whether it will be possible to quantitatively extract
	asymptotic normalization constants.

	\paragraph{Relation to L\"{u}scher-type formulas}

	We saw in Subsec.~\ref{subsec:lattice_theories} that lattice theories
	have an inherent IR and UV cutoff.
	Starting with the seminal work of L\"uscher~\cite{Luscher:1985dn}, a
	wide variety of formulas have been derived for the energy shift of
	bound states in finite-volume lattice calculations.  The usual
	application is to simulations that use periodic boundary conditions in
	cubic boxes (e.g., see Ref.~\cite{Konig:2011ti}).  The recent work by
	Pine and Lee~\cite{Lee:2010km,Pine:2012zv} extend the derivation to
	hard-wall boundary conditions using effective field theory for
	zero-range interactions and the method of images.  The result for
	$\Delta E_L$ in a three-dimensional cubic box has a different
	functional form than found here (the leading exponential is multiplied
	by $1/L$ with that geometry) and the subleading corrections are
	parametrically larger.

	However, because the HO truncation we consider is in partial waves,
	the one-dimensional analysis and formula from Ref.~\cite{Pine:2012zv}
	are applicable (because $\kinf$ and $\ANC$ are asymptotic quantities,
	the result for zero-range interaction is actually general for
	short-range interactions).  The method of images can be applied in a
	one-dimensional box of size $2L$ after specializing to a particular
	partial wave and then extending the space to odd solutions in $r$ from
	$-\infty$ to $+\infty$.  The leading-order finite-volume correction
	agrees with Eq.~\eqref{eq:complete_IR_scaling}, and the first omitted term
	is of the same order.
	The methods presented in \cite{Lee:2010km,Pine:2012zv}
	can be used to extend the present formulas to higher
	orders and more general cases, including coupled channels.  This area
	is ripe for investigation.

	Another area of investigation is how the trends for operator extrapolation
	carry over for $A > 2$.

	\medskip
	\subsubsection{Related developments}

	The results presented in this chapter have exclusively been for the two-body
	case.
	\begin{figure}[thbp]
	\centering
	\includegraphics[width=0.6\textwidth]%
	{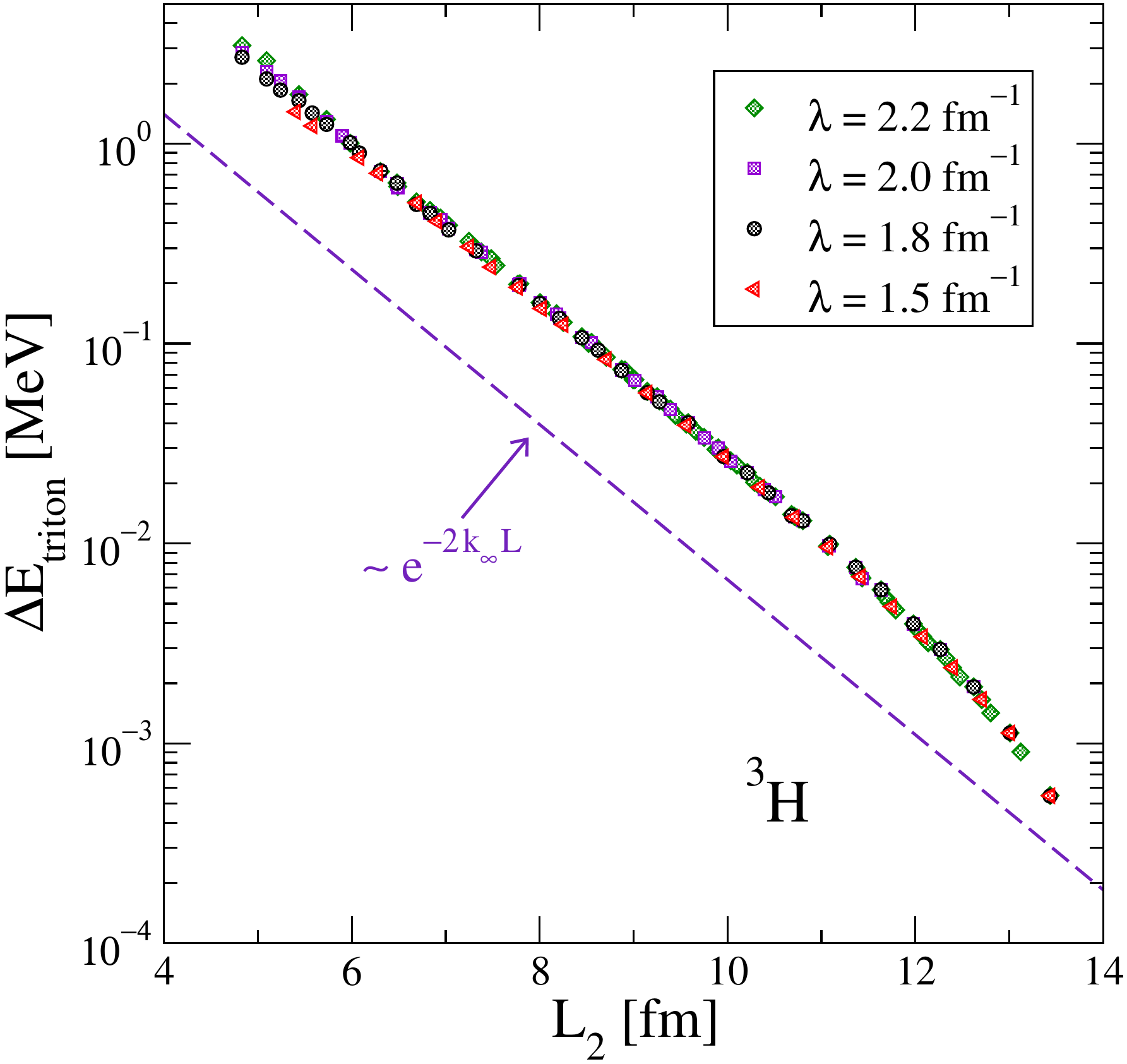}
	\caption{Residual error for triton plotted as a function of $L_2$ (here
 		calculated with the deuteron-neutron reduced mass) for the two- and
		three-nucleon	potential in Ref.~\cite{Jurgenson:2009qs} unitarily evolved
		by the SRG to four different resolutions (specified by $\lambda$) with the
		same binding energy~\cite{Jurgenson:2009qs,Jurgenson:2010wy}.
		$\kinf$ here is the lowest separation energy (triton breaking up
		into deuteron and neutron).}
	\label{fig:triton_vs_L2}
	\end{figure}
	Figure~\ref{fig:triton_vs_L2} shows $\Delta E = E_{\rm HO} - \Einf$ for triton
	plotted as a function of $L_2$.  Recall from Eq.~\eqref{eq:L2_def} that
	evaluating $L_2$ involves calculating the oscillator length $b$.  In
	Fig.~\ref{fig:triton_vs_L2}, we use the deuteron-neutron reduced mass,
	$\mu = 2/3 M_N$, to calculate $b$ and thereby $L_2$.  There are a few
	interesting observations to be made about Fig.~\ref{fig:triton_vs_L2}.
	Triton energies when plotted as a function of $L_2$ lie on a single line.
	Also, as in Fig.~\ref{fig:SRG_data_collapse}, triton energies from potentials
	evolved to various SRG $\lambda$'s, fall on the same line.  This indicates
	that $L_2$ is the correct length even for the three-body case and the
	three-body IR correction can also be written in terms of observables.
	Moreover the falloff is proportional to $e^{-2 \kinf L_2}$, with
	$\kinf$ being the lowest separation energy, as expected.

  There has been a lot of work on extending the results for IR energy corrections
	presented in this thesis to the many-body case.  This has been documented in
	Refs.~\cite{Furnstahl:2014hca, Wendt:2015nba}.  In
	Subsec.~\ref{subsec:radii_phase_shifts} we looked at the
	extrapolation of the radius-squared operator.  The authors of
	Ref.~\cite{Odell:2015xlw} extended this to the extrapolation of quadrupole
	moments and transitions for the deuteron.

	As mentioned previously, the approach of mapping the HO truncation into a
	hard-wall boundary condition (in both position as well as momentum space)
	can be used for any localized basis.  This has been explored for the case
	of Coulomb-Sturmian basis \cite{Caprio_Coulomb_Sturmian}.

	\subsection{UV front}
	\label{subsec:UV_front}

	The dependence of UV corrections on the number of nucleons $A$ is not
	yet established theoretically, but the tests in \cite{Konig:2014hma}
	seem to indicate that the cutoff dependence of $\Delta E$ for the
	many-body case is the same as in the two-body case, just scaled by an
	$A$-dependent overall constant.
	\begin{figure}[thbp]
	\centering
	\includegraphics[width=0.6\textwidth]%
	{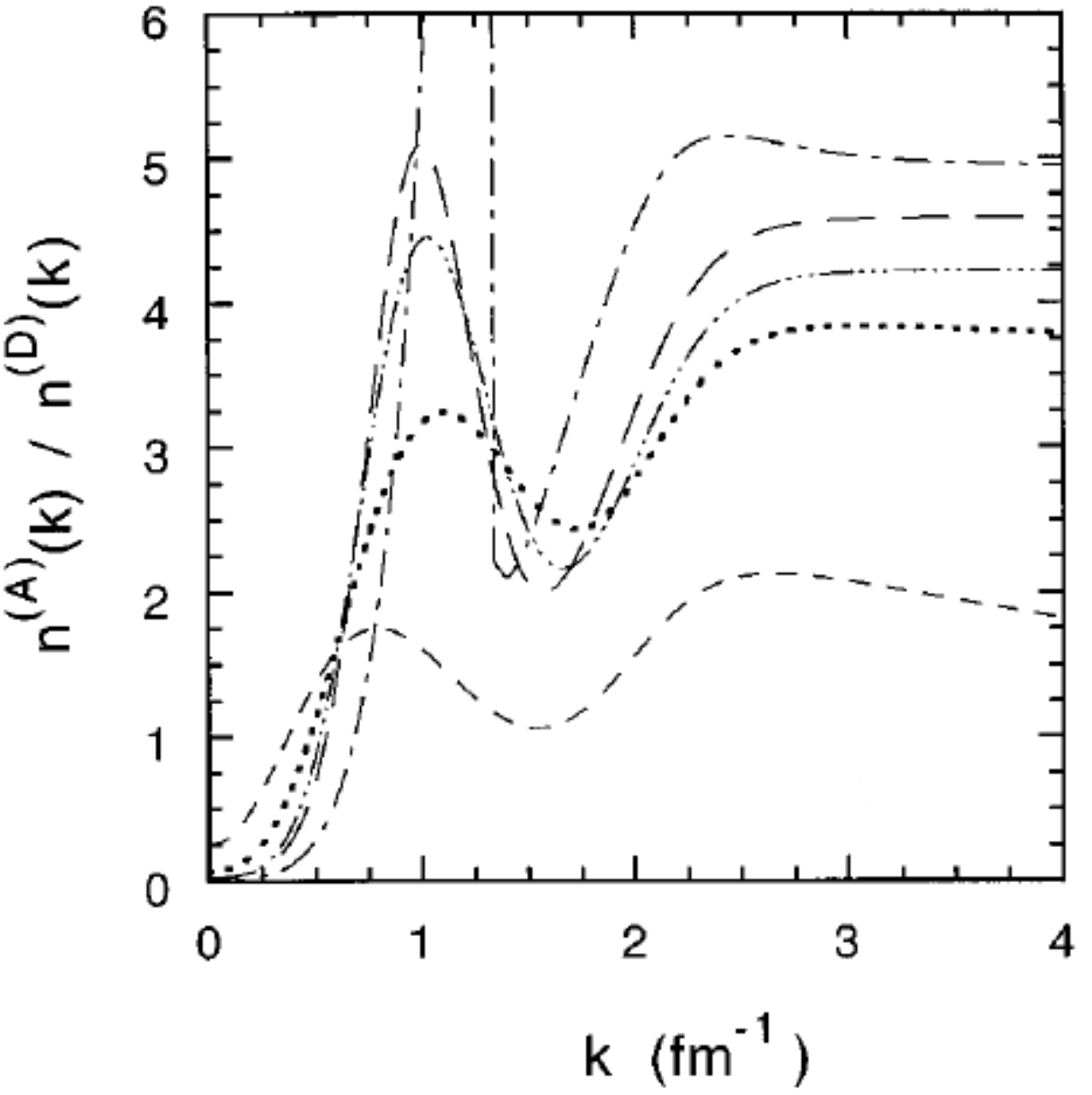}
	\caption{The ratio of the momentum distributions in nucleus to the deuteron
		momentum distributions.  The dashed, dotted, dot-dashed, long dashed,
		dot-long dashed lines correspond to $^3$He, $^4$He, $^{16}$O, $^{56}$Fe,
		and nuclear matter respectively.  Figure taken from
		\cite{CiofidegliAtti:1995qe}.}
	\label{fig:mom_distribution_ratio}
	\end{figure}
	This can be understood
	from general considerations of short-range correlations~\cite{Kimball:1973aa}
	or more systematically using the operator product
	expansion~\cite{Bogner:2012zm,Hofmann:2013aa}.  If there is a common two-body
	part, it may determine the dominant $\Lambda_2$ dependence
	with the rest providing the $A$-dependent scale factor.  This behavior would
	be consistent with the observation of a universal shape for high-momentum
	tails in momentum distributions in
	Fig.~\ref{fig:mom_distribution_ratio} (or the corresponding short-distance
	behavior)~\cite{CiofidegliAtti:1995qe,Feldmeier:2011qy}.  Connecting the
	two-body UV extrapolation results to the many-body case remains an open
	question.

	In this chapter we worked in the region where either the IR or the UV
	errors were dominant.  We saw in
	Fig.~\ref{fig:SRG_data_collapse_contamination} how UV contamination spoils
	the data collapse for IR extrapolation.  However, it is not always possible
	to isolate the IR and the UV contributions.  We therefore need reliable
	extrapolation schemes which can be employed when both the IR and UV errors
	are comparable.  Ref.~\cite{Jurgenson:2013yya} combined phenomenological
	UV errors \footnote{Phenomenological form for the UV error is found to be
	Gaussian in the cutoff $\Lambda_2$.  Ref.~\cite{Konig:2014hma} discusses how
	the Gaussian form arises.} with leading IR errors and used extrapolation of
	the	form
	\beq
	E(\Lambda_2, L_2) = \Einf + B_0 \ee^{-2 \Lambda_2^2/ B_1^2} + B_2
		\ee^{-2 \kinf L_2} \;.
	\label{eq:combine_IR_UV_phenomenological}
	\eeq
	\begin{figure}[thbp]
	\centering
	\includegraphics[width=0.6\textwidth]%
	{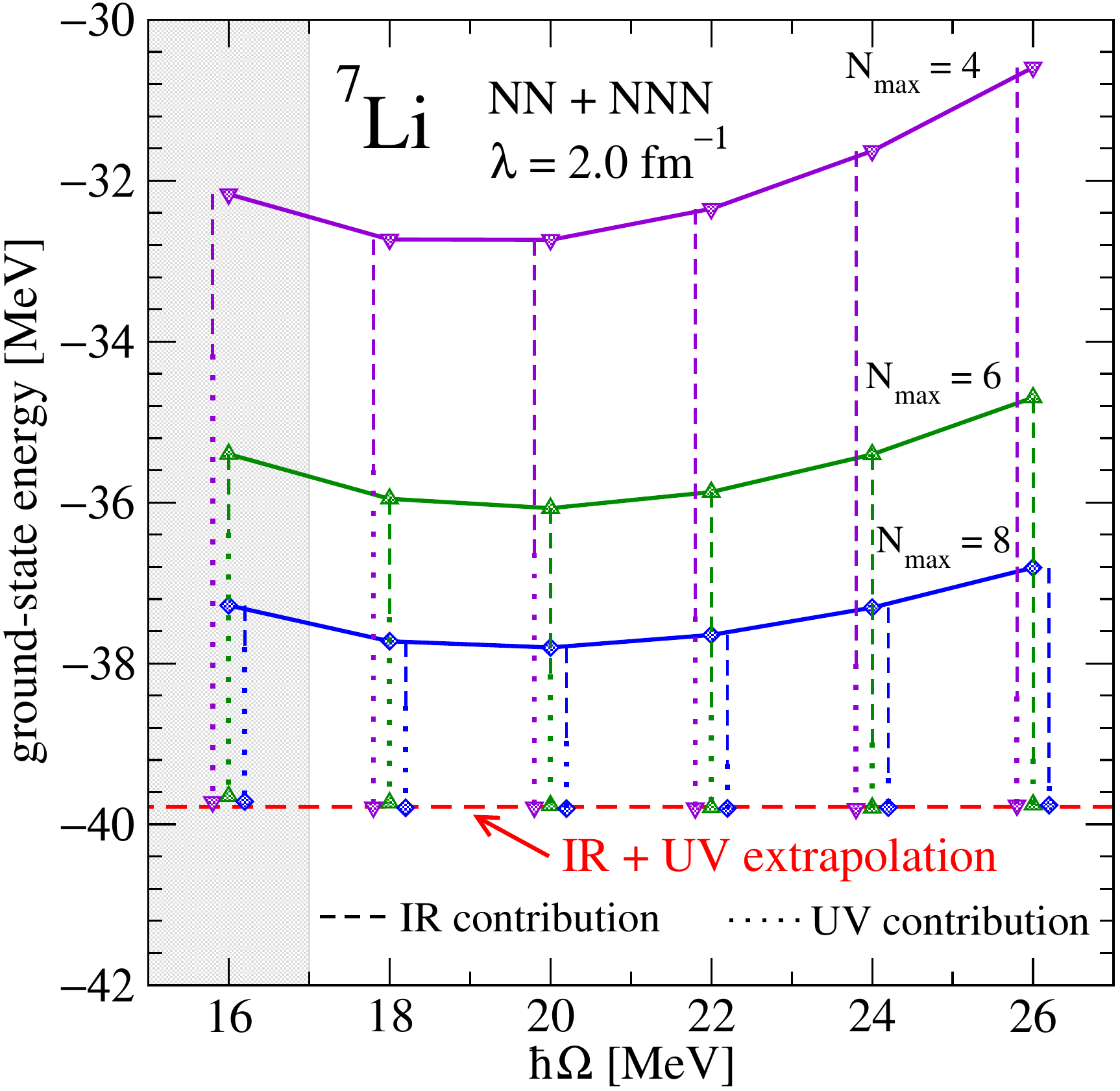}
	\caption{Ground-state energy for $^7$Li with IR (vertical dashed lines) and
		UV (vertical dotted lines) corrections from
		Eq.~\eqref{eq:combine_IR_UV_phenomenological} added to predict $\Einf$
		values.  The horizontal dashed line is the global $\Einf$.
		Figure taken from \cite{Jurgenson:2013yya}.}
	\label{fig:combine_IR_UV_Jurgenson}
	\end{figure}
	As seen in Fig.~\ref{fig:combine_IR_UV_Jurgenson} this simple addition of IR
	and UV errors seems to work well.  More work will be needed to place this
	on a sound theoretical foundation.

	The authors of Ref.~\cite{Binder:2015trg} developed interactions from chiral
	EFT that are tailored to the HO basis.  In their approach, the UV convergence
	with respect to the model space is implemented by construction (through
	refitting of LECs) and IR convergence is achieved by enlarging the model
	space for the kinetic energy.  This exhibited a fast convergence of
	ground-state energies and radii for nuclei up to $^{132}$Sn.  Thus,
	the development of reliable extrapolation schemes is indeed pushing the
	ab-initio frontier to heavier nuclei.

\cleardoublepage
\chapter[Factorization]{Factorization
	\footnote{Based on \cite{More:2015tpa}}}
	\label{chap:factorization}

	\section{Motivation}

	Most of the information we know about nuclear interactions and the properties
	of the nuclei comes from some kind of scattering experiments (either elastic
	or inelastic).  In such experiments we scatter a known probe off a nucleus
	and	extract information about nuclear interactions by looking at the final
	outcome of the scattering experiment.
	\begin{figure}[thbp]
	\centering
	\includegraphics[width=0.5\textwidth]%
	{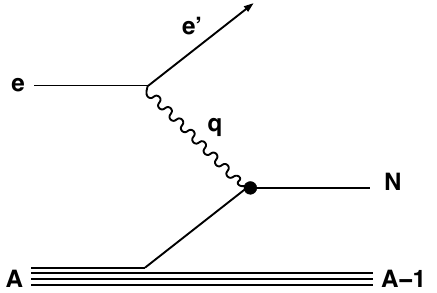}
	\caption{Schematic of a nucleon knockout reaction.}
	\label{fig:knock_out_schematic}
	\end{figure}
	\begin{figure}[thbp]
	\centering
	\includegraphics[width=0.6\textwidth]%
	{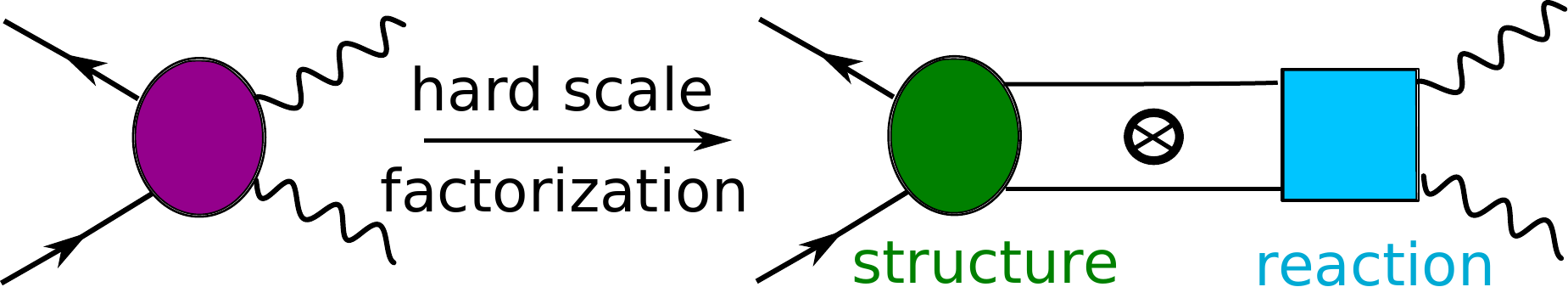}
	\caption{Schematic illustration of factorization between nuclear structure
		and reactions component.}
	\label{fig:factorization_schematic}
	\end{figure}
	Figure~\ref{fig:knock_out_schematic} shows a schematic for a nucleon knockout
	reaction where the probe is electrons, which interact with the nucleus
	by emitting virtual photons.

	The process of extracting nuclear properties from such experiments relies
	on the assumption that the effects of the probe are well understood and can
	be separated from the nuclear interactions we are trying to study.  This is
	the	\emph{factorization} between the nuclear structure and the nuclear
	reaction components illustrated schematically in
	Fig.~\ref{fig:factorization_schematic}.  The reaction component describes the
	probe and the structure includes the description of the initial and final
	states.
	This factorization between the structure and reaction components depends
	on the renormalization scale and scheme.  In some physical systems
	(e.g., in cold atoms near unitarity~\cite{Hoinka:2013fsa}), the scale and
	scheme dependence is very weak and can be safely neglected.  In some other
	physical systems such as in deep inelastic scattering (DIS) in high-energy
	QCD, the scale and scheme dependence is very manifest.
	\begin{figure}[thbp]
	\centering
	\includegraphics[width=0.55\textwidth]%
	{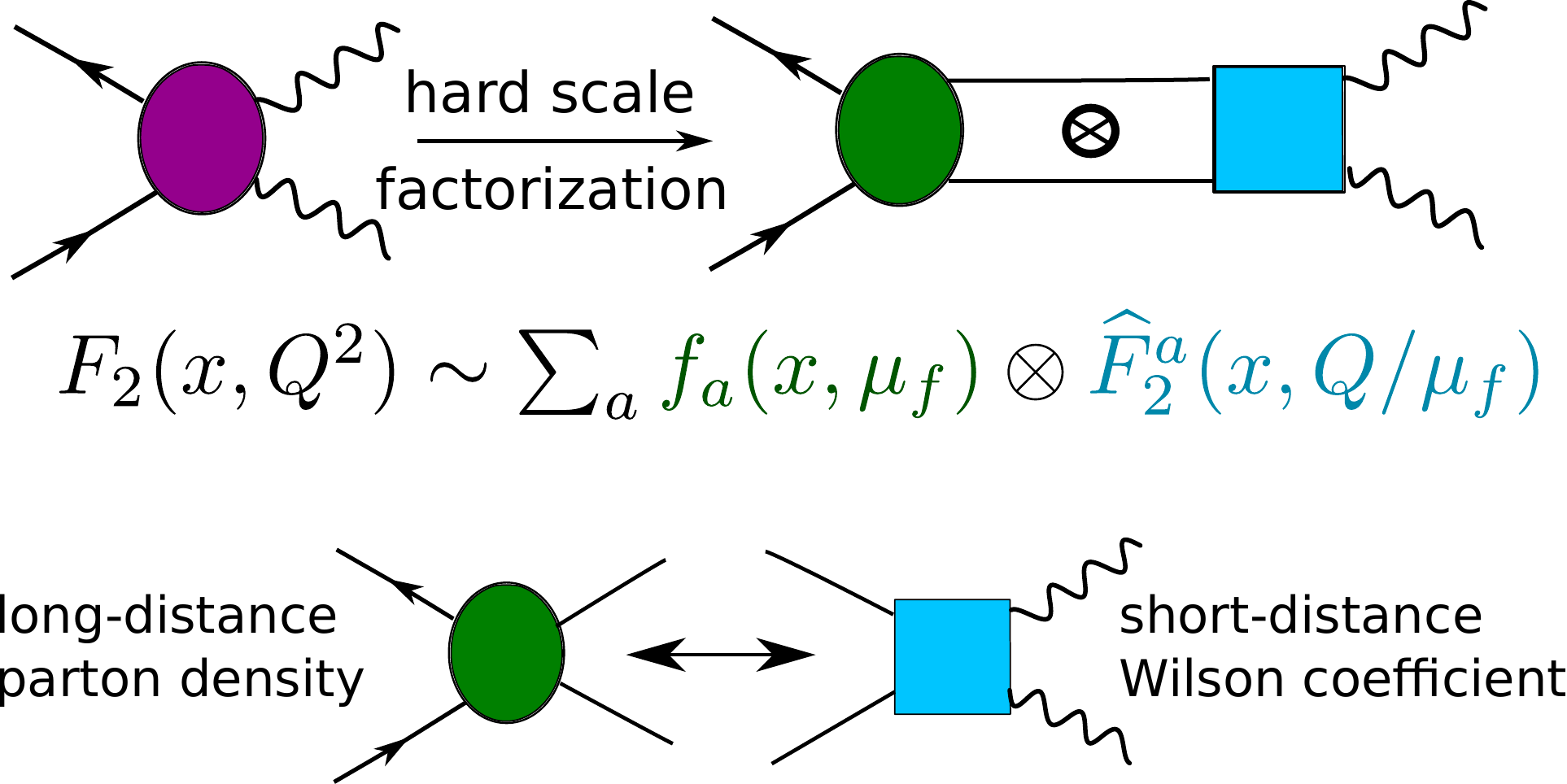}
	\caption{Factorization in high-energy QCD.  $x$ is the Bjorken-$x$ and it
		denotes the fraction of momentum of the nucleon carried by the parton
		under consideration.  $a$ denotes the parton flavor.}
	\label{fig:factorization_schematic_highE_QCD}
	\end{figure}
	Figure~\ref{fig:factorization_schematic_highE_QCD} illustrates the
	factorization in DIS.  The form factor $F_2$ of the nucleon
	(which up to some kinematic factors is the cross section) is given by the
	convolution of the long-distance parton density and the short-distance
	Wilson coefficient.  In this case, the parton density forms the structure
	part which is non-perturbative and the Wilson coefficient form the reactions
	part which can be calculated in perturbative QCD.  This separation
	between long- and short-distance physics is not unique, and is defined by
	the factorization scale $\mu_f$.  To minimize the contribution of logarithms
	that can disturb the perturbative expansion, $\mu_f$ is chosen to be equal to
	the	magnitude of the four-momentum transfer $Q$.
	The form factor $F_2$ (because it is
	related to the observable cross section) is independent of $\mu_f$, but
	the individual components are not.  As a consequence, the parton density
	(or distribution) function $f_a(x, Q^2)$ runs with $Q^2$.
	\begin{figure}[thbp]
	\centering
	\includegraphics[width=0.55\textwidth]%
	{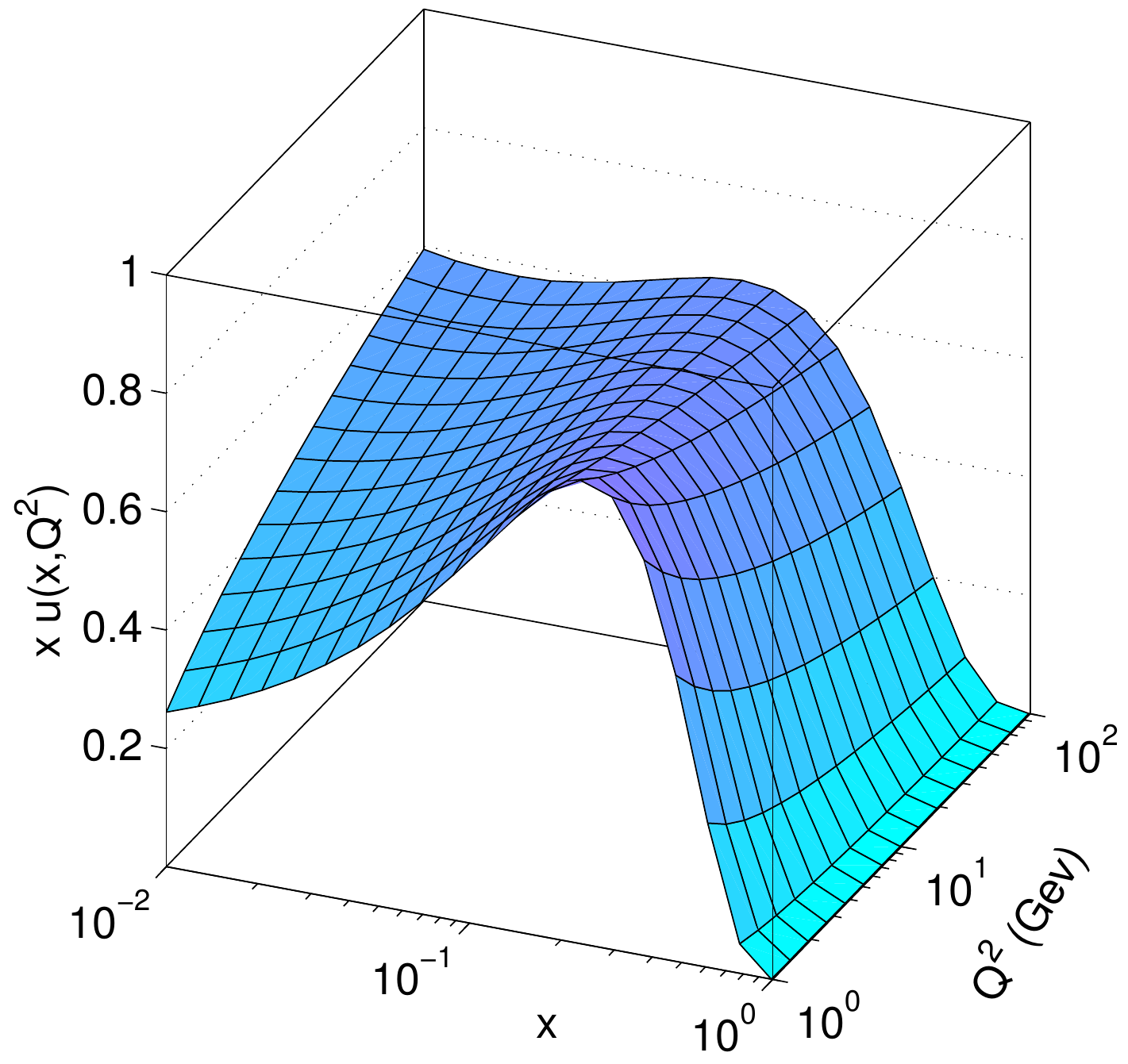}
	\caption{Parton distribution for the up quarks in the proton as a function
		of $x$ and $Q^2$.  Figure taken from \cite{Furnstahl:2013dsa}.}
	\label{fig:uquark_PDF_3D}
	\end{figure}
	This is demonstrated in Fig.~\ref{fig:uquark_PDF_3D}.  The parton
	distributions
	$f_a(x, Q^2)$ and $f_a(x, Q_0^2)$	at two different $Q^2$ are related by
	DGLAP evolution or the Altarelli-Parisi equations \cite{Altarelli:1977zs}.
	Thus the scale dependence of the structure and reaction components is well
	understood in high-energy QCD.

	The situation is far from well settled in low-energy nuclear physics.
	Nuclear
	structure has conventionally been treated largely separate from nuclear
	reactions
	(e.g., the two volumes of Feshbach's \emph{Theoretical Nuclear Physics} are
	divided
	this way).  The nuclear structure community usually dealt with calculating
	time-independent properties such as nuclear binding energies, excitation
	spectra, radii, so on whereas the nuclear reaction experts worked on
	disintegration,	knock-out, and
	transfer reactions.  However, both the communities invariably use inputs
	from the other side, and the consistency and universality of different
	components is not
	always guaranteed. 	To go back to the high-energy QCD analogy, the parton
	distribution functions (PDFs) $f_a(x, Q)$ extracted from the DIS are
	universal, in	the sense that they are process-independent.  For instance,
	the PDFs extracted
	from the DIS can be used for making prediction for the Drell-Yan process.
	The analogous process independence in the extracted quantities has not yet
	been demonstrated in low-energy nuclear physics.  This leads to ambiguous
	uncertainty	quantification when the nuclear properties extracted from one
	process	(cf.~Fig.~\ref{fig:factorization_schematic}) are used as an input to
	predict something	else.

\begin{figure}[thbp]
	\centering
	\includegraphics[width=0.6\textwidth]%
	{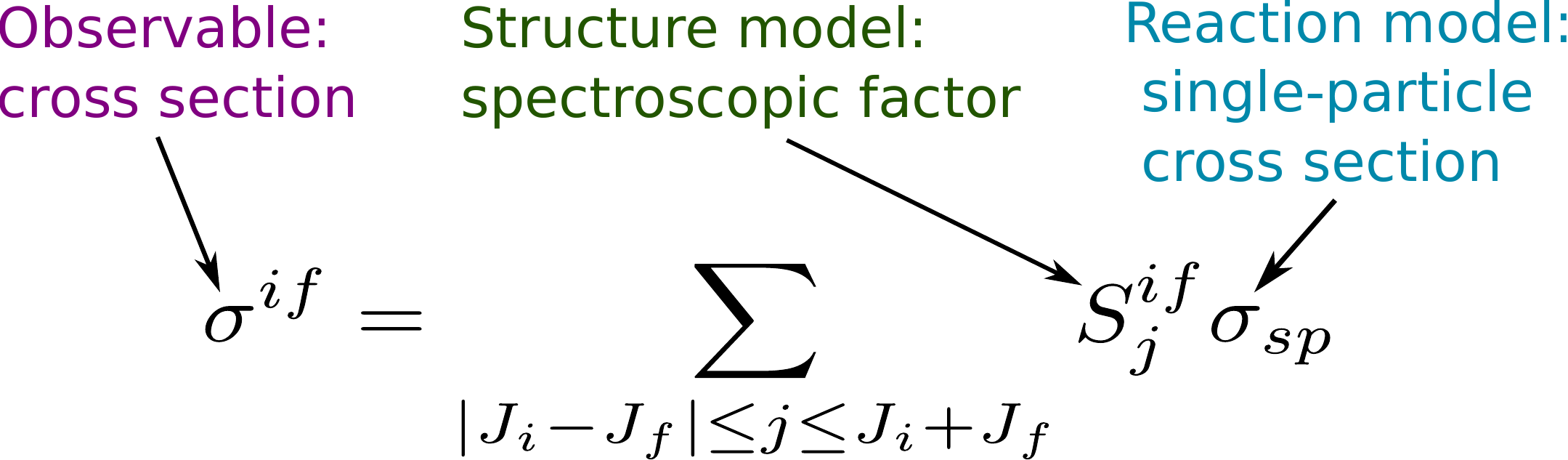}
	\caption{Schematic illustration of factorization in low-energy nuclear
		physics.}
	\label{fig:low_energy_factorization}
	\end{figure}

	The assumed factorization in low-energy nuclear physics is illustrated in
	Fig.~\ref{fig:low_energy_factorization}.
	The observable cross section in this case is written as a convolution of the
	spectroscopic factor
	and the single-particle cross section.  However, there are many open questions
	such when does this factorization hold and how can we justify it
	theoretically?  In cases that it does hold what are
	the	nuclear properties that we can extract and what is the
	scale/scheme dependence of these extracted properties?

	The Similarity Renormalization Group (or the SRG) transformations were
	introduced in	Subsec.~\ref{subsec:SRG_intro}.  We noted that SRG
	transformations are a class of unitary transformations that soften nuclear
	Hamiltonians and lead to accelerated convergence of observables.
	\begin{figure}[thbp]
	\centering
	\includegraphics[width=0.6\textwidth]%
	{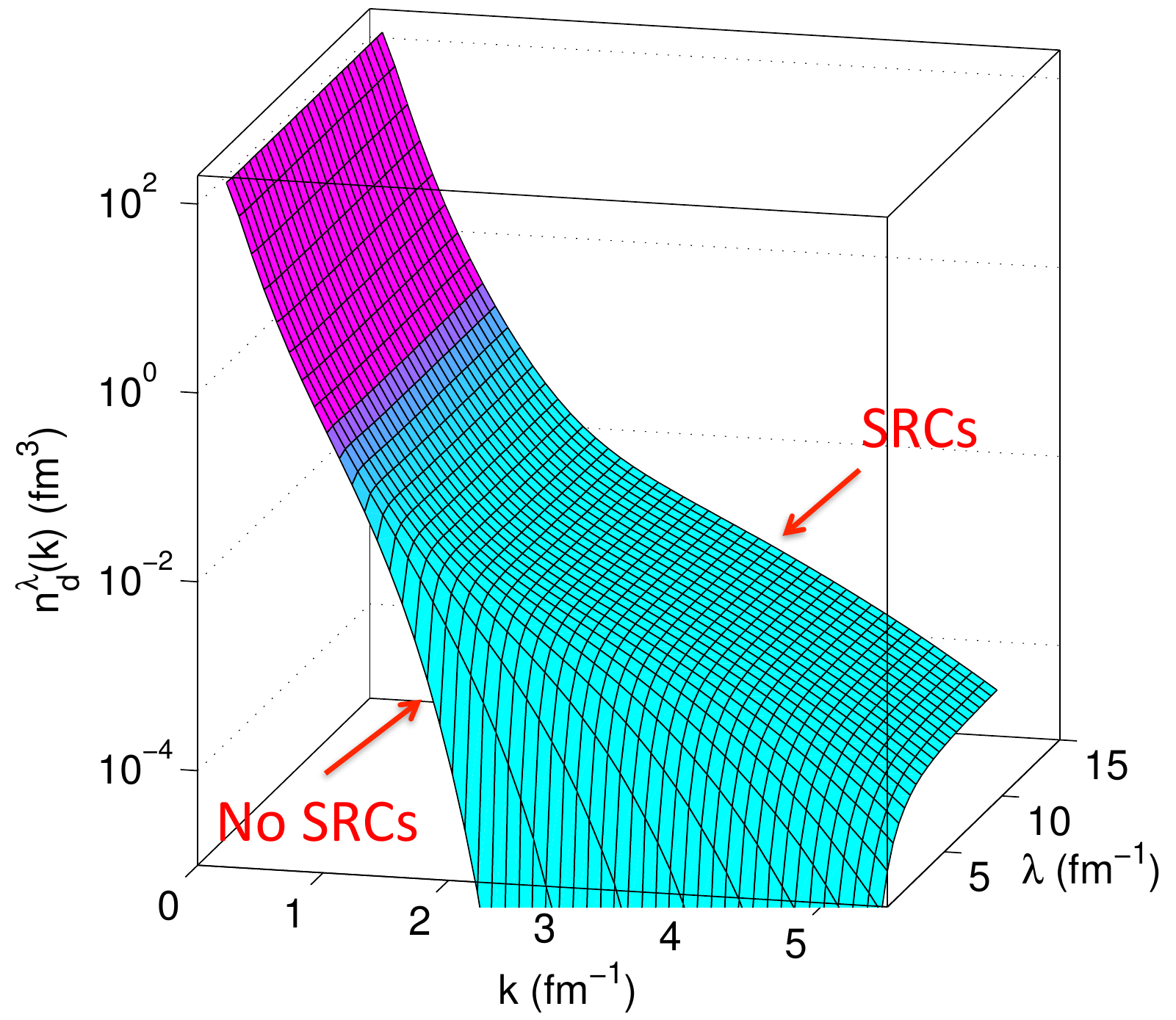}
	\caption{Deuteron momentum distribution at different SRG resolutions
		$\lambda$.  The evolved momentum distribution does not have the short-range
		correlations (SRCs).  Figure from \cite{Furnstahl:2013dsa}.}
	\label{fig:momentum_distribution}
	\end{figure}
	Figure~\ref{fig:momentum_distribution} shows the momentum distribution for
	the deuteron as a function of the SRG scale and momentum.  Note that
	Fig.~\ref{fig:momentum_distribution} is analogous to
	Fig.~\ref{fig:uquark_PDF_3D}.  The SRG evolution gets rid of the
	high-momentum components and therefore the evolved momentum distributions
	don't have the short-range correlations (SRCs).
	Figure~\ref{fig:momentum_distribution} makes it clear that the
	high-momentum tail of the momentum distribution is dramatically resolution
	dependent.  Yet it is common in the literature that
	high-momentum components are treated as measurable, at least
	implicitly~\cite{Frankfurt:2008zv,Arrington:2011xs,Rios:2013zqa,
	Boeglin:2015cha}.
	In fact, what can be extracted is the momentum distribution at some scale,
	and	with the specification of a scheme.  This makes momentum distributions
	model	dependent~\cite{Ford:2014yua, Sammarruca:2015hba}.

	\begin{figure}[thbp]
	\centering
	\includegraphics[width=0.6\textwidth]%
	{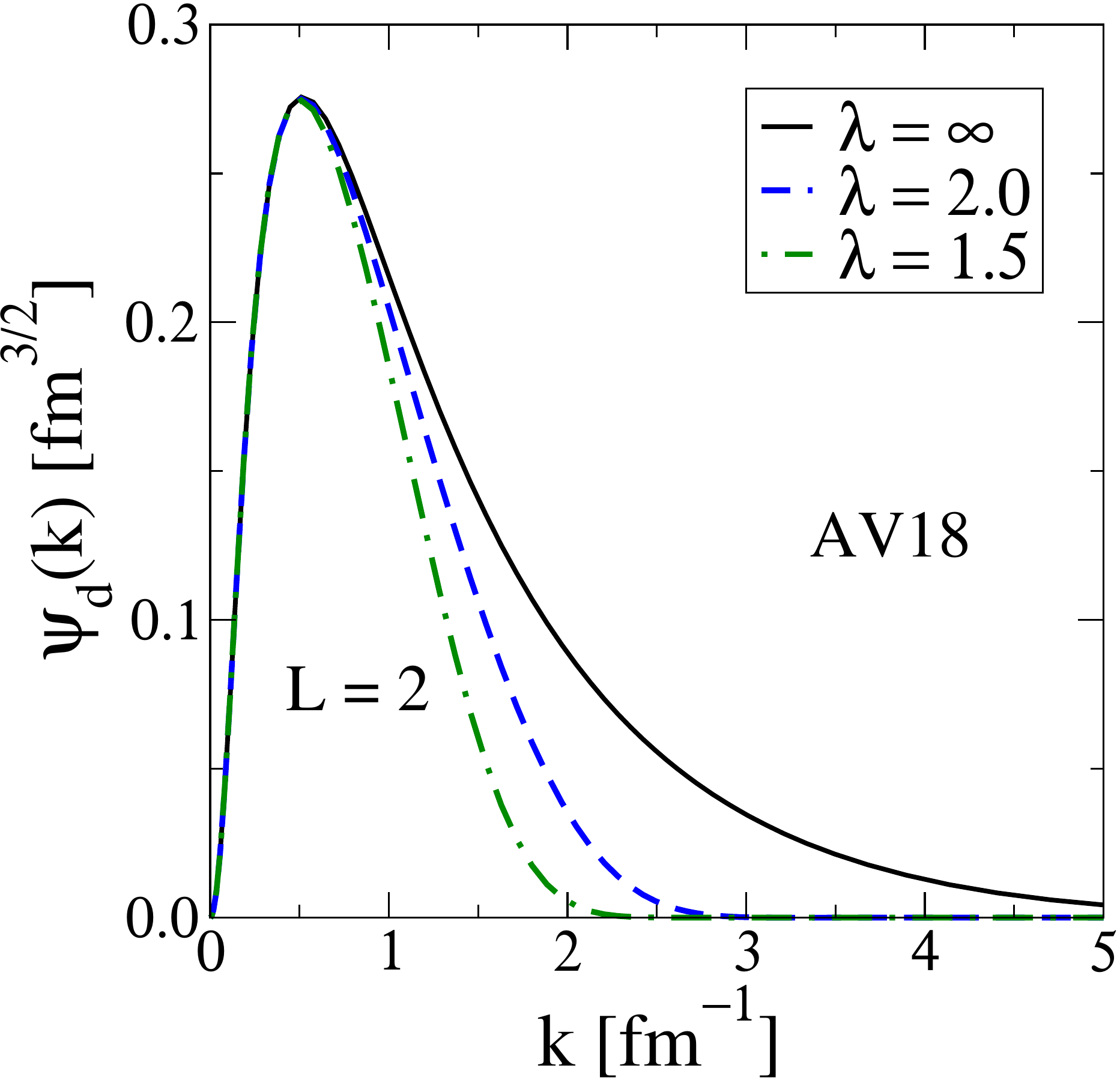}
	\caption{$D$-state wave functions for the deuteron for the AV18 potential and
		the AV18 potential evolved to two SRG $\lambda$'s.}
	\label{fig:wavefunction_evolution_deuteron_D_state}
	\end{figure}
\begin{figure}[thbp]
	\centering
	\includegraphics[width=0.55\textwidth]%
	{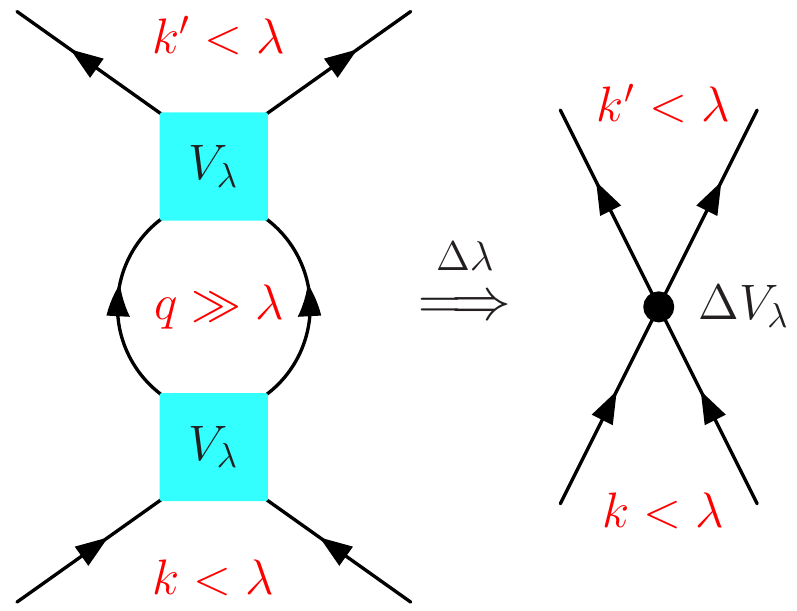}
	\caption{High-momentum modes in intermediate states replaced by contact
		interactions.  Figure from \cite{Furnstahl:2013dsa}.}
	\label{fig:replace_loop_contact}
	\end{figure}

	Figure~\ref{fig:wavefunction_evolution_deuteron_D_state} shows the $D$-state
	wave function for deuteron.  We see that just like the momentum distributions,
	SRG transformed wave functions do not have the high-momentum components.
	Therefore,
	if we use the SRG evolved wave function for calculating the cross section for
	a	process involving high-momentum probe, then the only way we get the same
	answer as with the unevolved wave function is if the relevant operator
	changed as well.  Thus, with SRG evolution, the high-momentum physics is
	shuffled from the wave function (nuclear structure) to the operator
	(nuclear reaction component).  This is reminiscent of chiral EFTs where the
	renormalization replaces the high-momentum modes in intermediate states
	by contact interactions (see Fig.~\ref{fig:replace_loop_contact}).
	The discussion so far shows how the SRG makes the scale dependence of
	factorization
	explicit.  SRG transformations come with the momentum scale $\lambda$ and
	\begin{figure}[thbp]
	\centering
	\begin{overpic}
   [width=0.6\textwidth]%
	 {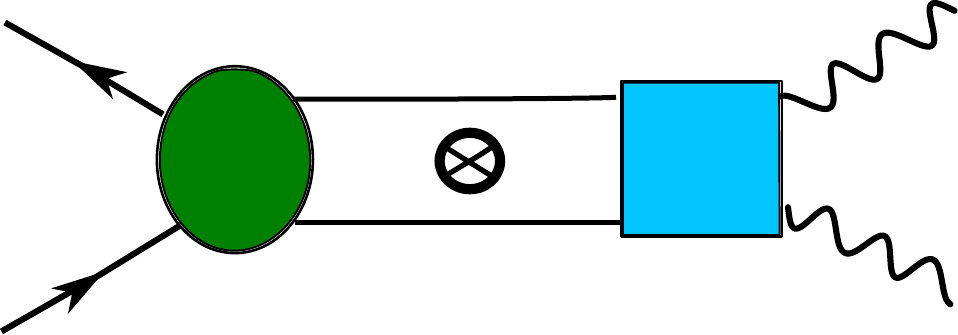}
   \put(20,33){\textcolor{black}{\large $p < \lambda$}}
   \put(67,33){\textcolor{black}{\large $p > \lambda$}}
   \end{overpic}
	\caption{The SRG scale $\lambda$ sets the natural scale for factorization.}
	\label{fig:SRG_factorization}
	\end{figure}
	this sets the scale for factorization.  As seen in the
	Fig.~\ref{fig:SRG_factorization}, we have a natural separation that the piece
	which involves momenta less than $\lambda$ forms the long-distance part and
	the piece that involves momenta greater than $\lambda$ forms the
	short-distance part.

	Consider the differential cross section given by the overlap matrix element
	of initial and final states.
	\beq
	\frac{d \sigma}{d \Omega} \propto \left|
	\mbraket{\psi_f}{\widehat{O}}{\psi_i} \right|^2 \;.
	\eeq
	The SRG evolved wave function is given by $\ket{\psi_i^\lambda} =
	U_{\lambda} \ket{\psi_i}$, where $U_{\lambda}$ is the unitary matrix
	associated with the SRG transformation.  The cross section is an
	experimental observable and should be independent of our choice of
	the SRG scale.  If we evolve all the components
	of the matrix element consistently
	\beq
	\mbraket{\psi_f}{\wh{O}}{\psi_i} = \mbraket{
	\underbrace{\psi_f U_{\lambda}^\dag}_{\psi_f^\lambda}}
	{\underbrace{U_\lambda \wh{O} U_{\lambda}^{\dag}}_{\wh{O}^\lambda}}
	{\underbrace{U_{\lambda} \psi_i}_{\psi_i^{\lambda}}} \;,
	\label{eq:matrix_element_invariance}
	\eeq
	then the evolved matrix element is same as the unevolved one and the
	observable cross section is unchanged.

	In general, to be consistent between structure and reactions one must
	calculate	cross sections or decay rates within a single framework.
	That is, one must	use
	the same Hamiltonian and consistent operators throughout the calculation
	(which means the same scale and scheme).  Such consistent calculations have
	existed	for some time for few-body nuclei (e.g.,
	see~\cite{Epelbaum:2008ga,Hammer:2012id,Carlson:2014vla,Marcucci:2015rca})
	and	are becoming increasingly feasible for heavier nuclei because of advances
	in reaction technology, such as using complex basis states to handle
	continuum	physics.  Recent examples in the literature include the
	No Core Shell Model Resonating Group Method
	(NCSM/RGM)~\cite{Quaglioni:2015via}, coupled cluster~\cite{Bacca:2013dma},
	and lattice EFT calculations~\cite{Pine:2013zja}.
	But there are many open	questions about constructing consistent currents and
	how to compare results from two such calculations.
	Some work along this direction which includes the evolution of the operator
	has recently been done.  Anderson \etal looked at the static properties
	of the deuteron such as momentum distributions, radii, and form factors under
	SRG evolution; they found no pathologies in the evolved operators, and
	the evolution effects were small for low-momentum observables
	\cite{Anderson:2010aq}.  Schuster \etal found in their work on radii and
	dipole transition matrix elements in light nuclei that the evolution effects
	are as important as three-body forces
	\cite{Schuster:2014lga,Schuster:2013sda}.
	Neff \etal looked at the SRG transformed density operators and concluded that
	it is essential to use evolved operators for observables sensitive to
	short-range physics \cite{Neff:2015xda}.  But all this work was done for
	expectation values of the operator, i.e, the state on the either side of the
	matrix element was the same.  In particular, there was no work which dealt
	with the issues related to operator evolution when we have a transition to
	continuum.  This is what we sought to address in \cite{More:2015tpa}.

	The electron scattering knock-out process is particularly interesting because
	of the connection to past, present, and planned
	experiments~\cite{Boffi:1996, LENP_white_paper2015}.
	The conditions for clean factorization
	of structure and reactions in this context is closely related to the impact of
	3N forces,
	two-body currents, and final-state interactions, which have not been cleanly
	understood as yet~\cite{Furnstahl:2010wd}.
	All of this becomes particularly relevant for high-momentum-transfer electron
	scattering.%
	\footnote{Note that high-momentum transfers imply high-resolution
	\emph{probes}, which is different from the resolution induced by the SRG
	scale.  How the latter should be chosen to best
	accommodate the former is a key unanswered question.}
	This physics is conventionally explained in terms of short-range correlation
	(SRC) phenomenology~\cite{Frankfurt:2008zv,Atti:2015eda}.  SRCs are two- or
	higher-body components of the nuclear wave function with high relative
	momentum and low center-of-mass momentum.  These explanations would seem to
	present a	puzzle for descriptions of nuclei with low-momentum Hamiltonians,
	for which	SRCs are essentially absent from the wave functions.

	This puzzle is resolved by the unitary transformations that mandate the
	invariance of the cross section (cf.~Eq.~\eqref{eq:matrix_element_invariance}).
	The physics that was described by SRCs in the
	wave functions must shift to a different component, such as a two-body
	contribution from the current (cf.~Fig.~\ref{fig:replace_loop_contact}).
	This may appear to complicate the reaction
	problem just as we have simplified the structure part, but past work and
	analogies to other processes suggests that factorization may in fact
	become cleaner~\cite{Anderson:2010aq,Bogner:2012zm}.  One of our goals is to
	elucidate this issue, although we have only begun to do so in
	\cite{More:2015tpa}.

	In particular, we take the first steps in exploring the interplay of
	structure and reaction as a function of kinematic variables and SRG decoupling
	scale $\lambda$ in a controlled calculation of a knock-out process.  There are
	various complications for such processes.  With RG evolution, a
	Hamiltonian---even with only a two-body potential initially---will develop
	many-body components as the decoupling scale decreases
	(cf.~Eq.~\eqref{eq:induced_forces_demonstration}).  Similarly, a one-body
	current will develop two- and higher-body components.

	Our strategy is to avoid dealing with all of these complications
	simultaneously
	by considering the cleanest knock-out process: deuteron electrodisintegration
	with only an initial one-body current.  With a two-body system, there are no
	three-body forces or three-body currents to contend with.  Yet it still
	includes several key ingredients to investigate: i) the wave function will
	evolve with changes in resolution; ii) at the same time, the one-body current
	develops two-body components, which are simply managed; and iii) there are
	final-state interactions (FSI).  It is these ingredients that will mix under
	the RG evolution.  We can focus on different effects or isolate parts of the
	wave function by choice of kinematics.  For example, we can examine when the
	impulse approximation is best and to what extent that is a
	resolution-dependent assessment.

	\section{Test ground: Deuteron disintegration}

	\subsection{Formalism}
	\label{subsec:formalism}

	Deuteron electrodisintegration is the simplest nucleon-knockout process
	and has been considered as a test ground for various $NN$ models for a
	long time (see, for example, Refs.~\cite{Arenhovel:2004bc,Boeglin:2015cha}).
	It has also been well studied
	experimentally~\cite{Gilad:1998wia,Egiyan:2007qj}.
	The absence of three-body currents and forces
	makes it an ideal starting point for studying the interplay with SRG evolution
	of the deuteron wave function, current, and final-state interactions.

\begin{figure}[htbp]
	 \centering
	 \includegraphics[width=0.6\textwidth]%
	 {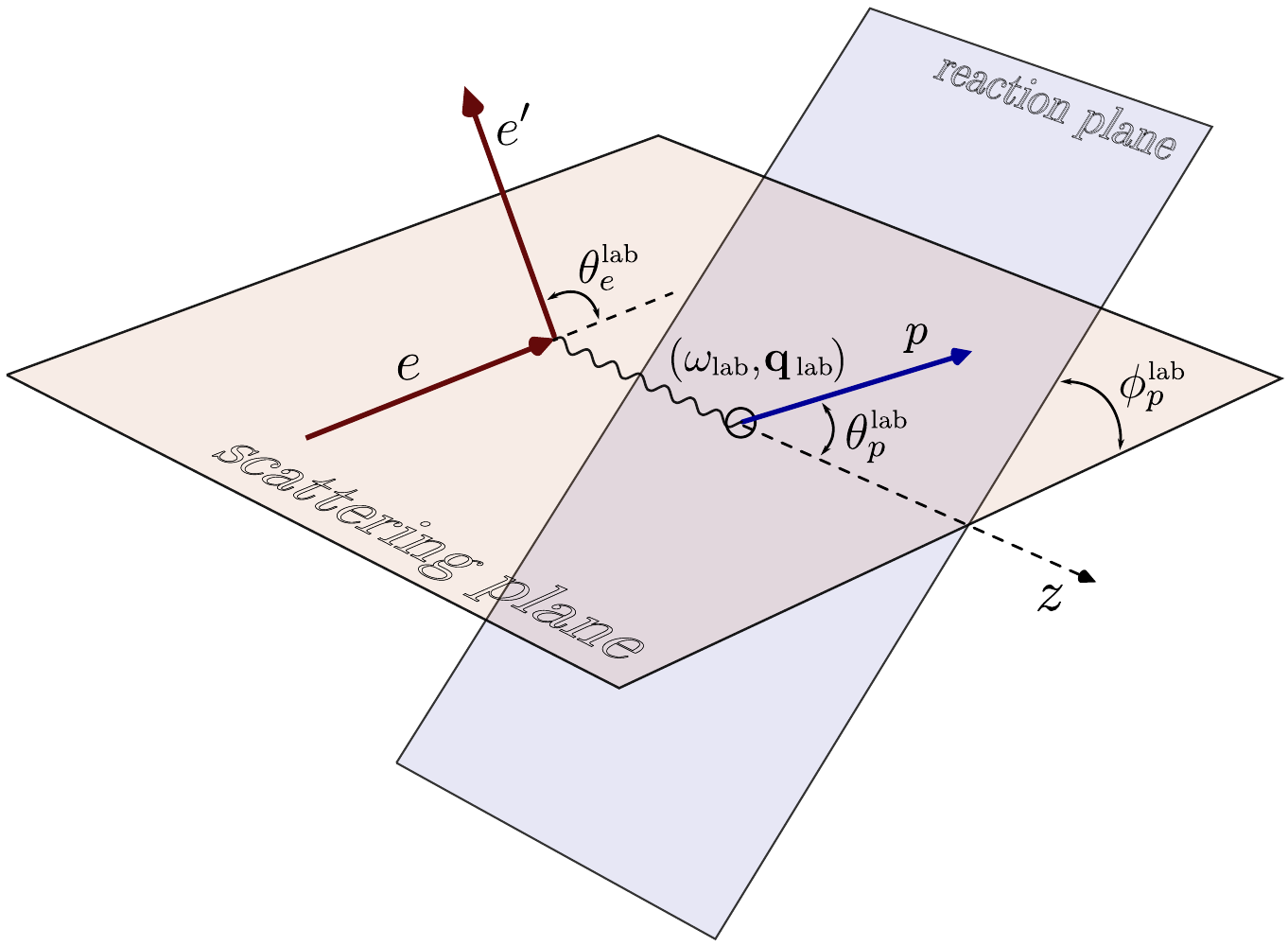}
	 \caption{The geometry of the electro-disintegration process in
	 the lab frame.  The virtual photon disassociates the deuteron into the proton
	 and the neutron (not shown in this figure).}
	 \label{fig:deut_dis_kinematics}
	\end{figure}

	We follow the approach of Ref.~\cite{Yang:2013rza}, which we briefly review.
	The kinematics for the process in the laboratory frame is shown in
	Fig.~\ref{fig:deut_dis_kinematics}.
	The virtual photon from electron scattering transfers enough energy and
	momentum to break up the deuteron into a proton and neutron.  The
	differential cross section for deuteron electrodisintegration for unpolarized
	scattering in the lab	frame is given by~\cite{Arenhovel:1988qh}
	\begin{multline}
	 \frac{\dd^3 \sigma}{\dd {k^\prime}^{\rm lab} \dd\Omega_e^{\rm lab}
	 \dd \Omega_p^{\rm lab}}
	 = \frac{\alpha}{6 \, \pi^2} \frac{{k^\prime}^{\rm lab}}
	 {k^{\rm lab} (Q^2)^2}
	 \Big[
	  v_L \fL + v_T f_T \\
	  \null + v_{TT} f_{TT} \cos 2 \phi_p^{\rm lab}
	  + v_{LT} f_{LT} \cos \phi_p^{\rm lab}
	 \Big] \,.
	\end{multline}
	Here $\Omega_e^{\rm lab}$ and $\Omega_p^{\rm lab}$ are the solid angles of the
	electron and the proton, $k^{\rm lab}$ and ${k^\prime}^{\rm lab}$ are the
	magnitude of incoming and outgoing electron 3-momenta,  $Q^2$ is the
	4-momentum-squared of the virtual photon, and $\alpha$ is the fine structure
	constant.  $\phi_p^{\rm lab}$ is the angle between the scattering plane
	containing the electrons and the plane spanned by outgoing nucleons.
	$v_L\,, v_T\,, \ldots$ are electron kinematic factors, and $\fL, f_T, \ldots$
	are the deuteron structure functions.  These structure functions contain all
	the dynamic information about the process.
	The four structure functions are independent and can be separated by combining
	cross-section measurements carried out with appropriate kinematic
	settings~\cite{Kasdorp:1997ba}.  Structure functions are thus cross sections
	up to kinematic factors and are independent of the SRG scale $\lambda$.
	They are analogous to form factor $F_2$ we saw in the DIS case
	(cf.~Fig.~\ref{fig:factorization_schematic_highE_QCD}).  In our
	work we focus on the longitudinal structure function $\fL$, following the
	approach of Ref.~\cite{Yang:2013rza}.

	\medskip

	\subsubsection{Calculating $\fL$}

	As in Ref.~\cite{Yang:2013rza}, we carry out the calculations in
	the center-of-mass frame of the outgoing proton-neutron pair.
	In this frame the photon
	four-momentum is $(\omega,\mbf{q})$, which can be obtained from the
	initial electron energy and $\theta_e$, the electron scattering angle.
	We denote the momentum of the outgoing proton by $\mathbf{\pp}$ and
	take $\mbf{q}$ to be along the $z$-axis.  The angles of $\mbf{\pp}$ are
	denoted by $\Omega_{\mbf{\pp}} = (\thetacm, \phicm)$.

	The longitudinal structure function can be written as
	\begin{equation}
	 \fL = \sum_{\substack{S_f, m_{s_f}\\ \mJd}}
	 \ampT_{S_f,m_{s_f},\mu = 0,\mJd}\!(\thetacm, \phicm) \,
	 \ampT_{S_f, m_{s_f},\mu = 0,\mJd}^\ast\!(\thetacm, \phicm) \,,
	\label{eq:f_L_from_T}
	\end{equation}
	where $S_f$ and $m_{s_f}$ are the spin quantum numbers of the final
	neutron-proton state, $\mu$ is the polarization index of the virtual photon,
	and $\mJd$ is the angular momentum of the initial deuteron
	state. The amplitude $\ampT$ is given by~\cite{Arenhoevel:1992xu}
	\begin{equation}
	 \ampT_{S,\msf,\mu,\mJd}
	 = -\pi \sqrt{2\alpha|\mathbf{\pp}|E_p E_d / M_d}
	 \,\la \psi_{f} \,|\, J_{\mu}(\mathbf{q}) \,| \psi_i \ra \,,
	\label{eq:T_definiton}
	\end{equation}
	where $\bra{\psi_f}$ is the final-state wavefunction of the outgoing
	neutron-proton pair, $\ket{\psi_i}$ is the initial deuteron state, and
	$J_{\mu} (\mbf{q})$ is the current operator that describes the momentum
	transferred by the photon.
	The variables in Eq.~\eqref{eq:T_definiton} are:
	\begin{itemize}
	\item fine-structure constant $\alpha$;
	\item outgoing proton (neutron) 3-momentum $\mbf{\pp} \, (-\mbf{\pp})$;
	\item proton energy $\displaystyle E_p = \sqrt{M^2 + \mbf{\pp}^2}$,
	where $M$ is the average of proton and neutron mass
	\item deuteron energy $\displaystyle E_d = \sqrt{M_d^2 + \mbf{q}^2} $,
	where $M_d$ is the mass of the deuteron.
	\end{itemize}
	As mentioned before, all of these quantities are in the center-of-mass
	frame of the outgoing nucleons.

	For $f_L$, $\mu = 0$ and therefore only $J_0$ contributes.
	The one-body current matrix element is given by
	\begin{multline}
	 \la \mbf{k}_1 \, T_1| \, J_0(\mbf{q}) \, | \,\mbf{k}_2 \, T\!=\!0 \ra
	 = \frac{1}{2} \big(G_E^p + (-1)^{T_1} G_E^n\big) \,
	 \delta(\mbf{k}_1 - \mbf{k}_2 - \mbf{q}/2) \\
	 \null + \frac{1}{2} \big((-1)^{T_1} G_E^p +  G_E^n\big) \, \delta(\mbf{k}_1
	 - \mbf{k}_2 + \mbf{q}/2) \,,
	\label{eq:J0_def}
	\end{multline}
	where $G_E^p$ and $G_E^n$ are the electric form factors of the proton and the
	neutron, and the deuteron state has isospin $T=0$.

	The final-state wave function of the outgoing proton-neutron pair can be
	written	as
	\begin{equation}
	 |\psi_f\ra = | \phi \ra + G_0 (E^\prime) \, t(E^\prime) \,| \phi \ra \,,
	\label{eq:psi_f_def}
	\end{equation}
	where $\ket{\phi}$ denotes a relative plane wave, $\GreensFn$ and $t$
	are the Green's function and the $t$-matrix respectively, and  $E^\prime =
	\mbf{\pp}^2/M$ is the energy of the outgoing nucleons.  The second term in
	Eq.~\eqref{eq:psi_f_def} describes the interaction between the outgoing
	nucleons.  As the momentum associated with the plane wave $\ket{\phi}$ is
	$\pp$, the $t$-matrix $t(E^\prime)$ that enters our calculation is always
	half on-shell.

	\begin{figure}[htbp]
	 \centering
	 \includegraphics[width=0.35\textwidth]%
	 {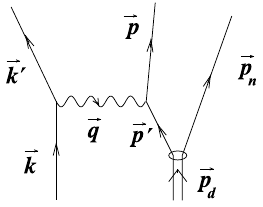}

	 \small{(a) Impulse Approximation (IA) \\[1.5 em]}

   \includegraphics[width=0.35\textwidth]%
	 {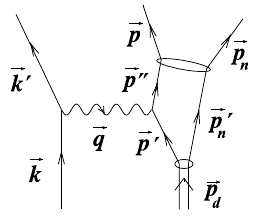}

   \small{(b) Final State Interaction (FSI)}

	 \caption{(a) The first term on the right side of Eq.~\eqref{eq:psi_f_def}.
	 The outgoing nucleons do not interact.
	 (b) The second term on the right side of Eq.~\eqref{eq:psi_f_def}.  The
	 outgoing nucleons interact through the NN potential.  Figures taken from
	 \cite{Ibrahim:2006lta}. }
	 \label{fig:IA_FSI_schematic}
	\end{figure}
	In the impulse approximation (IA) as defined here, the interaction between the
	outgoing nucleons is ignored and $|\psi_f \ra_{\rm IA} \equiv | \phi \ra$.
	A schematic of IA and FSI contributions is shown in
	Fig.~\ref{fig:IA_FSI_schematic}.
	The plane wave $|\phi\ra$ will have both isospin $0$ and $1$ components.
	The current $J_0$, $\GreensFn$, and the $t$-matrix are diagonal in spin space.
	The deuteron has spin $S=1$ and therefore the final state will also have
	$S=1$. 	Hence, we have
	\begin{eqnarray}
	 \ket{\phi} & \equiv & \ket{\mbf{\pp}\, S\!=\!1 \, \msf \psi_T} \nonumber \\
	 & = & \dfrac{1}{2} \sum_{T = 0,1}
	 \big(\ket{\mbf{\pp} \, S\!=\!1 \, \msf}
		+ (-1)^T \ket{{-}\mbf{\pp} \, S\!=\!1 \, \msf} \big) \,\ket{T} \,.
	\label{eq:phi_def}
	\end{eqnarray}
	Using Eqs.~\eqref{eq:J0_def} and~\eqref{eq:phi_def}, the overlap matrix
	element	in IA becomes
	\begin{multline}
	 \la \psi_f | \, J_0 \, |\psi_i \ra_{\rm IA}
	 = \sqrt{\frac{2}{\pi}} \sum_{L_{d} = 0, 2}
	 \CG{L_d}{\mJd - \msf}{S\!=\!1}{\msf}{J\!=\!1}{\mJd} \\
	 \null \times \Big[
	  G_E^p \, \psi_{L_d}(|\mbf{\pp} - \mbf{q}/2|)
	  \,Y_{L_d,\mJd-m_{s_f}}\!(\Omega_{\mbf{\pp} - \mbf{q}/2}) \\
	  \null + G_E^n \, \psi_{L_d}(|\mbf{\pp} + \mbf{q}/2|)
	  \,Y_{L_d,\mJd-m_{s_f}}\!(\Omega_{\mbf{\pp} + \mbf{q}/2})
	 \Big] \,,
	\label{eq:overlap_IA}
	\end{multline}
	where $\Omega_{\mbf{\pp} \pm \mbf{q}/2}$ is the solid angle between the
	unit vector $\hat{z}$ and $\mbf{\pp}\pm\mbf{q}/2$.  $\psi_{L_d}$ is the
	deuteron wave function in momentum space defined as
	\begin{equation}
	 \braket{k_1 \, J_1 \, m_{J_1} \, L_1 \, S_1 \, T_1}{\psi_i}
	 = \psi_{L_1}(k_1)
	  \delta_{J_1,1}\delta_{m_{J_1},\mJd}
	 \delta_{L_1,L_d}\delta_{S_1,1}\delta_{T_1,0} \,.
  \end{equation}
	The $S$-wave $(L=0)$ and $D$-wave $(L=2)$ components of the deuteron wave
	function satisfy the normalization condition
	\begin{equation}
	 \frac{2}{\pi} \int dp \, p^2 \, \big(\psi_0^2(p) + \psi_2^2(p)\big) = 1 \,.
	\end{equation}
	In deriving Eq.~\eqref{eq:overlap_IA} we have used the property of the
	spherical harmonics that
	\begin{equation}
	 Y_{lm}(\pi-\theta,\phi+\pi) = (-1)^l \, Y_{lm}(\theta,\phi) \,.
	\end{equation}
	In our work we follow the conventions of Ref.~\cite{Landau:1989}.
	Deriving Eq.~\eqref{eq:overlap_IA} also uses partial wave expansion
	\beq
	\ket{\mbf{k}} = \sqrt{\frac{2}{\pi}} \sum_{l, m} Y_{l \, m}^\ast
	(\Omega_{\mbf{k}}) \ket{k \, l \, m} \;,
	\eeq
	the normalization condition
	\beq
	\braket{p}{k} = \frac{\pi}{2} \frac{\delta(p - k)}{p^2} \;,
	\eeq
	and, the Clebsch-Gordan completeness relation
	\beq
	\ket{l \, m \, S\!=\!1 \, m_s} = \sum_{J, \, m_J} \ket{J \, m_J \, l \, S\!=\!1}
	\braket{J \, m_J \, l \, S\!=\!1}{l \, m \, S\!=\!1\, m_s} \;.
	\eeq
	Because
	$\thetacm$ and $\phicm$ are the angles of $\mbf{\pp}$, $\Omega_{\mbf{\pp} -
	\mbf{q}/2} \equiv \big(\thetacprime(\pp, \thetacm, q), \phicm \big)$
	and $\Omega_{\mbf{\pp} + \mbf{q}/2} \equiv
	\big(\thetacdoubleprime(\pp, \thetacm, q),
	\phicm \big)$, where
	\begin{equation}
	 \thetacprime (\pp, \thetacm, q) = \cos^{-1}
	 \!\left(
	  \frac{\pp \cos \thetacm - q/2}{\sqrt{{\pp}^2 - \pp q \cos\thetacm + q^2/4}}
	 \right)
	\label{eq:theta_c_prime_def}
	\end{equation}
	and
	\begin{equation}
	 \thetacdoubleprime (\pp, \thetacm, q) = \cos^{-1}
	 \!\left(
	  \frac{\pp \cos \thetacm + q/2}{\sqrt{{\pp}^2 + \pp q \cos\thetacm + q^2/4}}
	 \right) \,.
	\label{eq:theta_c_double_prime_def}
	\end{equation}
	The expressions for $\thetacprime$ and $\thetacdoubleprime$ can be obtained by
	elementary trigonometry.  Note that Eqs.~\eqref{eq:theta_c_prime_def} and
	\eqref{eq:theta_c_double_prime_def} reproduce the correct $\pp = 0$ and
	$q = 0$ limit.

	The overlap matrix element including the final-state interactions (FSI) is
	given by
	\begin{equation}
	 \la \psi_f| \, J_0 \,|\psi_i \ra
	 = \underbrace{\la \phi| \, J_0 \,|\psi_i \ra}_{\rm IA}
	 + \underbrace{\la \phi| t^\dag \, \GreensFn^\dag\, J_0 \,|\psi_i \ra}_
	 {\rm FSI} \,.
	\label{eq:overlap_IA_plus_FSI}
	\end{equation}
	The first term on the right side of Eq.~\eqref{eq:overlap_IA_plus_FSI}
	has already been evaluated in Eq.~\eqref{eq:overlap_IA}.  Therefore, the term
	we still need to evaluate is $\la \phi| t^\dag \, \GreensFn^\dag\, J_0
	\,|\psi_i \ra$.  The $t$-matrix is most conveniently calculated in a
	partial-wave basis.  Hence, the FSI term is evaluated by inserting complete
	sets of states in the form
	\begin{equation}
	 1 = \frac{2}{\pi} \sum_{\substack{L, S \\ J,  m_J }}
	 \sum_{T = 0, 1} \int dp \,p^2 \, |p \, J \, m_J \, L \, S \, T \ra
	 \, \la p \, J\, m_J \, L \, S \, T | \,.
	\label{eq:completeness_partial_wave}
	\end{equation}
	The outgoing plane-wave state in the partial-wave basis is given by
	\begin{multline}
	 \la \phi | \, k_1 \, J_1 \, m_{J_1} \, L_1 \, S\!=\!1 \, T_1 \ra
	  =  \frac{1}{2} \, {\sqrt\frac{2}{\pi}} \, \frac{\pi}{2} \,
	 \frac{\delta(\pp - k_1)}{k_1 ^2}
	 \CG{L_1}{m_{J_1} - \msf}{S\!=\!1}{\msf}{J_1}{m_{J_1}}
	 \, \\
	 \null \times \big(1 + (-1)^{T_1} (-1)^{L_1}\big)
	 \,Y_{L_1,m_{J_1} - \msf}\!(\thetacm, \phicm) \,.
	\label{eq:phi_def_pw}
	\end{multline}
	The Green's
	function is diagonal in $J$, $m_J$, $L$, $S$, and $T$, so we have
	\begin{equation}
	 \la k_1 | \, \GreensFn^\dag \,| k_2 \ra
	 = \frac{\pi}{2} \, \frac{\delta(k_1 - k_2)}{k_1^2}
	 \, \frac{M}{{\pp}^2 - k_1 ^2 - i\epsilon} \,.
	\label{eq:G0_def_pw}
	\end{equation}

	We also need to express the current in Eq.~\eqref{eq:J0_def} in the
	partial-wave basis.  To begin with, let us just work with first term in
	Eq.~\eqref{eq:J0_def}, which we denote by $J_0 ^-$.  In the partial-wave
	basis, it is written as
	\begin{multline}
	 \mbraket{k_1 \, J_1 \, \mJd \, L_1 \, S\!=\!1 \, T_1}{J_0^{-}}
	 {\,k_2 \, J\!=\!1 \, \mJd \, L_2 \, S\!=\!1 \, T\!=\!0}
	 = \frac{\pi^2}{2} \, \big(G_E ^p + (-1)^{T_1} \, G_E^n\big) \\
	 \null \times \sum_{\widetilde{m}_s = -1}^1 \int\!\dcostheta
	 \, \braket{J_1 \, \mJd}{L_1 \, \mJd - \wt{m}_s \, S\!=\!1 \,\wt{m}_s}
	 \,P_{L_1}^{\mJd - \widetilde{m}_s}\!(\cos\theta) \\
	 \null \times
	 \,P_{L_2}^{\mJd - \widetilde{m}_s}\!\big(\cos\thetacprime(k_1, \theta, q)
	 \big) \frac{\delta\big(k_2-\sqrt{k_1^2-k_1 q \cos\theta + q^2/4}\big)}
	 {k_2^2} \\
	 \null \times
	 \CG{L_2}{\mJd - \widetilde{m}_s}{S\!=\!1}{\widetilde{m}_s}{J\!=\!1}{\mJd} \,.
	\label{eq:J_0_minus_def_pw}
	\end{multline}
	Here $\mJd$ is the deuteron quantum number, which is preserved throughout.
	We have used the deuteron quantum numbers in the ket in anticipation
	that we will always evaluate the matrix element of $J_0$ with the deuteron
	wave function on the right.   $\thetacprime$ is as defined in
	Eq.~\eqref{eq:theta_c_prime_def}.  In deriving Eq.~\eqref{eq:J_0_minus_def_pw}
	we have also made use of the relation \cite{Jerry_thesis}
	\begin{equation}
	 \int Y_{lm} ^\ast(\theta, \varphi) \,
	 Y_{l^\prime m^\prime}(\alpha^\prime,\varphi)
	 \,\dd\!\cos \theta \, \dd \varphi
	 = 2\pi \delta_{m m^\prime}
	 \int\!\dcostheta
	 P_l^m(\cos\theta) \, P_{l^\prime}^m(\cos \alpha^\prime) \,.
	\end{equation}

	Equations~\eqref{eq:phi_def_pw}, \eqref{eq:G0_def_pw},
	and~\eqref{eq:J_0_minus_def_pw} can be combined to obtain
	\begin{multline}
	 \mbraket{\phi}{t^\dag \, \GreensFn^\dag \, J_0^-}{\psi_i}
	 = \sqrt{\frac{2}{\pi}} \,\frac{M}{\hbar c}
	 \sum_{T_1 = 0,1} \big(G_E ^p + (-1)^{T_1} \, G_E^n\big)
	 \sum_{L_1 = 0}^{L_{\rm max}} \big(1 + (-1)^{T_1} (-1)^{L_1}\big) \\
	 \null \times Y_{L_1 , \mJd - \msf}\!(\thetacm, \phicm)
	 \sum_{J_1 = |L_1 - 1|}^{L+1}
	 \CG{L_1}{\mJd - \msf}{S\!=\!1}{\msf}{J_1}{\mJd}
	 \sum_{L_2 = 0}^{L_{\rm max}} \int \dd k_2 \, k_2^2 \\
	 \null \times t^\ast(k_2, \pp, L_2, L_1, J_1, S\!=\!1, T_1)
	 \sum_{\widetilde{m}_s = -1}^1
	 \braket{J_1 \, \mJd}{L_2 \, \mJd - \wt{m}_s \, S\!=\!1 \, \wt{m}_s} \\
	 \null \times \sum_{L_d = 0,2}
	 \CG{L_d}{\mJd - \widetilde{m}_s}{S\!=\!1}{\widetilde{m}_s}{J\!=\!1}{\mJd}
	 \int \dcostheta
	 \, \frac{1}{{\pp}^2 - k_2^2 - i \epsilon} \\
	 \null \times
	 P_{L_2}^{\mJd - \widetilde{m}_s}(\cos\theta) \,
	 P_{L_d}^{\mJd - \widetilde{m}_s}\!\big(\cos\thetacprime(k_2, \theta, q)\big)
	 \, \psi_{L_d}\Big(\!\sqrt{{k_2}^2 - k_2 \, q \, \cos\theta + q^2/4}\Big) \,.
	\label{eq:phi_t_g0_J0_minus}
	\end{multline}
	Note that the matrix element of the $t$-matrix in
	Eq.~\eqref{eq:phi_t_g0_J0_minus} should strictly be written as
	${t^{\ast}}({E^\prime} = {\pp}^2/M; k_2, \pp, L_2, L_1, J_1, S\!=\!1, T_1)$.  However,
	keeping in mind that the $t$-matrix in this chapter is always evaluated
	half on-shell, we drop the $E^\prime$ index for the sake of brevity.
	To evaluate the hermitian conjugate, we use the property
	\beq
	t^\dag (\pp, k_2, L_1, L_2, J_1, S\!=\!1, T_1)
	= t^\ast (k_2, \pp, L_2, L_1, J_1, S\!=\!1, T_1) \;.
	\eeq

	We denote the second term in the one-body current Eq.~\eqref{eq:J0_def} by
	$J_0^+$.  The expression for $\mbraket{\phi}{t^\dag \, \GreensFn^\dag \,
	J_0^+}{\psi_i}$ is analogous to Eq.~\eqref{eq:phi_t_g0_J0_minus}, the only
	differences being that the form-factor coefficient is $(-1)^{T_1} G_E^p +
	G_E^n$ and the input arguments for the second associated Legendre polynomial
	and the deuteron wave function are different.  The two factors respectively
	become $\displaystyle P_{L_d}^{\mJd - \widetilde{m}_s}\!
	\big(\cos\thetacdoubleprime(k_2, \theta, q)\big)$
	and $\psi_{L_d}\big(\sqrt{{k_2}^2 + k_2 \,q \, \cos\theta + q^2/4}\big)$,
	where	$\thetacdoubleprime$ is defined in
	Eq.~\eqref{eq:theta_c_double_prime_def}.
	It can be shown that
	$\mbraket{\phi}{t^\dag \, \GreensFn^\dag \, J_0^+}{\psi_i}
	= \mbraket{\phi}{t^\dag \, \GreensFn^\dag \, J_0^-}{\psi_i}$.  Thus,
	\begin{equation}
	 \mbraket{\phi}{t^\dag \, \GreensFn^\dag \, J_0}{\psi_i}
	 = 2 \, \mbraket{\phi}{t^\dag \, \GreensFn^\dag \, J_0^-}{\psi_i} \,.
	\label{eq:J0_minus_twice_relation}
	\end{equation}
	Using this we can evaluate the overlap matrix element in
	Eq.~\eqref{eq:overlap_IA_plus_FSI}.  As outlined in Eqs.~\eqref{eq:f_L_from_T}
	and \eqref{eq:T_definiton}, this matrix element is related to the
	longitudinal structure function $\fL$.  Recall that the deuteron spin is
	conserved throughout and therefore $S_f = 1$ in Eq.~\eqref{eq:f_L_from_T}.

	In Subsec.~\ref{subsec:Results} we present results for $\fL$ both in
	the IA and including the FSI.  These results match those of
	Ref.~\cite{Yang:2013rza,Arenhoevel:1992xu}, verifying the accuracy of the
	calculations presented above.

	\subsection{Evolution setup}
	\label{subsec:evolution_setup}

	As outlined previously, we want to investigate the effect of
	unitary transformations on calculations of $\fL$.
	Let us start by looking at the IA matrix element:
	\begin{eqnarray}
	 \la \phi | J_0 | \psi_i \ra
	 &=& \la \phi | U^\dag \, U \, J_0 \, U^\dag \, U \, | \psi_i \ra \nonumber \\
	 &=& \underbrace{\la\phi|\widetilde{U}^\dag J_0^\lambda|\psi_i^\lambda\ra}_{A}
	 + \underbrace{\la \phi | \, J_0^\lambda | \psi_i^\lambda \ra}_{B} \,,
	\label{eq:A_B_split_up}
	\end{eqnarray}
	where we decompose the unitary matrix $U$ into the identity and a residual
	$\widetilde{U}$,
	\begin{equation}
	 U = I + \widetilde{U} \,.
	\label{eq:U_decomposition}
	\end{equation}
	The matrix $\widetilde{U}$ is smooth and therefore amenable to interpolation.
	The $U$ matrix is calculated following the approach in \cite{Anderson:2010aq}.
	The terms in Eq.~\eqref{eq:A_B_split_up} can be further split into
	\begin{equation}
	 \la \phi | \, J_0^\lambda | \psi_i^\lambda \ra
	 = \underbrace{\la \phi | \widetilde{U}\, J_0 \, \widetilde{U}^\dag |
	  \psi_i^\lambda \ra}_{B_1}
	 + \underbrace{\la \phi | \widetilde{U}\, J_0 \, | \psi_i^\lambda \ra}_{B_2}
	 + \underbrace{\la \phi | \, J_0 \, \widetilde{U}^\dag | \psi_i^\lambda
	  \ra}_{B_3}
	 + \underbrace{\la \phi |\, J_0 \, | \psi_i^\lambda \ra}_{B_4}
	\label{eq:B_split_up}
  \end{equation}
 	and
 	\begin{equation}
	 \la \phi | \widetilde{U}^\dag \, J_0^\lambda | \psi_i^\lambda \ra \\[0.25em]
	 = \underbrace{\la \phi | \widetilde{U}^\dag \, \widetilde{U}\, J_0 \,
	  \widetilde{U}^\dag | \psi_i^\lambda \ra}_{A_1}
	 + \underbrace{\la \phi | \widetilde{U}^\dag \, \widetilde{U}\, J_0 |
	  \psi_i^\lambda \ra}_{A_2}
	 + \underbrace{\la \phi | \widetilde{U}^\dag J_0 \,
	  \widetilde{U}^\dag | \psi_i^\lambda \ra}_{A_3}
	 + \underbrace{\la\phi | \widetilde{U}^\dag J_0 | \psi_i^\lambda\ra}_{A_4} \,.
	\label{eq:A_split_up}
 	\end{equation}
	The $B_4$ term is the same as in Eq.~\eqref{eq:overlap_IA}, but with the
	deuteron wave function replaced by the evolved version $\psi_{L_d}^\lambda$.
	Inserting complete sets of partial-wave basis states as in
	Eq.~\eqref{eq:completeness_partial_wave} and using Eqs.~\eqref{eq:phi_def_pw}
	and \eqref{eq:J_0_minus_def_pw}, we can obtain the expressions for $B_1$,
	$B_2$, $B_3$ and $A_1,\ldots,A_4$.  These expressions are given in
	Appendix~\ref{Appendix:sec:evolution_expressions}.

	Using the expressions for $A_1,\ldots, A_4$ and $B_1,\ldots, B_4$, we can
	obtain results for $\fL$ in the IA with one or more components of the overlap
	matrix element $\mbraket{\phi}{J_0}{\psi}$ evolved.  When calculated in IA,
	$\fL$ with all components evolved matches its unevolved counterpart, as shown
	later	in Subsec.~\ref{subsec:Results}.
	The robust agreement between the evolved and unevolved answers indicates that
	the expressions derived for $A_1, \ldots, B_4$ are correct and that there is
	no error in generating the $U$-matrices.  In
	Sec.~\ref{subsec:numerical_implementation} we provide some details about the
	numerical implementation of the equations presented here.

	Let us now take into account the FSI and study the effects
	of evolution.  The overlap matrix element should again be unchanged under
	evolution,
	\begin{equation}
	 \mbraket{\psi_f}{J_0}{\psi_i}
	 = \mbraket{\psi_f^\lambda}{J_0^\lambda}{\psi_i^\lambda} \,,
	\end{equation}
	where $\psi_f$ is given by Eq.~\eqref{eq:psi_f_def}.  Furthermore,
	\begin{equation}
	 |\psi_f ^\lambda \ra = | \phi \ra + G_0 \, t_\lambda |\phi \ra \,,
	\label{eq:psi_f_lam_def}
	\end{equation}
	where $t_\lambda$ is the evolved $t$-matrix, \ie, the $t$-matrix obtained
	by solving the Lippmann--Schwinger equation using the evolved potential, as
	discussed in Appendix~\ref{Appendix:sec:evolution_final_state}.  Thus
	\begin{equation}
	 \mbraket{\psi_f^\lambda}{J_0^\lambda}{\psi_i^\lambda}
	 = \underbrace{\mbraket{\phi}{J_0^\lambda}{\psi_i^\lambda}}_{B}
	 + \underbrace{\mbraket{\phi}{t_\lambda ^\dag \, G_0^\dag
	  \, J_0^\lambda}{\psi_i^\lambda}}_{F} \,.
	\end{equation}
	The term $B$ is the same that we already encountered in
	Eq.~\eqref{eq:A_B_split_up}.  The term $F$ can also be split up into four
	terms:
	\begin{multline}
	 \mbraket{\phi}{t_\lambda^\dag\,G_0^\dag\,J_0^\lambda}{\psi_i^\lambda}
	 = \underbrace{\mbraket{\phi}{t_\lambda ^\dag \, G_0^\dag \, \widetilde{U}
	  \, J_0 \, \widetilde{U}^\dag}{\psi_i^\lambda}}_{F_1}
	 + \underbrace{\mbraket{\phi}{t_\lambda ^\dag \, G_0^\dag \, \widetilde{U}
	  \, J_0 }{\psi_i^\lambda}}_{F_2} \\
	 \null + \underbrace{\mbraket{\phi}{t_\lambda ^\dag \, G_0^\dag
	  \, J_0 \, \widetilde{U}^\dag}{\psi_i^\lambda}}_{F_3}
	 + \underbrace{\mbraket{\phi}{t_\lambda ^\dag \, G_0^\dag
	  \, J_0}{\psi_i^\lambda}}_{F_4} \,.
	\label{eq:F_split_up}
	\end{multline}
	The expression for $F_4$ can easily be obtained from
	Eqs.~\eqref{eq:phi_t_g0_J0_minus} and~\eqref{eq:J0_minus_twice_relation} by
	replacing the deuteron wave function and the $t$-matrix by their evolved
	counterparts.  As before, we insert complete sets of partial-wave basis states
	using Eq.~\eqref{eq:completeness_partial_wave} and evaluate $F_3$, $F_2$,
	and $F_1$; see Eqs.~\eqref{eq:F3}, \eqref{eq:F2}, and \eqref{eq:F1}.
	Figures in Subsec.~\ref{subsec:Results} compare $\fL$ calculated
	from the matrix element with all components evolved to the unevolved $\fL$.
	We find an excellent
	agreement, validating the expressions for $F_1, \ldots, F_4$.

	\medskip

	\subsubsection{First-order analytical calculation}

	Recall that from Eqs.~\eqref{eq:f_L_from_T} and~\eqref{eq:T_definiton} we have
	\begin{equation}
	 \fL \propto \sum_{\msf, \mJd}\left|\mbraket{\psi_f}{J_0}{\psi_i}\right|^2 \,.
	\label{eq:fl_prop_matrix_element}
	\end{equation}
	When all three components---the final state, the current, and the initial
	state---are evolved consistently, then $\fL$ is unchanged.  However, if we
	miss evolving a component, then we obtain a different result.  It is
	instructive to
	illustrate this through a first-order analytical calculation.%
	\footnote{An analogous calculation based on field redefinitions appears
	in Ref.~\cite{Furnstahl:2001xq}.}

	Let us look at the effects due to the evolution of individual components for a
	general matrix element $\mbraket{\psi_f}{\widehat{O}}{\psi_i}$.  The evolved
	initial state is given by
	\begin{equation}
	 \ket{\psi_i^\lambda} \equiv U \, \ket{\psi_i}
	 = \ket{\psi_i} + \widetilde{U} \, \ket{\psi_i} \,,
	\label{eq:psi_i_evolution}
	\end{equation}
	where $\widetilde{U}$ is the smooth part of the $U$-matrix defined in
	Eq.~\eqref{eq:U_decomposition}.  Similarly, we can write down the expressions
	for the evolved final state and the evolved operator as
	\begin{equation}
	 \bra{\psi_f^\lambda} \equiv \bra{\psi_f} \, U^\dag
	 = \bra{\psi_f} - \bra{\psi_f}\,\widetilde{U}
	\label{eq:psi_f_evolution}
	\end{equation}
	and
	\begin{equation}
	 \widehat{O}^\lambda
	 \equiv U \, \widehat{O} \, U^\dag = \widehat{O}
	 + \widetilde{U}\, \widehat{O} - \widehat{O} \, \widetilde{U}
	 + \mathcal{O}(\widetilde{U}^2) \,.
	\label{eq:O_evolution}
	\end{equation}
	We assume here that $\widetilde{U}$ is small compared to $I$ (which can always
	be ensured by choosing the SRG $\lambda$ large enough) and therefore keep
	terms	only up to linear order in $\widetilde{U}$.  Using
	Eqs.~\eqref{eq:psi_i_evolution}, \eqref{eq:psi_f_evolution},
	and~\eqref{eq:O_evolution}, we get an expression for the evolved matrix
	element	in terms of the unevolved one and changes to individual components
	due to evolution:
	\begin{multline}
	 \mbraket{\psi_f^\lambda}{\widehat{O}^\lambda}{\psi_i^\lambda}
	 = \mbraket{\psi_f}{\widehat{O}}{\psi_i} - \underbrace{\mbraket{\psi_f}
	  {\widetilde{U} \, \widehat{O}}{\psi_i}}_{\delta \bra{\psi_f}} \\
	 \null + \underbrace{\mbraket{\psi_f}{\widetilde{U} \, \widehat{O}}{\psi_i}
	 - \mbraket{\psi_f}{\widehat{O} \, \widetilde{U}}{\psi_i}}_{\delta\widehat{O}}
	 + \underbrace{
	  \mbraket{\psi_f} {\widehat{O} \, \widetilde{U}}{\psi_i}
	 }_{\delta\ket{\psi_i}}
	\end{multline}
	\begin{equation}
	\Longrightarrow \mbraket{\psi_f^\lambda}{\widehat{O}^\lambda}{\psi_i^\lambda}
	 = \mbraket{\psi_f}{\widehat{O}}{\psi_i} + \mathcal{O}(\widetilde{U}^2) \,.
	\end{equation}
	We see that the change due to evolution in the operator is equal and opposite
	to the sum of changes due to the evolution of the initial and final states.
	We also find that changes in each of the components are of the same order,
	and	that they mix;  this feature persists to higher order.  Therefore, if one
	misses evolving an individual component, one will not reproduce the unevolved
	answer.  It is interesting to analyze how this is a function of kinematics
	and will be a subject of Subsec.~\ref{subsec:Results}.

	\subsection{Numerical implementation}
	\label{subsec:numerical_implementation}

	There are various practical issues in the calculation of evolved matrix
	elements that are worth detailing.  We use C++11 for our numerical
	implementation of	the expressions discussed in the previous section.  Matrix
	elements with a	significant	number of components evolved are computationally
	quite expensive due to a large number of nested sums and integrals (see in
	particular Appendix~\ref{Appendix:sec:evolution_expressions}).

	The deuteron wave function and $NN$ $t$-matrix are obtained by discretizing
	the	Schr\"{o}dinger and Lippmann--Schwinger equations, respectively; these
	equations
	are also used to interpolate the $t$-matrix and wave function to points not on
	the discretized mesh.  For example, if we write the momentum-space Schrödinger
	equation---neglecting channel coupling here for simplicity---as
	\begin{eqnarray}
	 \psi(p) =& \int \dd q\,q^2\,G_0(-E_B,q) V(p,q)\,\psi(q) \nonumber \\
	 \rightarrow& \sum_{i} w_i\,q_i^2\,G_0(-E_B,q_i) V(p,q_i)\,\psi(q_i) \,,
	\label{eq:SG-simple}
  \end{eqnarray}
	it can be solved numerically as a simple matrix equation by setting
	$p\in\{q_i\}$.  For any $p=p_0$ not on this mesh, the sum in
	Eq.~\eqref{eq:SG-simple} can then be evaluated to get $\psi(p_0)$.  This
	technique is based on what has been introduced in connection with
	contour-deformation methods in break-up scattering
	calculations~\cite{Hetherington:1965zza,Schmid:1974}.  For more details on
	interpolation of the $t$-matrix and wave function, please refer to
	Appendix~\ref{Appendix:t_matrix_details}.

	To interpolate the potential, which is stored on a momentum-space grid, we use
	the  two-dimensional cubic spline algorithm from ALGLIB~\cite{ALGLIB:0915}.
	In order to avoid unnecessary recalculation of expensive quantities---in
	particular of the off-shell $t$-matrix---while still maintaining an
	implementation very	close to the expressions given in this paper, we make use
	of transparent caching
	techniques.\footnote{This means that the expensive calculation is only carried
	out once, the first time the corresponding function is called for a given set
	of arguments, while subsequent calls with the same arguments return the result
	directly, using a fast lookup.  All this is done without the \emph{calling}
	code being aware of the caching details.}  For most integrations, in
	particular those involving a  principal value, we use straightforward nested
	Gaussian quadrature
	rules; only in a few cases did we find it more efficient to use adaptive
	routines for multi-dimensional integrals.

	With these optimizations, the calculations can in principle still be run on a
	typical laptop computer.  In practice, we find it more convenient to use a
	small cluster, with parallelization implemented using the TBB
	library~\cite{TBB:0915}.  On a node with 48 cores, generating data for a
	meaningful plot (like those shown in Subsec.~\ref{subsec:Results}) can then
	be done	in less than an hour.  For higher resolution and accuracy, we used
	longer runs	with a larger number of data and integration mesh points.

	\subsection{Results}
	\label{subsec:Results}

	For our analysis, we studied the effect of evolution of individual components
	on $\fL$ for selected kinematics in the ranges $E^\prime = 10$--$100~\MeV$ and
	$\mbf{q}^2 = 0.25$--$25~\fm^{-2}$, where $E^\prime$ is the energy of outgoing
	nucleons and $\mbf{q}^2$ is the three-momentum transferred by the virtual
	photon; both are taken in the center-of-mass frame of the outgoing nucleons.
	This range was chosen to cover a variety of kinematics and motivated by the
	set covered in Ref.~\cite{Yang:2013rza}.  We use the Argonne $v_{18}$
	potential
	(AV18)~\cite{Wiringa:1994wb} for our calculations.  It is one of the widely
	used potentials for nuclear few-body reaction calculations, particularly those
	involving large momentum transfers~\cite{Carlson:1997qn,Carlson:2014vla}.

	How strong the evolution of individual components (or a subset thereof)
	affects	the result for $\fL$ depends on the kinematics.  One kinematic
	configuration of particular interest is the so-called quasi-free ridge.  As
	discussed in Subsec.~\ref{subsec:formalism}, the four-momentum transferred by
	the	virtual photon in the center-of-mass frame is $(\omega, \mbf{q})$.  The
	criterion for a configuration to lie on the quasi-free ridge is $\omega = 0$.
	Physically, this means that the nucleons in the deuteron are on their mass
	shell.  As shown in Ref.~\cite{Yang:2013rza}, at the quasi-free ridge the
	energy of the outgoing nucleons ($E^\prime$) and the photon momentum transfer
	are related by
	\begin{equation}
	 \Ep = \sqrt{M_d^2 + \mbf{q}^2} - 2 M \,,
	 \label{eq:quasi_free_condition_exact}
	\end{equation}
	which reduces to
	\begin{equation}
	E^\prime~\text{(in~$\MeV$)} \approx 10 \, \mbf{q}^2~\text{(in~$\fm^{-2}$)} \,.
	\label{eq:quasi_free_condition}
	\end{equation}
	The quasi-free
	condition in the center-of-mass frame is the same as the quasi-elastic
	condition in the lab frame.  There, the quasi-elastic ridge is defined by $W^2
	= m_p^2 \Rightarrow Q^2 = 2 \, \omega_\text{lab} \, m_p$, where $W$ is
	the invariant mass.  On the quasi-elastic ridge, the so-called missing
	momentum%
	\footnote{The missing momentum is defined as the difference of
	the measured proton momentum and the momentum transfer, $\mbf{p}_\text{miss}
	\equiv \mbf{p}_\text{lab}^\text{proton} - \mbf{q}_\text{lab}$.}
	vanishes, $p_\text{miss} = 0$.

\begin{figure}[htbp]
	 \centering
	 \includegraphics[width=0.6\textwidth]%
	 {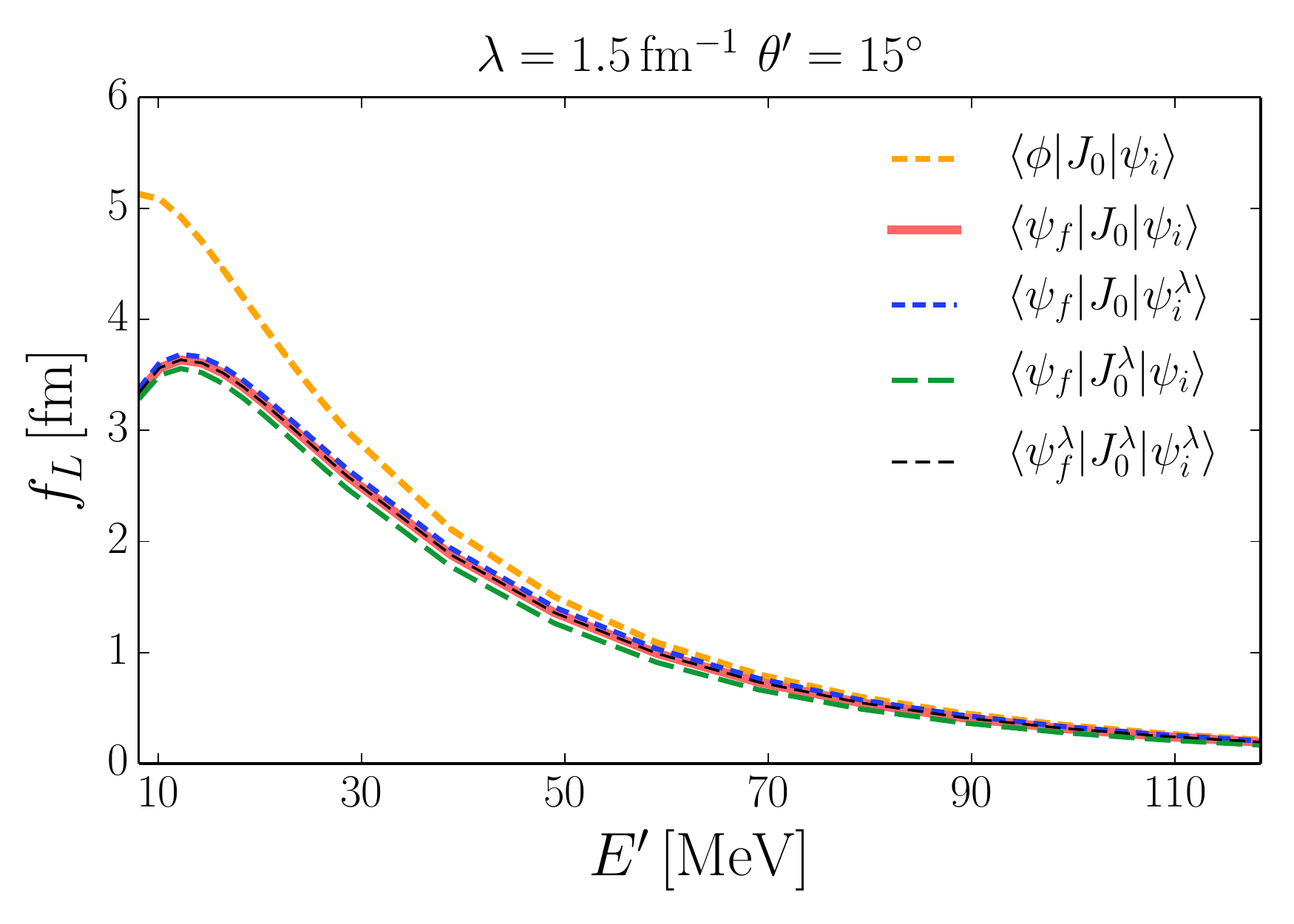}
	 \caption{ $\fL$ calculated at various points on the quasi-free ridge for
	   $\thetacm = 15^{\circ}$ for the AV18 potential.
	   Legends indicate which component of the matrix element in
	   Eq.~\eqref{eq:fl_prop_matrix_element} used to calculate $\fL$ is evolved.
	   There are no appreciable evolution effects all
	   along the quasi-free ridge.  The effect due to evolution of
	   the final state is small as well and is not shown here to avoid clutter.
	   $\fL$ calculated in the impulse
	   approximation is also shown for comparison.}
	 \label{fig:evolution_at_qfr}
	\end{figure}

	In Fig.~\ref{fig:evolution_at_qfr} we plot $\fL$ along the quasi-free ridge
	both in the impulse approximation (IA) and with the final-state interactions
	(FSI) included as a function of energy of the outgoing nucleons for a fixed
	angle, $\thetacm = 15^{\circ}$ of the outgoing proton.
	$\Ep$ and $\mbf{q}^2$ in Fig.~\ref{fig:evolution_at_qfr} are related by
	Eq.~\eqref{eq:quasi_free_condition_exact}.  Comparing the solid curve labeled
	$\mbraket{\psi_f}{J_0}{\psi_i}$ in the legend to the dashed curve (labeled
	$\mbraket{\phi}{J_0}{\psi_i}$) we find that FSI effects are minimal for
	configurations on the quasi-free ridge especially at large energies.

	In an intuitive picture, this is because after the initial photon is absorbed,
	both the nucleons in the deuteron are on their mass shell at the quasi-free
	ridge, and therefore no FSI are needed to make the final-state particles real.
	As we move away from the ridge, FSI become more important, as additional
	energy-momentum transfer is required to put the neutron and the proton on
	shell	in the final state.  The difference between full $\fL$ and $\fL$ in IA
	at small
	energies is also seen to hold for few-body nuclei \cite{Bacca:2014tla}.

	Figure~\ref{fig:evolution_at_qfr} also shows $\fL$ calculated from evolving
	only one of the components of the matrix element in
	Eq.~\eqref{eq:fl_prop_matrix_element}.  We note that the effects of SRG
	evolution of the individual components are minimal at the quasi-free ridge as
	well.  The kinematics at the quasi-free ridge are such that only the
	long-range (low-momentum) part of the deuteron wave function is probed, the
	FSI remains	small under evolution, and then unitarity implies minimal
	evolution of the current.  As one moves away from the quasi-free ridge, the
	effects of evolution
	of individual components become prominent.  Note that
	$\mbraket{\psi_f}{J_0}{\psi_i} =
	\mbraket{\psi_f^\lambda}{J_0^\lambda}{\psi_i^\lambda}$ and therefore the
	unevolved vs.\ all-evolved $\fL$ overlap in Fig.~\ref{fig:evolution_at_qfr}.

	\begin{figure}[htbp]
	 \centering
	 \includegraphics[width=0.6\textwidth]%
	 {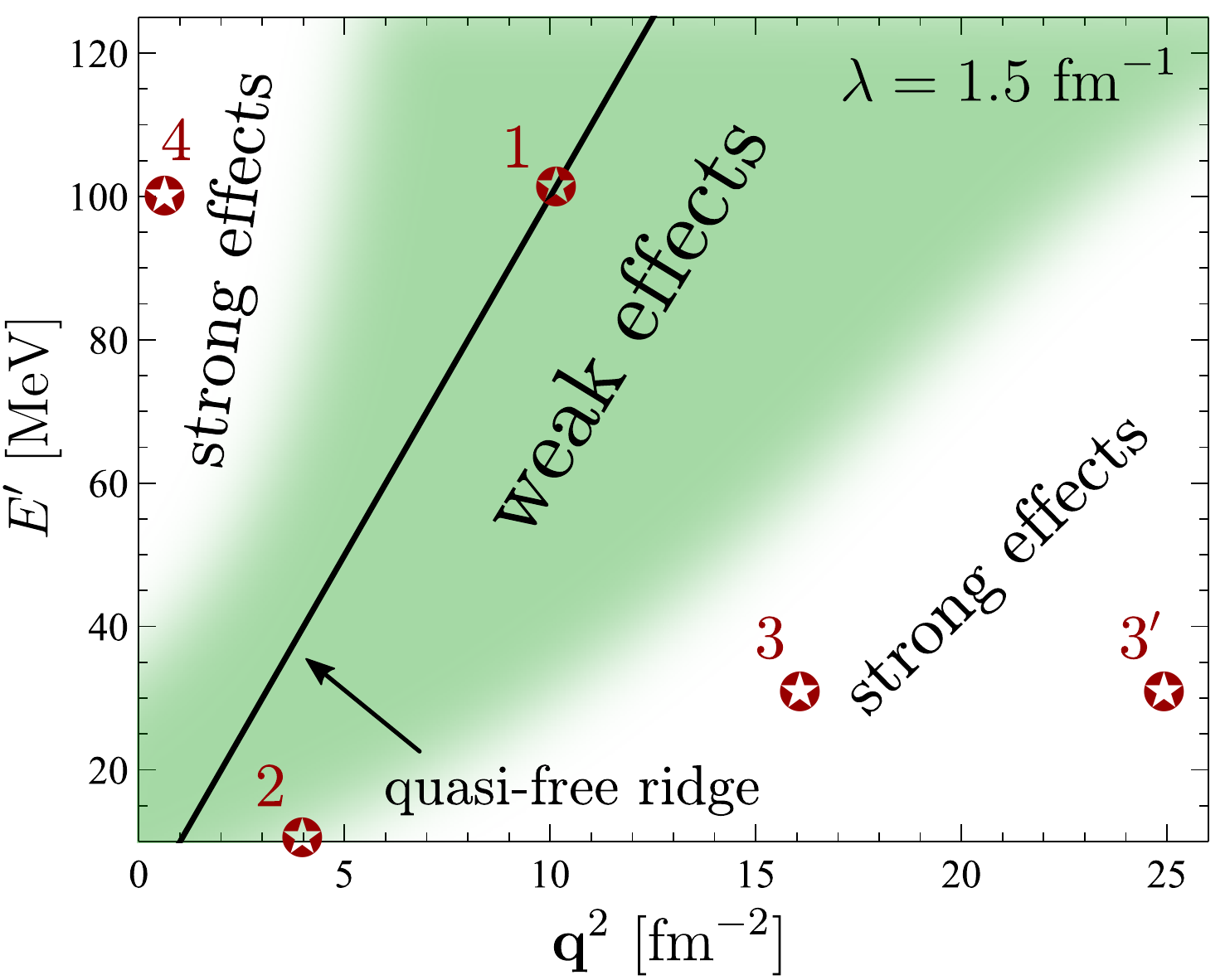}
	 \caption{`Phase space' of kinematics for
	 $\lambda = 1.5~\fm^{-1}$.  The effects of evolution
	 get progressively prominent as one moves further away from the quasi-free
	 ridge. The kinematics of the labeled points are considered later}
	 \label{fig:More-1p5}
	\end{figure}
	Figure~\ref{fig:More-1p5} shows the `phase space' of kinematics for SRG
	$\lambda = 1.5~\rm{fm^{-1}}$.  The quasi-free ridge is along the solid line
	in Fig.~\ref{fig:More-1p5}.  In the shaded region the effects generated by the
	evolution of individual components are weak (only a few percent
	relative difference).
	As one moves away from the quasi-free ridge, these differences get
	progressively	more prominent.
	The terms `small' and `weak' in Fig.~\ref{fig:More-1p5} are used in a
	qualitative
	sense.  In the shaded region denoted by `weak effects', the effects of evolution
	are not easily discernible on a typical $\fL$ versus $\thetacm$ plot, as
	seen in Fig.~\ref{fig:100_10_quasi_free_fsi}, whereas in the region
	labeled by `strong effects', the differences due to evolution are evident on
	such a plot (e.g., see Fig.~\ref{fig:30_16_fsi}).
	The size of the shaded region in Fig.~\ref{fig:More-1p5} depends
	on the SRG $\lambda$.  It is large for high $\lambda$'s and gets smaller as
	the	$\lambda$ is decreased (note that smaller SRG $\lambda$ means greater
	evolution).  Next, we look in detail at a few representative
	kinematics, indicated by points in Fig.~\ref{fig:More-1p5}.

  \medskip

	\subsubsection{At the quasifree ridge}

\begin{figure}[htbp]
	 \centering
	 \includegraphics[width=0.6\textwidth]%
	 {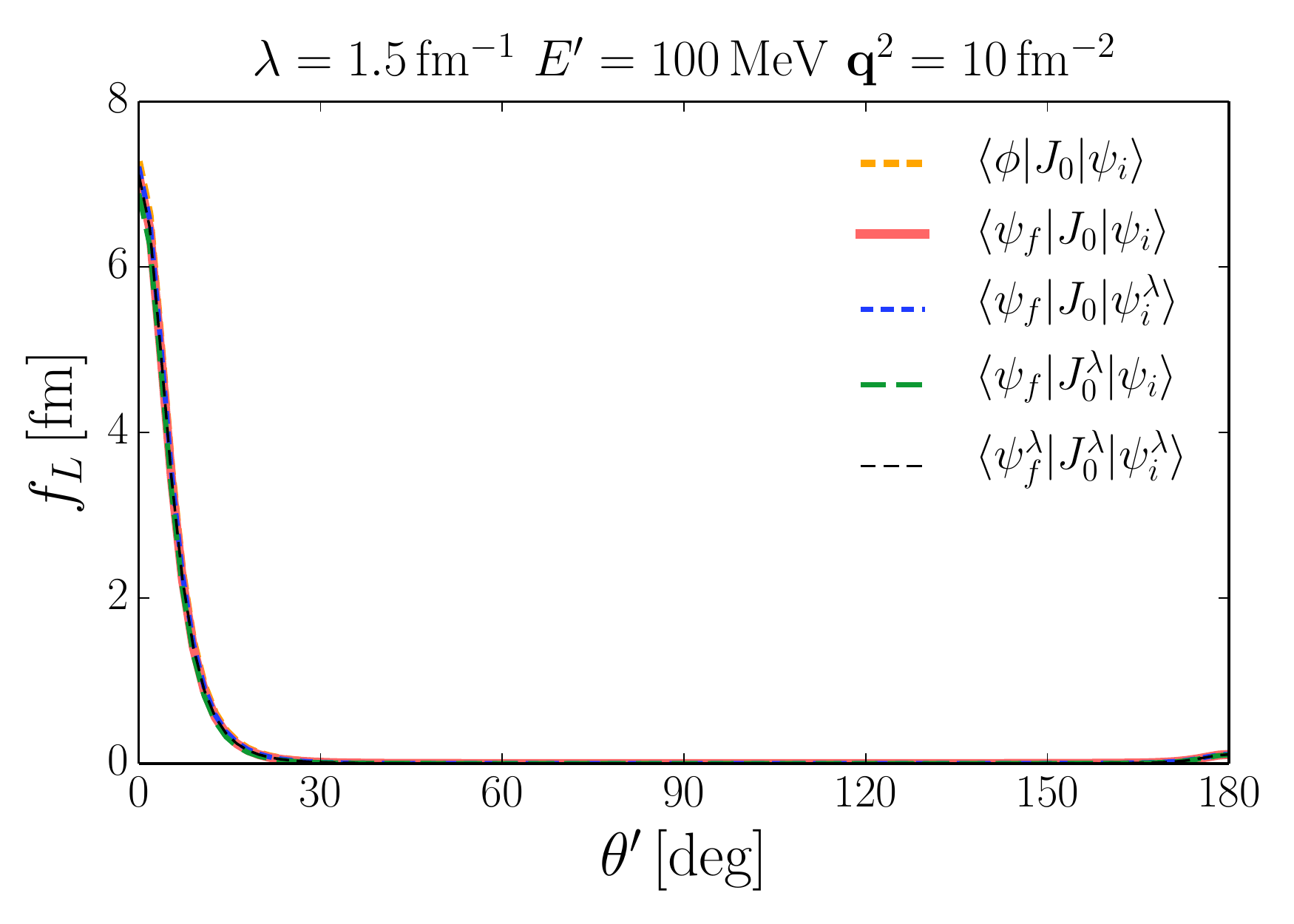}
	 \caption{
	   $\fL$ calculated for $E^\prime = 100~\MeV$ and $\mbf{q}^2 = 10~\fm^{-2}$
	   (point ``1'' in Fig.~\ref{fig:More-1p5}) for the AV18 potential.
	   Legends indicate which component of the matrix element in
	   Eq.~\eqref{eq:fl_prop_matrix_element} used to calculate $\fL$ is evolved.
	   $\thetacm$ is the angle of the outgoing proton in the center-of-mass frame.
	   There are no discernible evolution effects for all
	   angles.  The effect due to evolution
	   of the final state is small as well and is not shown here to avoid clutter.
	   $\fL$ calculated in the IA,
	   $\mbraket{\phi}{J_0}{\psi_i}$, is also shown for comparison.
	   }
	 \label{fig:100_10_quasi_free_fsi}
	\end{figure}

	As a representative of quasi-free kinematics, we choose
	$E^\prime = 100~\MeV$ and $\mbf{q}^2 = 10~\fm^{-2}$ and plot $\fL$ as a
	function of angle in Fig.~\ref{fig:100_10_quasi_free_fsi}.  The effect of
	including FSI is small for this configuration for all
	angles.  Also, the effects due to evolution of the individual components are
	too small to be discernible.  All this is consistent with the discussion in
	the	previous section.

	\medskip
	\subsubsection{Near the quasi-free ridge}

\begin{figure}[htbp]
	 \centering
	 \includegraphics[width=0.6\textwidth]%
	 {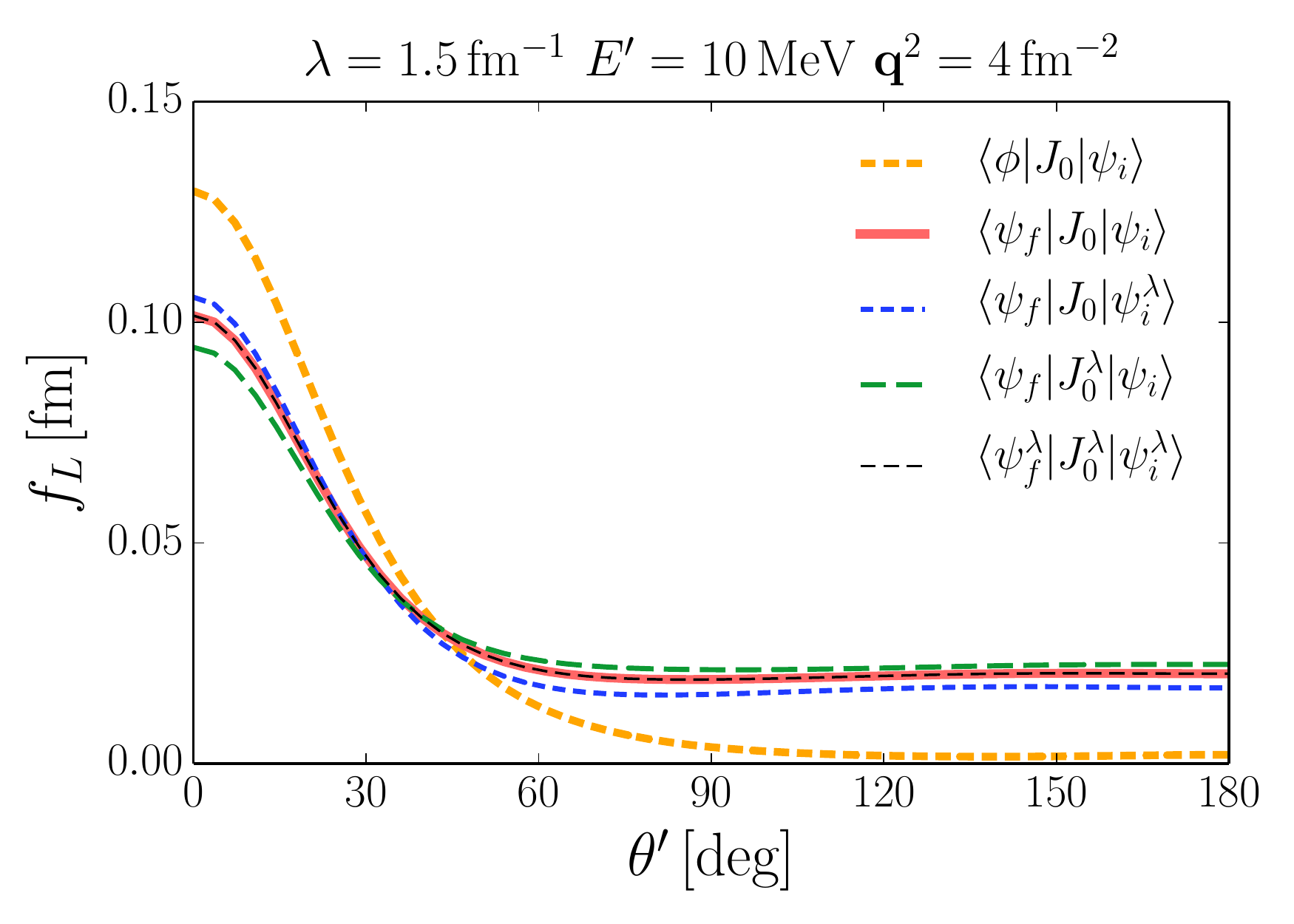}
	 \caption{
		 $\fL$ calculated for $E^\prime = 10~\MeV$ and $\mbfq^2 = 4~\fm^{-2}$
		 (point ``2'' in Fig.~\ref{fig:More-1p5}) for the AV18 potential.
		 Legends indicate which component of the matrix element in
		 Eq.~\eqref{eq:fl_prop_matrix_element} used to calculate $\fL$ is evolved.
		 $\fL$ calculated in the IA, $\mbraket{\phi}{J_0}{\psi_i}$,
		 is also shown for comparison.  The effects due to evolution of individual
		 components on $\fL$ are discernible, but still small (compared to the FSI
		 contribution).  The effect due to evolution of the final state is small as
		 well and is not shown here to avoid clutter. }
	 \label{fig:10_4_fsi}
	\end{figure}

	Next we look at the kinematics $E^\prime = 10~\MeV$ and $\mbf{q}^2 =
	4~\fm^{-2}$, which is near the quasi-free ridge.  This is the point ``2'' in
	Fig.~\ref{fig:More-1p5}.  As seen in Fig.~\ref{fig:10_4_fsi}, the different
	curves for $\fL$ obtained from evolving different components start to diverge.
	Figure~\ref{fig:10_4_fsi} also shows $\fL$ calculated in IA.  Comparing this
	to the full $\fL$ including FSI, we see that the effects due to evolution are
	small	compared to the FSI contributions.  This smallness prevents us from
	making any
	systematic observations about the effects due to evolution at this kinematics.
	We thus move on to kinematics which show more prominent effects.

	\medskip
	\subsubsection{Below the quasi-free ridge}

\begin{figure}[htbp]
	 \centering
	 \includegraphics[width=0.6\textwidth]
	 {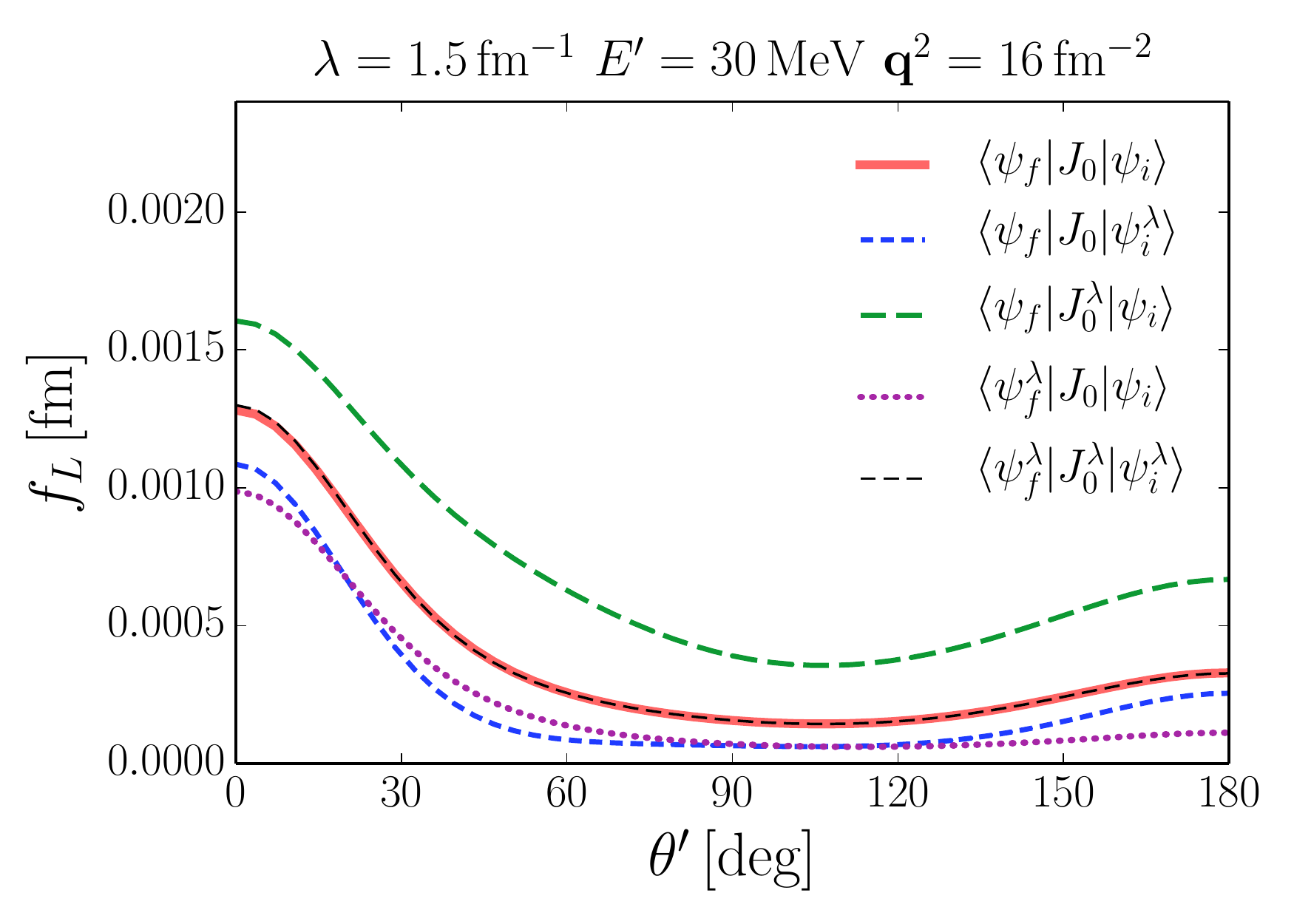}
	 \caption{$\fL$ calculated for
	 $E^\prime = 30~\MeV$ and $\mbfq^2 = 16~\fm^{-2}$
	 (point ``3'' in Fig.~\ref{fig:More-1p5})
	 for the AV18 potential.
	 Legends indicate which component of the matrix element in
	 Eq.~\eqref{eq:fl_prop_matrix_element} used to calculate $\fL$ is evolved.
	 Prominent enhancement with evolution of the
	 current only and suppression with evolution of the initial state and the final
	 state only, respectively.}
	 \label{fig:30_16_fsi}
	\end{figure}

\begin{figure}[htbp]
	 \centering
	 \includegraphics[width=0.6\textwidth]%
	 {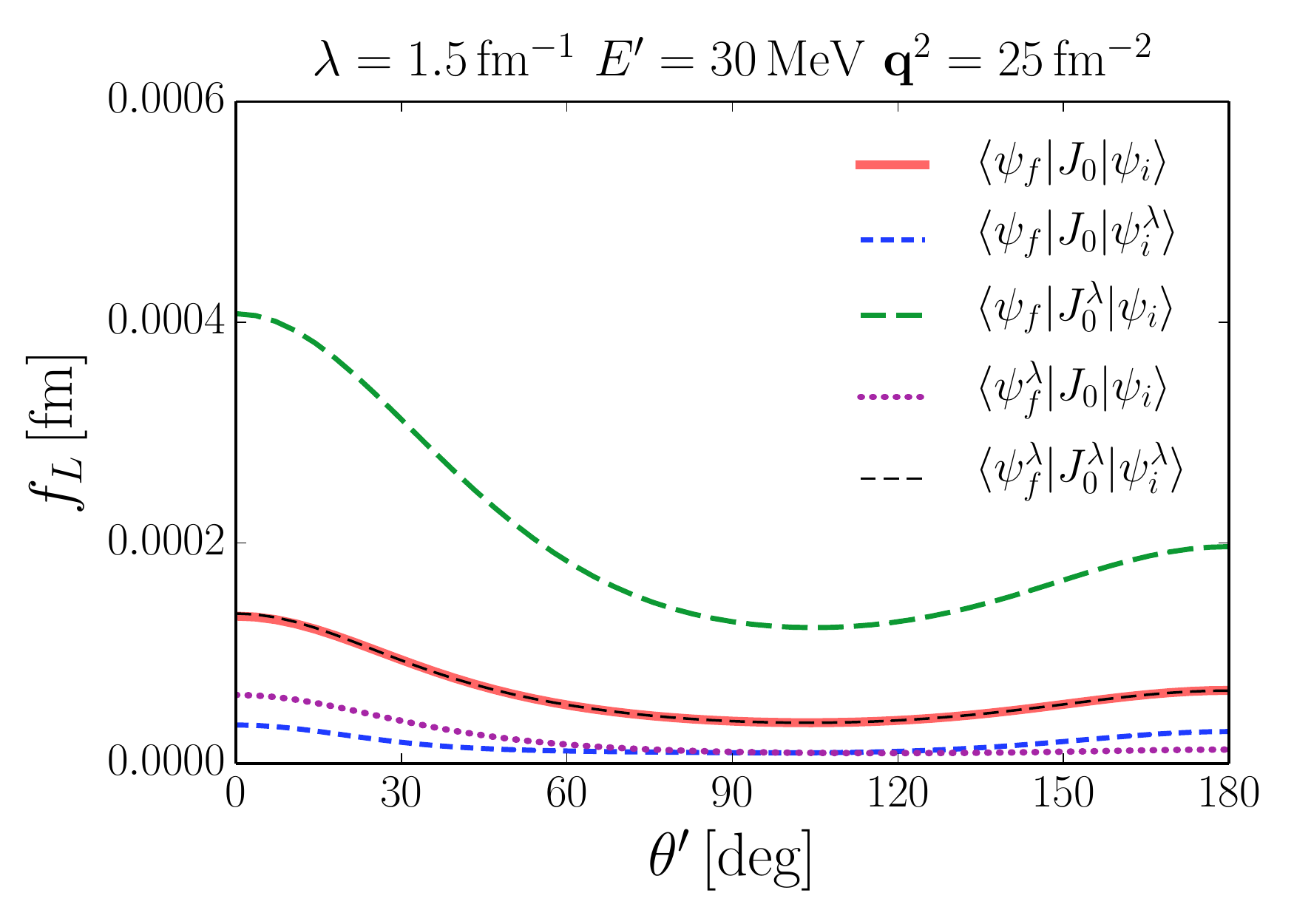}
	 \caption{
	 	 $\fL$ calculated for $E^\prime = 30~\MeV$ and $\mbfq^2 = 25~\fm^{-2}$
		 (point ``3$^{\prime}$'' in Fig.~\ref{fig:More-1p5})
		 for the AV18 potential.
		 Legends indicate which component of the matrix element in
		 Eq.~\eqref{eq:fl_prop_matrix_element} used to calculate $\fL$ is evolved.
		 Prominent enhancement with evolution of the
		 current only and suppression with evolution of the initial state and the
		 final state only, respectively.}
	 \label{fig:30_25_fsi}
	\end{figure}

	We next look in the region where $E^{\prime}~\text{(in~$\MeV$)}
	\ll 10\,\mbfq^2~\text{(in~$\fm^{-2}$)}$, \ie, below the quasi-free ridge in
	Fig.~\ref{fig:More-1p5}.  We look at two momentum transfers
	$\mbfq^2 = 16~\fm^{-2}$ and $\mbfq^2 = 25~\fm^{-2}$ for $\Ep = 30~\MeV$,
	which are points ``3'' and ``$3^\prime$'' in Fig.~\ref{fig:More-1p5}.
	Figures~\ref{fig:30_16_fsi} and~\ref{fig:30_25_fsi} indicate the effects on
	$\fL$	from evolving individual components of the matrix elements.
	It is noteworthy that in both cases evolution of the current gives a
	prominent enhancement, whereas evolution of the initial and final state gives
	a suppression.  When
	all	the components are evolved consistently, these changes combine and we
	recover	the unevolved answer for $\fL$.  This verifies the accurate
	implementation of	the equations derived in
	Subsec.~\ref{subsec:evolution_setup}.

	It is possible to qualitatively explain the behavior seen in
	Figs.~\ref{fig:30_16_fsi} and~\ref{fig:30_25_fsi}.  As noted in
	Eq.~\eqref{eq:overlap_IA_plus_FSI}, the overlap matrix element is given by the
	sum of the IA part and the FSI part.  Below the quasi-free ridge these two
	terms add constructively.  In this region, $\fL$ calculated in impulse
	approximation is smaller than $\fL$ calculated by including the final-state
	interactions.

\begin{figure}[htbp]
	 \centering
	 \includegraphics[width=0.6\textwidth]%
	 {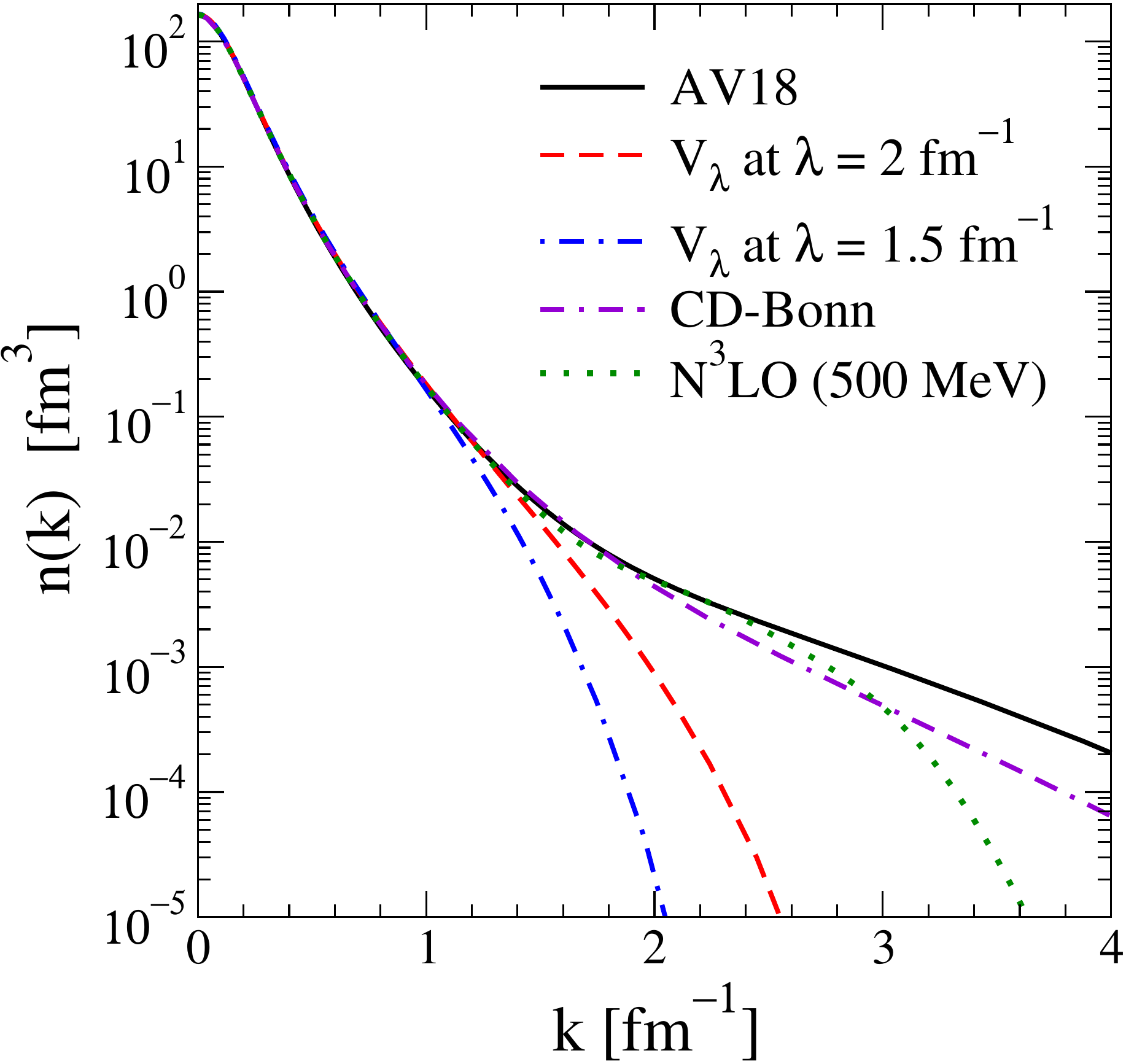}
	 \caption{
		 Momentum distribution for the deuteron for the AV18~\cite{Wiringa:1994wb},
		 CD-Bonn~\cite{Machleidt:2000ge},
		 and the Entem-Machleidt N$^3$LO chiral EFT~\cite{Entem:2003ft}
		 potentials, and for the AV18 potential evolved to two SRG $\lambda$'s.}
	 \label{fig:deut_md}
	\end{figure}

	\paragraph{\emph{(a) Evolving the initial state}} Let us first consider the
	effect of evolving the initial state only.  We have
	\begin{equation}
	 \mbraket{\psi_f}{J_0}{\psi_i^\lambda}
	 = \mbraket{\phi}{J_0}{\psi_i^\lambda}
	 + \mbraket{\phi}{t^\dag \, \GreensFn^\dag \, J_0}{\psi_i^\lambda} \,.
	\label{eq:overlap_IA_plus_FSI_evol_wf}
	\end{equation}
	As seen in Eq.~\eqref{eq:overlap_IA}, in the term
	$\mbraket{\phi}{J_0}{\psi_i^\lambda}$ the deuteron wave function is
	probed between $|\pp - q/2|$ and $\pp + q/2$.  These numbers are
	$(1.2, 2.9)~\fm^{-1}$ and $(1.7, 3.4)~\fm^{-1}$ for $\Ep = 30~\MeV$,
	$\mbfq^2 = 16~\fm^{-2}$ and $\Ep = 30~\MeV$, $\mbfq^2 = 25~\fm^{-2}$,
	respectively.  The evolved deuteron wave function is significantly suppressed
	at these high momenta.
	This behavior is reflected in the deuteron momentum distribution plotted in
	Fig.~\ref{fig:deut_md}.  The deuteron momentum distribution $n(k)$ is
	proportional to the sum of the squares of $S$- and $D$- state deuteron
	wave functions.  Thus, the first (IA) term in
	Eq.~\eqref{eq:overlap_IA_plus_FSI_evol_wf} is much smaller than its
	unevolved counterpart in Eq.~\eqref{eq:overlap_IA_plus_FSI}, for all angles.
	We note that even though we only use the AV18 potential to
	study changes due to evolution, these changes will be significant for other
	potentials as well.

	Evaluation of the second (FSI) term in
	Eq.~\eqref{eq:overlap_IA_plus_FSI_evol_wf} involves an integral over all
	momenta, as indicated in Eq.~\eqref{eq:phi_t_g0_J0_minus}.  We find that
	$|\mbraket{\phi}{t^\dag \, \GreensFn^\dag \, J_0}{\psi_i^\lambda}| <
	|\mbraket{\phi}{t^\dag \, \GreensFn^\dag \, J_0}{\psi_i}|$.  As mentioned
	before, because the terms $\mbraket{\phi}{J_0}{\psi_i}$ and
	$\mbraket{\phi}{t^\dag \, \GreensFn^\dag \, J_0}{\psi_i}$ add constructively
	below the quasi-free ridge and because the magnitude of both these terms
	decreases upon evolving the wave function, we have
	\begin{equation}
	 |\mbraket{\psi_f}{J_0}{\psi_i^\lambda}| < |\mbraket{\psi_f}{J_0}{\psi_i}|
	\label{eq:evolv_wf_below_qfr} \,.
	\end{equation}
	The above relation holds for most combinations of $\mJd$ and $\msf$.  For
	those	$\mJd$ and $\msf$ for which Eq.~\eqref{eq:evolv_wf_below_qfr} does not
	hold,	the absolute value of the matrix element is much smaller than for those
	for	which the Eq.~\eqref{eq:evolv_wf_below_qfr} \emph{does} hold, and
	therefore we have $\fL$ calculated from $\mbraket{\psi_f}{J_0}
	{\psi_i^\lambda}$ smaller than the
	$\fL$ calculated from $\mbraket{\psi_f}{J_0}{\psi_i}$, as seen in
	Figs.~\ref{fig:30_16_fsi} and~\ref{fig:30_25_fsi}.

	\paragraph{\emph{(b) Evolving the final state}}  As indicated in
	Eq.~\eqref{eq:psi_f_lam_def}, evolving the final state entails
	the evolution of the $t$-matrix.  The overlap matrix element therefore is
	\begin{equation}
	 \mbraket{\psi_f^\lambda}{J_0}{\psi_i}
	 = \mbraket{\phi}{J_0}{\psi_i}
	 + \mbraket{\phi}{t^\dag_\lambda \, \GreensFn^\dag \, J_0}{\psi_i} \,.
	\label{eq:evol_final_state_overlap}
	\end{equation}
	The IA term is the same as in the unevolved case.  The SRG evolution leaves the
	on-shell part of the $t$-matrix---which is directly related to
	observables---invariant.  The magnitude of the relevant off-shell $t$-matrix
	elements decreases
	on evolution, though.  As a result we have
	\begin{equation}
	|\mbraket{\psi_f^\lambda}{J_0}{\psi_i}| < |\mbraket{\psi_f}{J_0}{\psi_i}| \,.
	\end{equation}
	This is reflected in $\fL$ as calculated from the evolved final state, and
	seen in Figs.~\ref{fig:30_16_fsi} and~\ref{fig:30_25_fsi}.

	The effect of evolution of the initial state and the final state is to suppress
	$\fL$.  When all the three components are evolved, we reproduce the unevolved
	answer as indicated in Fig.~\ref{fig:30_16_fsi} and~\ref{fig:30_25_fsi}.  It is
	therefore required that we find a huge enhancement when just the current
	is evolved.

	The kinematics $\Ep = 30~\MeV$, $\mbfq^2 = 25~\fm^{-2}$ is further away from
	the quasi-free ridge than $\Ep = 30~\MeV$, $\mbfq^2 = 16~\fm^{-2}$.
	The evolution effects discussed above get
	progressively more prominent the further away one is from the quasifree
	ridge.  This can be verified by
	comparing the effects due to evolution of individual components in
	Figs.~\ref{fig:30_16_fsi} and~\ref{fig:30_25_fsi}.

\begin{figure}[htbp]
	 \centering
	 \includegraphics[width=0.6\textwidth]%
	 {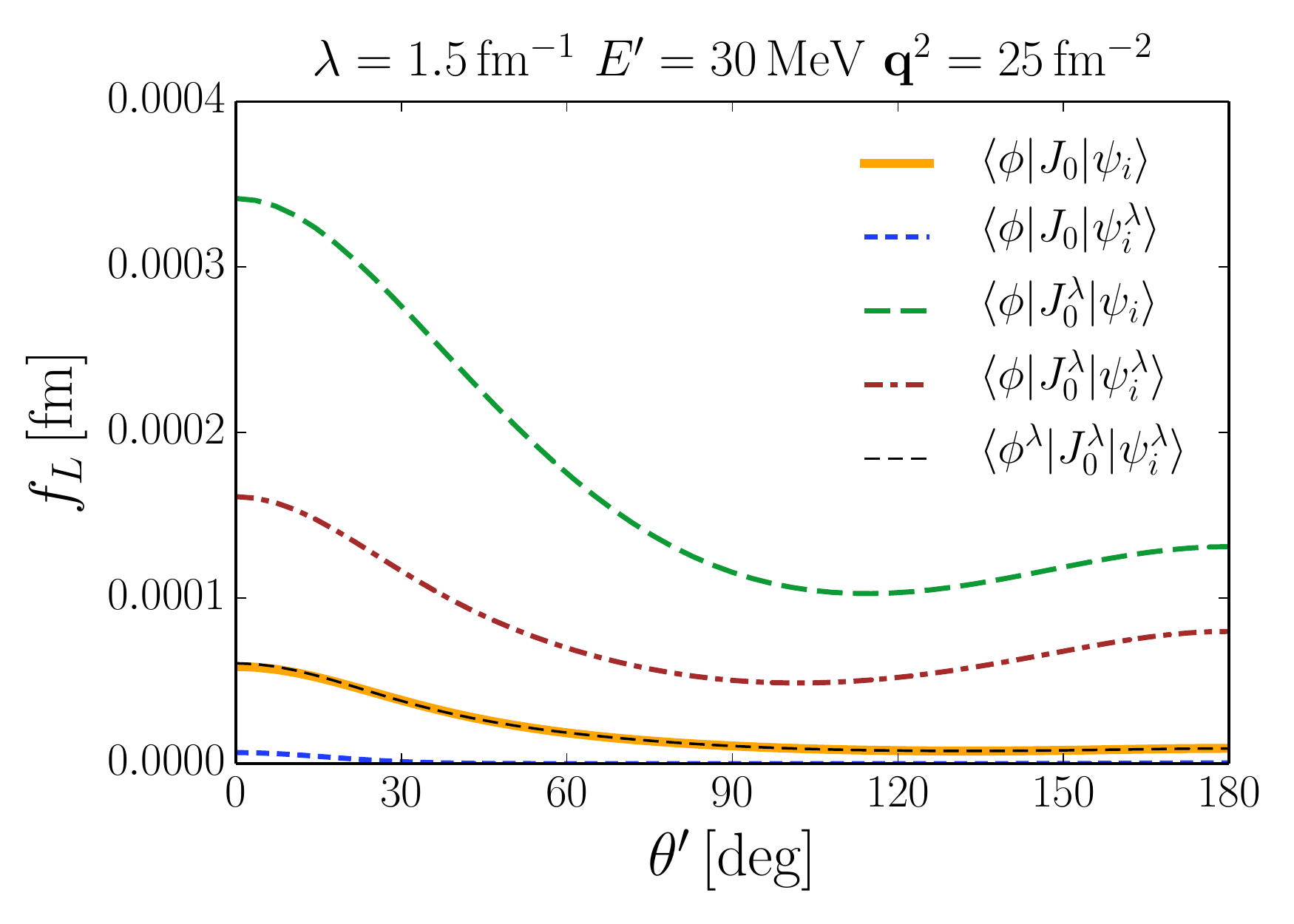}
	 \caption{
		 $\fL$ in IA $(\bra{\psi_f} \equiv \bra{\phi})$ calculated for
		 $E^\prime = 30~\MeV$ and $\mbfq^2 = 25~\fm^{-2}$ for the AV18 potential.
		 Legends indicate which component of the matrix element in
		 Eq.~\eqref{eq:fl_prop_matrix_element} used to calculate $\fL$ are evolved.}
	 \label{fig:30_25_ia}
	\end{figure}

	As remarked earlier, away from the quasi-free ridge the FSI
	become important.  Nonetheless, it is still instructive to look at $\fL$
	calculated in the IA at these kinematics.
	\begin{figure}[htbp]
	 \centering
	 \includegraphics[width=0.6\textwidth]%
	 {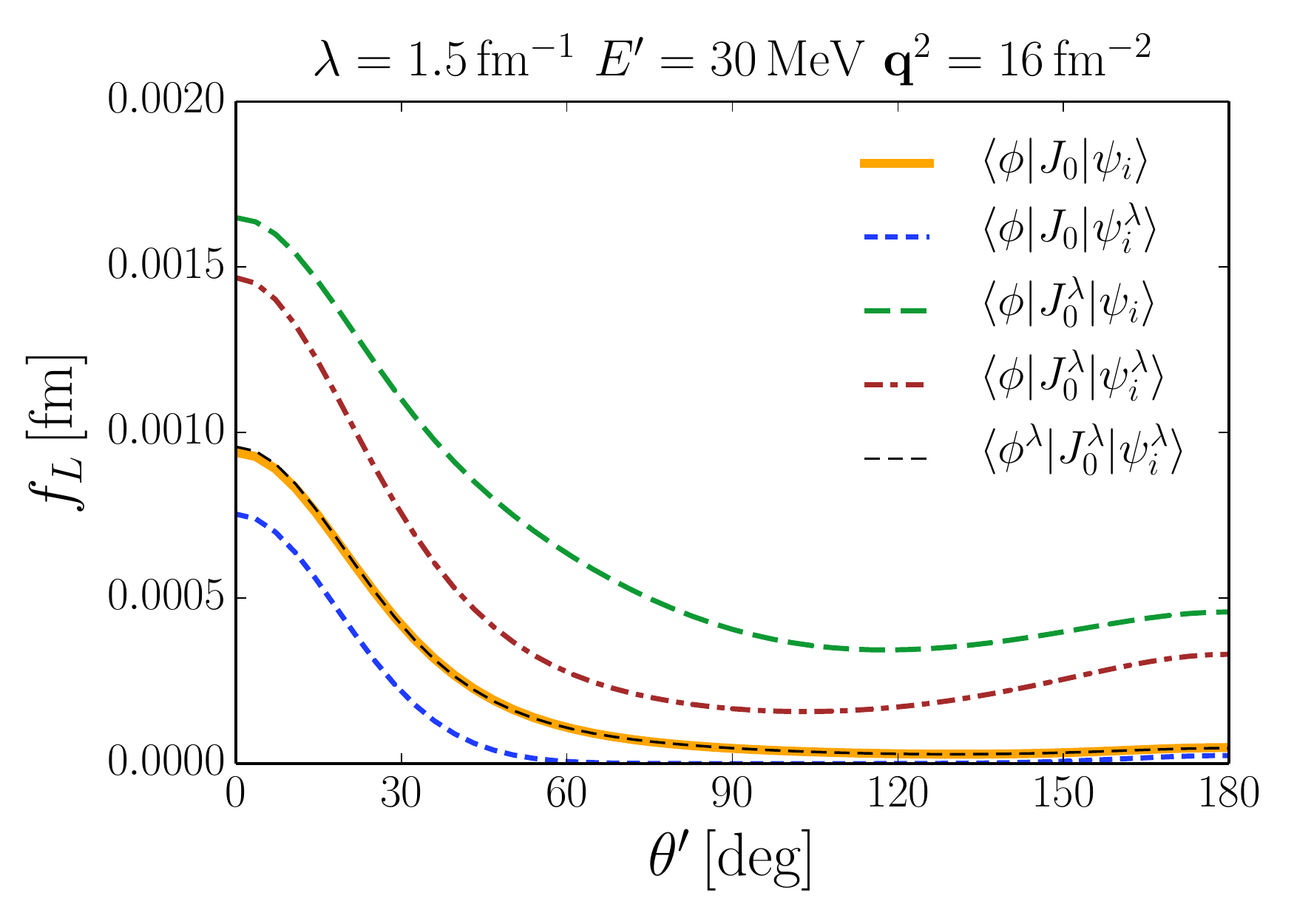}
	 \caption{
		 $\fL$ in IA $(\bra{\psi_f} \equiv \bra{\phi})$ calculated for
		 $E^\prime = 30~\MeV$ and $\mbfq^2 = 16~\fm^{-2}$ for the AV18 potential.
		 Legends indicate which component of the matrix element in
		 Eq.~\eqref{eq:fl_prop_matrix_element} used to calculate $\fL$ are evolved.}
	 \label{fig:30_16_ia}
	\end{figure}
	Note that the (unevolved) $\fL$ calculated in the IA, shown
	in Figs.~\ref{fig:30_16_ia} and~\ref{fig:30_25_ia}, is smaller than the full
	$\fL$ that takes into account the final state interactions (cf.~the
	corresponding curves in Figs.~\ref{fig:30_16_fsi} and~\ref{fig:30_25_fsi}).
	This is consistent with the claim made earlier that below the quasi-free ridge
	the two terms in Eq.~\eqref{eq:overlap_IA_plus_FSI} add constructively.

	The results in Figs.~\ref{fig:30_16_ia} and~\ref{fig:30_25_ia} can again be
	qualitatively explained based on our discussion above.  The evolution of the
	deuteron wave function leads to suppression as the evolved wave function does
	not have strength at high momentum.  The evolved current thus leads to
	enhancement.  Evolution of both the current and the initial state decreases
	$\fL$ from just the evolved current value, but it is not until we evolve all
	three components---final state, current, and the initial state---that we
	recover the unevolved answer.

	\begin{figure}[htbp]
		\centering
		\begin{subfigure}[c]{0.6\textwidth}
			\centering
			\includegraphics[width=\textwidth]
			{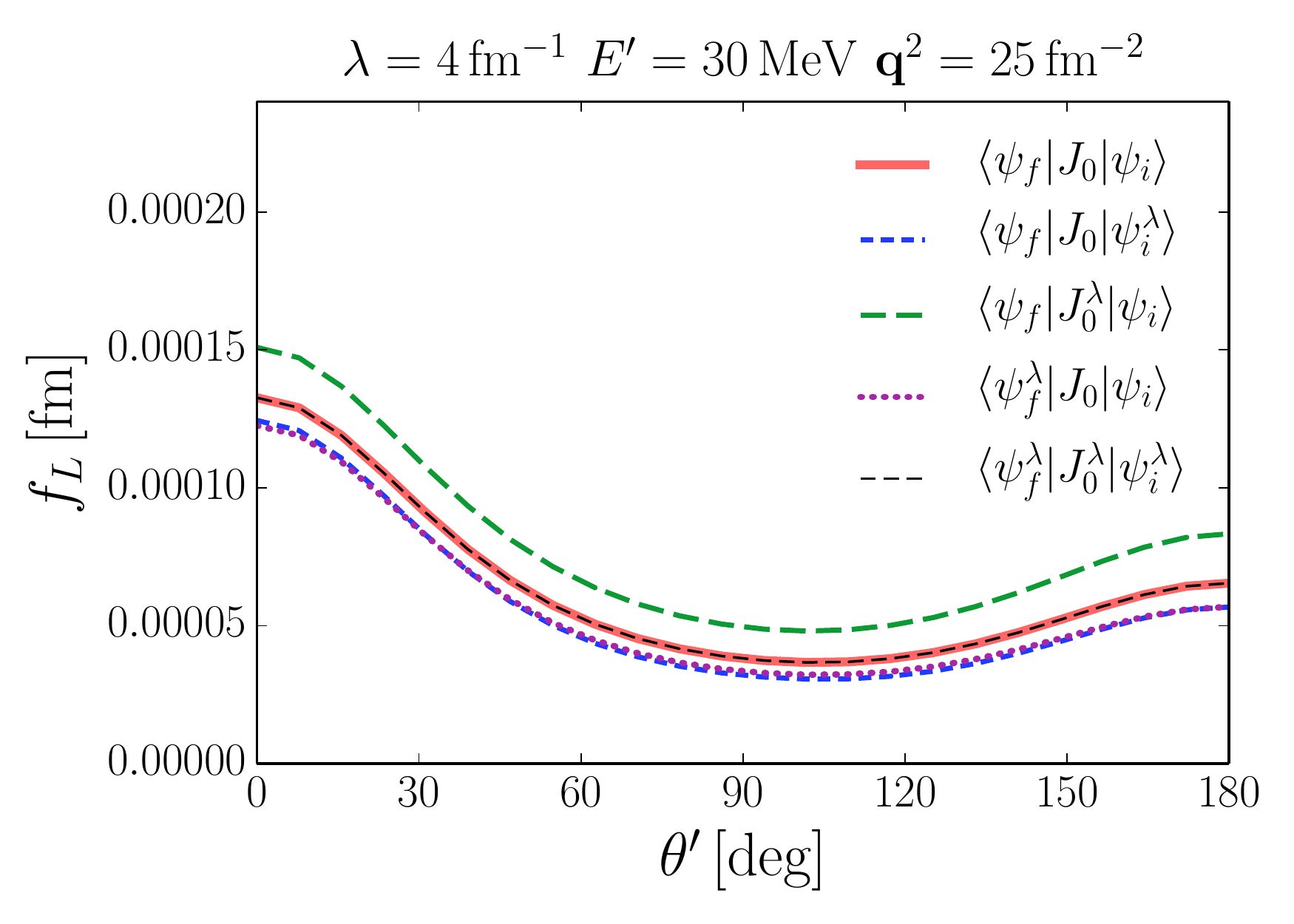}
			\caption{SRG $\lambda = 4 {\rm{~fm}}^{-1}$}
			\label{fig:30_25_lam4}
		\end{subfigure}
		\begin{subfigure}[c]{0.6\textwidth}
			\centering
			\includegraphics[width=\textwidth]
			{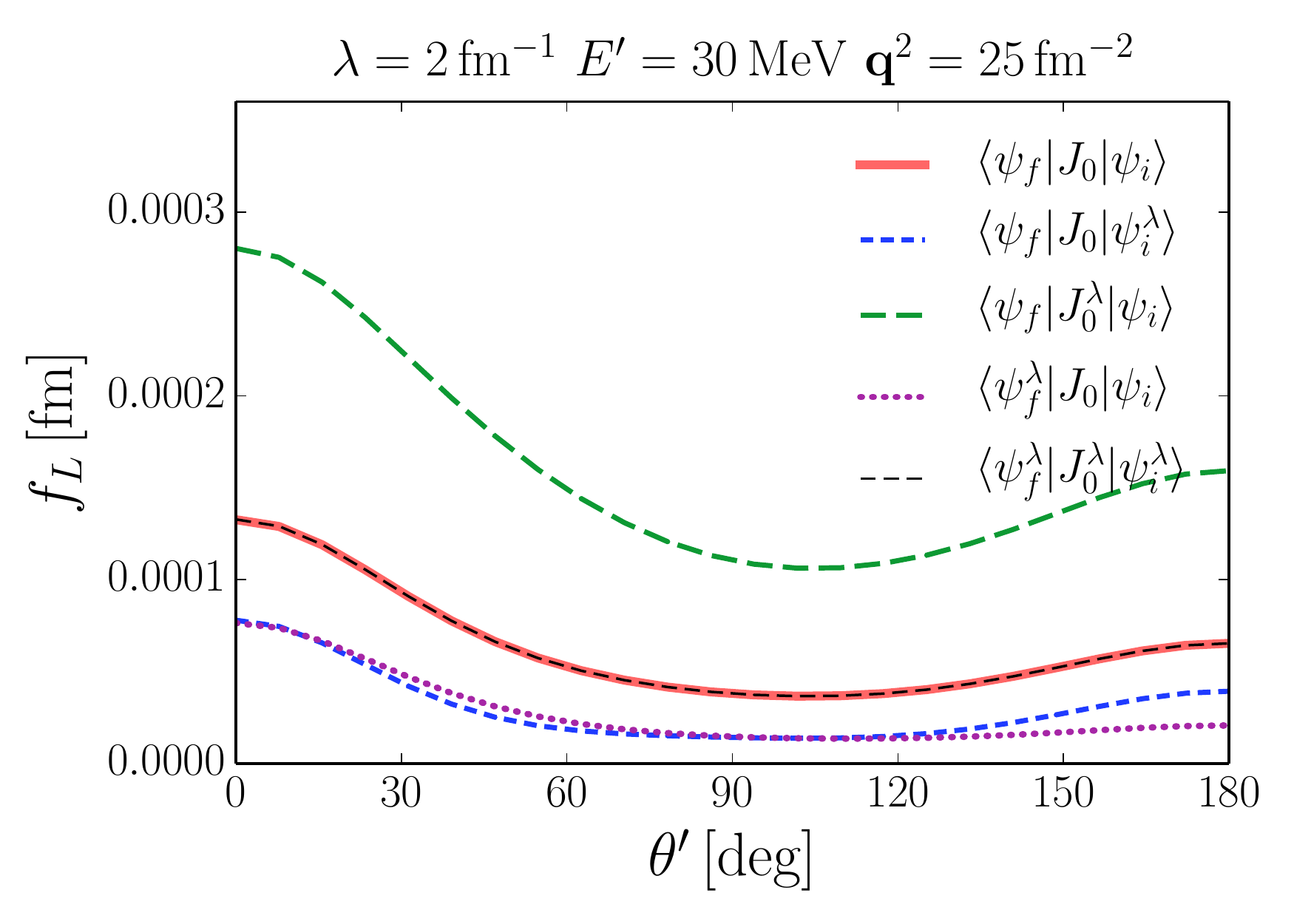}
			\caption{SRG $\lambda = 2 {\rm{~fm}}^{-1}$}
			\label{fig:30_25_lam2}
		\end{subfigure}
		\caption{$\fL$ calculated for $E^\prime = 30~\MeV$ and $\mbfq^2 = 25~\fm^{-2}$
			(point ``3$^{\prime}$'' in Fig.~\ref{fig:More-1p5})
			for the AV18 potential.
			Legends indicate which component of the matrix element in
			Eq.~\eqref{eq:fl_prop_matrix_element} used to calculate $\fL$ is evolved.
			The evolution is to SRG (a) $\lambda = 4 {\rm{~fm}}^{-1}$ and
			(b) $\lambda = 2 {\rm{~fm}}^{-1}$.}
		\label{fig:30_25_lam_fn}
	\end{figure}
	As expected, the effect due to evolution increases with further evolution.
	This can be seen by comparing the plots in Fig.~\ref{fig:30_25_lam_fn} to
	Fig.~\ref{fig:30_25_fsi}.
	\begin{figure}[htbp]
	 \centering
	 \includegraphics[width=0.6\textwidth]%
	 {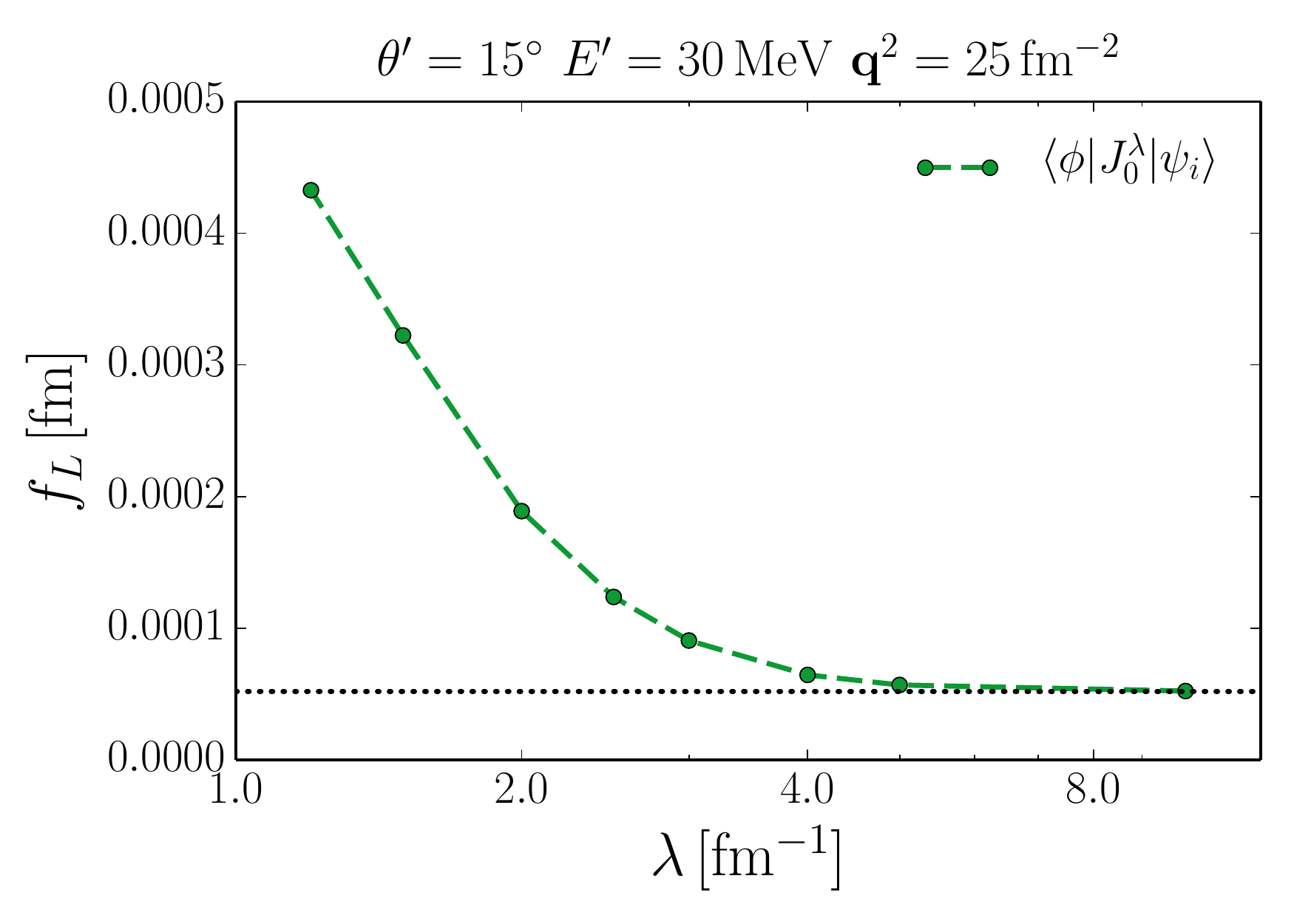}
	 \caption{
		 $\fL$ in IA calculated at $\thetacm = 15^{\circ}$ for
		 $E^\prime = 30~\MeV$ and $\mbfq^2 = 25~\fm^{-2}$
		 for the AV18 potential when the current operator in
		 Eq.~\eqref{eq:fl_prop_matrix_element} used to calculate $\fL$ is evolved to
		 various SRG $\lambda$'s.  The horizontal dotted line is the unevolved
		 answer.}
	 \label{fig:J0_evolution_vs_lambda}
	\end{figure}
	In Fig.~\ref{fig:J0_evolution_vs_lambda} we look at
	effects of the current-operator evolution on $\fL$ as a
	function of the SRG $\lambda$.  To isolate the effect of operator evolution,
	we only look at $\fL$ calculated in IA at a specific angle in
	Fig.~\ref{fig:J0_evolution_vs_lambda}.  Investigating details of the
	operator evolution forms the basis of ongoing work.  A few
	preliminary results along those lines are
	presented in Subsec.~\ref{subsec:operator_evolution}.

	\subsubsection{Above the quasi-free ridge}

\begin{figure}[htbp]
	 \centering
	 \includegraphics[width=0.6\textwidth]%
	 {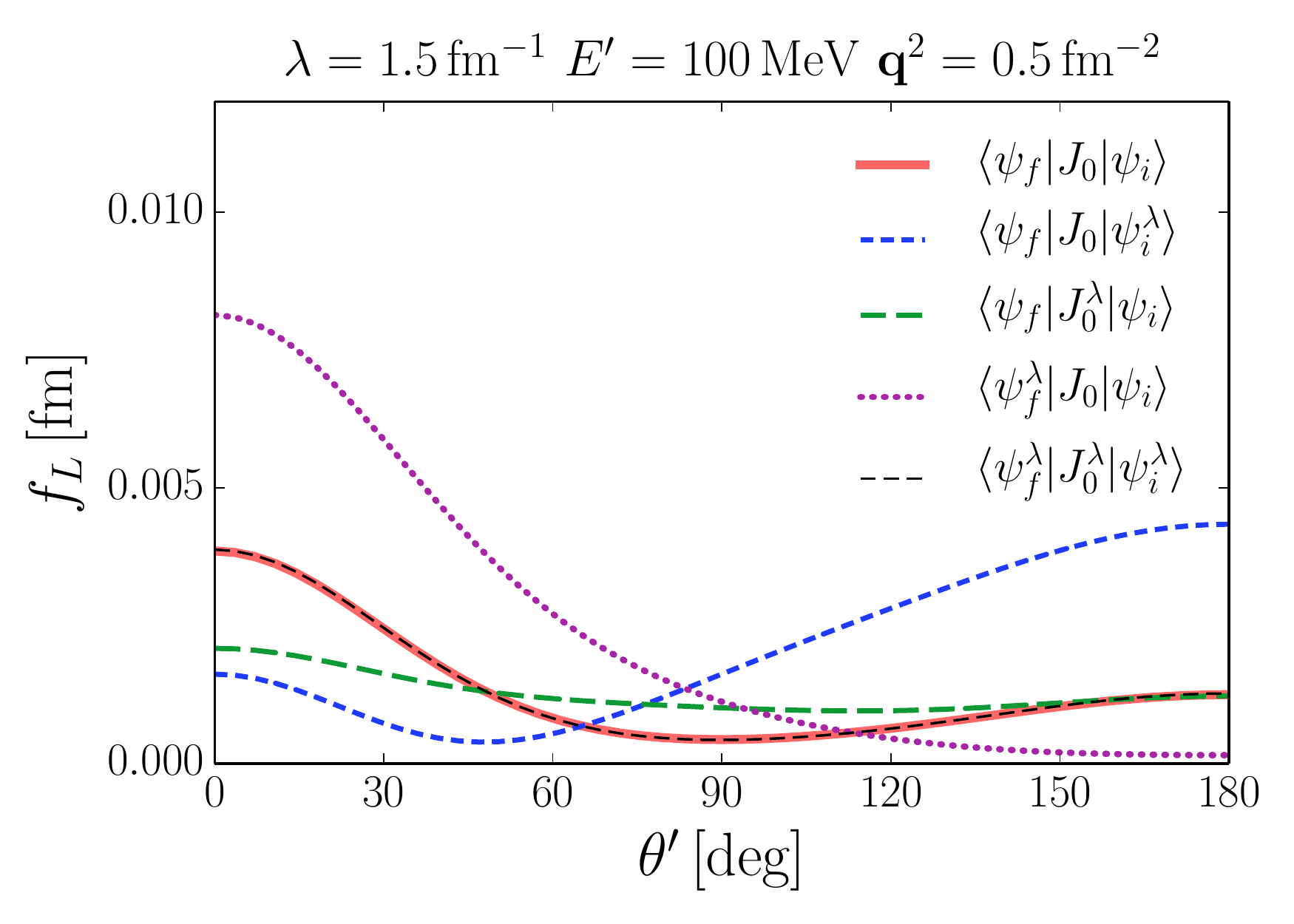}
	 \caption{
		 $\fL$ calculated for $E^\prime = 100~\MeV$ and
		 $\mbfq^2 = 0.5~\fm^{-2}$ (point ``4'' in Fig.~\ref{fig:More-1p5})
		 for the AV18 potential.
		 Legends indicate which component of the matrix element in
		 Eq.~\eqref{eq:fl_prop_matrix_element} used to calculate $\fL$ is evolved.
		 Opposite effects from the evolution of the
		 initial state and the final state.}
	 \label{fig:100_0p5_fsi}
	\end{figure}

	Finally, we look at an example from above the quasi-free ridge.
	Figure~\ref{fig:100_0p5_fsi} shows the effect of evolution of individual
	components on $\fL$ for $\Ep = 100~\MeV$ and $\mbfq^2 = 0.5~\fm^{-2}$, which
	is point ``4'' in Fig.~\ref{fig:More-1p5}.  The effects of evolution in this
	case are qualitatively different from those found below the quasi-free ridge.
	For instance, we see a peculiar suppression in $\fL$ calculated from the
	evolved	deuteron wave function at small angles, but an enhancement at large
	angles.  An	opposite behavior is observed for the final state.  It is again
	possible to	qualitatively explain these findings.

\begin{figure}[htbp]
	\centering
	\includegraphics[width=0.6\textwidth]
	{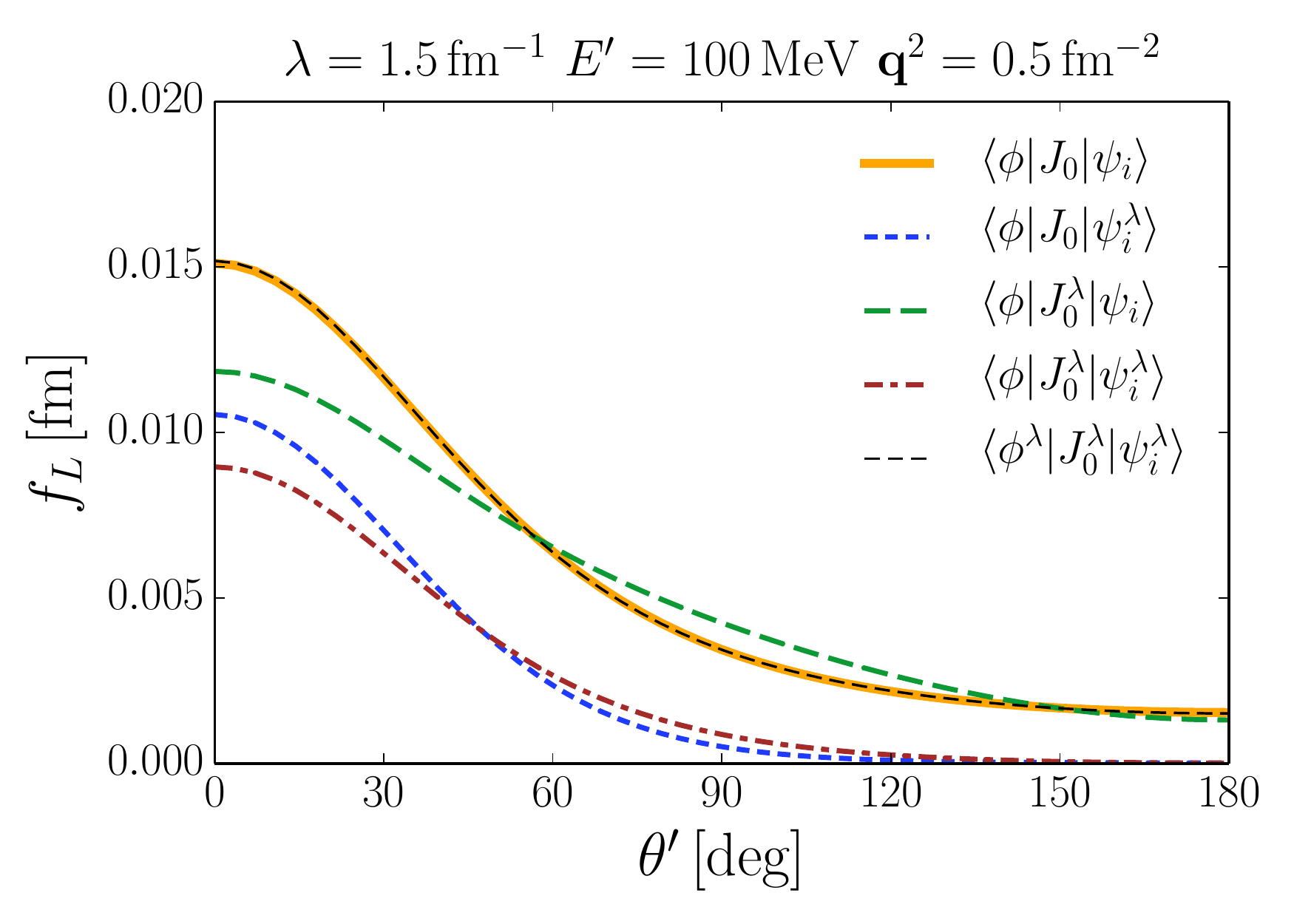}
	\caption{
	 $\fL$ in IA $(\bra{\psi_f} \equiv \bra{\phi})$ calculated for
	 $E^\prime = 100~\MeV$ and $\mbfq^2 = 0.5~\fm^{-2}$
	 for the AV18 potential.
	 Legends indicate which component of the matrix element in
	 Eq.~\eqref{eq:fl_prop_matrix_element} used to calculate $\fL$ are evolved.}
	\label{fig:100_0p5_ia}
	\end{figure}

	\paragraph{\emph{(a) Evolving the initial state}}	 Above the quasi-free ridge,
	the IA and FSI terms in
	Eq.~\eqref{eq:overlap_IA_plus_FSI} add destructively.  This can be seen by
	comparing the unevolved $\fL$ curves in Figs.~\ref{fig:100_0p5_fsi}
	and~\ref{fig:100_0p5_ia}.  Including the FSI brings down
	the value of $\fL$ when one is above the quasi-free ridge.

	At small angles, the magnitude of the IA term in
	Eq.~\eqref{eq:overlap_IA_plus_FSI} is larger than that of the FSI term.  The
	deuteron wave function for this kinematics is probed between $1.2$ and
	$1.9~\fm^{-1}$.  With the wave-function evolution, the magnitude of the IA
	term in Eq.~\eqref{eq:overlap_IA_plus_FSI_evol_wf} decreases, whereas the
	magnitude of the FSI term in that equation slightly increases compared to its
	unevolved counterpart.  Still, at small angles, we have
	$|\mbraket{\phi}{J_0}{\psi_i^\lambda}| > |\mbraket{\phi}{t^\dag \,
	\GreensFn^\dag \, J_0}{\psi_i^\lambda}|$, which leads to
	\begin{equation}
	 |\mbraket{\psi_f}{J_0}{\psi_i^\lambda}| < |\mbraket{\psi_f}{J_0}{\psi_i}| \,,
	 \label{eq:psi_i_evolution_small_angles_above_qfr}
	\end{equation}
	and thus to the suppression of $\fL$ at small angles observed in
	Fig.~\ref{fig:100_0p5_fsi}.

	At large angles, the magnitude of the IA term in
	Eq.~\eqref{eq:overlap_IA_plus_FSI} is smaller than that of the FSI term.
	With the wave-function evolution, the magnitude of IA term decreases
	substantially (large momenta in the deuteron wave function are probed at
	large	angles, cf.~Eq.~\eqref{eq:overlap_IA}), whereas the FSI term in
	Eq.~\eqref{eq:overlap_IA_plus_FSI} remains almost the same.  This results in
	increasing the difference between the two terms in
	Eq.~\eqref{eq:overlap_IA_plus_FSI} as the SRG $\lambda$ is decreased.  As
	mentioned before, above the quasi-free ridge, the IA and FSI terms
	in Eq.~\eqref{eq:overlap_IA_plus_FSI} add destructively and we therefore end
	up with $|\mbraket{\psi_f}{J_0}{\psi_i^\lambda}| >
	|\mbraket{\psi_f}{J_0}{\psi_i}|$, leading to the observed enhancement at large
	angles upon evolution of the wave function (see Fig.~\ref{fig:100_0p5_fsi}).

	\paragraph{\emph{(b) Evolving the final state}}

	The expression to consider is Eq.~\eqref{eq:evol_final_state_overlap}.  With
	the evolution of the $t$-matrix, the magnitude of the term
	$\mbraket{\phi}{t^\dag_\lambda \, \GreensFn^\dag \, J_0}{\psi_i}$
	decreases, and because of the opposite relative signs of the two terms in
	Eq.~\eqref{eq:evol_final_state_overlap}---and
	because at small angles the magnitude of the IA term is larger than the
	FSI term---the net
	effect is $|\mbraket{\psi_f^\lambda}{J_0}{\psi_i}| >
	|\mbraket{\psi_f}{J_0}{\psi_i}|$.  This leads to an enhancement of $\fL$ with
	evolved final state at small angles, as seen in Fig.~\ref{fig:100_0p5_fsi}.

	At large angles the magnitude of the IA term in
	Eq.~\eqref{eq:evol_final_state_overlap} is smaller than that of the FSI term.
	With the evolution of the $t$-matrix, the magnitude of the FSI term decreases
	and the difference between the IA and the FSI terms decreases as well.  This
	leads to the observed overall suppression in $\fL$ at large angles due to the
	evolution of the final state seen in Fig.~\ref{fig:100_0p5_fsi}.  For those
	few	($\msf$, $\mJd$) combinations for which the above general observations do
	not	hold, the value of individual components is too small to make any
	qualitative	difference.

	Figure~\ref{fig:100_0p5_ia} shows the effect of evolution of individual
	components on $\fL$ calculated in the IA for the kinematics
	under consideration.  Again the evolved deuteron wave function does not have
	strength at high momenta and therefore $\fL$ calculated from
	$\mbraket{\phi}{J_0}{\psi_i ^\lambda}$ has a lower value than its unevolved
	counterpart.

	Unitary evolution means that the effect of the evolved current is always
	such that it compensates the effect due to the evolution of the initial and
	final states.
	\begin{figure}[htbp]
	\centering
	\includegraphics[width=0.5\textwidth]
	{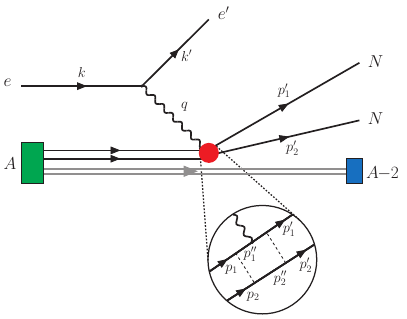}
	\caption{One-body current operator develops two- and higher-body
	components under SRG evolution. }
	\label{fig:current_blob}
	\end{figure}
	As mentioned before, varying $\lambda$ shuffles the physics between long-
	and short-distance parts (cf.~Fig.~\ref{fig:SRG_factorization}).
	As SRG $\lambda$ decreases the blob size in Fig.~\ref{fig:current_blob}
	increases, the high-momentum interaction between the nucleons can no longer
	be resolved, and thus the one-body current operator develops
	two- and higher-body components.
	Our ongoing work examines more directly the behavior of the
	current as it evolves to better understand how to carry over the results
	observed here to other reactions.

	\subsection{Operator evolution and $q$-factorization}
	\label{subsec:operator_evolution}

\begin{figure}[htbp]
	\centering
	\includegraphics[width=0.9\textwidth]{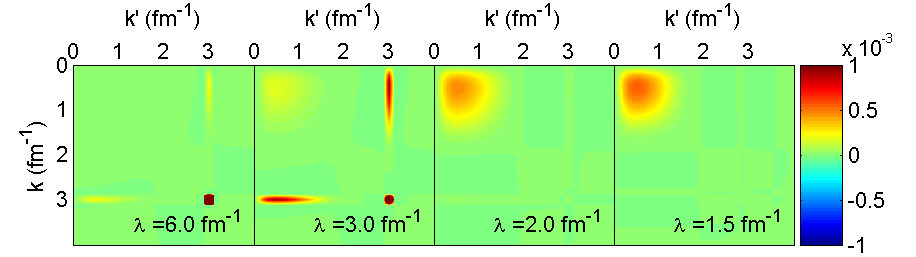}
	\caption{Integrand of $\mbraket{\psi_{\rm deut}^\lambda}
	{(a^\dag_q a_q)^{\lambda}}{\psi_{\rm deut}^\lambda}$ in the $^3S_1$ channel
	for $q = 3.02~{\rm fm^{-1}}$.
	The initial potential is the chiral $\rm{N^3LO}$ (500 MeV)
	potential \cite{Entem:2003ft}.  Figure from \cite{Anderson:2010aq}.}
	\label{fig:Anderson_mom_evolution}
	\end{figure}

	The first work on operator evolution via the SRG evolution was done
	in Ref.~\cite{Anderson:2010aq}.  Among other things the
	authors of Ref.~\cite{Anderson:2010aq} looked at the effect of evolution
	on the momentum distribution in the deuteron
	(cf.~Fig.~\ref{fig:Anderson_mom_evolution}).
	As seen in Fig.~\ref{fig:Anderson_mom_evolution}, the unevolved momentum
	operator which is peaked at large momentum develops strength at
	low momentum on evolution.  The high-momentum one-body part of the operator
	$a^\dag_q a_q$ is unchanged by the evolution, but it is
	suppressed by the evolved wave function, which is why only
	the two-body part survives in the matrix element
	$\mbraket{\psi_{\rm deut}^\lambda}
	{(a^\dag_q a_q)^{\lambda}}{\psi_{\rm deut}^\lambda}$ at lower $\lambda$.

	We follow the approach of \cite{Anderson:2010aq}, but focus instead on
	the current relevant to the deuteron disintegration.  The relevant current
	matrix element is given by Eq.~\eqref{eq:J_0_minus_def_pw}.  The $\delta$
	function in Eq.~\eqref{eq:J_0_minus_def_pw} can be used to do the integral
	analytically, giving a condition over the allowed momenta.
	\begin{align}
	\mbraket{k_1 \, J_1 \, \mJd \, L_1 \, S=1 \, T_1}{J_0^-}
	{k_2\, J=1 \, \mJd \, L_2 \, S=1 \, T=0} = \frac{\pi^2}{2}
	\big(G_E^p + (-1)^{T_1} G_E^n\big) \nonumber \\
	\times \sum_{\mst=-1}^{1} \braket{J_1 \, \mJd}
	{L_1 \, \mJd - \wt{m}_s \, S\!=\!1 \, \wt{m}_s}
	P_{L_1}^{\mJd - \mst} \left(\frac{k_1^2 - k_2^2 + q^2/4}{k_1 q}
	\right) \frac{2}{k_1 k_2 q} \nonumber \\
	\times
	P_{L_2}^{\mJd - \mst} \left(\frac{k_1^2 - k_2^2 - q^2/4}{k_2 q}
	\right)
	\CG{L_2}{\mJd - \mst}{S=1}{\mst}{J=1}{\mJd} \nonumber \\
	\cdots {\rm for~} k_2 \in (|k_1 - q/2|, k_1 + q/2) \nonumber \\
	= 0 {\rm ~otherwise~~~~~~~~~~}
	\label{eq:J0_minus_analytical_delta}
	\end{align}
	In deriving Eq.~\eqref{eq:J0_minus_analytical_delta} from
	Eq.~\eqref{eq:J_0_minus_def_pw} we have used the
	property of the $\delta$ function that
	\beq
	\delta\left(f(x)\right) = \frac{\delta(x - x_0)}{|f^\prime(x_0)|}\;,
	\eeq
	where $x_0$ is the zero of $f(x)$.

	Equation~\eqref{eq:J0_minus_analytical_delta} along with the
	expressions in Appendix~\ref{Appendix:sec:evolution_current}
	can be used to study the effects of evolution on the
	deuteron disintegration current operator.  In what follows,
	we will look at some representative graphs.
	For convenience we set $G_E^p = 1$ and $G_E^n = 0$; it
	should not qualitatively affect the results.

	\begin{figure}[htbp]
		\begin{subfigure}[c]{0.49\textwidth}
			\includegraphics[width=1.265\textwidth]
			{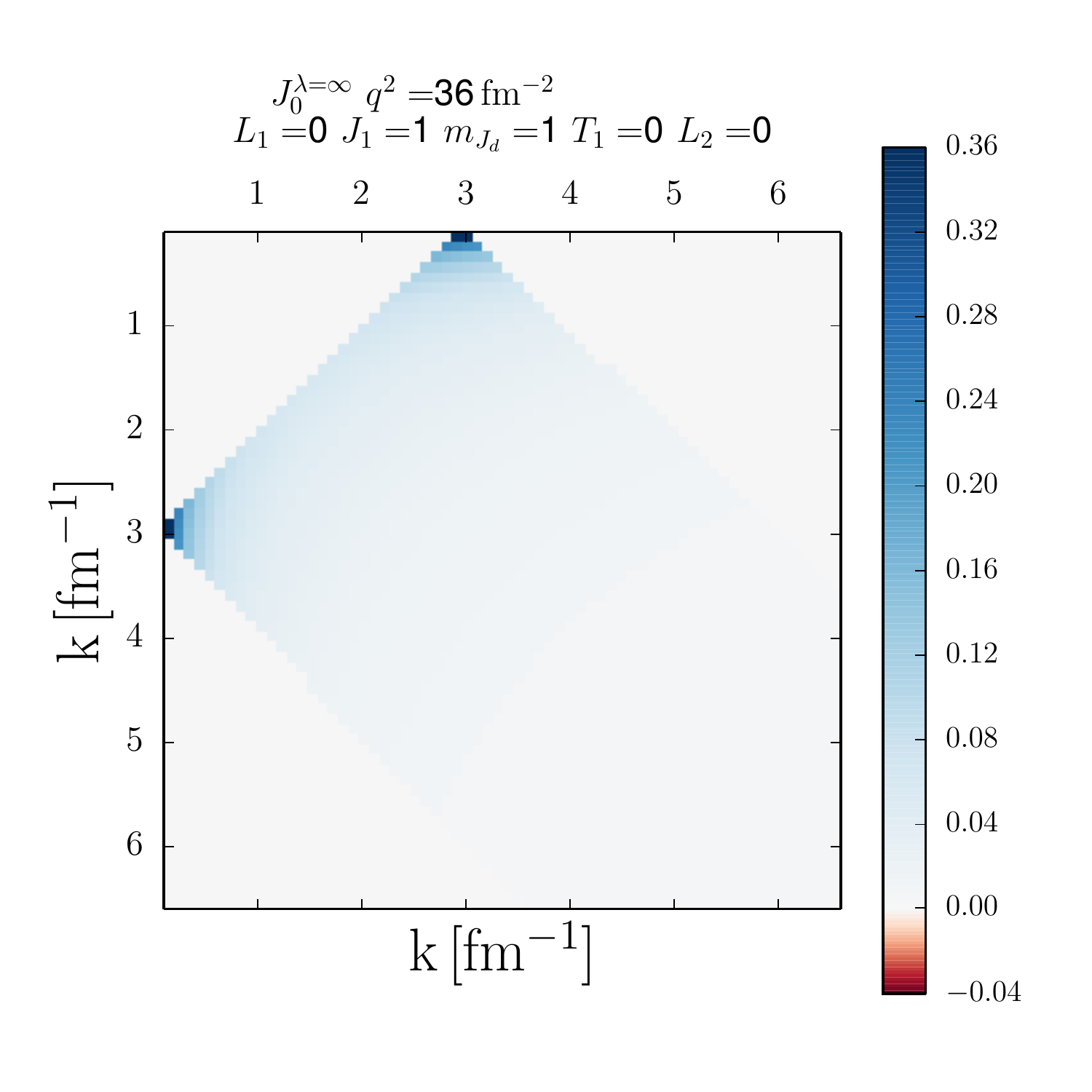}
		\end{subfigure}
		\begin{subfigure}[c]{0.49\textwidth}
			\includegraphics[width=1.265\textwidth]
			{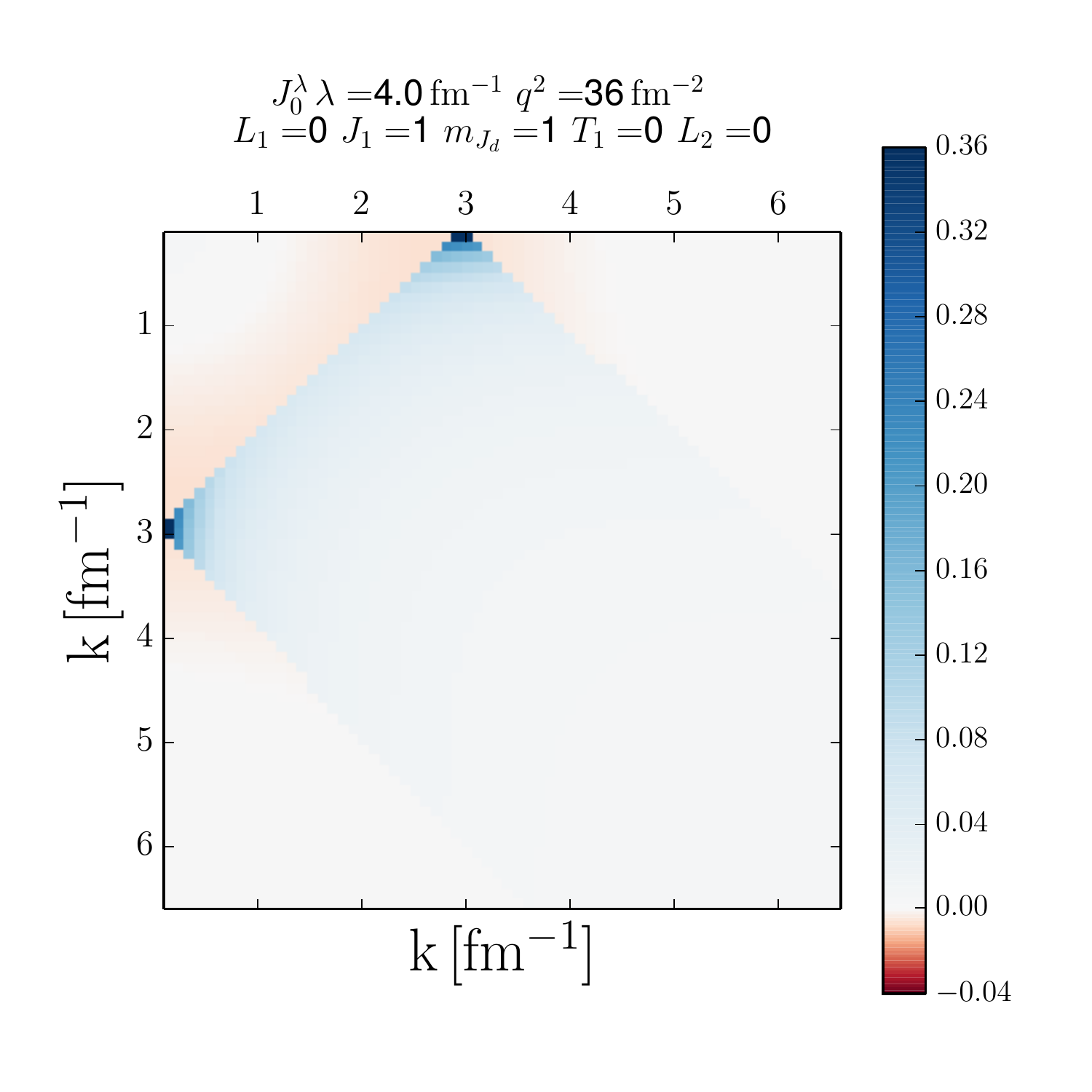}
		\end{subfigure}
		\caption{Contour plot for the matrix element
		$\mbraket{k_1 \, J_1 \, \mJd \, L_1 \, S=1 \, T_1}{J_0^{- \, \lambda}}
		{k_2\, J=1 \, \mJd \, L_2 \, S=1 \, T=0}$ for the quantum numbers
		indicated for $\lambda = \infty$ and $\lambda = 4 {\rm~fm^{-1}}$.  }
		\label{fig:current_evolv_unevol}
	\end{figure}
\begin{figure}[htbp]
		\begin{subfigure}[c]{0.49\textwidth}
			\includegraphics[width=1.265\textwidth]
			{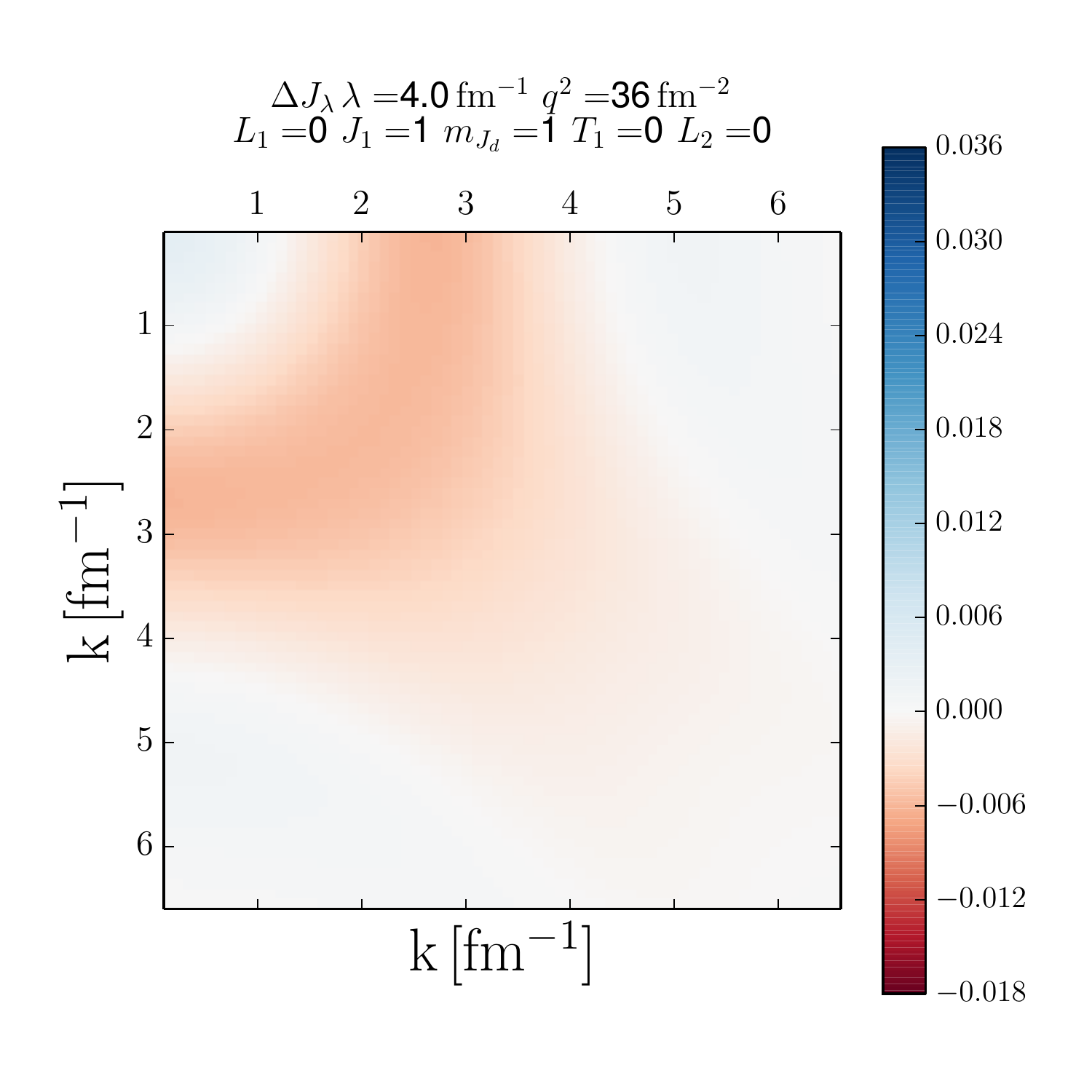}
		\end{subfigure}
		\begin{subfigure}[c]{0.49\textwidth}
			\centering
			\includegraphics[width=1.265\textwidth]
			{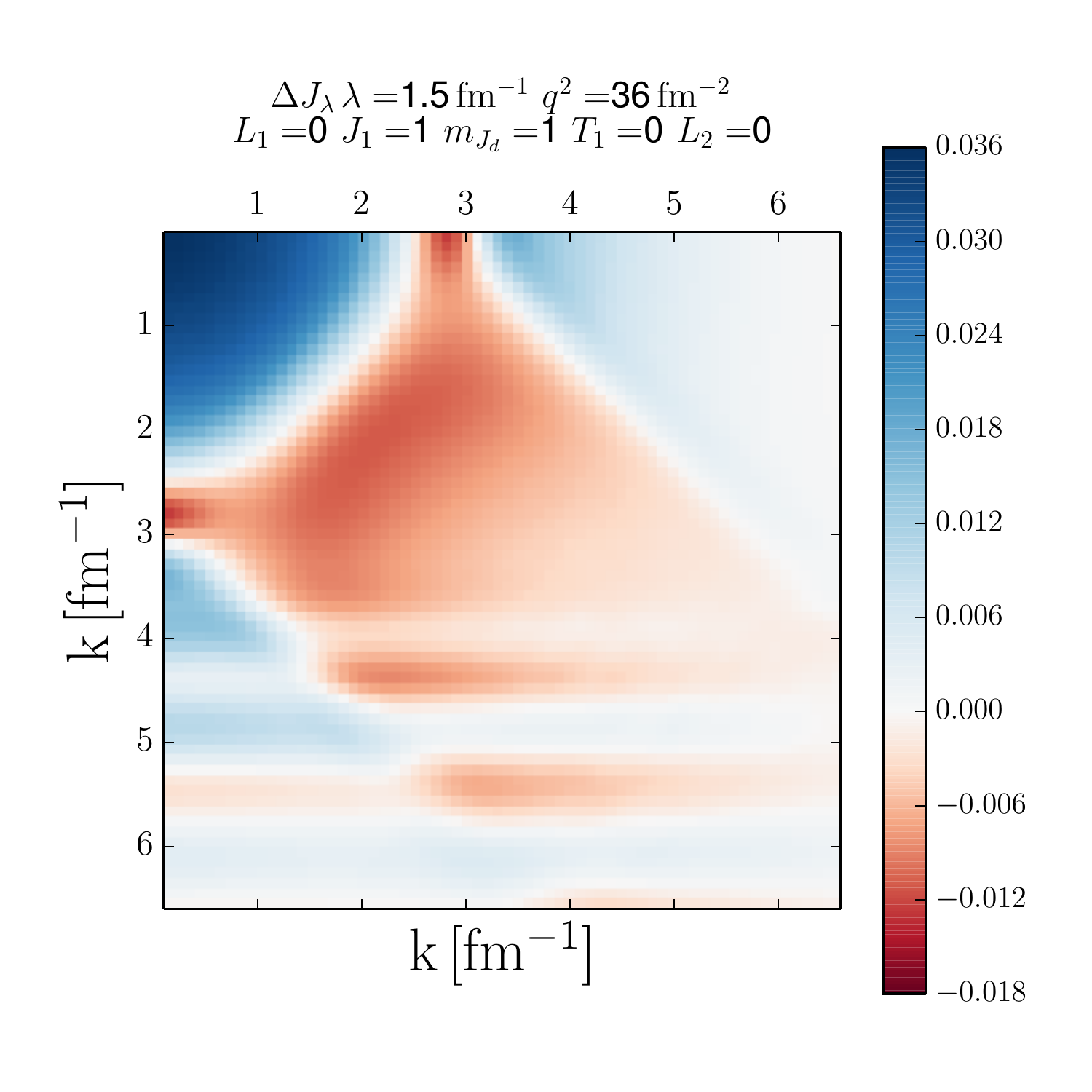}
		\end{subfigure}
		\caption{Contour plot for the matrix element
		$\mbraket{k_1 \, J_1 \, \mJd \, L_1 \, S=1 \, T_1}{\Delta J^{- \, \lambda}}
		{k_2\, J=1 \, \mJd \, L_2 \, S=1 \, T=0}$ for the quantum numbers
		indicated for $\lambda = 4$ and $\lambda = 1.5 {\rm~fm^{-1}}$.
		$\Delta J^{- \, \lambda} \equiv J_0^{- \, \lambda} - J_0$. }
		\label{fig:Delta_J_evolution}
	\end{figure}
\begin{figure}[htbp]
		\begin{subfigure}[c]{0.49\textwidth}
			\includegraphics[width=1.265\textwidth]
			{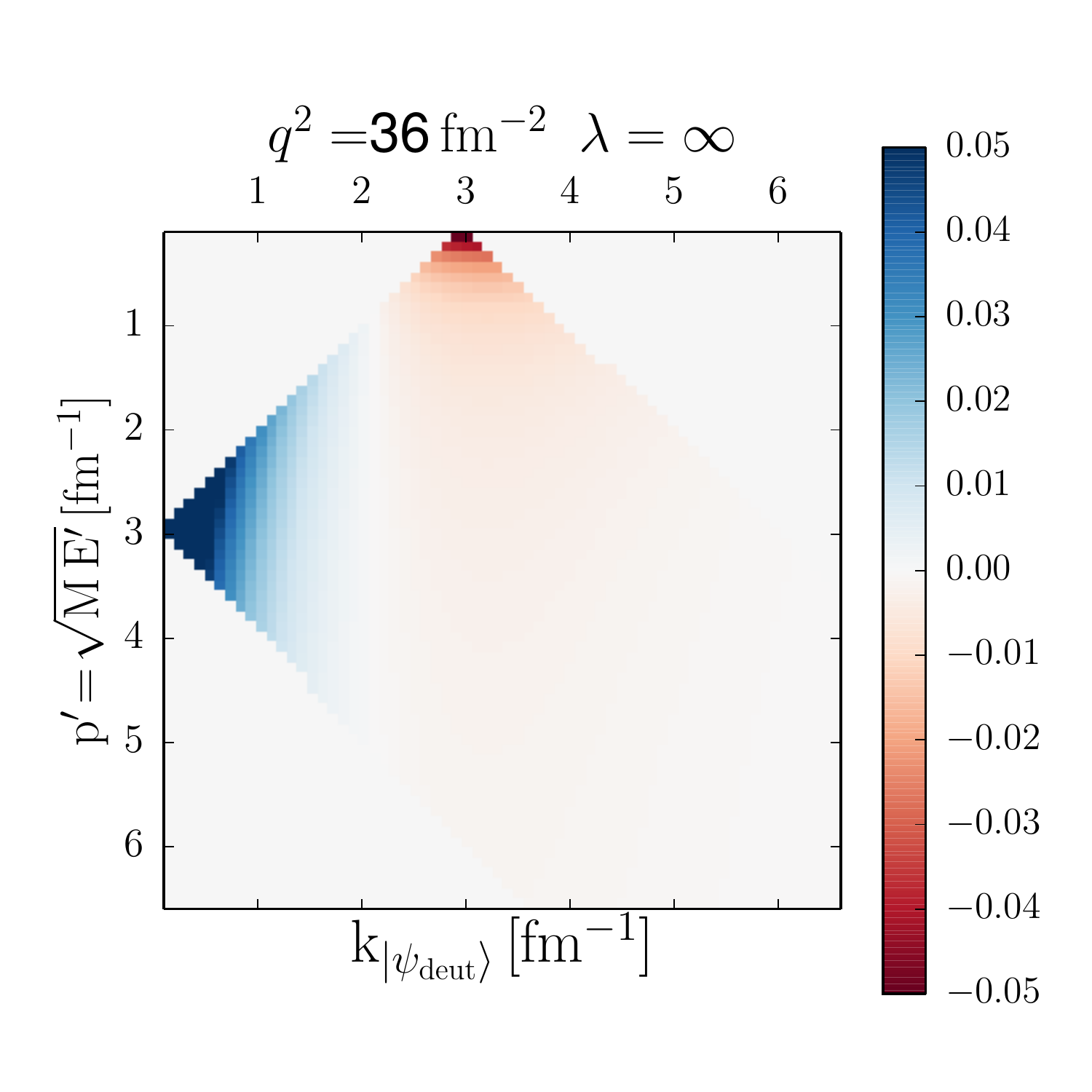}
		\end{subfigure}
		\begin{subfigure}[c]{0.49\textwidth}
			\centering
			\includegraphics[width=1.265\textwidth]
			{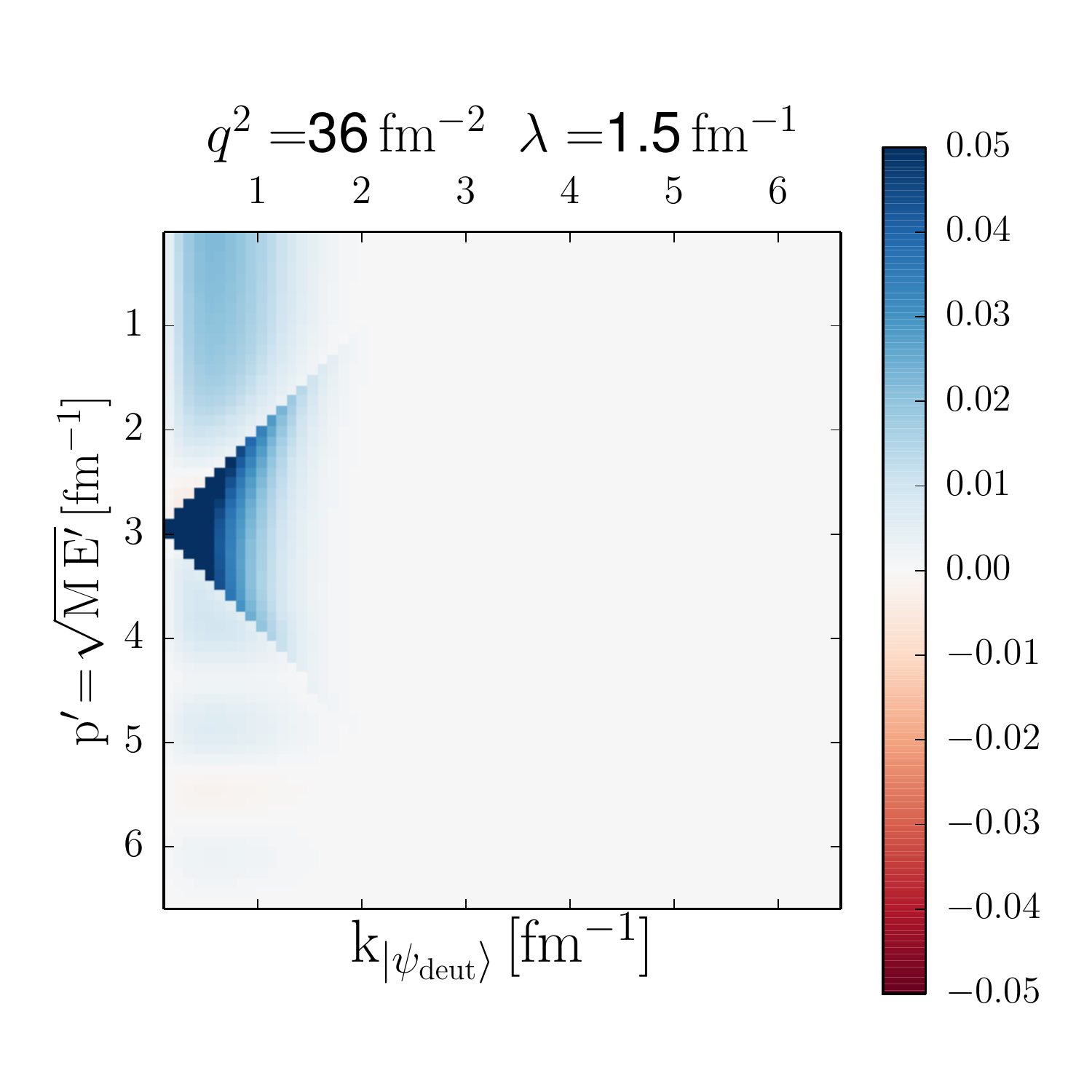}
		\end{subfigure}
		\caption{Contour plot for the integrand of
		$\thickmuskip=0mu\mbraket{p^\prime J_1 = 1 \mJd = 1
		L_1 = 0	S = 1	T_1 = 0}{J_0^{- \, \lambda}}
		{\psi_{\rm deut, \,^3S_1}^{\lambda}(k)}$ for $\lambda = \infty$ and
		$\lambda = 1.5 {\rm~fm^{-1}}$. }
		\label{fig:fold_J0_wf_evolution}
	\end{figure}

	Figure~\ref{fig:current_evolv_unevol} looks at the strength distribution
	of the unevolved and evolved current in a specific channel.  The unevolved
	current is a one-body operator and is peaked at $(0, q/2)$ and $(q/2, 0)$.
	With SRG evolution, the current develops two-body components.
	As seen in Figs.~\ref{fig:current_evolv_unevol} and
	\ref{fig:Delta_J_evolution}, the changes due to evolution are
	rather distributed.
	The evolved current doesn't become pathologically large
	at high momentum.  This is important because for practical calculations
	the evolved current will be used in conjunction with the evolved
	wave function.  The evolved wave functions have negligible strength
	at high momentum and the absence of pathologies in the evolved current
	make sure that the calculations with SRG in the reduced basis are possible.
	To illustrate this we turn to Fig.~\ref{fig:fold_J0_wf_evolution}.

	Figure~\ref{fig:fold_J0_wf_evolution} shows the contour plot for the
	integrand of $\mbraket{\pp; \, ^3S_1}{J_0^{\lambda}(q)}
	{\psi_{\rm deut, \,^3S_1}^{\lambda}}$.  The ket is the
	deuteron state and the bra state corresponds to the outgoing nucleons
	(up to factors of Spherical Harmonics and Clebsch-Gordan coefficients)
	without the final state interactions.
	We find that as we evolve to lower SRG $\lambda$, the strength in the
	integrand shifts to lower momenta in deuteron.  It is also possible to
	qualitatively
	explain the results in Subsec.~\ref{subsec:Results} on the basis of
	Fig.~\ref{fig:fold_J0_wf_evolution}.  For a given $\pp = p_0$, the
	value for $\mbraket{p_0; \, ^3S_1}{J_0^{\lambda}(q)}
	{\psi_{\rm deut, \,^3S_1}^{\lambda}}$ is obtained by adding all the
	points along the horizontal axis $\pp = p_0$ in
	Fig.~\ref{fig:fold_J0_wf_evolution}.

	The kinematics at the quasi-free ridge corresponds to
	$p_0 = q/2$ (can be derived from Eq.~\eqref{eq:quasi_free_condition}).
	We see from Fig.~\ref{fig:fold_J0_wf_evolution} that for
	$p_0 = q/2$, the contribution to the matrix element
	$\mbraket{p_0; \, ^3S_1}{J_0(q)}
	{\psi_{\rm deut, \,^3S_1}}$ comes from the low momentum components in the
	deuteron.  These low-momentum components are unchanged under SRG evolution
	and therefore we see hardly any scale dependence at the quasi-free ridge
	in the results presented in Subsec.~\ref{subsec:Results}.
	Next consider $p_0 = 1 {\rm~fm^{-1}}$.  For this case, the contribution
	to $\mbraket{p_0; \, ^3S_1}{J_0(q)}
	{\psi_{\rm deut, \,^3S_1}}$ comes from the high-momentum components in the
	deuteron, which change substantially under evolution.  Moreover, we see
	from Fig.~\ref{fig:fold_J0_wf_evolution} that the changes due to evolution
	are smooth low-momentum effects.  This indicates that the changes due to
	evolution are of the form of contact terms as expected from an EFT approach
	(cf.~Fig.~\ref{fig:replace_loop_contact}).

	\medskip
	\subsubsection{$q$-factorization of $f_L$}

	It was observed and explained in Refs.~\cite{Bogner:2007jb, Anderson:2010aq}
	that for $k < \lambda$ and $q \gg \lambda$ the unitary evolution operator
	factorizes: $U_{\lambda}(k, q) \rightarrow K_{\lambda}(k) Q_{\lambda}(q)$.
	\begin{figure}[htbp]
		\centering
		\includegraphics[width=0.55 \textwidth]
		{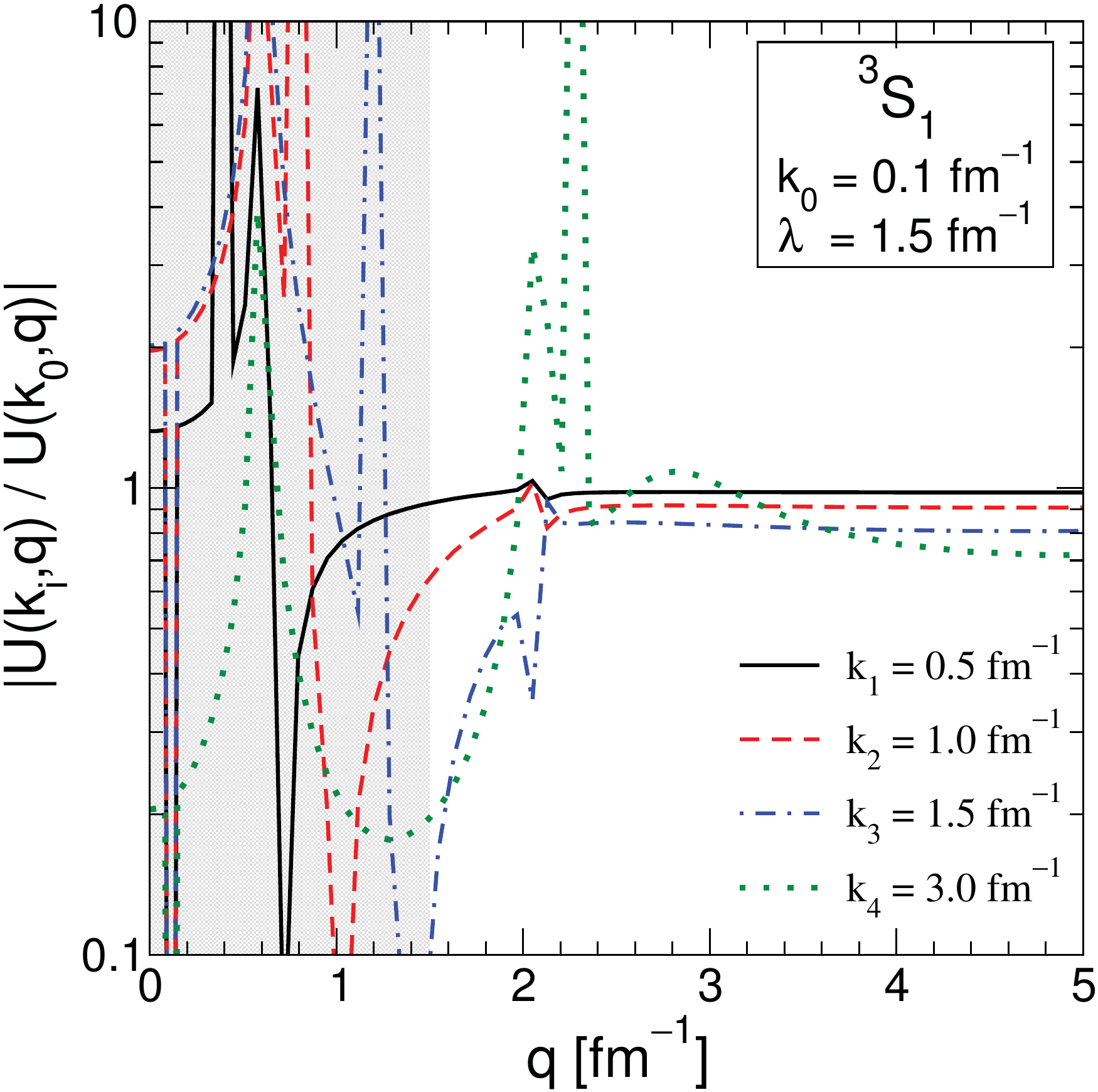}
		\caption{Factorization of $U$ matrices demonstrated by plateaus at high $q$.
		Figure from \cite{Anderson:2010aq}. }
		\label{fig:U_matrices_factorization}
	\end{figure}
	This factorization was observed by looking at the ratio of $U(k_i, q)/
	U(k_0, q)$ for small $k_i's$ and $k_0$, and noting that the ratio
	plateaus at high $q$ (cf.~Fig.~\ref{fig:U_matrices_factorization}).

	\begin{figure}[htbp]
		\centering
		\includegraphics[width=0.55 \textwidth]
		{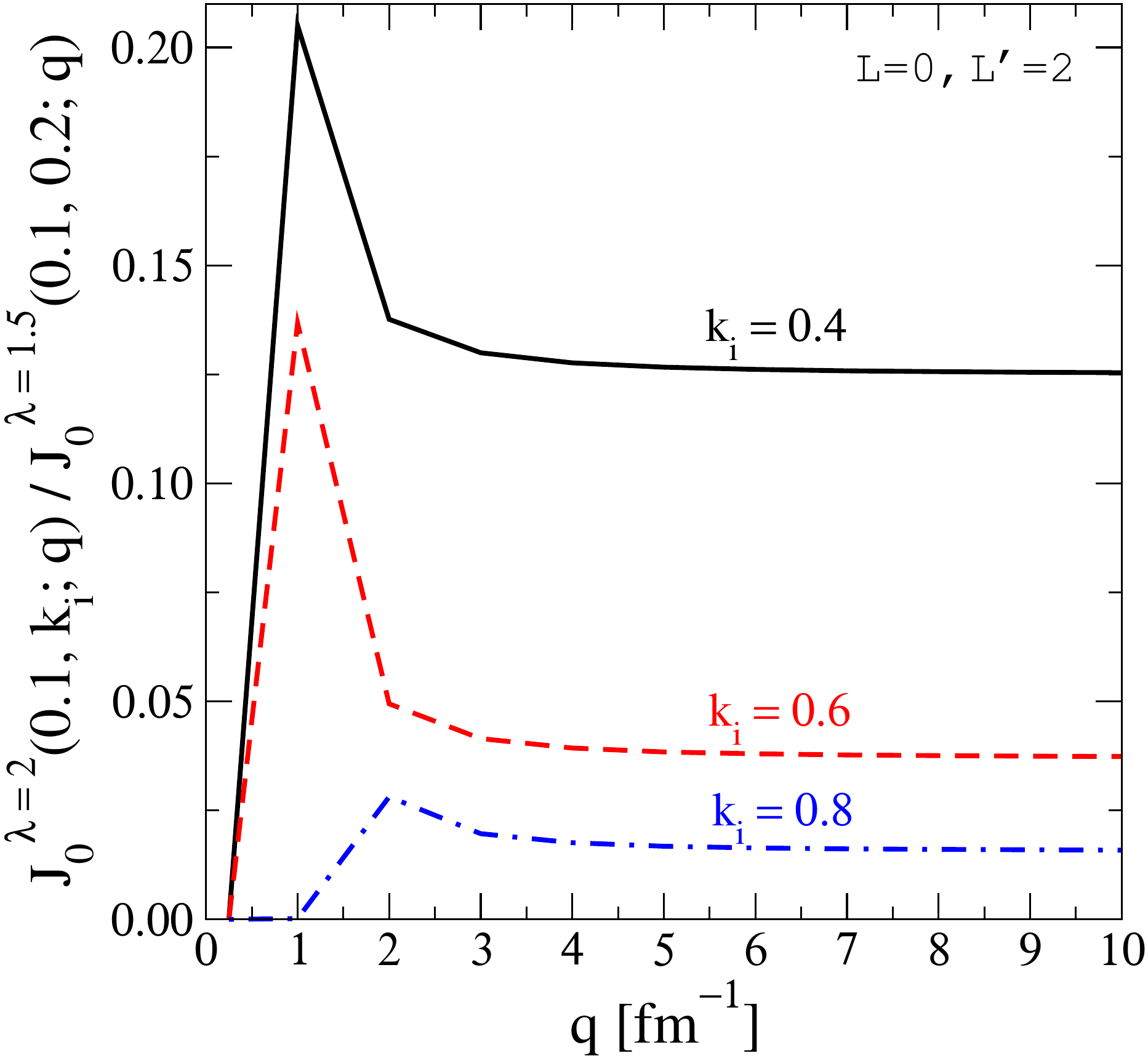}
		\caption{Factorization of the evolved current $J_0^\lambda$ demonstrated
		by plateaus at high $q$.}
		\label{fig:evolv_J0_factorization}
	\end{figure}
	Similar behavior is observed for the evolved current in
	Fig.~\ref{fig:evolv_J0_factorization}.  Preliminary analysis indicates that
	the scaling
	\beq
	J_0^{\lambda}(k, k^\prime; q) \rightarrow Z_{\lambda}(k, k^\prime) \, A(q)
	\eeq
	is a result of both the form of the unevolved current and
	the momentum factorization of $U$ matrices.

\begin{figure}[htbp]
		\centering
		\begin{subfigure}[b]{0.57\textwidth}
			\includegraphics[width=0.95\textwidth]
			{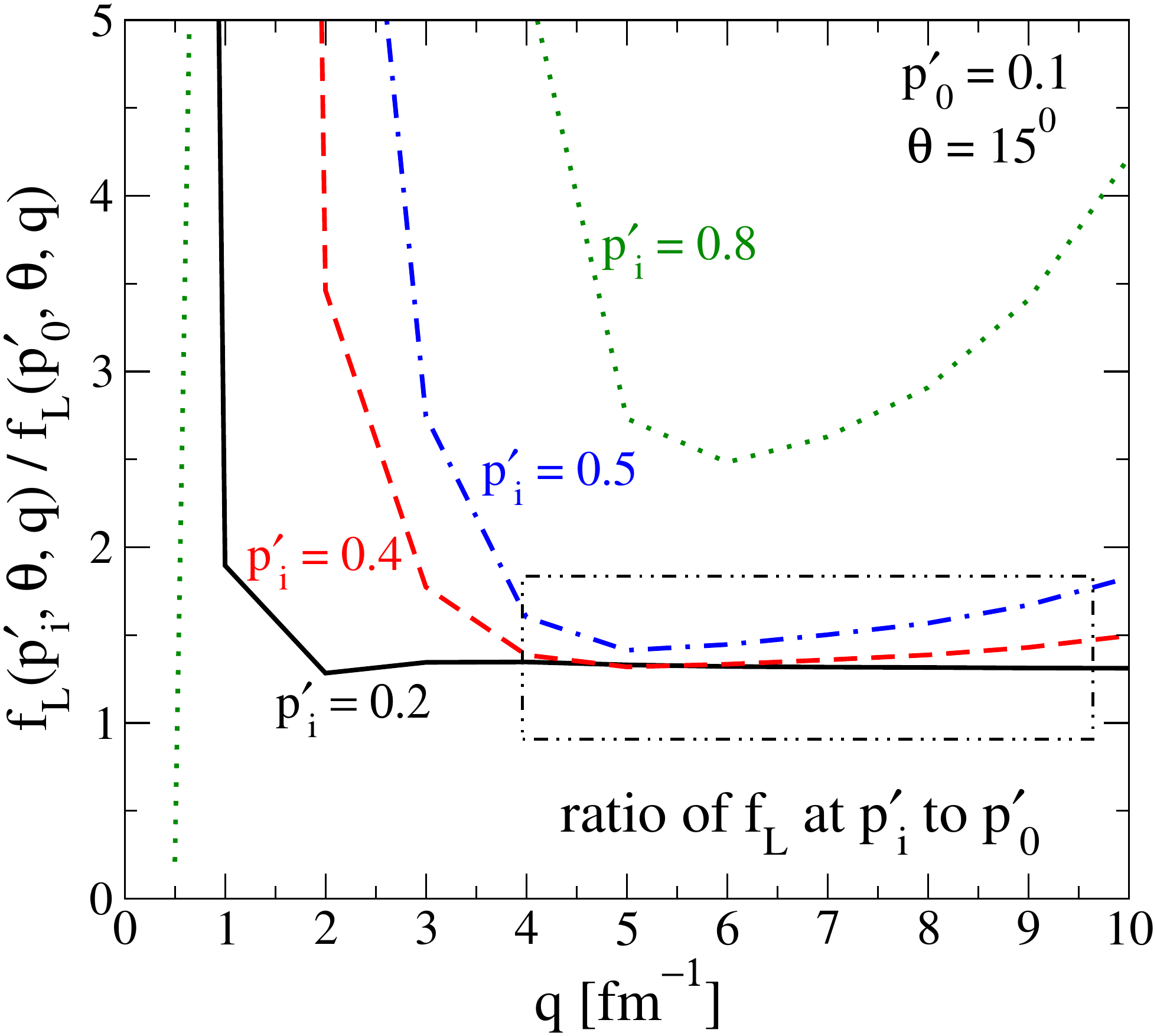}
			\caption{Demonstration that for $\pp \ll q$, the $q$ dependence of $\fL$
			factorizes.}
			\label{fig:q_factorization_fl}
		\end{subfigure}~~
		\begin{subfigure}[b]{0.37\textwidth}
			\centering
			\begin{overpic}[width= \textwidth]{Factorization/More-1p5}
			\put(68.45,11.2){\tikz\draw[red, thick] (0,0) rectangle (1.7,0.22);}
			\put(77,2.6){\textcolor{red}{\scriptsize \emph{plateau region}}}
			\end{overpic}
			\caption{The region of $q$-factorization of $\fL$ in the phase space
			covers the red box and extends further out along the $q^2$ axis.}
			\label{fig:q_factorization_fl_phase_space}
		\end{subfigure}
		\caption{$q$-factorization of $\fL$. }
		\label{fig:q_factorization_fl_plus_phase_space}
	\end{figure}

	Note that $\fL \sim \sum_{m_s, m_J} |\mbraket{\psi_f^\lambda}{J_0^\lambda}
	{\psi_i^\lambda}|^2$.  The ``q-factorization'' of $J_0^\lambda$ and $U$
	matrices indicates that the observable $f_L$ should scale with $q$ as well.
	We see in Fig.~\ref{fig:q_factorization_fl} that this indeed is the case.
	$\fL$ is a function of the outgoing nucleon momentum ($\pp$), the proton
	emission angle ($\theta^\prime$), and the momentum transferred by the
	photon ($q$).  For $\pp \ll q$, Fig.~\ref{fig:q_factorization_fl} tells us
	that
	\beq
	\fL(\pp, \theta^\prime; q) \rightarrow g(\pp, \theta) \, B(q) \;.
	\label{eq:fl_factorization}
	\eeq
	Figure~\ref{fig:q_factorization_fl_phase_space} shows the region in the
	phase space where Eq.~\eqref{eq:fl_factorization} holds.

	\begin{figure}[htbp]
		\centering
		\includegraphics[width=0.55 \textwidth]
		{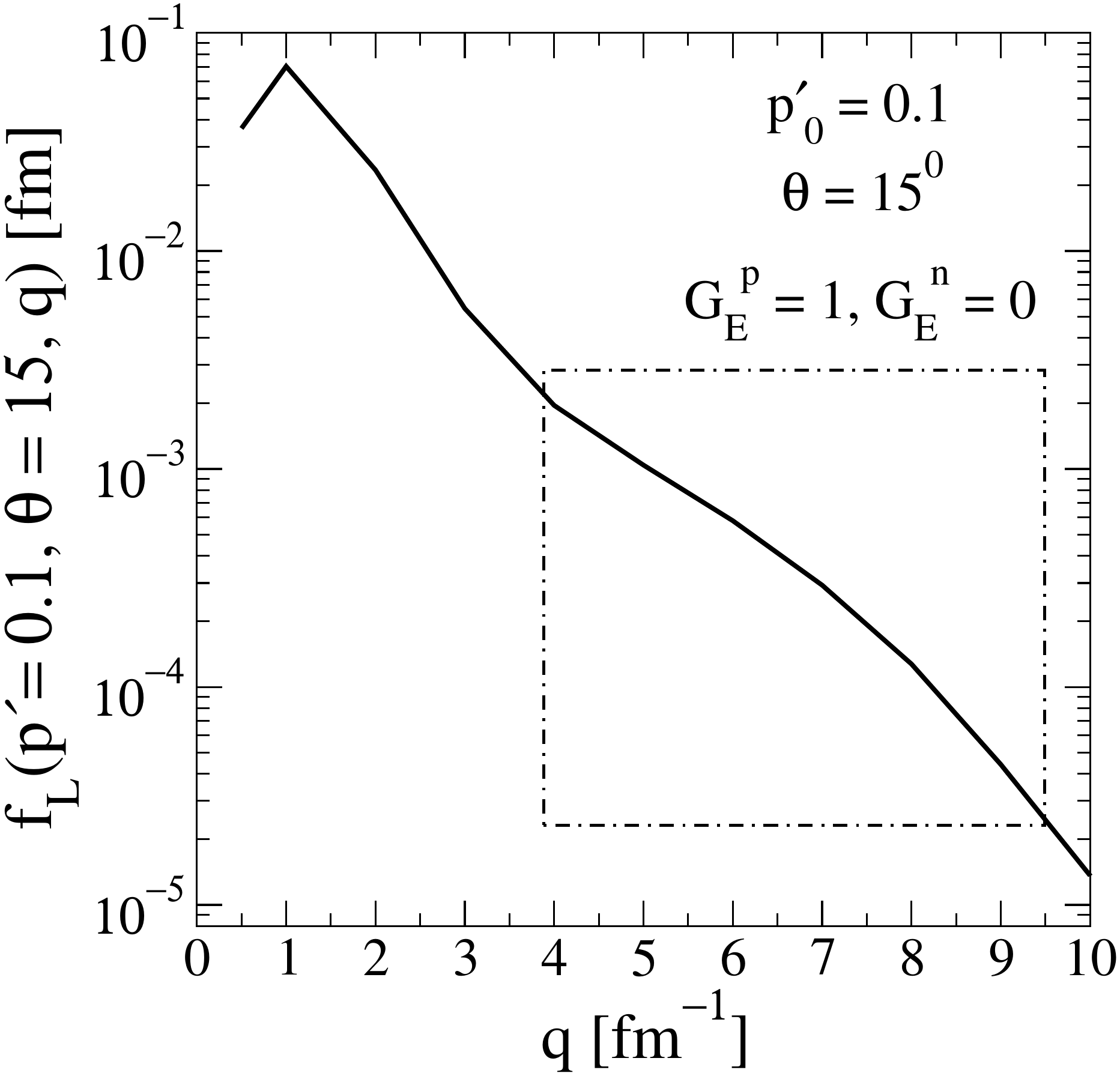}
		\caption{$\fL$ is a strong function of $q$.  The box indicates the plateau
		region in Fig.~\ref{fig:q_factorization_fl}.}
		\label{fig:fl_strong_q_fn}
	\end{figure}
	Figure~\ref{fig:fl_strong_q_fn} indicates that $\fL$ by itself is a strong
	function of $q$.  In the plateau region of Fig.~\ref{fig:q_factorization_fl},
	the denominator of the ratio varies by over two orders of magnitude.
	Given this and the complicated nature of $\fL$ calculations (results here
	include the FSIs), the scaling observed in Fig.~\ref{fig:q_factorization_fl}
	is non-trivial.

	It seems that the observed $q$-factorization of $\fL$ can be explained from
	the SRG perspective by invoking the factorization of $U$ matrices.
	It remains to be seen how the $q$-factorization could be explained starting
	from the unevolved matrix element.  This offers an interesting scenario,
	where the observation in Eq.~\eqref{eq:fl_factorization} can be explained
	by two different interpretations.

	\section{Summary and Outlook}
	\label{sec:factorization_summary}

	Nuclear properties such as momentum distributions are extracted from
	experiment by invoking the factorization of structure, which includes
	descriptions of	initial and final states, and reaction, which includes the
	description of the probe components.
	The factorization between reaction and structure depends on the scale and
	scheme chosen for doing calculations.  Unlike in high-energy QCD, this scale
	and scheme dependence of factorization is often not taken into account in
	low-energy nuclear physics calculations, but is potentially critical for
	interpreting experiment.
	In our work we investigated this issue by looking at the simplest knockout
	reaction: deuteron electrodisintegration.  We used SRG transformations to test
	the sensitivity of the longitudinal structure function~$\fL$ to evolution of
	its	individual components: initial state, final state, and the current.

	We find that the effects of evolution depend on kinematics,
	but in a \emph{systematic} way.  Evolution effects are negligible at the
	quasi-free ridge,	indicating that the
	scale dependence of individual components is minimal there.  This is
	consistent with the quasi-free ridge mainly probing the long-range part of
	the	wave function, which is largely invariant under SRG evolution.  This is
	also the region where contributions from FSI to $\fL$ are minimal.
	The effects get progressively more pronounced the further one moves away from
	the quasifree ridge.  The nature of these changes depends on whether one
	is above or below the quasifree ridge in the ‘“phase-space’” plot
	(Fig.~\ref{fig:More-1p5}).
	As indicated in Subsec.~\ref{subsec:Results}, these changes can also be
	explained qualitatively
	by looking at the overlap matrix elements.  This allows us to predict the
	effects due to evolution depending on kinematics.

	Our results demonstrate that scale dependence needs to be taken into account
	for low-energy nuclear calculations.  While we showed this explicitly only
	for	the case of the longitudinal structure function in deuteron
	disintegration,	we expect the results should qualitatively
	carry over for other knock-out reactions as well.
	An area of active investigation is the extension of the formalism presented
	here to hard scattering processes.

	SRG transformations are routinely used in nuclear structure calculations
	because	they lead to accelerated convergence for observables like binding
	energies.  We
	demonstrated that SRG transformations can be used for nuclear knock-out
	reactions as well as long as the operator involved is also consistently
	evolved. 	Naively, one would expect the evolved operator to be
	more complicated than the unevolved one.  However, as we saw in
	Subsec.~\ref{subsec:operator_evolution}, the SRG evolution makes
	interpretation of the high $q$-factorization of observables easy.
	It sets the stage for exploiting the operator product expansion as in
	Refs.~\cite{Anderson:2010aq,Bogner:2012zm}.  Moreover, we saw that the
	changes due to evolution are regulated contact terms as expected from
	an EFT perspective.

	We plan to use pionless EFT as a framework to quantitatively study the
	effects of operator evolution.  It should be a good starting point to
	understand in detail how a one-body operator develops strength in two- and
	higher-body sectors upon evolution.  This can give insight on the issue of
	power	counting of operator evolution.  Pionless EFT has been employed
	previously to
	study deuteron electrodisintegration in Ref.~\cite{Christlmeier:2008ye}, where
	it was used to resolve a discrepancy between theory and experiment.

	Extending our work to many-body nuclei requires inclusion of 3N forces and 3N
	currents.  Consistent evolution in that case would entail evolution in both
	two	and three-body sectors.  However, SRG transformations have proven to be
	technically feasible for evolving three-body
	forces~\cite{Jurgenson:2009qs,Jurgenson:2010wy,Hebeler:2012pr,Wendt:2013bla}.
	Thus, extending our calculations to many-body nuclei would be computationally
	intensive, but is feasible in the existing framework.  Including the effects
	of FSI is challenging for many-body systems and has been
	possible only recently for light nuclei \cite{Bacca:2014tla, Lovato:2015qka}.
	It would be interesting to investigate if the scale and scheme dependence of
	factorization allows us to choose a scale where the FSI effects are minimal.

\cleardoublepage
\chapter{Epilogue}

	This thesis presented the author's original work over the past four years.
	Most of the work has already been published.  The thesis offers more
	motivation for our work, added details about calculations, and
	presents new insights along with recent developments.

	The work in Chapter~\ref{chap:Extrapolation} started as author's warm-up
	problem during the summer following his first year.
	We kept finding
	interesting results, and the warm-up problem turned into a full-fledged
	project resulting in three publications
	\cite{More:2013rma,Furnstahl:2013vda,Konig:2014hma}.
	The key development was the mapping between the basis
	truncation and the hard-wall boundary condition.
	This mapping led to development of extrapolation schemes for energy, radii,
	and allowed extraction of phase shifts.
	Our work focused on two-body systems, though in principle, the
	two-body results usually don't need extrapolation.
	However, availability of exact answers allowed us to test our results.
	Our results pioneered
	the development of physically motivated extrapolation schemes in LENP.  
	The work in Refs.~\cite{Furnstahl:2014hca,Wendt:2015nba,Binder:2015trg}
	showed a way to extend our work to many-body nuclei.

	The uncertainty in the scale and scheme dependence of nuclear structure
	and reactions components made it difficult to make robust predictions
	for experiments (cf.~Fig.~\ref{fig:0nu_double_beta_matrix_elements}).
	To tackle this issue, we followed the same approach as in
	Chapter~\ref{chap:Extrapolation}, i.e., we started with a two-body
	system which is more tractable.  We found that the scale dependence
	depends strongly on kinematics, but in a systematic way and therefore
	can be understood.

	In Sec.~\ref{sec:factorization_summary}, we listed some of the direct
	extensions of our work.  Here we will discuss some of the broader topics
	relevant to our analysis in Chapter~\ref{chap:factorization}.  To
	begin with we looked at the deuteron disintegration reaction.  Its
	time-reversed version $n + p \rightarrow d + \gamma$ is appealing as
	well due to its relevance to the big bang nucleosynthesis.  It would be
	instructive to see how our analysis carries over to this reaction.

	One of the outstanding mysteries of nuclear physics is the \emph{EMC effect},
	named after the European Muon Collaboration that discovered it $32$
	years ago \cite{Aubert:1983xm}.  They observed that the probability of
	deep inelastic scattering (DIS) off a quark is significantly different from
	the same probability in a free nucleon.
	\begin{figure}[htbp]
	 \centering
	 \includegraphics[width=0.7\textwidth]%
	 {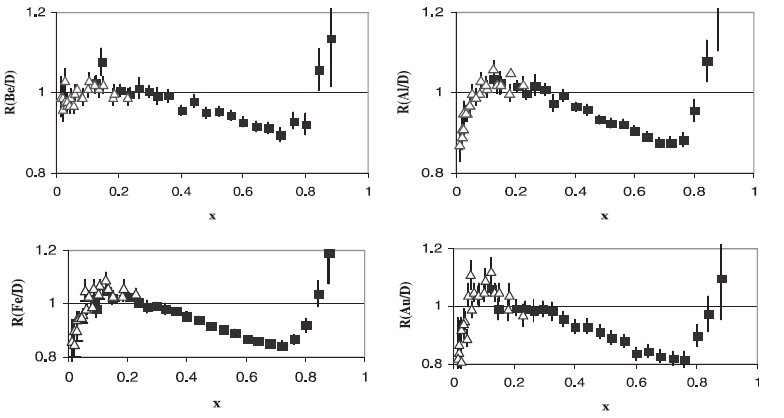}
	 \caption{The EMC effect in different nuclei \cite{Norton:2003cb}.
	 $x$ is the Bjorken-$x$. }
	 \label{fig:EMC_summary}
	\end{figure}
	Figure~\ref{fig:EMC_summary} shows the EMC effect for various nuclei.
	Given that nuclei are weakly bound (maximum of $8.8 {\rm~MeV}$ per nucleon)
	compared to the energy transfer in DIS (order of GeV), the deviation of
	ratio in Fig.~\ref{fig:EMC_summary} by up to 20\% from unity was
	unexpected.  A complete understanding of this curious EMC effect still
	remains elusive.

\begin{figure}[htbp]
	 \centering
	 \includegraphics[width=0.5\textwidth]%
	 {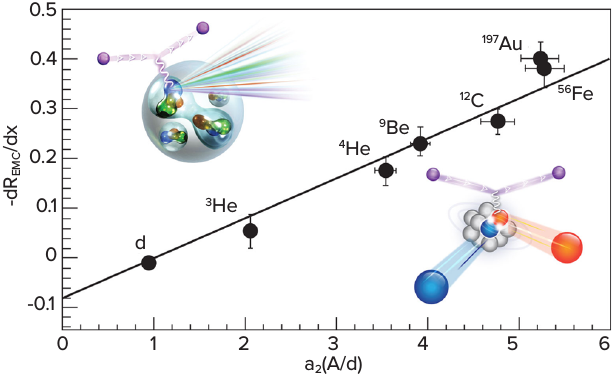}
	 \caption{The relationship between the number of two-nucleon correlated
	 pairs $a_2(A/d)$, and the strength of the EMC effect.  The later is
	 characterized by the slope of EMC effect in $0.3 < x < 0.7$.
	 Figure from \cite{Long_range_plan}.}
	 \label{fig:EMC_SRC_correlation}
	\end{figure}

	Experiments at Jefferson Lab indicate that there is a correlation
	between the two-nucleon short-range correlations and the EMC effect
	(cf.~Fig.~\ref{fig:EMC_SRC_correlation}).
	However, as we have already seen (cf.~Figs.~\ref{fig:SRG_evolution_wf}
	and \ref{fig:wavefunction_evolution_deuteron_D_state}), SRG evolved
	wave functions do not have the SRCs.  Instead the SRC physics is accounted
	for by the evolution of the operator.  Our results in
	Subsec.~\ref{subsec:operator_evolution} suggest that
	the relationship in Fig.~\ref{fig:EMC_SRC_correlation} might be explained
	by both the quantities being dominantly given by the leading
	two-body (contact) operator.  Studies analogous to the one
	presented in Chapter~\ref{chap:factorization} will help elucidate the
	model dependence of SRCs, and will be valuable for understanding the
	EMC effect.

	An on-going debate in the LENP community is the nature and interpretation
	of the spectroscopic factors \cite{Furnstahl:2010wd}.  Spectroscopic factors
	involve overlap of nuclear wave functions.  The short-range parts of the
	wave functions are scale and scheme dependent
	(cf.~Fig.~\ref{fig:SRG_evolution_wf}) and therefore so are the spectroscopic
	factors.
	Nonetheless, in experimental analysis they are often treated as observables
	with no scale/scheme dependence.  This is often because the dependence is
	unclear.  An exercise similar to our analysis in
	Chapter~\ref{chap:factorization} will shed a light on the scale/scheme
	dependence of the spectroscopic factors.  This in essence, will
	bridge the gap between theory and experiments resolving the issues
	associated with the extraction of nuclear properties from the experiments.

 \bibliographystyle{unsrt} 

\begin{thebibliography}{100}

\bibitem{Long_range_plan}
Long range plan for nuclear science.
\newblock {\url{http://science.energy.gov/np/nsac/}}, {September 2015}.

\bibitem{Aubert:1974js}
J.~J. Aubert et~al.
\newblock {Experimental Observation of a Heavy Particle J}.
\newblock {\em Phys. Rev. Lett.}, 33:1404--1406, 1974.

\bibitem{Gross:2005kv}
D.~J. Gross.
\newblock {The discovery of asymptotic freedom and the emergence of QCD}.
\newblock {\em Proc. Nat. Acad. Sci.}, 102:9099--9108, 2005.
\newblock [Rev. Mod. Phys.77,837(2005)].

\bibitem{Myhrer:2009uq}
F.~Myhrer and A.~W. Thomas.
\newblock {Understanding the proton's spin structure}.
\newblock {\em J. Phys.}, G37:023101, 2010.

\bibitem{Martinez:2013xka}
Gines Martinez.
\newblock {Advances in Quark Gluon Plasma}.
\newblock 2013.

\bibitem{Cyburt:2015mya}
Richard~H. Cyburt, Brian~D. Fields, Keith~A. Olive, and Tsung-Han Yeh.
\newblock {Big Bang Nucleosynthesis: 2015}.
\newblock {\em Rev. Mod. Phys.}, 88:015004, 2016.

\bibitem{PAS:9305903}
Amanda~I. Karakas and John~C. Lattanzio.
\newblock The dawes review 2: Nucleosynthesis and stellar yields of low- and
  intermediate-mass single stars.
\newblock {\em PASA - Publications of the Astronomical Society of Australia},
  31:e030 (62 pages), 2014.

\bibitem{Lattimer:2015eaa}
James~M. Lattimer.
\newblock {Introduction to neutron stars}.
\newblock {\em AIP Conf. Proc.}, 1645:61--78, 2015.

\bibitem{Feng:2010gw}
Jonathan~L. Feng.
\newblock {Dark Matter Candidates from Particle Physics and Methods of
  Detection}.
\newblock {\em Ann. Rev. Astron. Astrophys.}, 48:495--545, 2010.

\bibitem{Avignone:2007fu}
Frank~T. Avignone, III, Steven~R. Elliott, and Jonathan Engel.
\newblock {Double Beta Decay, Majorana Neutrinos, and Neutrino Mass}.
\newblock {\em Rev. Mod. Phys.}, 80:481--516, 2008.

\bibitem{Yukawa:1935xg}
Hideki Yukawa.
\newblock {On the Interaction of Elementary Particles I}.
\newblock {\em Proc. Phys. Math. Soc. Jap.}, 17:48--57, 1935.
\newblock [Prog. Theor. Phys. Suppl.1,1(1935)].

\bibitem{Wiringa:1994wb}
Robert~B. Wiringa, V.~G.~J. Stoks, and R.~Schiavilla.
\newblock {An Accurate nucleon-nucleon potential with charge independence
  breaking}.
\newblock {\em Phys. Rev.}, C51:38--51, 1995.

\bibitem{Stoks:1994wp}
V.~G.~J. Stoks, R.~A.~M. Klomp, C.~P.~F. Terheggen, and J.~J. de~Swart.
\newblock {Construction of high quality N N potential models}.
\newblock {\em Phys. Rev.}, C49:2950--2962, 1994.

\bibitem{Machleidt:1989tm}
R.~Machleidt.
\newblock {The Meson theory of nuclear forces and nuclear structure}.
\newblock {\em Adv. Nucl. Phys.}, 19:189--376, 1989.

\bibitem{Aoki:2008hh}
S.~Aoki, T.~Hatsuda, and N.~Ishii.
\newblock {Nuclear Force from Monte Carlo Simulations of Lattice Quantum
  Chromodynamics}.
\newblock {\em Comput. Sci. Dis.}, 1:015009, 2008.

\bibitem{Agashe:2014kda}
K.~A. Olive et~al.
\newblock {Review of Particle Physics}.
\newblock {\em Chin. Phys.}, C38:090001, 2014.

\bibitem{Anderson:1972pca}
P.~W. Anderson.
\newblock {More Is Different}.
\newblock {\em Science}, 177(4047):393--396, 1972.

\bibitem{Machleidt:1987hj}
R.~Machleidt, K.~Holinde, and C.~Elster.
\newblock {The Bonn Meson Exchange Model for the Nucleon Nucleon Interaction}.
\newblock {\em Phys. Rept.}, 149:1--89, 1987.

\bibitem{Machleidt:2000ge}
R.~Machleidt.
\newblock {The high-precision, charge-dependent Bonn nucleon-nucleon potential
  (CD-Bonn)}.
\newblock {\em Phys. Rev. C}, 63:024001, 2001.

\bibitem{LRP:2007}
{\em DOE/NSF Nuclear Science Advisory Committe, The Frontiers of Nuclear
  Science: A Long-Range Plan}, 2007.

\bibitem{Guth:1984rq}
Alan~H. Guth, K.~Huang, and R.~L. Jaffe, editors.
\newblock {\em {ASYMPTOTIC REALMS OF PHYSICS. ESSAYS IN HONOR OF FRANCIS E.
  LOW. PROCEEDINGS, SYMPOSIUM, CAMBRIDGE, USA, OCTOBER 16, 1981}}, 1984.

\bibitem{Weinberg:1978kz}
Steven Weinberg.
\newblock {Phenomenological Lagrangians}.
\newblock {\em Physica}, A96:327--340, 1979.

\bibitem{Peskin1995a}
Michael~E Peskin and Daniel~V Schroeder.
\newblock {\em {An Introduction to Quantum Field Theory}}.
\newblock Westview Press, 1995.

\bibitem{Machleidt:2011zz}
R.~Machleidt and D.R. Entem.
\newblock {Chiral effective field theory and nuclear forces}.
\newblock {\em Phys. Rept.}, 503:1--75, 2011.

\bibitem{Griesshammer:2015osb}
Harald~W. Griesshammer.
\newblock {Assessing Theory Uncertainties in EFT Power Countings from Residual
  Cutoff Dependence}.
\newblock In {\em {8th International Workshop on Chiral Dynamics (CD 2015)
  Pisa, Italy, June 29-July 3, 2015}}, 2015.

\bibitem{Epelbaum:2005pn}
Evgeny Epelbaum.
\newblock {Few-nucleon forces and systems in chiral effective field theory}.
\newblock {\em Prog. Part. Nucl. Phys.}, 57:654--741, 2006.

\bibitem{Jurgenson:2010wy}
E.~D. Jurgenson, P.~Navratil, and R.~J. Furnstahl.
\newblock {Evolving Nuclear Many-Body Forces with the Similarity
  Renormalization Group}.
\newblock {\em Phys. Rev.}, C83:034301, 2011.

\bibitem{Brown:2001zz}
B.~A. Brown.
\newblock {The nuclear shell model towards the drip lines}.
\newblock {\em Prog. Part. Nucl. Phys.}, 47:517--599, 2001.

\bibitem{Barrett:2013nh}
Bruce~R. Barrett, Petr Navratil, and James~P. Vary.
\newblock {Ab initio no core shell model}.
\newblock {\em Prog. Part. Nucl. Phys.}, 69:131--181, 2013.

\bibitem{Maris:2008ax}
P.~Maris, J.~P. Vary, and A.~M. Shirokov.
\newblock {Ab initio no-core full configuration calculations of light nuclei}.
\newblock {\em Phys. Rev. C}, 79:014308, 2009.

\bibitem{Roth:2009cw}
Robert Roth.
\newblock {Importance Truncation for Large-Scale Configuration Interaction
  Approaches}.
\newblock {\em Phys. Rev. C}, 79:064324, 2009.

\bibitem{Pieper:2002ne}
Steven~C. Pieper, K.~Varga, and Robert~B. Wiringa.
\newblock {Quantum Monte Carlo calculations of A=9, A=10 nuclei}.
\newblock {\em Phys. Rev.}, C66:044310, 2002.

\bibitem{Pieper:2007ax}
Steven~C. Pieper.
\newblock {Quantum Monte Carlo calculations of light nuclei}.
\newblock {\em Riv. Nuovo Cim.}, 31:709--740, 2008.
\newblock [,111(2007)].

\bibitem{Gandolfi:2007hs}
S.~Gandolfi, F.~Pederiva, S.~Fantoni, and K.~E. Schmidt.
\newblock {Auxiliary Field Diffusion Monte Carlo calculation of nuclei with A
  <= 40 with tensor interactions}.
\newblock {\em Phys. Rev. Lett.}, 99:022507, 2007.

\bibitem{Gezerlis:2014zia}
A.~Gezerlis, I.~Tews, E.~Epelbaum, M.~Freunek, S.~Gandolfi, K.~Hebeler,
  A.~Nogga, and A.~Schwenk.
\newblock {Local chiral effective field theory interactions and quantum Monte
  Carlo applications}.
\newblock {\em Phys. Rev.}, C90(5):054323, 2014.

\bibitem{Hagen:2013nca}
G.~Hagen, T.~Papenbrock, M.~Hjorth-Jensen, and D.~J. Dean.
\newblock {Coupled-cluster computations of atomic nuclei}.
\newblock {\em Rept. Prog. Phys.}, 77(9):096302, 2014.

\bibitem{Drut:2009ce}
J.~E. Drut, R.~J. Furnstahl, and L.~Platter.
\newblock {Toward ab initio density functional theory for nuclei}.
\newblock {\em Prog. Part. Nucl. Phys.}, 64:120--168, 2010.

\bibitem{Dobaczewski:2010gr}
Jacek Dobaczewski.
\newblock {Current Developments in Nuclear Density Functional Methods}.
\newblock {\em J. Phys. Conf. Ser.}, 312:092002, 2011.

\bibitem{Hergert:2015awm}
H.~Hergert, S.~K. Bogner, T.~D. Morris, A.~Schwenk, and K.~Tsukiyama.
\newblock {The In-Medium Similarity Renormalization Group: A Novel Ab Initio
  Method for Nuclei}.
\newblock {\em Phys. Rept.}, 621:165--222, 2016.

\bibitem{Savage:2015eya}
Martin~J. Savage.
\newblock {Nuclear Physics from Lattice Quantum Chromodynamics}.
\newblock In {\em {12th Conference on the Intersections of Particle and Nuclear
  Physics (CIPANP 2015) Vail, Colorado, USA, May 19-24, 2015}}, 2015.

\bibitem{Detmold:2003rq}
William Detmold, W.~Melnitchouk, and Anthony~William Thomas.
\newblock {Extraction of parton distributions from lattice QCD}.
\newblock {\em Mod. Phys. Lett.}, A18:2681--2698, 2003.

\bibitem{Detmold:2001jb}
William Detmold, W.~Melnitchouk, John~W. Negele, Dru~Bryant Renner, and
  Anthony~William Thomas.
\newblock {Chiral extrapolation of lattice moments of proton quark
  distributions}.
\newblock {\em Phys. Rev. Lett.}, 87:172001, 2001.

\bibitem{Durr:2010aw}
S.~Durr, Z.~Fodor, C.~Hoelbling, S.~D. Katz, S.~Krieg, T.~Kurth, L.~Lellouch,
  T.~Lippert, K.~K. Szabo, and G.~Vulvert.
\newblock {Lattice QCD at the physical point: Simulation and analysis details}.
\newblock {\em JHEP}, 08:148, 2011.

\bibitem{Furnstahl:2013oba}
R.~J. Furnstahl and K.~Hebeler.
\newblock {New applications of renormalization group methods in nuclear
  physics}.
\newblock {\em Rept. Prog. Phys.}, 76:126301, 2013.

\bibitem{Furnstahl:2012fn}
R.~J. Furnstahl.
\newblock {The Renormalization Group in Nuclear Physics}.
\newblock {\em Nucl. Phys. Proc. Suppl.}, 228:139--175, 2012.

\bibitem{Bogner:2009bt}
S.~K. Bogner, R.~J. Furnstahl, and A.~Schwenk.
\newblock {From low-momentum interactions to nuclear structure}.
\newblock {\em Prog. Part. Nucl. Phys.}, 65:94--147, 2010.

\bibitem{Bogner:2003wn}
S.~K. Bogner, T.~T.~S. Kuo, and A.~Schwenk.
\newblock {Model independent low momentum nucleon interaction from phase shift
  equivalence}.
\newblock {\em Phys. Rept.}, 386:1--27, 2003.

\bibitem{Bogner:2006pc}
S.~K. Bogner, R.~J. Furnstahl, and R.~J. Perry.
\newblock Similarity renormalization group for nucleon-nucleon interactions.
\newblock {\em Phys. Rev. C}, 75:061001, 2007.

\bibitem{Wegner:2000gi}
F.~J. Wegner.
\newblock {Flow equations for Hamiltonians}.
\newblock {\em Nucl. Phys. Proc. Suppl.}, 90:141--146, 2000.
\newblock [,141(2000)].

\bibitem{Li:2011sr}
W.~Li, E.~R. Anderson, and R.~J. Furnstahl.
\newblock {The Similarity Renormalization Group with Novel Generators}.
\newblock {\em Phys. Rev.}, C84:054002, 2011.

\bibitem{Epelbaum:2004fk}
E.~Epelbaum, W.~Glockle, and Ulf-G. Meissner.
\newblock {The Two-nucleon system at next-to-next-to-next-to-leading order}.
\newblock {\em Nucl. Phys.}, A747:362--424, 2005.

\bibitem{Hammer:2012id}
Hans-Werner Hammer, Andreas Nogga, and Achim Schwenk.
\newblock {Three-body forces: From cold atoms to nuclei}.
\newblock {\em Rev. Mod. Phys.}, 85:197, 2013.

\bibitem{Hebeler:2012pr}
K.~Hebeler.
\newblock {Momentum space evolution of chiral three-nucleon forces}.
\newblock {\em Phys. Rev. C}, 85:021002, 2012.

\bibitem{Wendt:2013bla}
Kyle~A. Wendt.
\newblock {Similarity Renormalization Group Evolution of Three-Nucleon Forces
  in a Hyperspherical Momentum Representation}.
\newblock {\em Phys. Rev. C}, 87:061001, 2013.

\bibitem{Garfagnini:2014nla}
Alberto Garfagnini.
\newblock {Neutrinoless Double Beta Decay Experiments}.
\newblock In {\em {12th Conference on Flavor Physics and CP Violation (FPCP
  2014) Marseille, France, May 26-30, 2014}}, 2014.

\bibitem{More:2013rma}
S.~N. More, A.~Ekstr{\"{o}}m, R.~J. Furnstahl, G.~Hagen, and T.~Papenbrock.
\newblock {Universal properties of infrared oscillator basis extrapolations}.
\newblock {\em Phys. Rev.}, C87(4):044326, 2013.

\bibitem{Furnstahl:2013vda}
R.~J. Furnstahl, S.~N. More, and T.~Papenbrock.
\newblock {Systematic expansion for infrared oscillator basis extrapolations}.
\newblock {\em Phys. Rev.}, C89(4):044301, 2014.

\bibitem{Konig:2014hma}
S.~K{\"{o}}nig, S.~K. Bogner, R.~J. Furnstahl, S.~N. More, and T.~Papenbrock.
\newblock {Ultraviolet extrapolations in finite oscillator bases}.
\newblock {\em Phys. Rev.}, C90(6):064007, 2014.

\bibitem{More:2015tpa}
S.~N. More, S.~K{\"{o}}nig, R.~J. Furnstahl, and K.~Hebeler.
\newblock {Deuteron electrodisintegration with unitarily evolved potentials}.
\newblock {\em Phys. Rev.}, C92(6):064002, 2015.

\bibitem{Furnstahl2012}
R.~Furnstahl, G.~Hagen, and T.~Papenbrock.
\newblock {Corrections to nuclear energies and radii in finite oscillator
  spaces}.
\newblock {\em Physical Review C}, 86(3):1--6, sep 2012.

\bibitem{Maris2009}
P.~Maris, J.~Vary, and A.~Shirokov.
\newblock {Ab initio no-core full configuration calculations of light nuclei}.
\newblock {\em Physical Review C}, 79(1):014308, jan 2009.

\bibitem{Hagen:2007hi}
G.~Hagen, D.~J. Dean, M.~Hjorth-Jensen, T.~Papenbrock, and A.~Schwenk.
\newblock {Benchmark calculations for 3H, 4He, 16O and 40Ca with ab- initio
  coupled-cluster theory}.
\newblock {\em Phys. Rev. C}, 76:044305, 2007.

\bibitem{Bogner:2007rx}
S.~K. Bogner, R.~J. Furnstahl, P.~Maris, R.~J. Perry, A.~Schwenk, and J.~P.
  Vary.
\newblock {Convergence in the no-core shell model with low-momentum two-nucleon
  interactions}.
\newblock {\em Nucl. Phys. A}, 801:21--42, 2008.

\bibitem{Forssen:2008qp}
C.~Forssen, J.P. Vary, E.~Caurier, and P.~Navratil.
\newblock {Converging sequences in the ab initio no-core shell model}.
\newblock {\em Phys. Rev. C}, 77:024301, 2008.

\bibitem{Coon:2012ab}
Sidney~A. Coon, Matthew~I. Avetian, Michael~K.G. Kruse, U.~van Kolck, Pieter
  Maris, et~al.
\newblock {Convergence properties of {\it ab initio} calculations of light
  nuclei in a harmonic oscillator basis}.
\newblock {\em Phys. Rev. C}, 86:054002, 2012.

\bibitem{Entem:2003ft}
D.~R. Entem and R.~Machleidt.
\newblock Accurate charge-dependent nucleon-nucleon potential at fourth order
  of chiral perturbation theory.
\newblock {\em Phys. Rev. C}, 68:041001, 2003.

\bibitem{gradshteyn}
L.~S. Gradshteyn and L.~M. Ryzhik.
\newblock {\em Tables of integrals, series, and products}.
\newblock Academic Press, San Diego, 6$^{\rm th}$ edition, 2000.

\bibitem{Stetcu:2006ey}
I.~Stetcu, B.~R. Barrett, and U.~van Kolck.
\newblock No-core shell model in an effective-field-theory framework.
\newblock {\em Phys. Lett. B}, 653:358--362, 2007.

\bibitem{Stetcu:2007ms}
I.~Stetcu, B.~R. Barrett, U.~van Kolck, and J.~P. Vary.
\newblock {Effective Theory for Trapped Few-Fermion Systems}.
\newblock {\em Phys. Rev. A}, 76:063613, 2007.

\bibitem{Luscher:1985dn}
M.~Luscher.
\newblock {Volume Dependence of the Energy Spectrum in Massive Quantum Field
  Theories. 1. Stable Particle States}.
\newblock {\em Commun. Math. Phys.}, 104:177, 1986.

\bibitem{Djajaputra:2000aa}
David Djajaputra and Bernard~R Cooper.
\newblock Hydrogen atom in a spherical well: linear approximation.
\newblock {\em European Journal of Physics}, 21(3):261, 2000.

\bibitem{taylor2006scattering}
J.R. Taylor.
\newblock {\em Scattering Theory: The Quantum Theory of Nonrelativistic
  Collisions}.
\newblock Dover, 2006.

\bibitem{Furnstahl:2012qg}
R.~J. Furnstahl, G.~Hagen, and T.~Papenbrock.
\newblock {Corrections to nuclear energies and radii in finite oscillator
  spaces}.
\newblock {\em Phys. Rev.}, C86:031301, 2012.

\bibitem{newton2002scattering}
R.G. Newton.
\newblock {\em Scattering theory of waves and particles}.
\newblock Dover, 2002.

\bibitem{wu2011scattering}
T.-Y. Wu and T.~Ohmura.
\newblock {\em Quantum Theory of Scattering}.
\newblock Dover, New York, 2011.

\bibitem{Phillips:1999hh}
Daniel~R. Phillips, Gautam Rupak, and Martin~J. Savage.
\newblock {Improving the convergence of N N effective field theory}.
\newblock {\em Phys. Lett. B}, 473:209--218, 2000.

\bibitem{Arteca1984}
Gustavo~Alberto Arteca, Francisco~M. Fern{\'a}ndez, and Eduardo~A. Castro.
\newblock {\em J. of Chem. Phys.}, 80:1569, 1984.

\bibitem{Fernandez1981}
Francisco~M. Fern{\'a}ndez and Eduardo~A. Castro.
\newblock Hypervirial analysis of enclosed quantum mechanical systems. i.
  dirichlet boundary conditions.
\newblock {\em Int. J. of Quantum Chem.}, 19(4):521--532, 1981.

\bibitem{Amado:1979zz}
Ralph~D. Amado.
\newblock {Problems in determining nuclear bound state wave functions}.
\newblock {\em Phys. Rev. C}, 19:1473--1481, 1979.

\bibitem{Tolle:2012cx}
S.~Tolle, H.~W. Hammer, and B.~Ch. Metsch.
\newblock {Convergence Properties of the Effective Theory for Trapped Bosons}.
\newblock {\em J. Phys.}, G40:055004, 2013.

\bibitem{Deano2013}
Alfredo Dea{\~n}o, Edmundo~J. Huertas, and Francisco Marcell{\'a}n.
\newblock {Strong and ratio asymptotics for Laguerre polynomials revisited}.
\newblock {\em J. Math. Anal. Appl.}, 403:477--486, 2013.

\bibitem{abramowitz1964}
Milton Abramowitz and Irene~A. Stegun.
\newblock {\em Handbook of Mathematical Functions}.
\newblock Dover, New York, 1972.

\bibitem{Konig:2011nz}
Sebastian Koenig, Dean Lee, and H.-W. Hammer.
\newblock {Volume Dependence of Bound States with Angular Momentum}.
\newblock {\em Phys. Rev. Lett.}, 107:112001, 2011.

\bibitem{Stetcu:2004wh}
Ionel Stetcu, Bruce~R. Barrett, Petr Navratil, and James~P. Vary.
\newblock {Effective operators within the ab initio no-core shell model}.
\newblock {\em Phys. Rev. C}, 71:044325, 2005.

\bibitem{bang2000}
J.~M. {Bang}, A.~I. {Mazur}, A.~M. {Shirokov}, Y.~F. {Smirnov}, and S.~A.
  {Zaytsev}.
\newblock {P-Matrix and J-Matrix Approaches: Coulomb Asymptotics in the
  Harmonic Oscillator Representation of Scattering Theory}.
\newblock {\em Annals of Physics}, 280:299--335, March 2000.

\bibitem{Luu:2010hw}
Thomas Luu, Martin~J. Savage, Achim Schwenk, and James~P. Vary.
\newblock {Nucleon-Nucleon Scattering in a Harmonic Potential}.
\newblock {\em Phys. Rev.}, C82:034003, 2010.

\bibitem{Stetcu:2009ic}
I.~Stetcu, J.~Rotureau, B.~R. Barrett, and U.~van Kolck.
\newblock {Effective interactions for light nuclei: An Effective (field theory)
  approach}.
\newblock {\em J. Phys.}, G37:064033, 2010.

\bibitem{busch1998}
Thomas Busch, Berthold-Georg Englert, Kazimierz Rzazewski, and Martin Wilkens.
\newblock Two cold atoms in a harmonic trap.
\newblock {\em Foundations of Physics}, 28:549--559, 1998.

\bibitem{Bhattacharyya:2006fg}
Anirban Bhattacharyya and T.~Papenbrock.
\newblock {Density functional theory for fermions close to the unitary regime}.
\newblock {\em Phys. Rev.}, A74:041602, 2006.

\bibitem{Harms:1970hd}
Edward Harms.
\newblock {Convenient Expansion for Local Potentials}.
\newblock {\em Phys. Rev. C}, 1:1667--1679, 1970.

\bibitem{Ernst:1973zzb}
D.J. Ernst, C.M. Shakin, and R.M. Thaler.
\newblock {Separable Representations of Two-Body Interactions}.
\newblock {\em Phys. Rev. C}, 8:46--52, 1973.

\bibitem{elgaroy1998}
\O{}. Elgar\o{}y and M.~Hjorth-Jensen.
\newblock Nucleon-nucleon phase shifts and pairing in neutron matter and
  nuclear matter.
\newblock {\em Phys. Rev. C}, 57:1174--1177, Mar 1998.

\bibitem{Lovelace:1964aa}
C.~Lovelace.
\newblock {Practical theory of three particle states. 1. Nonrelativistic}.
\newblock {\em Phys. Rev.}, 135:B1225--B1249, 1964.

\bibitem{Konig:2011ti}
Sebastian Koenig, Dean Lee, and H.-W. Hammer.
\newblock {Non-relativistic bound states in a finite volume}.
\newblock {\em Annals Phys.}, 327:1450--1471, 2012.

\bibitem{Lee:2010km}
Dean Lee and Michelle Pine.
\newblock {How quantum bound states bounce and the structure it reveals}.
\newblock {\em Eur. Phys. J. A}, 47:41, 2011.

\bibitem{Pine:2012zv}
Michelle Pine and Dean Lee.
\newblock {Effective Field Theory for Bound State Reflection}.
\newblock {\em Annals Phys.}, 331:24--50, 2013.

\bibitem{Jurgenson:2009qs}
E.~D. Jurgenson, P.~Navratil, and R.~J. Furnstahl.
\newblock {Evolution of Nuclear Many-Body Forces with the Similarity
  Renormalization Group}.
\newblock {\em Phys. Rev. Lett.}, 103:082501, 2009.

\bibitem{Furnstahl:2014hca}
R.~J. Furnstahl, G.~Hagen, T.~Papenbrock, and K.~A. Wendt.
\newblock {Infrared extrapolations for atomic nuclei}.
\newblock {\em J. Phys.}, G42(3):034032, 2015.

\bibitem{Wendt:2015nba}
K.~A. Wendt, C.~Forssén, T.~Papenbrock, and D.~Sääf.
\newblock {Infrared length scale and extrapolations for the no-core shell
  model}.
\newblock {\em Phys. Rev.}, C91(6):061301, 2015.

\bibitem{Odell:2015xlw}
D.~Odell, T.~Papenbrock, and L.~Platter.
\newblock {Infrared extrapolations of quadrupole moments and transitions}.
\newblock 2015.

\bibitem{Caprio_Coulomb_Sturmian}
Mark Caprio.
\newblock private communication.

\bibitem{CiofidegliAtti:1995qe}
Claudio Ciofi~degli Atti and S.~Simula.
\newblock {Realistic model of the nucleon spectral function in few and many
  nucleon systems}.
\newblock {\em Phys. Rev.}, C53:1689, 1996.

\bibitem{Kimball:1973aa}
J.C. Kimball.
\newblock {Short-Range Correlations and Electron-Gas Response Functions}.
\newblock {\em Phys. Rev. A}, 7:1648--1652, 1973.

\bibitem{Bogner:2012zm}
S.~K. Bogner and D.~Roscher.
\newblock {High-momentum tails from low-momentum effective theories}.
\newblock {\em Phys. Rev. C}, 86:064304, 2012.

\bibitem{Hofmann:2013aa}
Johannes Hofmann, Marcus Barth, and Wilhelm Zwerger.
\newblock {Short-distance properties of Coulomb systems}.
\newblock {\em Phys. Rev. B}, 87:235125, 2013.

\bibitem{Feldmeier:2011qy}
H.~Feldmeier, W.~Horiuchi, T.~Neff, and Y.~Suzuki.
\newblock {Universality of short-range nucleon-nucleon correlations}.
\newblock {\em Phys. Rev. C}, 84:054003, 2011.

\bibitem{Jurgenson:2013yya}
E.~D. Jurgenson, P.~Maris, R.~J. Furnstahl, P.~Navratil, W.~E. Ormand, and
  J.~P. Vary.
\newblock {Structure of $p$-shell nuclei using three-nucleon interactions
  evolved with the similarity renormalization group}.
\newblock {\em Phys. Rev.}, C87(5):054312, 2013.

\bibitem{Binder:2015trg}
S.~Binder, A.~Ekstr{\"{o}}m, G.~Hagen, T.~Papenbrock, and K.~A. Wendt.
\newblock {Effective field theory in the harmonic oscillator basis}.
\newblock 2015.

\bibitem{Hoinka:2013fsa}
Sascha Hoinka, Marcus Lingham, Kristian Fenech, Hui Hu, Chris~J. Vale,
  Joaquín~E. Drut, and Stefano Gandolfi.
\newblock {Precise determination of the structure factor and contact in a
  unitary Fermi gas}.
\newblock {\em Phys. Rev. Lett.}, 110(5):055305, 2013.

\bibitem{Furnstahl:2013dsa}
R.~J. Furnstahl.
\newblock {High-resolution probes of low-resolution nuclei}.
\newblock In {\em {Proceedings, International Conference on Nuclear Theory in
  the Supercomputing Era (NTSE-2013)}}, page 371, 2013.

\bibitem{Altarelli:1977zs}
Guido Altarelli and G.~Parisi.
\newblock {Asymptotic Freedom in Parton Language}.
\newblock {\em Nucl. Phys.}, B126:298, 1977.

\bibitem{Frankfurt:2008zv}
Leonid Frankfurt, Misak Sargsian, and Mark Strikman.
\newblock {Recent observation of short range nucleon correlations in nuclei and
  their implications for the structure of nuclei and neutron stars}.
\newblock {\em Int. J. Mod. Phys. A}, 23:2991--3055, 2008.

\bibitem{Arrington:2011xs}
J.~Arrington, D.W. Higinbotham, G.~Rosner, and M.~Sargsian.
\newblock {Hard probes of short-range nucleon-nucleon correlations}.
\newblock {\em Prog. Part. Nucl. Phys.}, 67:898--938, 2012.

\bibitem{Rios:2013zqa}
A.~Rios, A.~Polls, and W.~H. Dickhoff.
\newblock {Density and isospin asymmetry dependence of high-momentum
  components}.
\newblock {\em Phys. Rev.}, C89(4):044303, 2014.

\bibitem{Boeglin:2015cha}
Werner Boeglin and Misak Sargsian.
\newblock {Modern Studies of the Deuteron: from the Lab Frame to the Light
  Front}.
\newblock {\em Int. J. Mod. Phys.}, E24(03):1530003, 2015.

\bibitem{Ford:2014yua}
William~P. Ford, Sabine Jeschonnek, and J.~W. Van~Orden.
\newblock {Momentum distributions for $^2$H$(e,e'p)$}.
\newblock {\em Phys. Rev.}, C90(6):064006, 2014.

\bibitem{Sammarruca:2015hba}
Francesca Sammarruca.
\newblock {Short-range correlations in the deuteron: chiral effective field
  theory, meson-exchange, and phenomenology}.
\newblock {\em Phys. Rev.}, C92(4):044003, 2015.

\bibitem{Epelbaum:2008ga}
Evgeny Epelbaum, Hans-Werner Hammer, and Ulf-G. Mei{\ss}ner.
\newblock {Modern Theory of Nuclear Forces}.
\newblock {\em Rev. Mod. Phys.}, 81:1773--1825, 2009.

\bibitem{Carlson:2014vla}
J.~Carlson, S.~Gandolfi, F.~Pederiva, Steven~C. Pieper, R.~Schiavilla, K.~E.
  Schmidt, and R.~B. Wiringa.
\newblock {Quantum Monte Carlo methods for nuclear physics}.
\newblock 2014.

\bibitem{Marcucci:2015rca}
L.~E. Marcucci, F.~Gross, M.~T. Pena, M.~Piarulli, R.~Schiavilla, I.~Sick,
  A.~Stadler, J.~W. Van~Orden, and M.~Viviani.
\newblock {Electromagnetic Structure of Few-Nucleon Ground States}.
\newblock {\em J. Phys.}, G43:023002, 2016.

\bibitem{Quaglioni:2015via}
S.~Quaglioni, G.~Hupin, A.~Calci, P.~Navratil, and R.~Roth.
\newblock {Ab initio calculations of reactions with light nuclei}.
\newblock {\em EPJ Web Conf.}, 113:01005, 2016.

\bibitem{Bacca:2013dma}
Sonia Bacca, Nir Barnea, Gaute Hagen, Giuseppina Orlandini, and Thomas
  Papenbrock.
\newblock {First Principles Description of the Giant Dipole Resonance in
  $^{16}$O}.
\newblock {\em Phys. Rev. Lett.}, 111(12):122502, 2013.

\bibitem{Pine:2013zja}
Michelle Pine, Dean Lee, and Gautam Rupak.
\newblock {Adiabatic projection method for scattering and reactions on the
  lattice}.
\newblock {\em Eur. Phys. J.}, A49:151, 2013.

\bibitem{Anderson:2010aq}
E.~R. Anderson, S.~K. Bogner, R.~J. Furnstahl, and R.~J. Perry.
\newblock {Operator Evolution via the Similarity Renormalization Group I: The
  Deuteron}.
\newblock {\em Phys. Rev.}, C82:054001, 2010.

\bibitem{Schuster:2014lga}
Micah~D Schuster, Sofia Quaglioni, Calvin~W. Johnson, Eric~D. Jurgenson, and
  Petr Navratil.
\newblock {Operator evolution for ab initio nuclear theory}.
\newblock {\em Phys. Rev.}, C90(1):011301, 2014.

\bibitem{Schuster:2013sda}
Micah~D. Schuster, Sofia Quaglioni, Calvin~W. Johnson, Eric~D. Jurgenson, and
  Petr Navratil.
\newblock {Operator evolution for ab initio electric dipole transitions of
  4He}.
\newblock {\em Phys. Rev.}, C92(1):014320, 2015.

\bibitem{Neff:2015xda}
Thomas Neff, Hans Feldmeier, and Wataru Horiuchi.
\newblock {Short-range correlations in nuclei with similarity renormalization
  group transformations}.
\newblock {\em Phys. Rev.}, C92(2):024003, 2015.

\bibitem{Boffi:1996}
Sigfrido Boffi, Carlotta Giusti, Franco~Davide Pacati, and Marco Radici.
\newblock {\em Electromagnetic Response of Atomic Nuclei}.
\newblock Clarendon Press, Oxford, 1996.

\bibitem{LENP_white_paper2015}
White paper on nuclear astrophysics and low-energy nuclear physics.
\newblock {\url{www.lecmeeting.org}}, {January 2015}.

\bibitem{Furnstahl:2010wd}
R.~J. Furnstahl and A.~Schwenk.
\newblock {How should one formulate, extract, and interpret `non- observables'
  for nuclei?}
\newblock {\em J. Phys. G}, 37:064005, 2010.

\bibitem{Atti:2015eda}
Claudio Ciofi~degli Atti.
\newblock {In-medium short-range dynamics of nucleons: Recent theoretical and
  experimental advances}.
\newblock {\em Phys. Rept.}, 590:1--85, 2015.

\bibitem{Arenhovel:2004bc}
Hartmuth Arenh{\"{o}}vel, Winfried Leidemann, and Edward~L. Tomusiak.
\newblock {General survey of polarization observables in deuteron
  electrodisintegration}.
\newblock {\em Eur. Phys. J.}, A23:147--190, 2005.

\bibitem{Gilad:1998wia}
S.~Gilad, W.~Bertozzi, and Z.~L. Zhou.
\newblock {New observables in 2H(e,e′p)n}.
\newblock {\em Nucl. Phys.}, A631:276--295, 1998.

\bibitem{Egiyan:2007qj}
K.~S. Egiyan et~al.
\newblock {Experimental study of exclusive H-2(e,e-prime p)n reaction
  mechanisms at high Q**2}.
\newblock {\em Phys. Rev. Lett.}, 98:262502, 2007.

\bibitem{Yang:2013rza}
C.-J. Yang and Daniel~R. Phillips.
\newblock {The longitudinal response function of the deuteron in chiral
  effective field theory}.
\newblock {\em Eur. Phys. J. A}, 49:122, 2013.

\bibitem{Arenhovel:1988qh}
H.~Arenhovel, W.~Leidemann, and E.~L. Tomusiak.
\newblock {The Role of the Neutron Electric Form-factor in $D (e$, $e^\prime
  N$) $N$ Including Polarization Observables}.
\newblock {\em Z. Phys.}, A331:123--138, 1988.

\bibitem{Kasdorp:1997ba}
Willem-Jan Kasdorp.
\newblock {\em {Deuteron electrodisintegration at large momentum values}}.
\newblock PhD thesis, Utrecht U., 1997.

\bibitem{Arenhoevel:1992xu}
Hartmuth Arenh{\"{o}}vel, Winfried Leidemann, and Edward~L. Tomusiak.
\newblock {Exclusive deuteron electrodisintegration with polarized electrons
  and a polarized target}.
\newblock {\em Phys. Rev.}, C46:455--470, 1992.

\bibitem{Ibrahim:2006lta}
Hassan~F. Ibrahim.
\newblock {\em {The $^{2}H(e, e^\prime \,p) n$ reaction at high four-momentum
  transfer}}.
\newblock PhD thesis, Old Dominion U., 2006-12-01.

\bibitem{Landau:1989}
Rubin~H. Landau.
\newblock {\em Quantum Mechanics II}.
\newblock John Wiley \& Sons, New York, 1989.

\bibitem{Jerry_thesis}
C.-J. Yang.
\newblock {\em Subtractive Renormalization of the NN interaction in Chiral
  Effective Theory and the Deuteron Electro-disintegration Calculation}.
\newblock Ph.d thesis, Ohio University, 2010.

\bibitem{Furnstahl:2001xq}
R.~J. Furnstahl and H.-W. Hammer.
\newblock Are occupation numbers observable?
\newblock {\em Phys. Lett. B}, 531:203--208, 2002.

\bibitem{Hetherington:1965zza}
J.~H. Hetherington and L.~H. Schick.
\newblock {Exact Multiple-Scattering Analysis of Low-Energy Elastic K--d
  Scattering with Separable Potentials}.
\newblock {\em Phys. Rev.}, 137:B935--B948, 1965.

\bibitem{Schmid:1974}
E.~W. Schmid and H.~Ziegelmann.
\newblock {\em {The Quantum Mechanical Three-Body Problem}}.
\newblock Vieweg, Braunschweig, 1974.

\bibitem{ALGLIB:0915}
Sergey Bochkanov.
\newblock {ALGLIB}.
\newblock {\url{www.alglib.net}}.

\bibitem{TBB:0915}
Intel®.
\newblock {Threading Building Blocks (TBB)}.
\newblock {\url{www.threadingbuildingblocks.org}}, {September 2015}.

\bibitem{Carlson:1997qn}
J.~Carlson and R.~Schiavilla.
\newblock {Structure and dynamics of few nucleon systems}.
\newblock {\em Rev. Mod. Phys.}, 70:743--842, 1998.

\bibitem{Bacca:2014tla}
Sonia Bacca and Saori Pastore.
\newblock {Electromagnetic reactions on light nuclei}.
\newblock {\em J. Phys.}, G41(12):123002, 2014.

\bibitem{Bogner:2007jb}
S.~K. Bogner, R.~J. Furnstahl, R.~J. Perry, and A.~Schwenk.
\newblock {Are low-energy nuclear observables sensitive to high-energy phase
  shifts?}
\newblock {\em Phys. Lett.}, B649:488--493, 2007.

\bibitem{Christlmeier:2008ye}
Stefan Christlmeier and Harald~W. Griesshammer.
\newblock {Pion-less Effective Field Theory on Low-Energy Deuteron
  Electro-Disintegration}.
\newblock {\em Phys. Rev.}, C77:064001, 2008.

\bibitem{Lovato:2015qka}
A.~Lovato, S.~Gandolfi, J.~Carlson, Steven~C. Pieper, and R.~Schiavilla.
\newblock {Electromagnetic and neutral-weak response functions of $^4$He and
  $^{12}$C}.
\newblock {\em Phys. Rev.}, C91(6):062501, 2015.

\bibitem{Aubert:1983xm}
J.~J. Aubert et~al.
\newblock {The ratio of the nucleon structure functions $F_2^N$ for iron and
  deuterium}.
\newblock {\em Phys. Lett.}, B123:275--278, 1983.

\bibitem{Norton:2003cb}
P.~R. Norton.
\newblock {The EMC effect}.
\newblock {\em Rept. Prog. Phys.}, 66:1253--1297, 2003.

\bibitem{Stapp:1956mz}
H.~P. Stapp, T.~J. Ypsilantis, and N.~Metropolis.
\newblock {Phase shift analysis of 310-MeV proton proton scattering
  experiments}.
\newblock {\em Phys. Rev.}, 105:302--310, 1957.

\bibitem{Koenig:2013}
Sebastian K\"{o}nig.
\newblock {\em Effective quantum theories with short- and long-range forces}.
\newblock PhD thesis, Bonn University, 2013.

\end{thebibliography}


% ...


\appendix

\cleardoublepage
\chapter{$t$-matrix details}
\label{Appendix:t_matrix_details}

  \section{Solving the Lippmann-Schwinger equation}

  The Lippmann-Schwinger equation (LSE), which is essentially the
  Schr\"{o}dinger equation for scattering states is given in operator form by
  \beq
  t = V + V \, G \, t \;,
  \label{eq:LSE_operator_form}
  \eeq
  where $t$ is the $t$-matrix, $V$ is the potential, and $G$ is the Green's
  function.  In momentum space, Eq.~\eqref{eq:LSE_operator_form} becomes
  \beq
  t_{\lp l} (E_0 \!=\! p_0^2/M ; \pp, p) = V_{\lp l} + \sum_{\lpp}
  \frac{2}{\pi} M \int \frac{\dd \ppp \, {\ppp}^2 V_{l \lpp}(\pp, \ppp) \,
    t_{\lpp l}(E_0; \ppp, p)}{p_0^2 - {\pp}^2 + i \epsilon} \;.
  \label{eq:t_matrix_mom_space}
  \eeq
  Derivation of Eq.~\eqref{eq:t_matrix_mom_space} makes use of the completeness
  relation in Eq.~\eqref{eq:completeness_partial_wave} and the definition of
  Green's function in Eq.~\eqref{eq:G0_def_pw}.   The indices which are the
  same on both sides of Eq.~\eqref{eq:t_matrix_mom_space} and are not summed
  over are suppressed.  For deuteron disintegration calculations, we need only
  the half on-shell $t$-matrix.  But here we will look at the more general case
  of evaluating the fully off-shell $t$-matrix.

  For the sake of pedagogy, let us consider that we are evaluating the
  $t$-matrix for uncoupled channels.  Suppressing the angular momentum indices
  and putting in the limits of integration,
  the Eq.~\eqref{eq:t_matrix_mom_space} becomes
  \beq
  t(E_0; \pp, p) = V(\pp, p) + \frac{2}{\pi} M \int_0 ^\Lambda
  \frac{\dd \ppp \, {\ppp}^2
  V(\pp, \ppp)\, t(E_0; \ppp, p)}{p_0^2 - {\ppp}^2 + i \epsilon} \;.
  \label{eq:t_uncoupled_channels}
  \eeq
  Next we outline the steps involved in solving
  Eq.~\eqref{eq:t_uncoupled_channels} numerically.  We follow the approach of
  Ref.~\cite{Landau:1989}.

  The integrals are efficiently evaluated numerically using a Gauss-Legendre
  quadrature.  However, the integrand in Eq.~\eqref{eq:t_uncoupled_channels} has
  a pole at $\ppp = p_0$, and that needs to be accounted for properly.
  Let's consider the expression
  \bea
  \label{eq:separating_pole_first_step}
  \int_0 ^\Lambda \dd p \frac{p^2 f(p)}{p_0^2 - p^2 + i \epsilon}
  &=& \int_0 ^\Lambda \dd p \frac{p^2 f(p)}{(p_0 + p)(p_0 - p + i \epsilon)}
  \\ [0.4 em]
  &\equiv& \int_0 ^\Lambda \dd p \frac{\wt{f}(p)}{(p_0 - p + i \epsilon)} \;,
  \label{eq:separating_pole}
  \eea
  where we have defined $\wt{f}(p)$ as
  \beq
  \wt{f}(p) = \frac{p^2 \, f(p)}{p_0 + p} \;.
  \eeq
  In principle, we can work without separating the singular and non-singular
  factors of $p_0^2 - p^2 + i \epsilon$.  However, we find better numerical
  convergence when the pole term is factorized as in
  Eq.~\eqref{eq:separating_pole_first_step}.

  Using Sokhotsky's formula
  \beq
  \frac{1}{x \pm i \epsilon} = \mathcal{P} \left(\frac{1}{x}\right) \mp i \,
  \pi \, \delta(x) \;,
  \eeq
  we have
  \beq
  \int_0 ^\Lambda \dd p \frac{\wt{f}(p)}{(p_0 - p + i \epsilon)} =
  \mathcal{P} \int_0 ^\Lambda \dd p \frac{\wt{f}(p)}{p_0 - p}  - i \, \pi \,
  \wt{f}(p_0)\;.
  \label{eq:f_tilde_simple_pole}
  \eeq
  Let's first evaluate the principal value integration in the
  Eq.~\eqref{eq:f_tilde_simple_pole}.
  \bea
  \mathcal{P} \int_0 ^\Lambda \dd p \frac{\wt{f}(p)}{p_0 - p}
  & = &  \mathcal{P} \int_0 ^\Lambda \dd p \frac{\wt{f}(p) - \wt{f}(p_0) +
  \wt{f}(p_0)}{p_0 - p} \nonumber \\ [0.4 em]
  & = & \int_0 ^\Lambda \dd p \frac{\wt{f}(p) - \wt{f}(p_0)}{p_0 - p} +
  \wt{f}(p_0) \, \mathcal{P} \int_0 ^\Lambda \dd p \frac{1}{p_0 - p} \;.
  \label{eq:add_subtract_pole}
  \eea
  The integrand of the first term on the right side of
  Eq.~\eqref{eq:add_subtract_pole} is zero at the pole $p = p_0$ and therefore
  non-singular.  We can therefore drop the principal value for that term and
  evaluate it as a normal integral.  The second term on the right side of
  Eq.~\eqref{eq:add_subtract_pole} can be evaluated analytically.
  \bea
  \mathcal{P} \int_0 ^\Lambda \dd p \frac{1}{p_0 - p}
  & = & \int_0 ^{p_0 - \epsilon} \dd p \frac{1}{p_0 - p} +
  \int_{p_0 + \epsilon} ^\Lambda \dd p \frac{1}{p_0 - p} \nonumber \\ [0.5 em]
  & = &  \null - {\rm{ln}}(p_0 - p) \Big\vert_0 ^{p_0 - \epsilon} +
         \null - {\rm{ln}}(p_0 - p) \Big\vert_{p_0 + \epsilon} ^\Lambda
         \nonumber \\ [0.5 em]
  & = & - {\rm{ln}}\left(\frac{\Lambda - p_0}{p_0}\right)\;.
  \label{eq:PV_1_over_x}
  \eea
  From Eqs.~\eqref{eq:PV_1_over_x}, \eqref{eq:add_subtract_pole}, and
  \eqref{eq:f_tilde_simple_pole}, we have
  \begin{equation}
  \int_0 ^\Lambda \dd p \frac{\wt{f}(p)}{p_0 - p + i \epsilon}
  = \int_0 ^\Lambda \dd p \frac{\wt{f}(p) - \wt{f}(p_0)}{p_0 - p} -
  \wt{f}(p_0) \, {\rm{ln}}\left(\frac{\Lambda - p_0}{p_0}\right)
  \null - i \, \pi \, \wt{f}(p_0)\;.
  \end{equation}
  Discretizing this on the Gauss-Legendre mesh we have
  \beq
  \int_0 ^\Lambda \dd p \frac{\wt{f}(p)}{p_0 - p + i \epsilon}
  = \sum_{j = 1}^N \frac{\wt{f}(p_j)}{p_0 - p_j} w_j - \wt{f}(p_0)
  \left[ i \pi + {\rm{ln}}\big(\frac{\Lambda - p_0}{p_0}\big) + \sum_{j=1}^N
  \frac{w_j}{p_0 - p_j}\right] \;.
  \label{eq:pole_GL}
  \eeq
  $p_j$'s are the momentum mesh points, $N$ is the number of mesh points,
  and $w_j$'s are the associated weights.

  Comparing Eqs.~\eqref{eq:t_uncoupled_channels} and
  Eq.~\eqref{eq:separating_pole}, the corresponding $\wt{f}$ function for
  the LSE is
  \beq
  \wt{f}(\ppp) = \frac{2}{\pi} M \frac{{\ppp}^2 V(\pp, \ppp) \, t(E_0; \ppp, p)}
  {(p_0 + \ppp)} \;.
  \eeq
  Using the result of Eq.~\eqref{eq:pole_GL}, the LSE from
  Eq.~\eqref{eq:t_uncoupled_channels} on the Gauss-Legendre mesh becomes
  \begin{multline}
  t(E_0; \pp, p) = V(\pp, p) + \frac{2}{\pi} M \sum_{j = 1}^N
  \frac{k_j^2 \, V(\pp, k_j) \, t(E_0; k_j, p)}{p_0^2 - k_j^2} w_j \\
  \null - \frac{2}{\pi} M \frac{p_0^2 \, V(\pp, p_0) \, t(E_0; p_0, p)}{2 p_0}
  \left[ i \pi + {\rm{ln}}\big(\frac{\Lambda - p_0}{p_0}\big)
  + \sum_{j = 1}^N \frac{w_j}{p_0 - k_j} \right] \;.
  \label{eq:LSE_GL}
  \end{multline}
  Note that $k_j$ are Gauss-Legendre momentum mesh points and $w_j$ are the
  corresponding weights.

  Let's define an array $D$ such that
  \beq
  D_j = \left\{ \begin{array}{ll}
          \dfrac{2}{\pi} M \dfrac{k_j^2 \, w_j}{p_0^2 - k_j^2}
          \mbox{~~~~for $j = 1, \cdots, N$} \\ [0.7em]
          -\dfrac{2}{\pi} M \dfrac{p_0^2}{2 p_0} \Big( i \pi + {\rm{ln}}\big(
          \dfrac{\Lambda - p_0}{p_0}\big) + \sum_{j = 1}^N \dfrac{w_j}{p_0 - k_j}
          \Big) \mbox{~~for $j = N+ 1$}.
        \end{array} \right.
  \label{eq:D_def}
  \eeq
  Using this definition of $D$ and with the identification that
  $k_{j = N+1} = p_0$, Eq.~\eqref{eq:LSE_GL} can be written as
  \beq
  t(E_0; \pp, p) - \sum_{j = 1}^{N+1} V(\pp, k_j) \, D_j \, t(E_0; k_j, p)
  = V(\pp, p) \;.
  \label{eq:t_D}
  \eeq

  To solve Eq.~\eqref{eq:t_D} in matrix form, we let $\pp \to \{p_i\}$,
  where $i = 1, \cdots, N$ are the Gauss-Legendre mesh points and $i = N+1$
  is the on-shell point $p = p_0$ \footnote{$k_i$'s and $p_i$'s are actually
  the same set of momentum points.  To avoid confusion, we keep the notation
  separate.}.  Eq.~\eqref{eq:t_D} can then be written as
  a matrix multiplication equation.
  \beq
  t(E_0; \pp_i, p) - \sum_{j = 1}^{N+1} V(\pp_i, k_j) \, D_j \, t(E_0; k_j, p)
  = V(\pp_i, p)
  \label{eq:t_D_p_i}
  \eeq
  \beq
  \sum_{j = 1}^{N+1} \underbrace{\big(\delta_{i j} - V(p_i, k_j) \, D_j\big)}
  _{\equiv \, F_{ij}} \, t(E_0, k_j, p)  =  V(p_i, p)
  \label{eq:F_def}
  \eeq
  \beq
  [F]_{(N+1)\times (N+1)} [t]_{N+1} = [V]_{N+1}
  \label{eq:F_t_matrix}
  \eeq
  Note that $[V]$ in Eq.~\eqref{eq:F_t_matrix} is an array whose $N+1^{\rm{th}}$
  element is $V(p_0, p)$ and the first $N$ elements are $V(p_i, p)$, where as
  mentioned before $p_i$'s are the Gauss-Legendre mesh points.  The same
  indexing holds for $[t]$.

  $[F]$ and $[V]$ in Eq.~\eqref{eq:F_t_matrix} are
  known.  Eq.~\eqref{eq:F_t_matrix} can be solved using standard matrix equation
  solving subroutines to get the $t$-matrix array $[t]$.  Recall that
  $\displaystyle {[t] = \big( t(E_0; p_{j = 1,\cdots, N}, p), \, t(E_0; p_0, p)
  \big)}$.  Thus, for a given $p$ and $E_0$, we have the $t$-matrix
  $t(E_0; p_j, p)$ for any point $p_j$ on the Gauss-Legendre mesh, and also
  have it at the half on-shell point $t(E_0; p_0, p)$.  In principle, we can
  use any standard interpolation routine to get the $t$-matrix at a point
  not on the Gauss-Legendre mesh.  However, it turns out that we can use the
  LSE itself for interpolation.  Consider a point $\wt{p}$ not on the momentum
  mesh.  From Eq.~\eqref{eq:t_D}, we have
  \beq
  t(E_0; \wt{p}, p) = V(\wt{p}, p) + \sum_{j = 1}^{N+1}
  V(\wt{p}, k_j) \, D_j \, t(E_0, k_j, p) \;.
  \label{eq:t_interpolation}
  \eeq
  As mentioned before $\{k_j\} = \{p_j\}$, and therefore all the terms on the
  right side of Eq.~\eqref{eq:t_interpolation} are known allowing us to
  evaluate $t(E_0; \wt{p}, p)$.  To interpolate the potential, which is stored
  on a momentum-space grid, we use
	the two-dimensional cubic spline algorithm from ALGLIB~\cite{ALGLIB:0915}.

  \medskip
  \subsubsection{Coupled channels}

  The neutron-proton system has both spin $S = 0$ and $S = 1$ channels.  The
  uncoupled channels have $S = 0$, whereas coupled channels have $S = 1$.
  The angular momentum numbers of the coupled channel pairs differ by $2$.
  For example, some of the coupled channel pairs are $^3 S_1$ - $^3 D_1$,
  $^3 P_2$ - $^3 F_2$, $^3 D_3$ - $^3 G_3$, so on.

  For coupled channels Eq.~\eqref{eq:LSE_operator_form} becomes
  \beq
  \left( \begin{array}{cc}
    t_{00} & t_{02} \\
    t_{20} & t_{22}
  \end{array} \right) =  \left( \begin{array}{cc}
    V_{00} & V_{02} \\
    V_{20} & V_{22}
  \end{array} \right) + \left( \begin{array}{cc}
    V_{00} & V_{02} \\
    V_{20} & V_{22}
  \end{array} \right)  \left( \begin{array}{cc}
    G_{0} & 0 \\
    0 & G_{0}
  \end{array} \right)  \left( \begin{array}{cc}
    t_{00} & t_{02} \\
    t_{20} & t_{22}
  \end{array} \right) \,
  \eeq
  where the subscripts $00$, $02$ etc.\ indicate the coupled channels.
  Following steps similar to the uncoupled case, the analog of
  Eq.~\eqref{eq:t_D_p_i} is
  \begin{multline}
    \left( \begin{array}{cc}
      t_{00}(\pp_i, p) & t_{02}(\pp_i, p) \\
      t_{20}(\pp_i, p) & t_{22}(\pp_i, p)
    \end{array} \right) - \sum_{j = 1}^{N+1} \left( \begin{array}{cc}
      V_{00}(\pp_i, k_j) & V_{02}(\pp_i, k_j) \\
      V_{20}(\pp_i, k_j) & V_{22}(\pp_i, k_j)
    \end{array} \right) \left( \begin{array}{cc}
      \mathcal{D}(k_j) & 0 \\
      0 & \mathcal{D}(k_j)
    \end{array} \right) \\  \left( \begin{array}{cc}
      t_{00}(k_j, p) & t_{02}(k_j, p) \\
      t_{20}(k_j, p) & t_{22}(k_j, p)
    \end{array} \right) =  \left( \begin{array}{cc}
      V_{00}(\pp_i, p) & V_{02}(\pp_i, p) \\
      V_{20}(\pp_i, p) & V_{22}(\pp_i, p)
    \end{array} \right) \;.
  \end{multline}
  $\mathcal{D}$ is a $(N+1) \times (N+1)$ diagonal matrix element with the
  diagonal matrix elements given by $D$ from Eq.~\eqref{eq:D_def}.
  Each of $V_{0,2 \, 0,2}(\pp_i, k_j)$ are a $(N+1) \times (N+1)$ matrix,
  whereas $t_{0,2 \, 0,2}(\pp_i, p)$ and $V_{0,2 \, 0,2}(\pp_i, p)$ have
  dimensions of $(N+1) \times 1$.

  Analogous to Eqs.~\eqref{eq:F_def} and \eqref{eq:F_t_matrix}, we now have
  \begin{multline}
  \sum_{j = 1}^{N+1} \bigg[\delta_{i j} \left( \begin{array}{cc}
  \mathds{1}_{(N+1) \times (N+1)} & 0 \\
  0 & \mathds{1}_{(N+1) \times (N+1)}
  \end{array} \right) -
  \left( \begin{array}{cc}
  V_{00}(\pp_i, k_j) D_j & V_{02}(\pp_i, k_j) D_j \\
  V_{20}(\pp_i, k_j) D_j & V_{22}(\pp_i, k_j) D_j
  \end{array} \right)
  \bigg] \\
  \left( \begin{array}{cc}
  t_{00}(k_j, p) & t_{02}(k_j, p) \\
  t_{20}(k_j, p) & t_{22}(k_j, p)
  \end{array} \right)
  = \left( \begin{array}{cc}
    V_{00}(\pp_i, p) & V_{02}(\pp_i, p) \\
    V_{20}(\pp_i, p) & V_{22}(\pp_i, p)
    \end{array} \right) \;,
  \end{multline}
  and
  \beq
  [F]_{(2 N+2)\times (2N+2)} [t]_{(2 N+ 2) \times 2 } =
  [V]_{(2 N+2) \times 2} \;.
  \label{eq:F_t_coupled}
  \eeq
  Solving Eq.~\eqref{eq:F_t_coupled} for a given $E_0$ and $p$ yields
  $t_{0,2 \, 0,2}(E_0; k_j, p)$ for points $k_j$ on the mesh.  Analogous to
  Eq.~\eqref{eq:t_interpolation}, we can again use the LSE to interpolate
  the $t$-matrix in coupled channel.  Below, we note interpolation for one
  of the components for the point $p = \wt{p}$ not on the mesh.
  \beq
  t_{02}(E_0; \wt{p}, p) = V_{02}(\wt{p}, p) + \sum_{j = 1}^{N+1}
  V_{00}(\wt{p}, k_j) \, D_j \, t_{02}(E_0, k_j, p) + \sum_{j = 1}^{N+1}
  V_{02}(\wt{p}, k_j) \, D_j \, t_{22}(E_0, k_j, p)
  \label{eq:t_interpolation_coupled}
  \eeq
  Note that all the quantities on the right side of
  Eq.~\eqref{eq:t_interpolation_coupled} are known allowing us to evaluate
  $t_{02}(E_0; \wt{p}, p)$.  Similarly, we can write down equations for
  interpolation of other components of the $t$-matrix.

  \section{$t$-matrix checks}

  We checked the accuracy of our $t$-matrix by calculating the phase shifts and
  verifying them against standard values (such as from NN-online).
  For uncoupled channels, the on-shell part of the $t$-matrix is related to
  the phase shift as follows \footnote{Be aware that the factors of $M$
  (nucleon mass) and $\hbar$ might differ based on the conventions and units
  used.}
  \beq
  t_{l} (E_k; k, k) = \frac{\ee^{i \delta_l} \, \sin \delta_l}{- M \, k} \;.
  \label{eq:t_phase_shift}
  \eeq
  Thus, argument of the $t$-matrix gives the phase shift.
  \beq
  \delta_l(k) = {\rm{arg}} (t_{l} (E_k; k, k))
  \label{eq:delta_l_uncoupled}
  \eeq

  The coupled channel calculation involves an additional parameter called
  mixing angle denoted by $\bar{\epsilon}$.  In the ``Stapp'' or the ``bar''
  phase shift parametrization \cite{Stapp:1956mz}, the $S$-matrix is written
  as
  \beq
  S = \left( \begin{array}{cc}
  \cos 2 \bar{\epsilon} \, \ee^{2 i \bar{\delta}_1} &
  i \, \sin 2 \bar{\epsilon} \, \ee^{i (\bar{\delta}_1 + \bar{\delta}_2)} \\
  i \, \sin 2 \bar{\epsilon} \, \ee^{i (\bar{\delta}_1 + \bar{\delta}_2)} &
  \cos 2 \bar{\epsilon} \, \ee^{2 i \bar{\delta}_2}
  \end{array} \right) \;.
  \label{eq:S_stapp}
  \eeq
  We calculate $S$ in terms $t$-matrix using (see for instance Eq.~(8.70) in
  \cite{Landau:1989})
  \beq
  S = \left( \begin{array}{cc}
  1 - 2 \, i \, k \, t_{00}(E_k; k, k )  &
  - 2 \, i \, k \, t_{02}(E_k; k, k ) \\
  - 2 \, i \, k \, t_{20}(E_k; k, k ) &
  1 - 2 \, i \, k \, t_{22}(E_k; k, k )
  \end{array} \right) \;.
  \eeq
  From Eq.~\eqref{eq:S_stapp}, we can work out that the phase shifts and the
  mixing angles are given as follows.
  \bea
  \bar{\delta}_1 = \dfrac{1}{2} \tan^{-1} \left( \dfrac{
  {\rm{Im}}\big[S[1,1]\big]}
  {{\rm{Re}}\big[S[1,1]\big]} \right) \\ [0.5 em]
  \bar{\delta}_2 = \dfrac{1}{2} \tan^{-1} \left( \dfrac{
  {\rm{Im}}\big[S[2,2]\big]}
  {{\rm{Re}}\big[S[2,2]\big]} \right) \\ [0.5 em]
  \bar{\epsilon} = \dfrac{1}{2} \sin^{-1} \left( \dfrac{
  {\rm{Im}}\big[S[1,2]\big]}
  {{\rm{Re}}\big[\sqrt{{\rm{det}}(S)}\big]} \right)
  \eea
  The phase shifts and the mixing angles calculated using these formulas match
  the results on NN-online.  This indicates that $t$-matrix we have is correct.

  \subsubsection{Checking the imaginary part of the $t$-matrix}

  The formulas for phase shifts check the ratio of real and imaginary parts of
  $t$-matrix.  This is particularly evident in Eq.~\eqref{eq:delta_l_uncoupled}.
  But it is also possible to check the imaginary part of the $t$-matrix.
  From Eq.~\eqref{eq:t_phase_shift}, we have (suppressing the arguments of the
  $t$-matrix)
  \bea
  t & = &  \frac{\sin \delta}{-M \, k \, (\cos \delta - i \, \sin \delta)}
  \;. \\
  \Rightarrow \dfrac{1}{-M \, t} & = & k \, \cot \delta - i \, k \;. \\
  \Rightarrow {\rm{Im}}[1/t] & \propto & k \;.
  \label{eq:Im_t_behavior}
  \eea
  Thus, the imaginary part of $1/t_l(E_k; k, k)$ when plotted as a function of
  $k$ should be a straight line.  Our $t$-matrix satisfies this condition.
  The slope of the line in this case is $M$, but in general depends on the
  units chosen.

  \subsubsection{Symmetric property of $t$-matrix}

  The phase shifts, mixing angles, and the behavior of the imaginary part of
  $t$-matrix described in Eq.~\eqref{eq:Im_t_behavior}, all check the
  on-shell part of the $t$-matrix.  To get some confidence about the off-shell
  part of $t$-matrix, we can check if it has the right symmetries.
  The $t$-matrix is symmetric under the angular momentum and momentum
  interchange, i.e.,
  \beq
  t(E_0; k, k^\prime, L, L^\prime, J, S , T) = t(E_0; k^\prime, k, L^\prime, L,
  J, S, T) \;.
  \eeq

  Note that the $t$-matrix is not Hermitian as one would naively expect.

  \section{Using the LSE for the wave function interpolation}

  Solving Schr\"{o}dinger's equation we obtain the deuteron wave function
  on the momentum mesh on which our potential is stored.  We can use the LSE
  to obtain the wave function at any intermediate momentum point.
  We use the property that near the bound state pole, the $t$-matrix
  factorizes as (see Appendix of Ref.~\cite{Koenig:2013} and references therein)
  \beq
  \lim_{E \to -E_B} (E + E_B) \, t(E; k, k^\prime) = \mathcal{B}^\ast (k) \,
  \mathcal{B} (k^\prime) \;,
  \label{eq:t_near_pole}
  \eeq
  where $E_B$ is the bound state energy, and the factor $\mathcal{B}$ is
  the wave function apart from a factor of propagator
  \beq
  \mathcal{B}(q) = \dfrac{-\pi (k_B^2 + q^2)}{2 \, M} \psi(q) \;,
  \label{eq:B_psi_relation}
  \eeq
  where $k_B$ is the bound state momentum.

  Multiplying the LSE equation for the $t_{22}$ channel by $(E + E_B)$ and
  taking the limit $E \to -E_B$, we have
  \begin{multline}
  \lim_{E \to -E_B} (E+E_B) \, t_{22} (E; k, k^\prime)
  = \lim_{E \to -E_B} (E+E_B) \, V_{22} (k, k^\prime) \\
  \null + \frac{2}{\pi} M
  \lim_{E \to -E_B}  \int \dd p \, p^2 \, V_{20} (k, p)
  \frac{t_{02}(E; p, k^\prime)}{k_E^2 - p^2 + i \epsilon} (E+E_B) \\
  \null + \frac{2}{\pi} M \lim_{E \to -E_B}  \int \dd p \, p^2 \, V_{22} (k, p)
  \frac{t_{22}(E; p, k^\prime)}{k_E^2 - p^2 + i \epsilon} (E+E_B)
  \label{eq:lim_LSE} \;.
  \end{multline}
  The first term on the right side of Eq.~\eqref{eq:lim_LSE} vanishes as the
  potential does not have a singular part.
  Using Eq.~\eqref{eq:t_near_pole}, we get
  \begin{equation}
  \mathcal{B}_2^\ast(k) = \frac{2}{\pi} M \int \dd p \, p^2
  \frac{V_{20}(k, p) \, \mathcal{B}_0^\ast (p)}{-k_B^2 - p^2} +
  \frac{2}{\pi} M \int \dd p \, p^2
  \frac{V_{22}(k, p) \, \mathcal{B}_2^\ast (p)}{-k_B^2 - p^2} \;.
  \label{eq:eq_for_B2}
  \end{equation}
  In deriving Eq.~\eqref{eq:eq_for_B2}, we have dropped the factor of
  $\mathcal{B}_2(k^\prime)$ which is common on both sides.
  Substituting Eq.~\eqref{eq:B_psi_relation} gives
  \beq
  \frac{-\pi}{2 \, M} \, (k_B^2 + k^2) \, \psi_2^\ast(k) =
  \int \dd p \, p^2 \, V_{20}(k, p) \psi_0^\ast(p) +
  \int \dd p \, p^2 \, V_{22}(k, p) \psi_2^\ast(p) \;.
  \eeq
  The wave function in our case is real and therefore the complex conjugation
  can be dropped.  Writing the integral in terms of sum, we have
  \beq
  \psi_2(k) = \frac{-2 \, M}{\pi \, (k_B^2 + k^2)} \left[
  \sum_{j = 1}^N w_j \, p_j^2 \, V_{20}(k, p_j) \, \psi_0(p_j)
  + \sum_{j = 1}^N w_j \, p_j^2 \, V_{22}(k, p_j) \, \psi_2(p_j) \right] \;.
  \label{eq:psi_2_interpolation}
  \eeq
  The wave functions on the mesh---$\psi_0(p_j)$ and $\psi_2(p_j)$---are
  already known from solving the Schr\"{o}dinger equation and therefore
  Eq.~\eqref{eq:psi_2_interpolation} allows us to obtain $\psi_2(k)$ for any
  desired momentum $k$.  We also checked that if we choose $k$ in
  Eq.~\eqref{eq:psi_2_interpolation} to be one of the on mesh points, then we
  get back the expected answer.

  Similarly for the $S$-state wave function, we have
  \beq
  \psi_0(k) = \frac{-2 \, M}{\pi \, (k_B^2 + k^2)} \left[
  \sum_{j = 1}^N w_j \, p_j^2 \, V_{00}(k, p_j) \, \psi_0(p_j)
  + \sum_{j = 1}^N w_j \, p_j^2 \, V_{02}(k, p_j) \, \psi_2(p_j) \right] \;.
  \label{eq:psi_0_interpolation}
  \eeq
  A word of caution---the $M$ in Eqs.~\eqref{eq:psi_2_interpolation} and
  \eqref{eq:psi_0_interpolation} depends on conventions.  In some cases, the
  mass factor is absorbed in the potential.  Same goes for the factor of
  $\hbar$'s.  Therefore, it is important to check that the units are consistent.

  The interpolation techniques for the wave functions and for the $t$-matrix
  (Eqs.~\eqref{eq:t_interpolation} and \eqref{eq:t_interpolation_coupled})
  keep the numerical errors minimal.  In our case, the only source of error is
  from interpolation of the potential.

\cleardoublepage
\chapter{Evolution details for deuteron disintegration}
\label{Appendix:evolution_details}

  \section{Expressions for the evolved matrix elements}
  \label{Appendix:sec:evolution_expressions}

  Here we document the expressions used in Subsec.~\ref{subsec:evolution_setup}.
  As seen in Eq.~\eqref{eq:B_split_up}, in order to evaluate the term $\la \phi|
  \, J_0^\lambda | \psi_i^\lambda \ra$ we split it into four terms: $B_1$,
  $B_2$, $B_3$, and $B_4$.  $B_4$ is obtained from Eq.~\eqref{eq:overlap_IA} by
  using the evolved deuteron wave function instead of the unevolved one.  The
  expressions for the terms $B_3$, $B_2$, and $B_1$ are as follows:
  \begin{multline}
   B_3 \equiv \la \phi | \, J_0 \, \widetilde{U}^\dag | \psi_i^\lambda \ra
   = 2 \, \sqrt{\frac{2}{\pi}} \,
   \sum_{T_1 = 0,1} \big(G_E ^p + (-1)^{T_1} \, G_E^n\big)
   \sum_{L_1 = 0}^{L_{\rm max}} \big(1 + (-1)^{T_1} (-1)^{L_1}\big) \\
   \null \times Y_{L_1 , \mJd - \msf} (\thetacm, \phicm)
   \sum_{J_1 = |L_1 - 1|}^{L+1}
   \CG{L_1}{\mJd - \msf}{S\!=\!1}{\msf}{J_1}{\mJd} \\
   \null \times \sum_{\widetilde{m}_s = -1}^1
   \braket{J_1 \, \mJd}{L_1 \, \mJd - \wt{m}_s \, S\!=\!1 \, \wt{m}_s}
   \sum_{L_2 = 0}^{L_{\rm max}}
   \CG{L_2}{\mJd-\widetilde{m}_s}{S\!=\!1}{\widetilde{m}_s}{J\!=\!1}{\mJd} \\
   \null \times \sum_{L_d = 0, 2}
   \int \! \dd k_3 \, \psi_{L_d}^\lambda(k_3) \, k_3^2 \int \! \dcostheta \,
    P_{L_1}^{\mJd-\widetilde{m}_s}(\cos \theta)
    P_{L_2}^{\mJd-\widetilde{m}_s}\!\big(\cos\thetacprime(\pp, \theta, q)\big)
    \\ \null \times
    \widetilde{U}\!\left(
     k_3, \sqrt{{\pp}^2 - \pp q \cos\theta + q^2/4}, L_d, L_2, J\!=\!1, S\!=\!1,
     T\!=\!0 \right) \,,
  \end{multline}
  \begin{multline}
   B_2 \equiv \la \phi | \,\widetilde{U}\, J_0 \, | \psi_i^\lambda \ra
   = 2 \, \sqrt{\frac{2}{\pi}} \,
   \sum_{T_1 = 0,1} \big(G_E ^p + (-1)^{T_1} \, G_E^n\big)
   \sum_{L_1 = 0}^{L_{\rm max}} \big(1 + (-1)^{T_1} (-1)^{L_1}\big) \\
   \null \times Y_{L_1 , \mJd - \msf} (\thetacm, \phicm)
   \sum_{J_1 = |L_1 - 1|}^{L+1}
   \CG{L_1}{\mJd - \msf}{S\!=\!1}{\msf}{J_1}{\mJd} \\
   \null \times
   \sum_{L_2, \widetilde{m}_s}
   \braket{J_1 \, \mJd}{L_1 \, \mJd - \wt{m}_s \, S\!=\!1 \, \wt{m}_s}
   \sum_{L_d = 0, 2}
   \CG{L_d}{\mJd - \widetilde{m}_s}{S\!=\!1}{\widetilde{m}_s}{J\!=\!1}{\mJd} \\
   \null \times
   \int \! \dd k_2 \, k_2^2 \, \widetilde{U}(\pp, k_2, L_1, L_2, J_1, S\!=\!1,
   T_1)  \int \! \dcostheta \,
   P_{L_2}^{\mJd-\widetilde{m}_s}(\cos\theta) \\
   \null \times
   P_{L_d}^{\mJd-\widetilde{m}_s}\!\big(\cos\thetacprime(k_2, \theta, q)\big)
   \psi_{L_d}^\lambda\!\left(\sqrt{{k_2}^2 - k_2 q \cos\theta + q^2/4}\right)
   \,,
  \end{multline}
  \begin{multline}
   B_1 \equiv \la \phi | \,\widetilde{U}\, J_0 \,\widetilde{U}^\dag \, |
   \psi_i^\lambda \ra = \frac{4}{\pi} \, \sqrt{\frac{2}{\pi}} \,
   \sum_{T_1 = 0,1} \big(G_E ^p + (-1)^{T_1} \, G_E^n\big)
   \sum_{L_1 = 0}^{L_{\rm max}} \big(1 + (-1)^{T_1} (-1)^{L_1}\big) \\
   \null \times
   Y_{L_1 , \mJd - \msf} (\thetacm, \phicm)
   \sum_{J_1 = |L_1 - 1|}^{L+1}
   \CG{L_1}{\mJd-\msf}{S\!=\!1}{\msf}{J_1}{\mJd} \\
   \null \times
   \sum_{L_2, \widetilde{m}_s}
   \braket{J_1 \, \mJd}{L_2 \, \mJd - \wt{m}_s \, S\!=\!1 \, \wt{m}_s}
   \sum_{L_3 = 0}^{L_{\rm max}}
   \CG{L_3}{\mJd-\widetilde{m}_s}{S\!=\!1}{\widetilde{m}_s}{J\!=\!1}{\mJd} \\
   \null \times
   \int \! \dd k_2 \, k_2^2 \, \widetilde{U} (\pp, k_2, L_1, L_2, J_1, S\!=\!1,
   T_1)  \sum_{L_d = 0, 2} \int \! \dd k_4 \, k_4^2 \, \psi_{L_d}^\lambda (k_4) \\
   \null \times
   \int \! \dcostheta \, P_{L_2}^{\mJd - \widetilde{m}_s}(\cos\theta)
   P_{L_3}^{\mJd - \widetilde{m}_s}\!\big(\cos \thetacprime(k_2, \theta, q)\big)
   \\ \null \times
   \widetilde{U}\!\left(
     k_4, \sqrt{{k_2}^2 - k_2 q \cos\theta + q^2/4}, L_d, L_3, J\!=\!1, S\!=\!1,
     T\!=\!0   \right) \,.
  \end{multline}
  In deriving the equations for $B_1$, $B_2$, and $B_3$ we have made use of
  the fact that the matrix elements with $J_0$ are twice the matrix elements
  with $J_0^-$, i.e., $\mbraket{\phi}{J_0 \, \widetilde{U}^\dag}{\psi_i^\lambda}
  = 2 \,\mbraket{\phi}{J_0^- \, \widetilde{U}^\dag}{\psi_i^\lambda}$, and
  similarly for $B_2$ and $B_1$ (cf.~Eq.~\ref{eq:J0_minus_twice_relation}).

  Evaluating Eq.~\eqref{eq:A_split_up} involves calculating the individual terms
  $A_1$, $A_2$, $A_3$, and $A_4$.  The expressions for $A_4$ and $A_3$ can be
  obtained from expressions for $B_2$ and $B_1$, respectively, by replacing
  $\widetilde{U}$ with $\widetilde{U}^\dag$.  The $U$-matrices are real.
  Therefore, $\widetilde{U}^\dag$ is obtained from $\widetilde{U}$ by
  interchanging momentum and angular momentum indices.  The expressions for
  $A_2$ and $A_1$ are
  \begin{multline}
   A_2 \equiv \la \phi | \widetilde{U}^\dag \, \widetilde{U}\, J_0 |
   \psi_i^\lambda \ra  = \frac{4}{\pi} \, \sqrt{\frac{2}{\pi}} \,
   \sum_{T_1 = 0,1} \big(G_E ^p + (-1)^{T_1} \, G_E^n\big)
   \sum_{L_1 = 0}^{L_{\rm max}} \big(1 + (-1)^{T_1} (-1)^{L_1}\big) \\
   \null \times  Y_{L_1 , \mJd - \msf} (\thetacm, \phicm)
   \sum_{J_1 = |L_1 - 1|}^{L+1}
   \CG{L_1}{\mJd-\msf}{S\!=\!1}{\msf}{J_1}{\mJd} \\
   \null \times
   \sum_{L_3, \widetilde{m}_s}
   \braket{J_1 \, \mJd}{L_3 \, \mJd - \wt{m}_s \, S\!=\!1 \, \wt{m}_s}
   \sum_{L_2 = 0}^{L_{\rm max}} \int \! \dd k_2 \, k_2^2 \,
   \widetilde{U} (k_2, \pp, L_2, L_1, J_1, S\!=\!1, T_1) \\
   \null \times
   \sum_{L_d = 0, 2}
   \CG{L_d}{\mJd-\widetilde{m}_s}{S\!=\!1}{\widetilde{m}_s}{J\!=\!1}{\mJd}
   \int \! \dd k_3 \, k_3^2 \,
   \widetilde{U} (k_2, k_3, L_2, L_3, J_1, S\!=\!1, T_1) \\
   \null \times
   \int \! \dcostheta \,P_{L_3}^{\mJd - \widetilde{m}_s}(\cos\theta)
   P_{L_d}^{\mJd - \widetilde{m}_s}\!\big(
   \cos\thetacprime(k_3, \theta, q)\big) \,
   \psi_{L_d}^\lambda\!\left(\sqrt{{k_3}^2 - k_3 q \cos\theta + q^2/4} \right)
  \end{multline}
  and
  \begin{multline}
   A_1 \equiv \la \phi | \widetilde{U}^\dag \, \widetilde{U}\, J_0 |
   \psi_i^\lambda \ra  = \frac{8}{\pi^2} \,\sqrt{\frac{2}{\pi}} \,
   \sum_{T_1 = 0,1} \big(G_E ^p + (-1)^{T_1} \,G_E^n\big)
   \sum_{L_1 = 0}^{L_{\rm max}} \big(1 + (-1)^{T_1} (-1)^{L_1}\big) \\
   \null \times
   Y_{L_1 , \mJd - \msf} (\thetacm, \phicm)
   \sum_{J_1 = |L_1 - 1|}^{L+1}
   \CG{L_1}{\mJd - \msf}{S\!=\!1}{\msf}{J_1}{\mJd} \\
   \null \times \sum_{L_3, \widetilde{m}_s}
   \braket{J_1 \, \mJd}{L_3 \, \mJd - \wt{m}_s \, S\!=\!1 \, \wt{m}_s}
   \sum_{L_4 = 0}^{L_{\rm max}}
   \CG{L_4}{\mJd - \widetilde{m}_s}{S\!=\!1}{\widetilde{m}_s}{J\!=\!1}{\mJd} \\
   \null \times
   \sum_{L_2 = 0}^{L_{\rm max}} \int \! \dd k_2 \, k_2^2 \,
    \widetilde{U} (k_2, \pp, L_2, L_1, J_1, S\!=\!1, T_1)
   \int \! \dd k_3 \, k_3^2 \,
    \widetilde{U} (k_2, k_3, L_2, L_3, J_1, S\!=\!1, T_1) \\
   \null \times
   \sum_{L_d = 0, 2} \int \! \dd k_5 \, k_5^2 \, \psi_{L_d}^\lambda (k_5)
   \int \! \dcostheta \, P_{L_3}^{\mJd - \widetilde{m}_s}(\cos \theta)
   P_{L_4}^{\mJd - \widetilde{m}_s}\!\big(\cos\thetacprime(k_3, \theta, q)\big)
   \\  \null \times
   \,\widetilde{U}\!\left(
    k_5, \sqrt{{k_3}^2 - k_3 q \cos\theta  + q^2/4}, L_d, L_4, J\!=\!1, S\!=\!1,
    T\!=\!0  \right) \,.
  \end{multline}

  Evaluating the evolved current while including the final-state interactions
  involves computing the terms $F_1$, $F_2$, $F_3$, and $F_4$, as indicated in
  Eq.~\eqref{eq:F_split_up}.  $F_4$ is obtained from
  Eqs.~\eqref{eq:phi_t_g0_J0_minus} and~\eqref{eq:J0_minus_twice_relation} by
  replacing the deuteron wave function and the $t$-matrix by their evolved
  counterparts.  The expressions for the terms $F_3$, $F_2$, and $F_1$ are then
  as follows:
  \begin{multline}
   F_3 \equiv \mbraket{\phi}{t_\lambda ^\dag \, G_0^\dag \,
   J_0 \, \widetilde{U}^\dag}{\psi_i^\lambda}
   = \frac{4}{\pi} \, \sqrt{\frac{2}{\pi}} \, \frac{M}{\hbar c}
   \int \! \frac{\dd k_2 \, k_2^2}{(\pp + k_2)(\pp - k_2 - i \epsilon)}
   \sum_{T_1 = 0,1} \big(G_E ^p + (-1)^{T_1} \,G_E^n\big) \\
   \null \times
   \sum_{L_1 = 0}^{L_{\rm max}} \big(1 + (-1)^{T_1} (-1)^{L_1}\big)
   Y_{L_1 , \mJd - \msf}(\thetacm, \phicm)
   \sum_{J_1 = |L_1 - 1|}^{L+1}
    \CG{L_1}{\mJd - \msf}{S\!=\!1}{\msf}{J_1}{\mJd} \\
    \null \times
   \sum_{L_2 = 0}^{L_{\rm max}}
    t^\ast_\lambda(k_2, \pp, L_2, L_1, J_1, S\!=\!1, T_1)
    \sum_{\widetilde{m}_s = -1}^1
    \braket{J_1 \, \mJd}{L_2 \, \mJd - \wt{m}_s \, S\!=\!1 \, \wt{m}_s} \\
    \null \times
   \sum_{L_3 = 0}^{L_{\rm max}}
    \CG{L_3}{\mJd - \widetilde{m}_s}{S\!=\!1}{\widetilde{m}_s}{J\!=\!1}{\mJd}
   \int \! \dcostheta \, P_{L_2}^{\mJd - \widetilde{m}_s}(\cos \theta)
   P_{L_3}^{\mJd - \widetilde{m}_s}\!\big(\cos \thetacprime(k_2, \theta, q)
   \big) \\
   \null \times
   \int \! \dd k_5 \, k_5^2 \sum_{L_d = 0, 2}
   \widetilde{U}\!\left(
    k_5, \sqrt{{k_2}^2 - k_2 q \cos\theta + q^2/4}, L_d, L_3, J\!=\!1, S\!=\!1,
    T\!=\!0  \right) \psi_{L_d}^\lambda (k_5) \,,
  \label{eq:F3}
  \end{multline}
  \begin{multline}
   F_2 \equiv \mbraket{\phi}{t_\lambda ^\dag \, G_0^\dag \, \widetilde{U} \,
   J_0}{\psi_i^\lambda}
   = \frac{4}{\pi} \, \sqrt{\frac{2}{\pi}} \,\frac{M}{\hbar c}
   \int \! \frac{\dd k_2 \, k_2^2}{(\pp + k_2)(\pp - k_2 - i \epsilon)}
   \sum_{T_1 = 0,1} \big(G_E ^p + (-1)^{T_1} \,G_E^n\big) \\
   \null \times
   \sum_{L_1 = 0}^{L_{\rm max}} \big(1 + (-1)^{T_1} (-1)^{L_1}\big)
   Y_{L_1 , \mJd - \msf} (\thetacm, \phicm)
   \sum_{J_1 = |L_1 - 1|}^{L+1}
   \CG{L_1}{\mJd - \msf}{S\!=\!1}{\msf}{J_1}{\mJd} \\
   \null \times
   \sum_{L_2 = 0}^{L_{\rm max}}
    t^\ast_\lambda(k_2, \pp, L_2, L_1, J_1, S\!=\!1,T_1)
   \sum_{L_3 = 0}^{L_{\rm max}} \int \! \dd k_4 \, k_4^2 \,
   \widetilde{U}(k_2, k_4, L_2, L_3, J_1, S\!=\!1, T_1) \\
   \null \times
   \sum_{\widetilde{m}_s = -1}^1
   \braket{J_1 \, \mJd}{L_3 \, \mJd - \wt{m}_s \, S\!=\!1 \, \wt{m}_s}
   \sum_{L_d = 0, 2}
   \CG{L_d}{\mJd - \widetilde{m}_s}{S\!=\!1}{\widetilde{m}_s}{J\!=\!1}{\mJd} \\
   \null \times
   \int \! \dcostheta \, P_{L_3}^{\mJd - \widetilde{m}_s}(\cos \theta)
   P_{L_d}^{\mJd - \widetilde{m}_s} \big(\cos \thetacprime(k_4, \theta, q)\big)
   \, \psi_{L_d}^\lambda\!\left(\sqrt{{k_4}^2 - k_4 q \cos\theta + q^2/4}\right)
   \,,
  \label{eq:F2}
  \end{multline}
  \begin{multline}
   F_1 \equiv \mbraket{\phi}{t_\lambda ^\dag \, G_0^\dag \, \widetilde{U} \,
   J_0 \, \widetilde{U}^\dag}{\psi_i^\lambda}
   = \frac{8}{\pi^2} \, \sqrt{\frac{2}{\pi}} \, \frac{M}{\hbar c}
   \int \! \frac{\dd k_2 \, k_2^2}{(\pp + k_2)(\pp - k_2 - i \epsilon)}
   \sum_{T_1 = 0,1} \big(G_E ^p + (-1)^{T_1} \,G_E^n\big) \\
   \null \times
   \sum_{L_1 = 0}^{L_{\rm max}} \big(1 + (-1)^{T_1} (-1)^{L_1}\big)
   Y_{L_1 , \mJd - \msf} (\thetacm, \phicm)
   \sum_{J_1 = |L_1 - 1|}^{L+1}
    \CG{L_1}{\mJd - \msf}{S\!=\!1}{\msf}{J_1}{\mJd} \\
    \null \times
   \sum_{L_2 = 0}^{L_{\rm max}}
    t^\ast_\lambda(k_2, \pp, L_2, L_1, J_1, S\!=\!1, T_1)
    \sum_{L_3 = 0}^{L_{\rm max}} \, \sum_{\widetilde{m}_s = -1}^1
    \braket{J_1 \, \mJd}{L_3 \, \mJd - \wt{m}_s \, S\!=\!1 \, \wt{m}_s} \\
   \null \times \sum_{L_4 = 0}^{L_{\rm max}}
    \CG{L_4}{\mJd - \widetilde{m}_s}{S\!=\!1}{\widetilde{m}_s}{J\!=\!1}{\mJd}
    \int \! \dd k_4 \, k_4^2 \,
    \widetilde{U}(k_2, k_4, L_2, L_3, J_1, S\!=\!1, T_1) \\
    \null \times
   \int \! \dcostheta \, P_{L_3}^{\mJd - \widetilde{m}_s}(\cos \theta) \,
   P_{L_4}^{\mJd - \widetilde{m}_s}\!\big(\cos \thetacprime(k_4, \theta, q)\big)
   \int \! \dd k_6 \, k_6^2
   \sum_{L_d = 0, 2} \psi_{L_d}^\lambda (k_6) \\
   \null \times
   \widetilde{U}\!\left(
    k_6, \sqrt{{k_4}^2 - k_4 q \cos\theta + q^2/4}, L_d, L_4, J\!=\!1, S\!=\!1,
    T\!=\!0  \right)  \,.
  \label{eq:F1}
  \end{multline}

  \section{Evolution of the final state}
  \label{Appendix:sec:evolution_final_state}

  The interacting final neutron-proton state $\ket{\psi_f}$ as defined in
  Eq.~\eqref{eq:psi_f_def} is the formal solution of the Lippmann--Schwinger
  (LS) equation for the scattering wave function,
  \begin{spliteq}
   \ket{\psi_f}
   &= \ket{\phi} + G_0 (E^\prime) \, V \ket{\psi_f} \\
   &= \ket{\phi} + G_0 (E^\prime) \, t(E^\prime) \ket{\phi} \,.
   \label{eq:LS-psi}
  \end{spliteq}
  The $t$-matrix, in turn, is defined by the LS equation
  \begin{equation}
   t(E^\prime) = V + V \, G_0(E') \, t(E') \,.
  \label{eq:LS-t}
  \end{equation}
  The subsitution $E' \to E' + \ii\epsilon$ and the limit $\epsilon\to0$
  are implied to select outgoing boundary conditions.  We want to show now
  that the SRG-evolved final state can be obtained directly by using the
  solution $t^\lambda$ of Eq.~\eqref{eq:LS-t} with $V \to V_\lambda$ in the
  second line of
  Eq.~\eqref{eq:LS-psi}, which is the same as Eq.~\eqref{eq:psi_f_def} in
  Subsec.~\ref{subsec:formalism}, i.e.,
  \begin{equation}
   U_\lambda \ket{\psi_f} = \ket{\psi_f^\lambda} \,,
  \label{eq:U-psi-final}
  \end{equation}
  where
  \begin{equation}
   \ket{\psi_f^\lambda} = \ket{\phi} +
   G_0 (E^\prime) \, t_\lambda(E^\prime) \ket{\phi} \,.
  \end{equation}
  In this section, we suppress all spin and isospin degrees of freedom, and only
  denote the (arbitrary) energy parameter as $E^\prime$ for consistency with
  Subsec.~\ref{subsec:formalism}.

  First, it is important to recall that by definition the free Hamiltonian
  $H_0$ does not evolve, so that for $H = H_0 + V$ we have
  \begin{equation}
   H_\lambda = U_\lambda \;\! H \, U_\lambda^\dagger
   \equiv H_0 + V_\lambda \,.
  \label{eq:H-lambda}
  \end{equation}
  In other words, the evolved potential $V_\lambda$ is defined such that it
  absorbs the evolution of the initial free Hamiltonian (kinetic energy) as well.

  In order to prove Eq.~\eqref{eq:U-psi-final}, it is convenient to consider the
  evolved and unevolved full Green's functions $G_\lambda(E^\prime)$ and
  $G(E^\prime)$, defined via
  \begin{subalign}[eq:G-inv]
   G_\lambda(z)^{-1} &= z - H_\lambda = G_0^{-1}(z)^{-1} - V_\lambda \,,
   \label{eq:G-lambda-inv} \\
   G(z)^{-1} &= z - H_{\phantom{\lambda}} = G_0^{-1}(z)^{-1} - V \,.
   \label{eq:G-bare-inv}
  \end{subalign}
  Here, $G_0^{-1}(z)^{-1} = z - H_0$ is the free Green's function (which does
  not change under the SRG evolution because $H_0$ does not), and $z$ is an
  arbitrary complex energy parameter that is set to $E^\prime + \ii\epsilon$ to
  recover the
  physically relevant case.  The Green's functions can be expressed in terms of
  the $t$-matrix as
  \begin{equation}
   G(z) = G_0(z) + G_0(z) \, t(z) \, G_0(z) \,,
  \end{equation}
  and analogously for the evolved version.  Furthermore, the Green's functions
  can be written in their spectral representations
  \begin{subalign}[eq:G-spectral]
   G_\lambda(z)^{-1} &\simeq \int\!\dd^3 k
   \, \frac{\ket{\psi_f^\lambda(k)}\bra{\psi_f^\lambda(k)}}{z - k^2/M}
   + \text{bound states} \,,
   \label{eq:G-lambda-spectral} \\
   G(z)^{-1} &\simeq \int\!\dd^3 k
   \, \frac{\ket{\psi_f(k)}\bra{\psi_f(k)}}{z - k^2/M}
   + \text{bound states} \,.
   \label{eq:G-bare-spectral}
  \end{subalign}
  Here, $\ket{\psi_f^{(\lambda)}(k)}$ denotes the (evolved) continuum states
  with momentum $k$, and we have $\ket{\psi_f^{(\lambda)}}
  = \ket{\psi_f^{(\lambda)}(\sqrt{M E^\prime})}$

  From Eqs.~\eqref{eq:H-lambda} and~\eqref{eq:G-inv} it now follows that
  \begin{eqnarray}
   G_\lambda(z)^{-1} &=& z - H_\lambda
   = z - U_\lambda \;\! H \, U_\lambda^\dagger \nonumber\\
   &=& U_\lambda (z - H) \:\! U_\lambda^\dagger
   = U_\lambda \, G(z)^{-1} \, U_\lambda^\dagger \,.
 \end{eqnarray}
  Combining this with Eqs.~\eqref{eq:G-spectral} and matching residues at $z =
  E^\prime + \ii\epsilon$, we find that indeed $\ket{\psi_f^{\lambda}} =
  U_\lambda \ket{\psi_f}$, as stated in Eq.~\eqref{eq:U-psi-final}.

  \section{Evolution of the current}
  \label{Appendix:sec:evolution_current}

  Here we document the equations used to obtain the results in
  Subsec.~\ref{subsec:operator_evolution}.  The evolved current $J_0^{\lambda}$
  is given by
  \beq
  J_0^{\lambda} = \underbrace{\wt{U} J_0 \wt{U}^\dag}_{J_1} +
  \underbrace{\wt{U} J_0}_{J_2} + \underbrace{J_0 \wt{U}^\dag}_{J_3} + J_0\;.
  \eeq

  \begin{multline}
    J_3 \equiv \mbraket{{\thickmuskip=0mu k_1 \, J_1 \, \mJd \,
    L_1 \, S\!=\!1 \, T_1}}
    {J_0^-(q) \, \wt{U}^\dag}{{\thickmuskip=0mu k_2 \, J\!=\!1 \,
    \mJd \, L_2 \, S\!=\!1 \,
    T\!=\!0}} = \frac{2}{\pi} \frac{\pi^2}{2} \big(G_E^p + (-1)^{T_1} G_E^n
    \big) \\
    \times
    \sum_{L_p} \int_{|k_1 - q/2|}^{k_1 + q/2} \dd p \, p^2 \sum_{\mst = -1}^1
    \braket{J_1 \, \mJd}{L_1 \, \mJd \!- \!\mst \, {S\!=\!1}
    \, \mst} \, \frac{2}{k_1 \, p \, q} \\
    \times
    P_{L_1}^{\mJd - \mst}\left(\frac{k_1^2 - p^2 + q^2/4}{k_1 q}
    \right)
    P_{L_1}^{\mJd - \mst}\left(
    \frac{k_1^2 - p^2 - q^2/4}{p q}
    \right) \\
    \times
    \CG{L_p}{\mJd\!-\!\mst}{S\!=\!1}{\mst}{J\!=\!1}{\mJd} \,
    \wt{U}^\dag({\thickmuskip=0mu p_1, k_2, L_p, L_2, J\!=\!1, S\!=\!1,
    T = 0}) \;.
  \end{multline}
  Evaluation of $J_2$ requires writing the current $J_0$ such
  that the integration over the bra state can be done.  We get the
  same expression as in Eq.~\eqref{eq:J0_minus_analytical_delta},
  but on a different domain.
  \begin{align}
  \mbraket{k_1 \, J_1 \, \mJd \, L_1 \, S=1 \, T_1}{J_0^-}
  {k_2\, J=1 \, \mJd \, L_2 \, S=1 \, T=0} = \frac{\pi^2}{2}
  \big(G_E^p + (-1)^{T_1} G_E^n\big) \nonumber \\
  \times \sum_{\mst=-1}^{1} \braket{J_1 \, \mJd}
  {L_1 \, \mJd - \wt{m}_s \, S\!=\!1 \, \wt{m}_s}
  P_{L_1}^{\mJd - \mst} \left(\frac{k_1^2 - k_2^2 + q^2/4}{k_1 q}
  \right) \frac{2}{k_1 k_2 q} \nonumber \\
  \times
  P_{L_2}^{\mJd - \mst} \left(\frac{k_1^2 - k_2^2 - q^2/4}{k_2 q}
  \right)
  \CG{L_2}{\mJd - \mst}{S=1}{\mst}{J=1}{\mJd} \nonumber \\
  \cdots {\rm for~} k_1 \in (|k_2 - q/2|, k_2 + q/2) \nonumber \\
  = 0 {\rm ~otherwise~~~~~~~~~~}
  \label{eq:J0_minus_analytical_delta2}
  \end{align}

  Equation~\eqref{eq:J0_minus_analytical_delta2} is gotten from the
  following expression for the current:
  \begin{align}
   \mbraket{k_1 \, J_1 \, \mJd \, L_1 \, S\!=\!1 \, T_1}{J_0^{-}}
 	 {\,k_2 \, J\!=\!1 \, \mJd \, L_2 \, S\!=\!1 \, T\!=\!0}
 	 = \frac{\pi^2}{2} \, \big(G_E ^p + (-1)^{T_1} \, G_E^n\big) \\
 	 \null \times \sum_{\widetilde{m}_s = -1}^1 \int\!\dcostheta
 	 \, \braket{J_1 \, \mJd}{L_1 \, \mJd - \wt{m}_s \, S\!=\!1 \,\wt{m}_s}
 	 \,P_{L_2}^{\mJd - \widetilde{m}_s}\!(\cos\theta) \\
 	 \null \times
 	 \,P_{L_1}^{\mJd - \widetilde{m}_s}\!\big(\cos\thetacdoubleprime(k_1,
   \theta, q)
 	 \big) \frac{\delta\big(k_1-\sqrt{k_2^2 + k_2 q \cos\theta + q^2/4}\big)}
 	 {k_1^2} \\
 	 \null \times
 	 \CG{L_2}{\mJd - \widetilde{m}_s}{S\!=\!1}{\widetilde{m}_s}{J\!=\!1}{\mJd} \,.
 	\label{eq:J_0_minus_def_pw2}
  \end{align}
  Note that Eq.~\eqref{eq:J_0_minus_def_pw2} differs from
  Eq.~\eqref{eq:J_0_minus_def_pw} in terms of the arguments of the
  $\delta$ function and the Legendre polynomials.

  Using Eq.~\eqref{eq:J0_minus_analytical_delta2}, we can derive the expression
  for $J_2$.
  \begin{multline}
    J_2 \equiv \mbraket{{\thickmuskip=0mu k_1 \, J_1 \, \mJd \,
    L_1 \, S\!=\!1 \, T_1}}
    {\wt{U} \, J_0^-(q)}{{\thickmuskip=0mu k_2 \, J\!=\!1 \,
    \mJd \, L_2 \, S\!=\!1 \,
    T\!=\!0}} = \frac{2}{\pi} \frac{\pi^2}{2} \big(G_E^p + (-1)^{T_1} G_E^n
    \big) \\
    \times
    \sum_{L_p} \int_{|k_2 - q/2|}^{k_2 + q/2} \dd p \, p^2 \sum_{\mst = -1}^1
    \braket{J_1 \, \mJd}{L_p \, \mJd \!- \!\mst \, {S\!=\!1}
    \, \mst} \, \frac{2}{k_2 \, p \, q} \\
    \times
    P_{L_p}^{\mJd - \mst}\left(\frac{p^2 - k_2^2 + q^2/4}{p q}
    \right)
    P_{L_2}^{\mJd - \mst}\left(
    \frac{p^2 -k_2^2 - q^2/4}{k_2 q}
    \right) \\
    \times
    \CG{L_2}{\mJd\!-\!\mst}{S\!=\!1}{\mst}{J\!=\!1}{\mJd} \,
    \wt{U}({\thickmuskip=0mu k_1, p, L_1, L_p, J\!=\!1, S\!=\!1,
    T = 0}) \;.
  \end{multline}

  Finally, the expression for $J_1$ is
  \begin{multline}
    J_1 \equiv \mbraket{{\thickmuskip=0mu k_1 \, J_1 \, \mJd \,
    L_1 \, S\!=\!1 \, T_1}}
    {\wt{U} J_0^-(q) \wt{U}^\dag}{{\thickmuskip=0mu k_2 \, J\!=\!1 \,
    \mJd \, L_2 \, S\!=\!1 \,
    T\!=\!0}} = \big(\frac{2}{\pi}\big)^2 \frac{\pi^2}{2} \\
    \times
    \big(G_E^p + (-1)^{T_1} G_E^n \big)
    \sum_{L_{p_1}, L_{p_2}, \mst}
    \int \dd p_1 \int_{p_2 = |p_1 - q/2|}^{p_1 + q/2} \dd p_2 \, p_1^2
    \, p_2^2 \, \frac{2}{p_1 p_2 q} \\
    \times \wt{U}({\thickmuskip=0mu k_1, p_1, L_1, L_{p_1}, J_1, S\!=\!1,
      T_1}) \, \wt{U}^\dag({\thickmuskip=0mu p_2, k_2, L_{p_2}, L_2, J\!=\!1,
      S\!=\!1, T\!=\!0}) \\
    \times \braket{J_1 \, \mJd}{L_{p_1} \, \mJd \!- \!\mst \, {S\!=\!1}
      \, \mst} \, \CG{L_{p_2}}{\mJd\!-\!\mst}{S\!=\!1}{\mst}{J\!=\!1}{\mJd} \\
    \times P_{L_{p_1}}^{\mJd - \mst}\left(\frac{p_1^2 - p_2^2 + q^2/4}{p_1 q}
      \right) \,
      P_{L_{p_2}}^{\mJd - \mst}\left(
      \frac{p_1^2 -p_2^2 - q^2/4}{p_2 q}
      \right)
  \end{multline}

\end{document}